\documentclass[11pt,a4paper]{article}
\usepackage{jcappub}
\usepackage{longtable}
\pdfoutput=1

\newcommand{\bB}{\mathbf{b}}
\newcommand{\dB}{\mathbf{d}}
\newcommand{\hB}{\mathbf{h}}
\newcommand{\nB}{\mathbf{n}}
\newcommand{\pB}{\mathbf{p}}
\newcommand{\qB}{\mathbf{q}}
\newcommand{\sB}{\mathbf{s}}
\newcommand{\uB}{\mathbf{u}}
\newcommand{\vB}{\mathbf{v}}
\newcommand{\yB}{\mathbf{y}}
\newcommand{\bm}[1]{\mbox{\boldmath{$#1$}}}
\newcommand{\bms}[1]{\mbox{{\tiny\boldmath{$#1$}}}}
\DeclareMathOperator*{\argmin}{arg\,min}

\title{Reconstruction of the primordial power spectrum of curvature
  perturbations using multiple data sets} 
\author[a]{Paul Hunt}
\author[b,c]{\& Subir Sarkar}

\affiliation[a]{Institute of Theoretical Physics, Warsaw University,
  ul Ho\.za 69, 00-681 Warsaw, POLAND}
\affiliation[b]{Rudolf Peierls Centre for Theoretical Physics,
  University of Oxford, Oxford OX1 3NP, UK}
\affiliation[c]{Niels Bohr Institute, Blegdamsvej 17, 2100 Copenhagen,
  DENMARK}
\emailAdd{Paul.Hunt@fuw.edu.pl}
\emailAdd{s.sarkar@physics.ox.ac.uk}

\abstract{Detailed knowledge of the primordial power spectrum of
  curvature perturbations is essential both in order to elucidate the
  physical mechanism (`inflation') which generated it, and for
  estimating the cosmological parameters from observations of the
  cosmic microwave background and large-scale structure.  Hence it
  ought to be extracted from such data in a model-independent manner,
  however this is difficult because relevant cosmological observables
  are given by a convolution of the primordial perturbations with some
  smoothing kernel which depends on both the assumed world model and
  the matter content of the universe. Moreover the deconvolution
  problem is ill-conditioned so a regularisation scheme must be
  employed to control error propagation.  We demonstrate that
  `Tikhonov regularisation' can robustly reconstruct the primordial
  spectrum from multiple cosmological data sets, a significant
  advantage being that both its uncertainty and resolution are then
  quantified. Using Monte Carlo simulations we investigate several
  regularisation parameter selection methods and find that generalised
  cross-validation and Mallow's $C_p$ method give optimal results. We
  apply our inversion procedure to data from the Wilkinson Microwave
  Anisotropy Probe, other ground-based small angular scale CMB
  experiments, and the Sloan Digital Sky Survey. The reconstructed
  spectrum (assuming the standard $\Lambda$CDM cosmology) is
  \emph{not} scale-free but has an infrared cutoff at $k \lesssim 5
  \times 10^{-4}\; \mathrm{Mpc}^{-1}$ (due to the anomalously low CMB
  quadrupole) and several features with $\sim 2 \sigma$ significance
  at $k/\mathrm{Mpc}^{-1} \sim$ 0.0013--0.0025, 0.0362--0.0402 and
  0.051--0.056, reflecting the `WMAP glitches'. To test whether these
  are indeed real will require more accurate data, such as from the
  Planck satellite and new ground-based experiments.}


\arxivnumber{1308.2317}
\begin{document}
\maketitle
\flushbottom

\section{Introduction \label{intro}}

The primordial curvature perturbations which seeded the growth of
large-scale structure (LSS) is presently our only observational window
on the very early universe. Precision observations of anisotropies in
the cosmic microwave background (CMB) by the Wilkinson Microwave
Anisotropy Probe (WMAP) satellite have greatly enhanced our knowledge
of these primordial perturbations. The pattern of acoustic peaks in
the CMB angular power spectra measured by WMAP, notably the
anticorrelation in the temperature-electric (TE) polarisation spectrum
on degree scales, indicate that the initial fluctuations are coherent,
predominantly adiabatic and were generated on superhorizon scales
\cite{Peiris:2003ff,Dodelson:2003ip}. The perturbations also obey
Gaussian statistics to a high level of accuracy
\cite{Komatsu:2008hk}. Together this provides strong support for an
early quasi-de Sitter phase of `inflation', usually assumed to be
driven by a scalar field with potential $V(\phi)$, as the origin of
the primordial perturbations \cite{Linde:2005ht,Lyth:1998xn}. However
the gravitational dynamics of vacuum energy remains a complete mystery
--- the `Cosmological Constant problem'
\cite{Weinberg:1988cp,Nobbenhuis:2004wn}. Hence it is essential to
gain further insights concerning the underlying \textit{physical}
mechanism which generated the primordial curvature perturbations.

Cosmologists have so far attempted only to discriminate between
various toy models of $V(\phi)$ on the basis of simple modelling of
the primordial power spectrum (PPS) of curvature perturbations ---
$\mathcal{P_R}(k)$ --- usually as a featureless power-law
spectrum. This has partly been because of practical limitations in
extracting the PPS from CMB and LSS data (to be discussed below), and
also because of the prejudice that `simple' models of inflation
predict just such a power-law form (with logarithmic
corrections). However these are all severely fine-tuned `toy models'
so e.g. favouring $V=m^2\phi^2$ over $V=\lambda\phi^4$
\cite{Peiris:2003ff} does not in fact provide any useful insight into
how inflation actually occurred.\footnote{Moreover, these are actually
  the \textit{same} model in the sense that in the tiny range in
  $\phi$ space which corresponds to fluctuations on all observable
  scales, any monomial potential $\phi^n$ can be Taylor-expanded as
  $V_0 + \alpha\phi + \ldots$, so both quartic $\lambda\phi^4$ and
  quadratic $m^2\phi^2$ potentials are in fact `linear inflation'
  models \cite{German:2001tz}.} While detection of tensor
perturbations (gravitational waves) would provide important new input,
these may be observable through $B$-mode polarisation of the CMB only
if the energy scale of inflation is in a narrow window around the
Grand Unified scale of ${\cal O}(10^{16})$~GeV, which is certainly not
mandatory \cite{Hotchkiss:2008sa,Bird:2008cp}. Hence scalar density
perturbations are likely to remain the only available probe of the
early universe, and we must therefore determine their properties in as
much detail as is possible, in order to establish how they were
generated.

Inhomogenities after the primordial era depend upon both the initial
fluctuations and their evolution. The latter is governed by the
cosmological parameters describing the unperturbed background
cosmology such as the baryon density $\Omega_\mathrm{b}$, the Hubble
constant $h$, \textit{etc}. Consequently it is difficult to
disentangle the PPS and background parameters using observables such
as CMB anisotropies and galaxy clustering which effectively only
sample one time slice. For example, varying a PPS consisting of 75
wavenumber bins can mimic the effect of changes in the background
parameters on the CMB temperature angular power spectrum to within the
limit of cosmic variance \cite{Kinney:2001js}. Therefore the PPS must
be assumed to have a simple form in order to avoid the degeneracy
between the spectrum and the background parameters. It is usually
taken to be a \textit{featureless} power-law, $\mathcal{P_R}(k)\propto
k^{n_\mathrm{s}-1}$, in the wavenumber $k$, with the spectral index
$n_\mathrm{s}$ close to unity (the `Harrison-Zeldovich'
scale-invariant spectrum), as is expected in toy inflationary models
based on monomial potentials e.g. $V(\phi) \propto \phi^n$
\cite{Linde:2005ht}. Then the WMAP data favour the so-called
`concordance' $\Lambda$CDM cosmology and constrain its parameters to
within a few per cent \cite{Komatsu:2010fb}. However the results are
very sensitive to the assumed PPS. Indeed by incorporating power
spectra with `bumps', broken power-laws, or a strongly `running'
spectral index, an Einstein-de Sitter model with a $\sim 10\%$ hot
dark matter component can still fit both CMB and LSS data
\cite{Blanchard:2003du,Hunt:2007dn,Hunt:2008wp}. Moreover all
geometrical measures, i.e. based on an assumed (in this case
Freidmann-Robertson-Walker) metric, such as the luminosity distance to
Type Ia supernovae or the baryon acoustic oscillation angular scale
can also be fitted in such models by adopting a different metric
e.g. the radially inhomogeneous Lemaitre-Tolman-Bondi form
\cite{Alexander:2007xx,Nadathur:2010zm}. Given that the observational
situation concerning \textit{dynamical} evidence for dark energy such
as the late integrated Sachs-Wolfe (ISW) effect is still confused
(e.g. \cite{Nadathur:2011iu}), an accurate estimate of the PPS is
clearly essential.

Knowledge of the PPS would also discriminate between models of
inflation and probe physics at very high energy scales, particularly
if the spectrum is distinctive. Single field slow-roll models predict
values of the spectral index and its running
$\mathrm{d}n_\mathrm{s}/\mathrm{d}{\ln}k$ so these parameters can be
constrained by fitting to the WMAP and other complementary data (see
e.g. \cite{Komatsu:2010fb,Finelli:2009bs,Alabidi:2010sf}). However
there are physically better motivated models which produce fully
adiabatic perturbations with \emph{broken} scale-invariance: `steps',
`bumps' and localised oscillations in the PPS can arise from
interruptions to slow-roll evolution caused by phase transitions
\cite{Adams:1997de,Hunt:2004vt, Hotchkiss:2009pj}, resonant particle
production \cite{Chung:1999ve,
  Mathews:2004vu,Romano:2008rr,Barnaby:2009mc,Barnaby:2010sq} or
features in the effective inflaton potential such as a `kink'
\cite{Starobinsky:1992ts,Leach:2001zf,Gong:2005jr}, `step'
\cite{Adams:2001vc,Chen:2006xjb,Chen:2008wn,Lerner:2008ad,
  Dvorkin:2009ne,Liu:2010dh,Adshead:2011jq} and others
\cite{Hodges:1989dw,Leach:2000yw,Ashoorioon:2006wc,Joy:2007na,
  Jain:2007au,Bean:2008na,Saito:2008em,Tye:2009ff,
  Achucarro:2010da,Goswami:2010qu,Brax:2011si,Arroja:2011yu,Liu:2011cw}. Such
PPS features also occur in models with modulated preheating
\cite{Chambers:2007se,Bond:2009xx}, a limited duration of inflation
\cite{Burgess:2002ub,Piao:2003zm,Powell:2006yg,Nicholson:2007by} and
other nonstandard scenarios \cite{Dvali:2003us,Lasenby:2003ur,
  Langlois:2004px}. Undamped PPS oscillations can be generated by
trans-Planckian physics \cite{Brandenberger:2000wr,
  Danielsson:2002kx,Schalm:2004xg,Greene:2005aj,Easther:2005yr} and in
other models that seek to incorporate Planck scale physics
\cite{Wang:2002hf,Kaloper:2003nv,Flauger:2009ab,
  Biswas:2010si,Jackson:2011qg,Chen:2011zf}. Clearly a
model-\textit{independent} method of recovering the PPS from
observational data is required.

Interest in broken scale-invariant spectra was stimulated by the first
WMAP data release which showed a lack of power on large scales in the
temperature (TT) angular power spectrum, suggesting a cutoff in the
PPS on the present horizon scale.  Although the octupole moment was
higher in the 3-year data release, the quadrupole is still
unexpectedly low in the WMAP 9-year data and outliers or `glitches'
(exceeding the cosmic variance) persist around the $\ell=22$ and
$\ell=40$ multipoles, despite the improved control of experimental
systematics and the additional integration time. In addition, a
possible anomaly around $\ell=120$ has been identified
\cite{Ichiki:2009xs,Kumazaki:2012mx}. The overall $\chi^2$ assuming a
power-law spectrum equals 3336.4 for 3115 degrees of freedom, the
probability to exceed this being only 0.3\% (stated to be mainly due
to the polarized likelihood \cite{Bennett:2012fp}). Many models of
inflation with broken scale-invariance have been compared to the low
\cite{Contaldi:2003zv,Cline:2003ve,
  Feng:2003zua,Kawasaki:2003dd,BasteroGil:2003bv,Hannestad:2004ts,
  Liguori:2004fa,Sinha:2005mn,Joy:2008qd,Jain:2008dw} and/or high
\cite{Martin:2003sq,Martin:2004yi,Kawasaki:2004pi,Covi:2006ci,
  Hamann:2007pa,Hoi:2007sf,Hazra:2010ve,Nakashima:2010sa,Aich:2011qv,
  Benetti:2011rp,Meerburg:2011gd} multipole anomalies.

Since the PPS is a free function it must be given a parameterisation.
In the model-independent approach the parameterisation is designed to
be as general as possible and to fit all conceivable forms of the PPS.
Model-independent methods of estimating the PPS fall into two classes,
depending on the number of power spectrum parameters. In parametric
methods the number of parameters, which is typically around 20 but can
be up to 50 \cite{Ichiki:2009zz}, is much less than the number of data
points.  The PPS has been described using wavelets
\cite{Mukherjee:2003cz,
  Mukherjee:2003yx,Mukherjee:2003ag,Mukherjee:2005dc}, principal
components \cite{Leach:2005av} and smoothing splines
\cite{Sealfon:2005em,Verde:2008zz,Peiris:2009wp,Bird:2010mp,Gauthier:2012aq},
in addition to bins in wavenumber with no interpolation (i.e. `tophat'
bins) \cite{Wang:1998gb}, linear interpolation \cite{Wang:2000js,
  Bridle:2003sa,Hannestad:2003zs,Bridges:2005br,Spergel:2006hy,
  Bridges:2006zm,Chantavat:2008nu,Bridges:2008ta}, cubic spline
interpolation
\cite{Ichiki:2009zz,Hlozek:2011pc,Guo:2011re,Guo:2011hy,Guo:2012jn}
and power-law bins \cite{Hannestad:2000pm}. The small number of
parameters means that accurate confidence limits for the cosmological
parameters can be set using Markov Chain Monte Carlo (MCMC) analysis
to draw samples from the posterior distribution of the parameters
given the data. However from the outset a filter is effectively
imposed on the recovered PPS by the choice of parameterisation, which
limits the resolution with which the PPS can be recovered. For
example, when the spectrum is parameterised using wavenumber bins or
smoothed with a spline, this will obviously miss features in the PPS
narrower than the bin width or the spine scale.

In non-parametric methods the number of parameters describing the
power spectrum is comparable to, and often greater than, the number of
data points. Since the CMB angular power spectrum is given by a
convolution of the PPS with a radiative transport kernel, this
approach is essentially an exercise in deconvolution. Recovery of the
PPS by deconvoluting the CMB angular power spectrum was first
discussed in \cite{Berera:1999rt} but only the Sachs-Wolfe effect was
considered. In the `cosmic inversion' method
\cite{Matsumiya:2001xj,Matsumiya:2002tx,Kogo:2003yb,Kogo:2004vt,
  Kogo:2005qi,Nagata:2008tk} a differential equation for the PPS
derived from cosmological perturbation theory was solved iteratively.
Regularisation schemes can be employed in non-parametric methods which
act as variable filters, governed by a regularisation parameter
$\lambda$ which can be tuned to optimally extract the PPS in the
presence of noise. Regularisation methods include truncated singular
value decomposition \cite{Nicholson:2009zj}, Richardson-Lucy iteration
\cite{Shafieloo:2003gf,
  Shafieloo:2006hs,Shafieloo:2007tk,Nicholson:2009pi,
  Hamann:2009bz,Gibelyou:2010qe,Hazra:2013xva,Hazra:2013eva} and
maximum entropy deconvolution \cite{Goswami:2013uja}. Tikhonov
regularisation was previously adopted by e.g.
\cite{Tegmark:2002cy,Tocchini-Valentini:2004ht,
  Tocchini-Valentini:2005ja,Nagata:2008zj,Ichiki:2009xs}.

We too adopt Tikhonov regularisation because it has notable advantages
over other non-parametric techniques. There is a relatively simple
relationship between the input data and the output estimated PPS
(unlike in iterative regularisation methods) which means the results
are easy to interpret. Furthermore Tikhonov regularisation is rapid
and computationally inexpensive, so that extensive testing of the
method on simulated data is feasible.

We improve over previous work in several significant respects.
Non-parametric techniques have mainly been applied to CMB TT power
spectra data alone (with the exception of
\cite{Kogo:2004vt,Kogo:2005qi, Nicholson:2009pi,Nicholson:2009zj}
where CMB polarisation data was also used).  By using additional data
sets it should be possible to recover the PPS over a wider wavenumber
range with increased accuracy.  In the pre-WMAP era, separate
estimates of the PPS were produced from a number of data sets
including non-CMB data \cite{Tegmark:2002cy}.  Here we extend the
Tikhonov regularisation method and show how it can be used to obtain a
\textit{single} combined high-resolution estimate of the PPS from
multiple data sets. We illustrate the method using TT and TE data from
WMAP and other CMB experiments, in addition to measurements of the
clustering of luminous red galaxies (LRG) in the Sloan Digital Sky
Survey (SDSS).
 
Simply constructing an estimate of the PPS is insufficient; it is
equally important to understand the relationship of the estimate to
the true PPS.\footnote{This is known as the `appraisal' of the
  estimate in the inverse theory literature \cite{Scales:2000ja}.} The
estimate will inevitably differ from the true PPS for at least three
reasons. The first is that the data are contaminated by noise. We
quantify the effects of error propagation on our recovered PPS using
Monte Carlo simulation in addition to estimating both Bayesian and
frequentist covariance matrices. The second is that the background
cosmological model parameters used to calculate the convolution
kernels are uncertain. While the effect on the reconstructed spectrum
of varying these has been studied earlier
\cite{Kogo:2004vt,Nicholson:2009zj}, we quantify the propagated error
from this effect for the first time, using covariance matrices. The
third reason is that a finite number of data points contains only a
limited amount of information about the true PPS.  In practice this
means the recovered PPS has limited resolution --- it would be unable
to distinguish fine features in the true PPS, even with noise-free
data. Also for the first time we quantify the resolution of the
estimated PPS, using the so-called `resolution kernel'.

Nearly all non-parametric inversion algorithms, including Tikhonov
regularisation, contain at least one implicit or explicit adjustable
regularisation parameter which depends on the signal-to-noise ratio of
the data and whose value must be chosen correctly for optimal
results.\footnote{For example, in truncated singular value
  decomposition and Richardson-Lucy iteration the regularisation
  parameter is identified with the truncation level and the iteration
  count respectively.} In all previous work the parameter values have
either been set by hand or chosen using techniques untested on
simulated cosmological data. We pay particular attention to this
issue, and study several different objective methods for selecting the
Tikhonov regularisation parameter using Monte Carlo simulation.

Almost all of the formal mathematical theory developed on Tikhonov
regularisation concerns its application to a single data set, usually
with uncorrelated Gaussian errors (see e.g. \cite{groetsch}). However
CMB data sets usually have \textit{correlated non-}Gaussian
errors. Moreover, the noise distributions at low multipoles depend on
the true PPS due to cosmic variance. While Tikhonov regularisation can
readily be adapted to multiple data sets with complicated noise
properties, it leads to some novel results not previously reported in
the literature.

The plan of the paper is as follows. In Section~\ref{method} we
describe our method, referring to Appendix~\ref{error} where we
discuss frequentist error analysis and comment on how it can be
adapted to Bayesian inference (\ref{bayesian}). Appendix~\ref{data}
presents the data used --- from WMAP (\ref{wmap}), ground-based CMB
experiments (\ref{groundbased}) and SDSS (\ref{sdss}) --- and the
corresponding likelihood function derivatives (\ref{matrix}). In
Appendix~\ref{validation}, we examine the performance of our method
using mock data (\ref{mockdata}) for several inflationary model test
spectra (\ref{spectest}) and study how the value of the regularisation
parameter affects the inversion. In Appendix~\ref{regularisation},
methods of choosing the optimum regularisation parameter are
investigated. We present our main results in Section~\ref{results}.
In addition to the estimated PPS (\ref{spectra}) we include analysis
of the uncorrelated bandpowers (\ref{uncorr}), uncertainties in the
extracted cosmological parameters (\ref{bgpe}) and the statistical
significance of the PPS features (\ref{statsig}). In
Section~\ref{conclusions} we present our conclusions.  A glossary of
symbols used is provided in Appendix~\ref{glossary}.

\section{Inversion method \label{method}}

\subsection{Tikhonov regularisation \label{tikreg}}

The two-power correlation function of the primordial comoving
curvature perturbation $\mathcal{R}$ is related to the PPS
through:~\footnote{ Our notation/definition of the PPS is that of
  \cite{Lyth:1998xn}. In \cite{Komatsu:2008hk} the PPS is denoted by
  $\Delta^2_\mathcal{R}\left(k\right)$.}
\begin{equation}
\label{2ptcorr}
 \langle \mathcal{R}^{*}\left(\mathbf{k}\right)
 \mathcal{R}\left(\mathbf{k}^{\prime}\right) \rangle =
 \delta^3\left(\mathbf{k} - \mathbf{k}^{\prime}\right) 
 \frac{2\pi^2}{k^3}\mathcal{P_R}
 \left(k\right).
\end{equation}
Let us assume there are $N$ available cosmological data sets which
probe the PPS. The data points are denoted
$\mathrm{d}_a^{(\mathbb{Z})}$ where the superscript labels the data
set ($\mathbb{Z}$) and the subscript $a$ runs from $1$ up to the
number $N_\mathbb{Z}$ of points in the set.\footnote{ For ease of
  notation a data set superscript is promoted to a subscript when
  labelling an entire vector or matrix rather than one of its elements
  e.g. the vector $\mathrm{d}_a^{(\mathbb{Z})}$ is simply
  $\dB_{\mathbb{Z}}$ and the matrix $W_{ai}^{(\mathbb{Z})}$ is simply
  $\mathsf{W}_\mathbb{Z}$.} The data points extracted from
measurements of CMB anisotropy, galaxy clustering, Lyman-$\alpha$
forest, cluster abundance and weak lensing obey a first-order integral
equation of the form \cite{Tegmark:2002cy}:
\begin{equation}
  \mathrm{d}_a^{(\mathbb{Z})} = \int^{\infty}_0
  \mathcal{K}_a^{(\mathbb{Z})}\left(\bm{\theta},k\right)\mathcal{P_R}
  \left(k\right)\,\mathrm{d}k + \mathrm{n}_a^{(\mathbb{Z})},
\label{int1}
\end{equation}
where the integral kernels $\mathcal{K}_a^{(\mathbb{Z})}$ depend on
the background parameters $\bm{\theta}$, and the noise vectors
$\mathbf{n}_\mathbb{Z}$ have zero mean and covariance matrices
$\mathsf{N}_\mathbb{Z}\equiv\langle \nB_\mathbb{Z}
\nB_\mathbb{Z}^\mathrm{T}\rangle$, where $^\mathrm{T}$ denotes the
transpose matrix. Eq.(\ref{int1}) is quite general: it holds on large
scales where \emph{linear} cosmological perturbation theory is valid
and also applies in the nonlinear regime if a suitable transformation
or linearisation can be performed. We assume an estimate
$\hat{\bm{\theta}}$ of the background parameters exists which is
\emph{independent} of the $N$ data sets, and has a zero mean
uncertainty $\uB$, with elements $\mathrm{u}_\alpha$.  Then $\langle
\mathrm{u}_\alpha \mathrm{n}_a^{(\mathbb{Z})}\rangle=0$ for all
elements of the uncertainty and noise vectors as these are
uncorrelated by assumption. The covariance matrix for the estimated
background parameters is just $\mathsf{U}\equiv\langle \uB
\uB^\mathrm{T}\rangle$. Given our estimate of the background parameter
set $\hat{\bm{\theta}}$, the goal is to obtain an estimate
$\hat{\mathcal{P}}_\mathcal{R}\left(k\right)$ of the PPS from the data
sets $\dB_{\mathbb{Z}}$.

This is an example of an inverse problem (as opposed to the forward
problem of calculating the expected data for a given PPS) which
generally involve the determination of some unknown model input from
indirect measurements. Inverse theory concerns both parameter and
function estimation. A `well-posed' problem is defined
\cite{hadamard:1923tg} as having an unique solution which depends
continuously on the data, \emph{i.e.} as the error in the data tends
to zero, the induced error in the solution also tends to zero. If this
is not so, the problem is said to be `ill-posed'. Inverse problems are
usually ill-posed, and the reconstruction of the PPS is no
exception. Ill-posed inverse problems occur also for example in
medical and geophysical imaging, remote sensing and image restoration,
and an extensive literature is devoted to their solution (for a
textbook introduction see e.g. \cite{hansen:1998dc,engl:2000we}). In
cosmology, there have been applications in e.g. gravitational wave
detection \cite{Rakhmanov:2006qm} and in cosmography
\cite{Kitaura:2007pe}, as well as of course in the present context
\cite{Tegmark:2002cy,Tocchini-Valentini:2004ht,Tocchini-Valentini:2005ja}.

For any finite number of data points there is an infinite-dimensional
null space of functions $\mathcal{N}_\mathcal{R}\left(k\right)$ such that
\begin{equation}
\label{null}
 0 = \int^{\infty}_0 \mathcal{K}_a^{(\mathbb{Z})}\left(\bm{\theta},k\right) 
     \mathcal{N}_\mathcal{R}\left(k\right)\,\mathrm{d}k
\end{equation} 
for all of the integral kernels \cite{parker:1994dc}. The available
data cannot distinguish between two power spectra that differ by a
member of the null space (e.g. because of finite resolution or even
because the difference would be obvious only where there is no
data). Clearly some additional criterion must be used to narrow down
the infinite number of power spectra consistent with the data, the
vast majority of which may well be unphysical. From our present
(admittedly crude) understanding of inflation the PPS is expected to
be smooth in some sense with little fine detail. Hence we assume that
the simplest solution is the preferred one i.e. the optimal estimate
of the PPS is the \textit{smoothest} of the spectra matching the
data. While other spectra may fit the data equally well, they will
inevitably be more complex and likely possess misleading
features. Thus our prior knowledge (prejudice) concerning the PPS is
combined with the information contained in the data to obtain the
optimal estimate.

We approximate the PPS as a piecewise function given by a sum of $N_j$
basis functions $\phi_i\left(k\right)$, weighted by coefficients
$\mathrm{p}_i$ giving the power in each bin:
\begin{equation}
 \mathcal{P_R}\left(k\right) = \sum_{i=1}^{N_j}\mathrm{p}_i\phi_i\left(k\right).
\label{basis} 
\end{equation}
The basis functions are defined using a grid of wavenumbers
$\left\{k_i\right\}$ as
\begin{equation}
\label{phidef}
 \phi_i\left(k\right) \equiv \left\{\begin{array}{ll}
 1,& k_{i}<k\leq k_{i+1},  \\
 0,&  \mbox{elsewhere.} 
 \end{array}\right.
\end{equation}
We set $N_j=2000$ and use a logarithmically spaced grid between
$k_1=7\times 10^{-6}\; \mathrm{Mpc}^{-1}$ and $k_{N_j+1}=0.7\;
\mathrm{Mpc}^{-1}$. Substituting eq.(\ref{basis}) into eq.(\ref{int1})
yields
\begin{equation}
 \mathrm{d}_a^{(\mathbb{Z})} = \sum_i
 W_{ai}^{(\mathbb{Z})}\left(\bm{\theta}\right)\mathrm{p}_i +
 \mathrm{n}_a^{(\mathbb{Z})},
\label{rel}
\end{equation}  
where the $N_\mathbb{Z}\times N_j$ matrices
$W_{ai}^{(\mathbb{Z})}\left(\bm{\theta}\right)$
($\mathsf{W}_\mathbb{Z}$ for short) depend on the background
parameters:
\begin{equation}
\label{rel1}
 W_{ai}^{(\mathbb{Z})}\left(\bm{\theta}\right) = \int^{k_{i+1}}_{k_i}
 \mathcal{K}_a^{(\mathbb{Z})}\left(\bm{\theta},k\right)\,
 \mathrm{d}k.
\end{equation}
Vectors of length $N_d\equiv\sum_\mathbb{Z} N_\mathbb{Z}$ assembled
from the individual data and noise vectors are denoted by
$\dB\equiv\left(\dB_1^\mathrm{T},\dB_2^\mathrm{T}, \ldots,
\dB_N^\mathrm{T}\right)^\mathrm{T}$ and
$\nB\equiv\left(\nB_1^\mathrm{T},\nB_2^\mathrm{T}, \ldots,
\nB_N^\mathrm{T}\right)^\mathrm{T}$ respectively. We also define the
$N_d\times N_j$ matrix
$\mathsf{W} \equiv \left(\mathsf{W}_1^\mathrm{T},\mathsf{W}_2^\mathrm{T},\ldots,
\mathsf{W}_N^\mathrm{T}\right)^\mathrm{T}$ so that
\begin{equation}
 \dB = \mathsf{W}\left(\bm{\theta}\right)\pB+\nB,
\label{rel2}
\end{equation} 
where $\pB$ is the vector with elements $\mathrm{p}_i$ defining the PPS.

Standard methods of statistical inference such as maximum likelihood
analysis perform poorly when applied to ill-posed inverse problems.
Instead, specialist techniques must be employed. To illustrate the
difficulties involved, we use a superficially appealing estimate of
the PPS labelled $\hat{\pB}_0$ defined in the following manner. Recall
that the rank of a matrix is equal to the number of nonzero singular
values, and is also equal to the number of linearly independent column
and row vectors. If the rank $\rho$ of $\mathsf{W}$ is equal to $N_j$
then $\hat{\pB}_0$ is the unique maximum of the likelihood function of
the data given the PPS and the background parameters. In general the
matrix $\mathsf{W}$ has a $\left(N_j-\rho\right)$-dimensional null
space of vectors $\mathbf{v}_\mathrm{nl}$ such that $\mathsf{W}
\mathbf{v}_\mathrm{nl}=0$. Hence if $\rho<N_j$ there are infinitely
many solutions which maximise the likelihood and differ by members of
the null space, so that the likelihood function has no well-defined
peak.  We introduce a function $\mathrm{R}\left(\pB\right)$ which
quantifies the `roughness' of $\pB$. Then if $\rho<N_j$ the estimate
$\hat{\pB}_0$ is taken as the vector with the minimum $\mathrm{R}$
value (i.e. the `smoothest') of those that maximise the likelihood.

For multiple data sets with Gaussian noise the likelihood function
$\mathcal{L}\left(\pB,\bm{\theta}|\dB\right)$ is given by
\begin{equation}
 -2\ln\mathcal{L}\left(\pB,\bm{\theta}|\dB\right) = 
\left[\mathsf{W}\left(\bm{\theta}\right)\pB - 
\dB\right]^\mathrm{T}\mathsf{N}^{-1} 
\left[\mathsf{W}\left(\bm{\theta}\right)\pB - \dB\right]
\label{gauslike}
\end{equation}
up to an unimportant normalisation constant, where
$\mathsf{N}\equiv\mathrm{diag}\left(\mathsf{N}_1,\mathsf{N}_2,\dots,
\mathsf{N}_N\right)$ is a $N_d\times N_d$ block diagonal matrix
assembled from the individual data covariance matrices. It is diagonal
because the individual data sets are assumed to be
\emph{independent}. If in addition the roughness function has the
quadratic form $\mathrm{R}\left(\pB\right)=\pB^\mathrm{T}
\mathsf{\Gamma} \pB$ where $\mathsf{\Gamma}$ is a positive definite
matrix, then
$\hat{\pB}_0 = \mathsf{W}_{\mathsf{N}\mathsf{\Gamma}}^{\dagger}\dB$ for
$\rho\leq N_j$. Here
$\mathsf{W}_{\mathsf{N}\mathsf{\Gamma}}^{\dagger}$ is the weighted
Moore-Penrose inverse of $\mathsf{W}$, also known as the weighted
generalised inverse \cite{benisrael:2003we,wangwei:2004we} and can be
calculated using the weighted singular value decomposition of
$\mathsf{W}$ developed in \cite{vanloan:1976ja}. This is written in
terms of the $N_d\times N_d$ matrix $\mathsf{J}$ and the $N_j\times
N_j$ matrix $\mathsf{K}$ which satisfy
$\mathsf{J}^\mathrm{T}\mathsf{N}\mathsf{J}=\mathsf{I}$ and
$\mathsf{K}^\mathrm{T}\mathsf{\Gamma}^{-1}\mathsf{K} = \mathsf{I}$ and
has the form
\begin{equation}
\mathsf{W} = \mathsf{J}\left(\begin{array}{cc}
\mathsf{\Lambda} & \mathsf{0} \\
\mathsf{0} & \mathsf{0} \end{array} \right)\mathsf{K}^\mathrm{T},
\end{equation}  
where
$\mathsf{\Lambda} \equiv \mathrm{diag} 
\left(\sigma_1,\sigma_2,\dots,\sigma_\rho\right)$
with $\sigma_1\geq\sigma_2\geq\ldots\geq\sigma_\rho>0$.  The
$\left\{\sigma_i\right\}$ are called the weighted singular values of
$\mathsf{W}$ and are just the square-roots of the nonzero eigenvalues of
$\mathsf{\Gamma}^{-1}\mathsf{W}^\mathrm{T}\mathsf{N}\mathsf{W}$. The
weighted Moore-Penrose inverse is then given by:
\begin{equation}
\mathsf{W}_{\mathsf{N}\mathsf{\Gamma}}^{\dagger} = 
\mathsf{\Gamma}^{-1}\mathsf{K}\left(\begin{array}{cc}
  \mathsf{\Lambda}^{-1} & \mathsf{0} \\ \mathsf{0} &
  \mathsf{0} \end{array} \right)\mathsf{J}^\mathrm{T}\mathsf{N}.
\label{mp}
\end{equation}
The significance of this will be discussed shortly. Note that the
ordinary Moore-Penrose inverse $\mathsf{W}^{\dagger}$ is recovered
when $\mathsf{N}$ and $\mathsf{\Gamma}$ are both equal to the
identity matrix.

Unfortunately $\hat{\pB}_0$ is `ill-conditioned', i.e. excessively
sensitive to small perturbations of the data. This is because
convolving the PPS with an integral kernel acts as a smoothing
operation, due to the finite width of the kernel. High frequency
components of the PPS are damped more than low frequency
ones. Conversely, high frequency noise is amplified to produce wild
excursions in $\hat{\pB}_0$, rendering the estimate useless. Roughly
speaking, the condition number of a system measures how it amplifies
errors. The condition number of the weighted Moore-Penrose is
$\sigma_1/\sigma_\rho$, which indicates that the ill-conditioning of
$\hat{\pB}_0$ is associated with $\mathsf{W}$ possessing small
weighted singular values, as can be seen from eq.(\ref{mp}).

So-called regularisation schemes are designed to produce a
well-behaved approximate solution to an ill-posed inverse
problem. They effectively filter out the troublesome small singular
values of $\mathsf{W}$.  Regularisation reduces the propagation of
error from the data to the solution at the cost of inducing a bias;
the former is intended to offset the latter. Since the optimum
solution depends on the level of noise in the data, a regularisation
scheme actually produces a family of solutions characterised by a
`regularisation parameter'.  Obtaining the optimum solution then
amounts to choosing the appropriate value of the regularisation
parameter according to the noise level. Common regularisation schemes
include spectral cut-off methods such as truncated singular value
decomposition, iterative methods such as Landweber iteration, and the
Tikhonov regularisation scheme --- which we use in this work.

Tikhonov regularisation is based upon the notion of a `correctness
set' \cite{tikhonov}. For an inverse problem this is a subset of
solution space (i.e. the space of all possible solutions) known
\textit{a priori} to contain the true solution. Moreover, the solution
is continuous for data such that the solution remains in the
correctness set.\footnote{Indeed regularisation schemes need to
  incorporate \emph{some} prior expectation concerning the solution in
  order to overcome the generally `ill-posed' nature of the problem
  (as is obvious in e.g. applications to remote geophysical sensing or
  medical imaging where the physically plausible answer is easily
  recognised) \cite{hansen:1998dc,engl:2000we}.} Since the PPS is
believed to be smooth, there is effectively an upper bound
$\mathrm{R}_0$ on its roughness: spectra rougher than $\mathrm{R}_0$
may be regarded as physically implausible. The compact set of spectra
which satisfy $\mathrm{R}\left(\mathbf{p}\right)\le \mathrm{R}_0$ form
a natural correctness set. The estimated PPS is chosen as the member
of the correctness set which \textit{maximises} the likelihood
function. In practice the roughest elements of the correctness set
give the best fit to the data and so the estimate satisfies
$\mathrm{R}\left(\hat{\mathbf{p}}\right)= \mathrm{R}_0$.  Therefore to
obtain $\hat{\mathbf{p}}$ we maximise the likelihood subject to this
constraint. This is equivalent to the more convenient unconstrained
minimisation of the quantity
\begin{equation} 
\label{minq}
 Q\left(\pB,\dB,\hat{\bm{\theta}},\lambda\right)\equiv
 L\left(\pB,\hat{\bm{\theta}},\dB\right) + \lambda \mathrm{R}\left(\pB\right),
\end{equation}
where
$L\left(\pB,\bm{\theta},\dB\right)\equiv-2\ln\mathcal{L}
\left(\pB,\bm{\theta}|\dB\right)$
and the regularisation parameter $\lambda$ plays the role of a
Lagrange multiplier. Thus the estimated PPS is given by
\begin{equation}
 \hat{\mathbf{p}}\left(\dB,\hat{\bm{\theta}},\lambda\right) = 
 {\argmin}_\pB\,Q\left(\pB,\dB,\hat{\bm{\theta}},\lambda\right).
\end{equation}
For a quadratic roughness function, the estimate $\hat{\pB}$ is a
unique continuous function of the data.  It is defined as the
best-fitting spectrum for a particular roughness value, but it can
equally well be thought of as the \emph{smoothest} spectrum for a
given value of the likelihood.

The regularisation parameter thus controls the trade-off between the
smoothness of the estimated PPS and its fidelity to the data.
Increasing $\lambda$ reduces the correctness set so that $\hat{\pB}$
becomes smoother and less spiky but fits the data less well. As
$\lambda$ tends to zero, $\hat{\pB}$ tends to the smoothest spectrum
which maximises the likelihood, viz. the ill-conditioned
$\hat{\pB}_0$. If the noise level of the data and the error in the
estimated background parameters also tend to zero, then $\hat{\pB}$
approaches the true PPS. In the opposite limit $\hat{\pB}$ approaches
the best-fitting spectrum which minimises $\mathrm{R}$ as $\lambda$
tends to infinity. Thus we attempt to find a smooth PPS which provides
a reasonable fit to the data, in a similar manner to nonparametric
regression.

While the regularisation parameter determines the \textit{amount} by
which $\hat{\pB}$ is smoothed, the roughness function governs the
\textit{manner} in which $\hat{\pB}$ is smoothed. In this work we employ
the roughness function $\mathrm{R}\left(\mathbf{p}\right)=\mathbf{p}^\mathrm{T}
\mathsf{\Gamma} \mathbf{p}$, where
\begin{equation}
\label{firstl}
\mathsf{\Gamma}\equiv\mathsf{L}^\mathrm{T} \mathsf{L} =
\left(\begin{array}{ccccc} 1 & -1 & & & \\ -1 & 2 & -1 & & \\ & \ddots
  & \ddots & \ddots & \\ & & -1 & 2 & -1 \\ & & & -1 & 1
\end{array}\right),
\end{equation} 
and $\mathsf{L}$ is a discrete version of the first-order derivative
operator. For a logarithmically spaced wavenumber grid, $\mathrm{R}$
represents a discretisation of the integral of
$\left(\mathrm{d}\mathcal{P_R}/\mathrm{d}\ln k\right)^2$ over $\ln
k$. This choice of roughness function is known as first-order Tikhonov
regularisation \cite{aster}. Similar roughness functions involving
first-order derivatives were used previously in
\cite{Tocchini-Valentini:2004ht,Tocchini-Valentini:2005ja}. Any vector
with equal components minimises $\mathrm{R}$, corresponding to a
Harrison-Zeldovich (H-Z) PPS with $n_\mathrm{s}=1$. Note that for
wavenumbers where the corresponding data is incomplete or does not
constrain the PPS, $\hat{\pB}$ is determined mainly by the roughness
function. Hence with the above choice of the roughness function we
ensure that the recovered PPS is generally positive. This would not be
the case if we chose instead a second-order derivative operator
corresponding to a `tilted' spectrum with $n_\mathrm{s} \lesssim 1$
(even though this is in fact the best-fit to data for a $\Lambda$CDM
cosmology). Note that the procedure used in \cite{Tegmark:2002cy} is
equivalent to setting $\mathsf{\Gamma}$ equal to the identity matrix,
which corresponds to zeroth-order Tikhonov regularisation.

We minimise $Q$ using the Broyden-Fletcher-Goldfarb-Shanno (BFGS)
algorithm --- a quasi-Newton method \cite{press}. In the usual 
Newton-Raphson method the function to be minimised is assumed to be
quadratic near its minimum and at each iteration a better approximation
for the minimum is obtained by solving a linear system of equations
involving the gradient vector and the Hessian matrix of second partial
derivatives of $Q$. The BFGS algorithm avoids the computationally
expensive evaluation of the Hessian; instead an accurate approximation
to the inverse Hessian is built up using changes in the gradient
vector between successive iterations, in a multidimensional
generalisation of the secant method. Expressions for the gradients of
the various likelihood functions are listed in
Appendix~\ref{matrix}, together with the initial guesses for the
Hessian.

\section{Results \label{results}}

In Appendix~\ref{validation} we validate our inversion method on
simulated data. Below we apply it to real cosmological data.

\subsection{Recovered spectra \label{spectra}}

\begin{figure*}
\includegraphics*[angle=0,width=0.5\columnwidth,trim = 32mm 171mm 23mm
  15mm, clip]{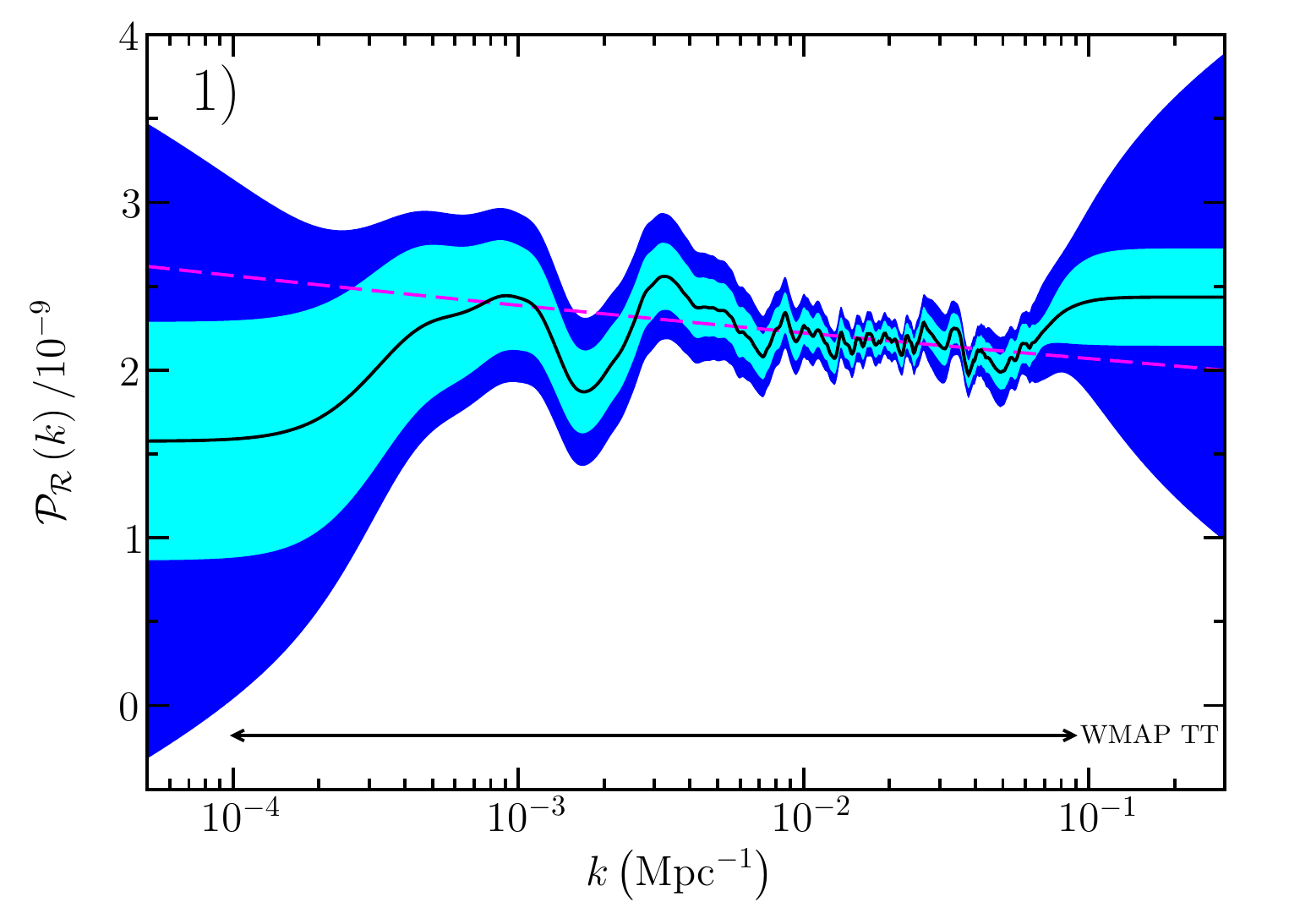}
\includegraphics*[angle=0,width=0.5\columnwidth,trim = 32mm 171mm 23mm
  15mm, clip]{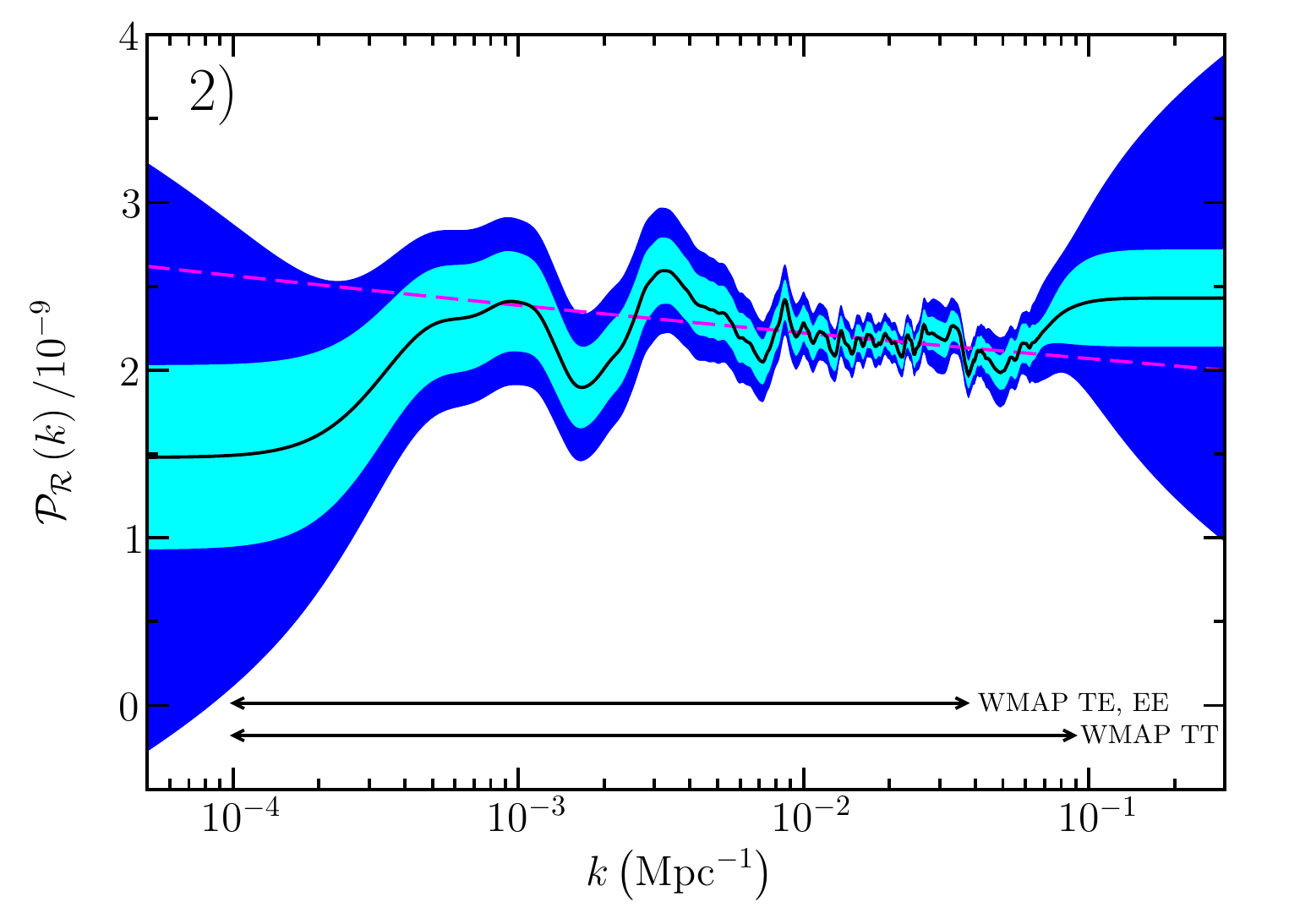}
\includegraphics*[angle=0,width=0.5\columnwidth,trim = 32mm 171mm 23mm
  15mm, clip]{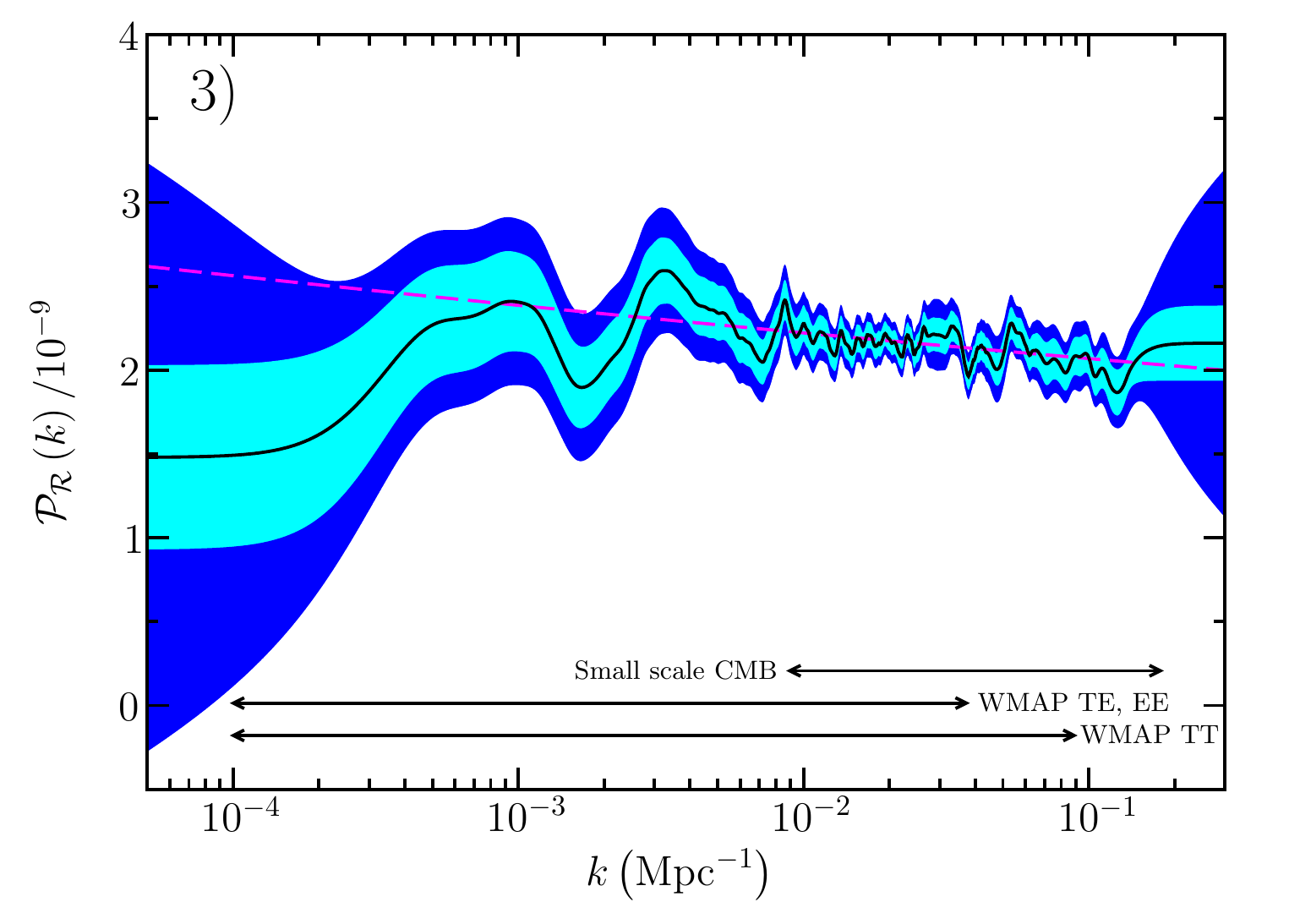}
\includegraphics*[angle=0,width=0.5\columnwidth,trim = 32mm 171mm 23mm
  15mm, clip]{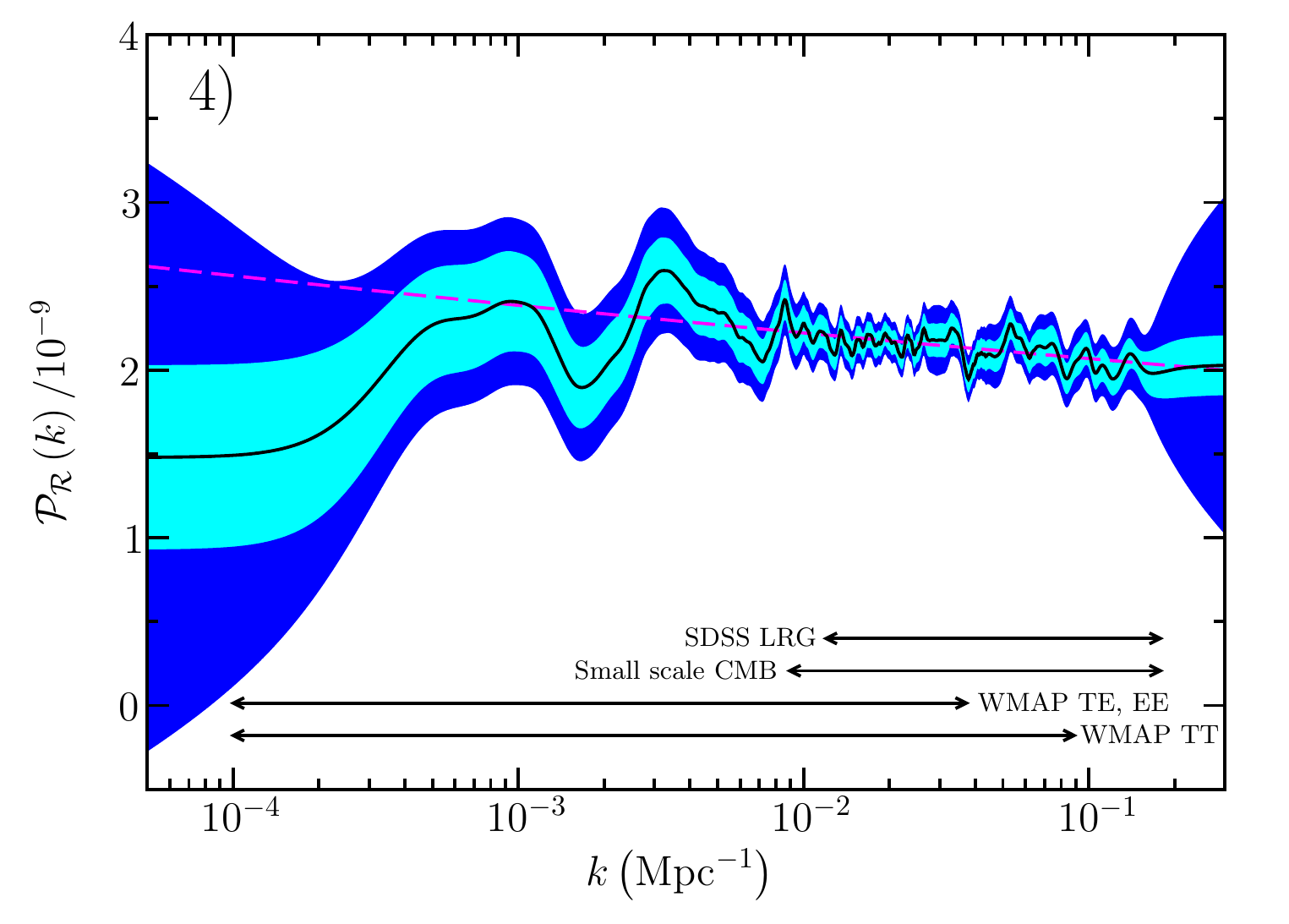}
\caption{\label{lolam} The PPS recovered with $\lambda=100$ from the
  1) WMAP-9 temperature data alone, 2) WMAP-9 temperature and
  polarisation data, 3) WMAP-9 plus small angular scale CMB data, 4)
  WMAP-9, small-scale CMB and SDSS-4 LRG data. In each plot the
  central black line is the reconstructed PPS and the dark band is the
  $1\sigma$ error given by the square root of the diagonal elements of
  the Bayesian covariance matrix $\mathsf{\Sigma}_\mathrm{B}$
  (eq.\,\ref{sigb}), while the overlaying light band is similarly
  obtained from the frequentist covariance matrix
  $\mathsf{\Sigma}_\mathrm{F}$ (eq.\,\ref{sigmaf}). The dashed line is
  our `best-fit' power-law spectrum with slope $n_\mathrm{s}=0.969$
  assuming the standard $\Lambda$CDM model parameters. Also indicated
  are the approximate wavenumber ranges over which the different
  datasets have the most impact.}
\end{figure*}

\begin{figure*}
\begin{minipage}{150mm}
\includegraphics*[angle=0,width=0.5\columnwidth,trim = 32mm 171mm 23mm
  15mm, clip]{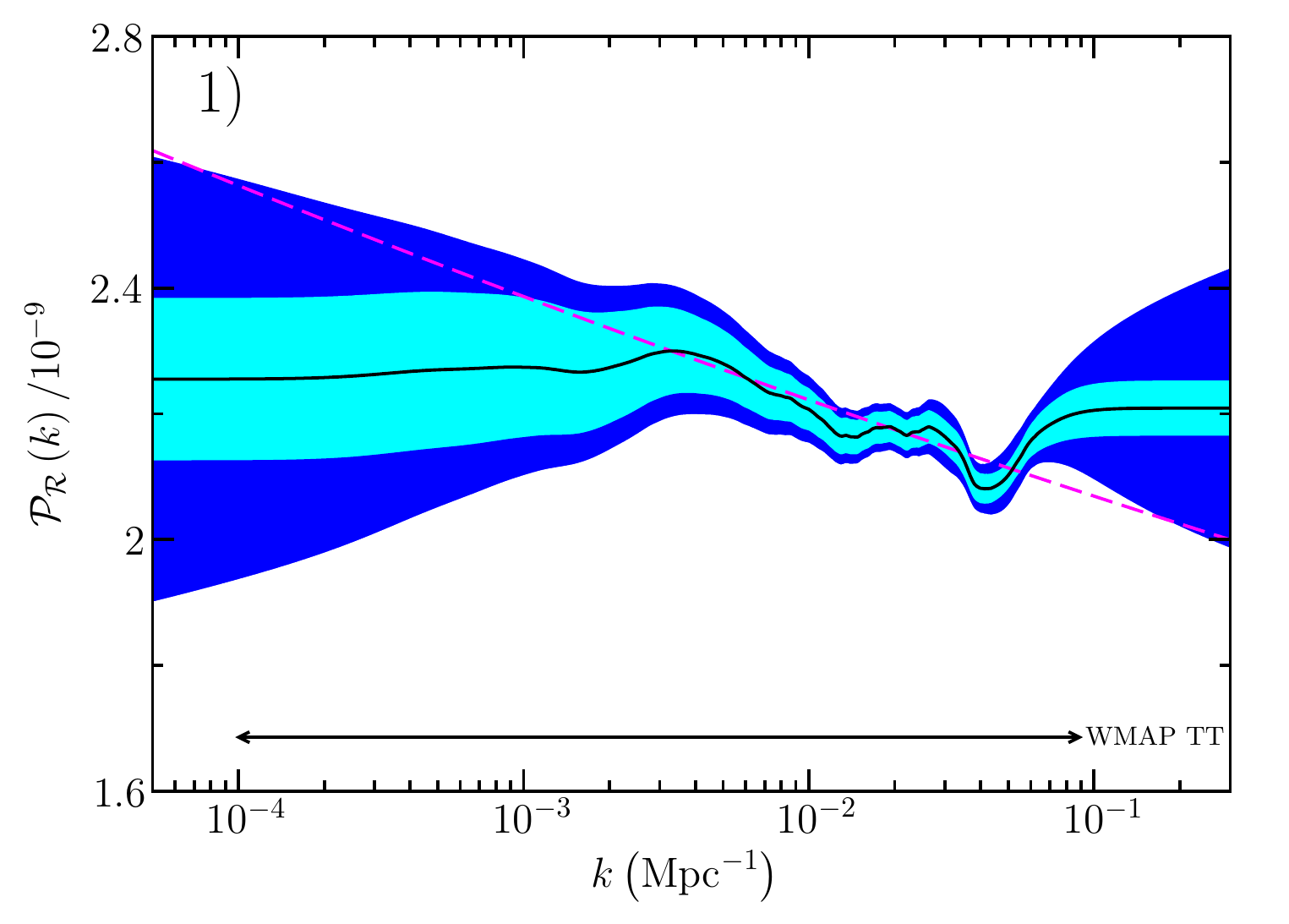}
\includegraphics*[angle=0,width=0.5\columnwidth,trim = 32mm 171mm 23mm
  15mm, clip]{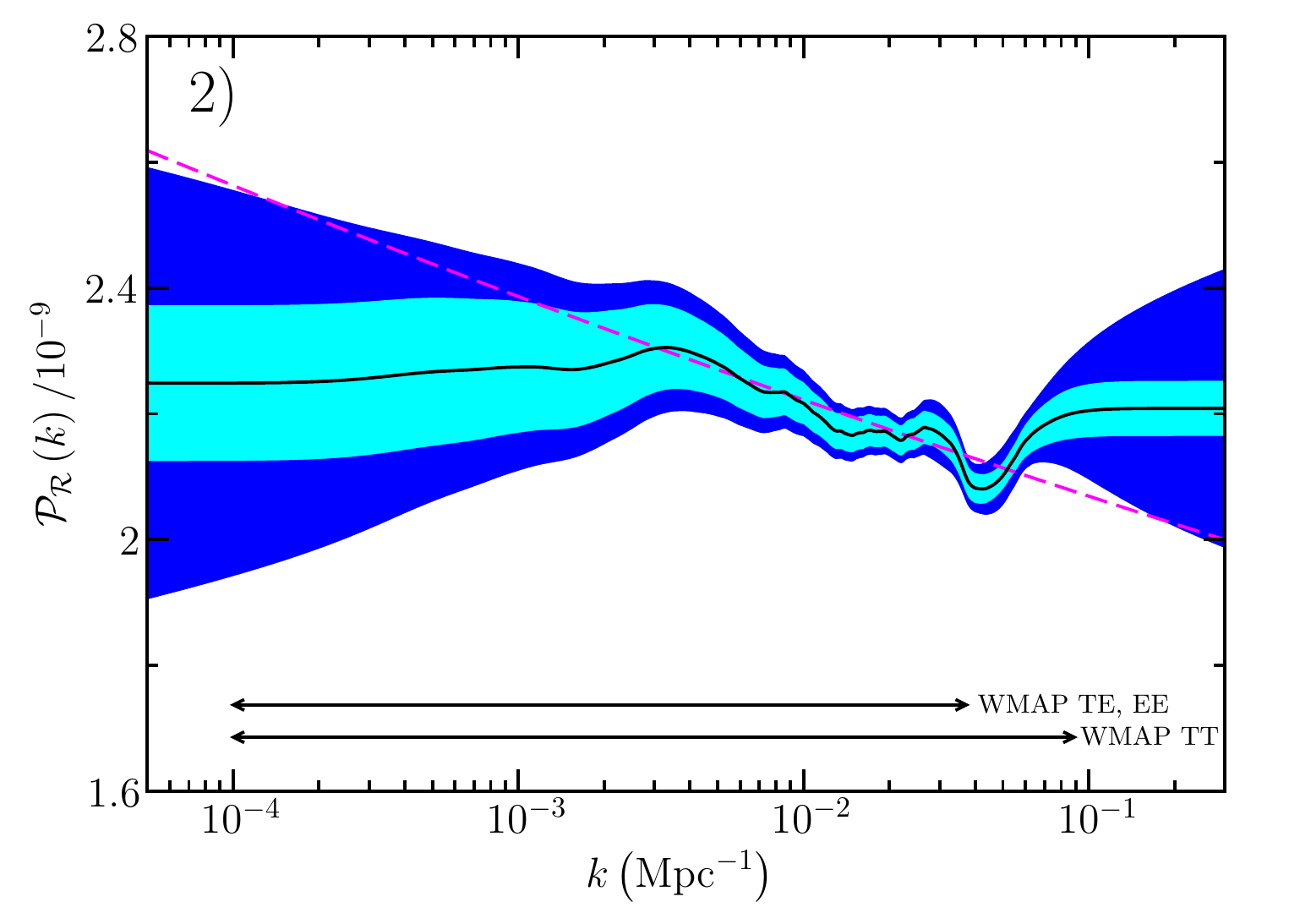}
\includegraphics*[angle=0,width=0.5\columnwidth,trim = 32mm 171mm 23mm
  15mm, clip]{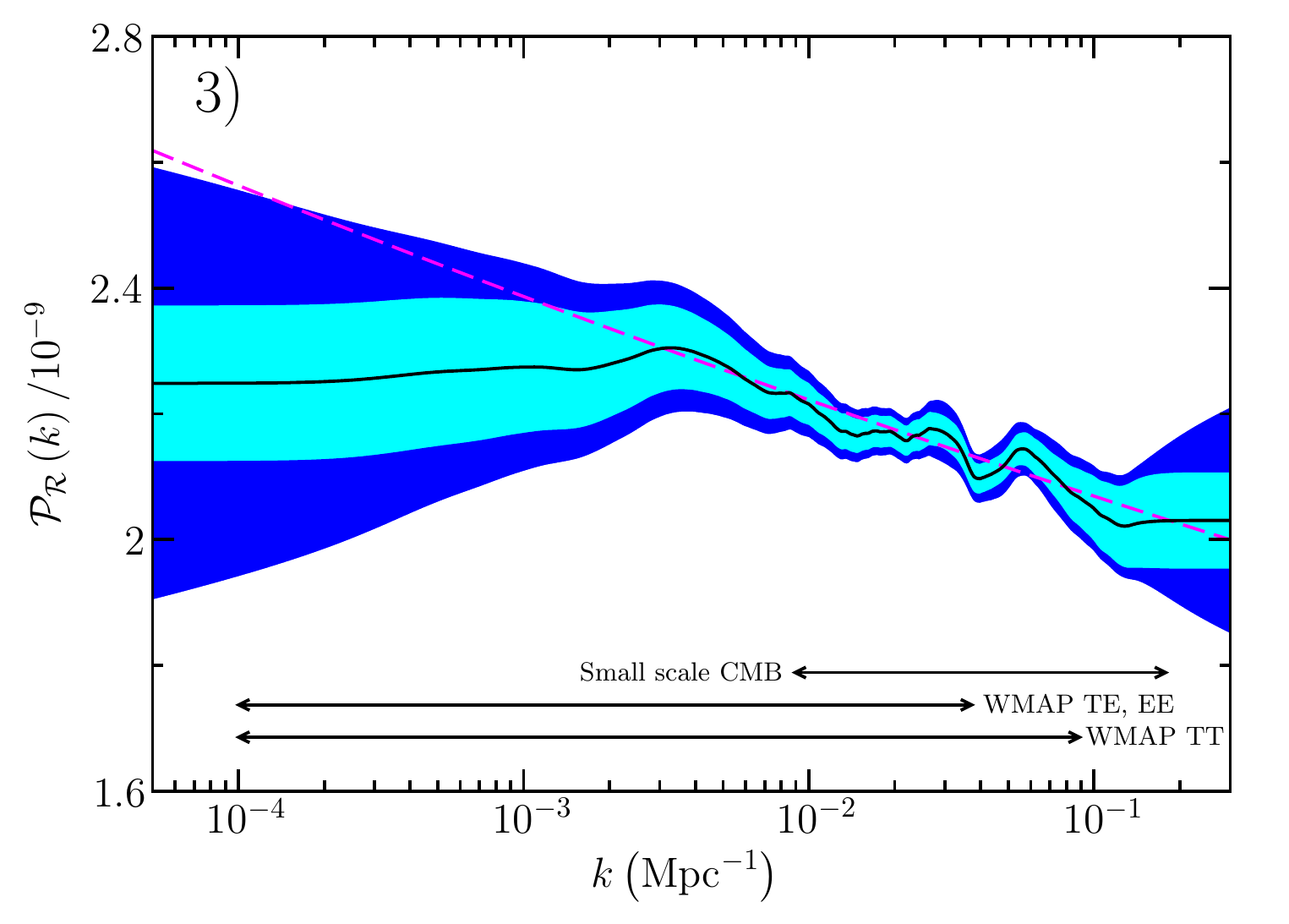}
\includegraphics*[angle=0,width=0.5\columnwidth,trim = 32mm 171mm 23mm
  15mm, clip]{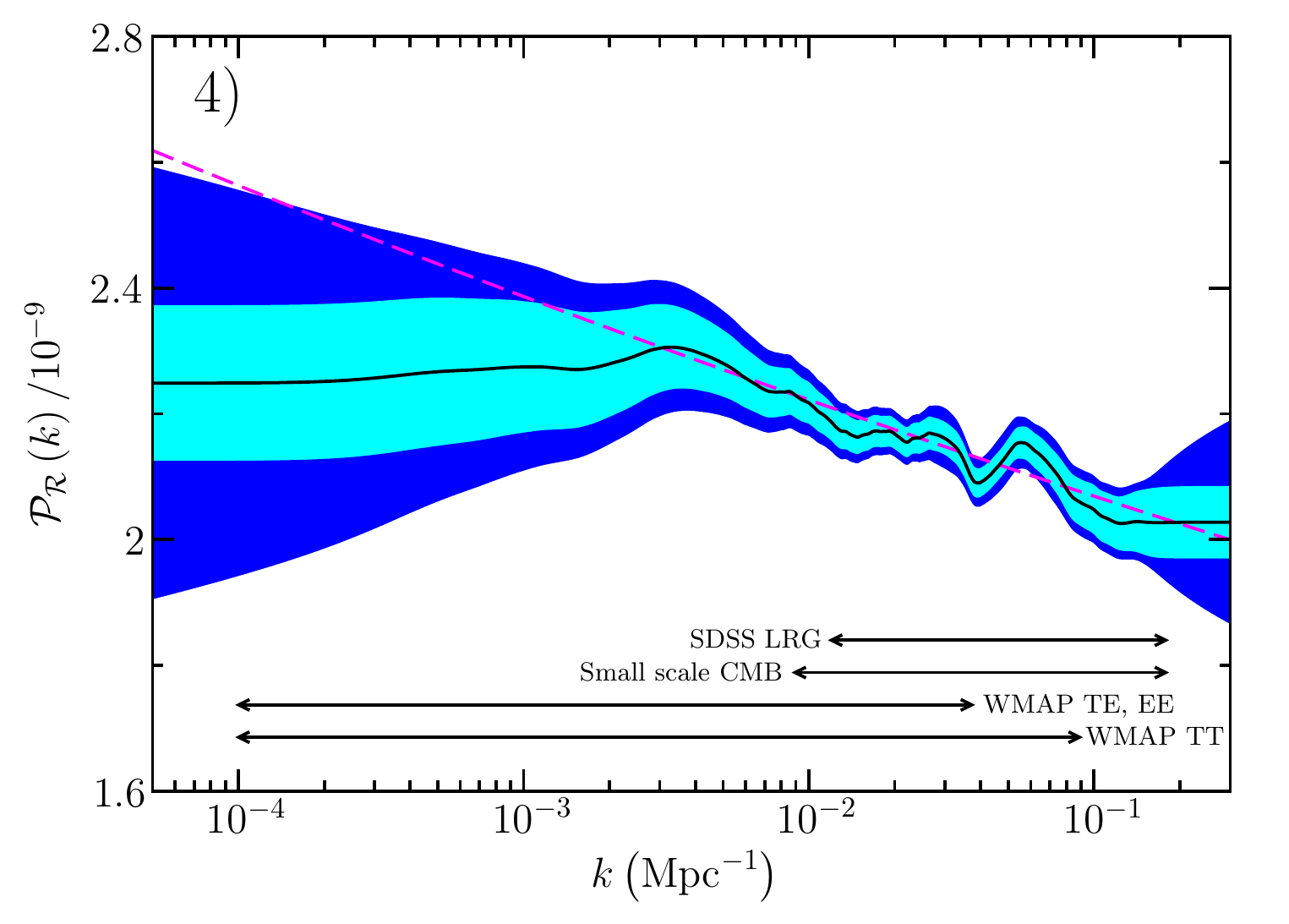}
\caption{\label{hilam} Same as Fig.\,\ref{lolam}, but for inversions
  with $\lambda=5000$.}
\end{minipage}
\end{figure*}

We refer to 4 data set combinations of CMB and LSS data as follows:

\begin{itemize}

\item Data combination 1: WMAP TT 
\item Data combination 2: WMAP all (TT + TE + EE)
\item Data combination 3: All CMB (WMAP all + ground-based: ACBAR +
  BOOMERanG (TT + TE + EE) + CBI + QUAD (TT + TE + EE) + VSA)
\item Data combination 4: All CMB + SDSS-4 LRG

\end{itemize}
Obviously these can be updated with data from other experiments as and
when necessary.

We adopt typical parameter values for the standard $\Lambda$CDM model:
$\omega_\mathrm{b}\equiv\Omega_\mathrm{b}h^2=0.0224$,
$\omega_\mathrm{c}\equiv\Omega_\mathrm{c}h^2=0.102$, $h=0.73$, optical
depth to last scattering $\tau=0.095$, and bias of luminous red
galaxies $b_\mathrm{LRG}=1.9$ \cite{Tegmark:2006az}. This model is
spatially flat, hence $\Omega_\Lambda \equiv 1 - \Omega_\mathrm{b} -
\Omega_\mathrm{c}$.

 As shown in Appendix~\ref{testspec}, if the PPS is an exact power-law
 then deconvoluting with $\lambda\simeq 5000$ would be most likely to
 minimise the squared-error (Panel A, Fig.~\ref{mse}). However if the
 PPS has features then a lower value like $\lambda\simeq 100$ would be
 more appropriate (Panels B and C, Fig.~\ref{mse}). Therefore we
 perform reconstructions with both $\lambda=100$ (Fig.\,\ref{lolam})
 and $\lambda= 5000$ (Fig.\,\ref{hilam}) but are not able to advocate
 any particular value of $\lambda$ in this range.

The recovered spectra oscillate about a power-law with $n_s=0.969$,
consistent with the spectral index found by the WMAP team. However,
there are several interesting features. The $\lambda=100$ spectra have
the well-known infrared cutoff from the low WMAP-9 TT
quadrupole. There are also `bumps' at $k\simeq 0.0032, 0.0086$ and
$0.033\;\mathrm{Mpc}^{-1}$ due to the excess power in the WMAP-9
temperature angular power spectrum around the $\ell\simeq 40, 117$ and
470 multipoles. Similarly `dips' at $k\simeq 0.0017, 0.0072, 0.013,
0.015$ and $0.038\;\mathrm{Mpc}^{-1}$ are caused by the lack of power
around the $\ell\simeq 24, 95, 181, 209$ and 540 multipoles. Adding
the WMAP-9 polarisation data slightly increases the infrared
suppression of the reconstructed PPS, boosts the bump at $k\simeq
0.0086\;\mathrm{Mpc}^{-1}$ and reduces the error bands on large
scales. Including the small-scale CMB data improves the reconstruction
for $k\gtrsim 0.05\;\mathrm{Mpc}^{-1}$. The small scale observations
at $\ell\simeq 750$ create an additional bump at
$k\simeq0.053\;\mathrm{Mpc}^{-1}$. Adding the SDSS-4 LRG data further
sharpens the estimated PPS on small scales and introduces further dips
at $k\simeq0.084\;\mathrm{Mpc}^{-1}$ and
$k\simeq0.12\;\mathrm{Mpc}^{-1}$.  For $\lambda= 5000$ however much of
the structure in the $\lambda=100$ reconstruction is smoothed away,
particularly on large scales where the infrared cutoff disappears ---
this is due to our prior of a H-Z spectrum combined with the increased
cosmic variance on large scales. Prominent features remaining in the
PPS estimated from the WMAP-9, small-scale CMB and SDSS-4 LRG data
include a peak at $k\simeq0.055\;\mathrm{Mpc}^{-1}$ and dips at
$k\simeq0.014\;\mathrm{Mpc}^{-1}$ and
$k\simeq0.038\;\mathrm{Mpc}^{-1}$.
 
\begin{figure*}
\includegraphics*[angle=0,width=0.5\columnwidth,trim = 32mm 171mm 23mm
  15mm, clip]{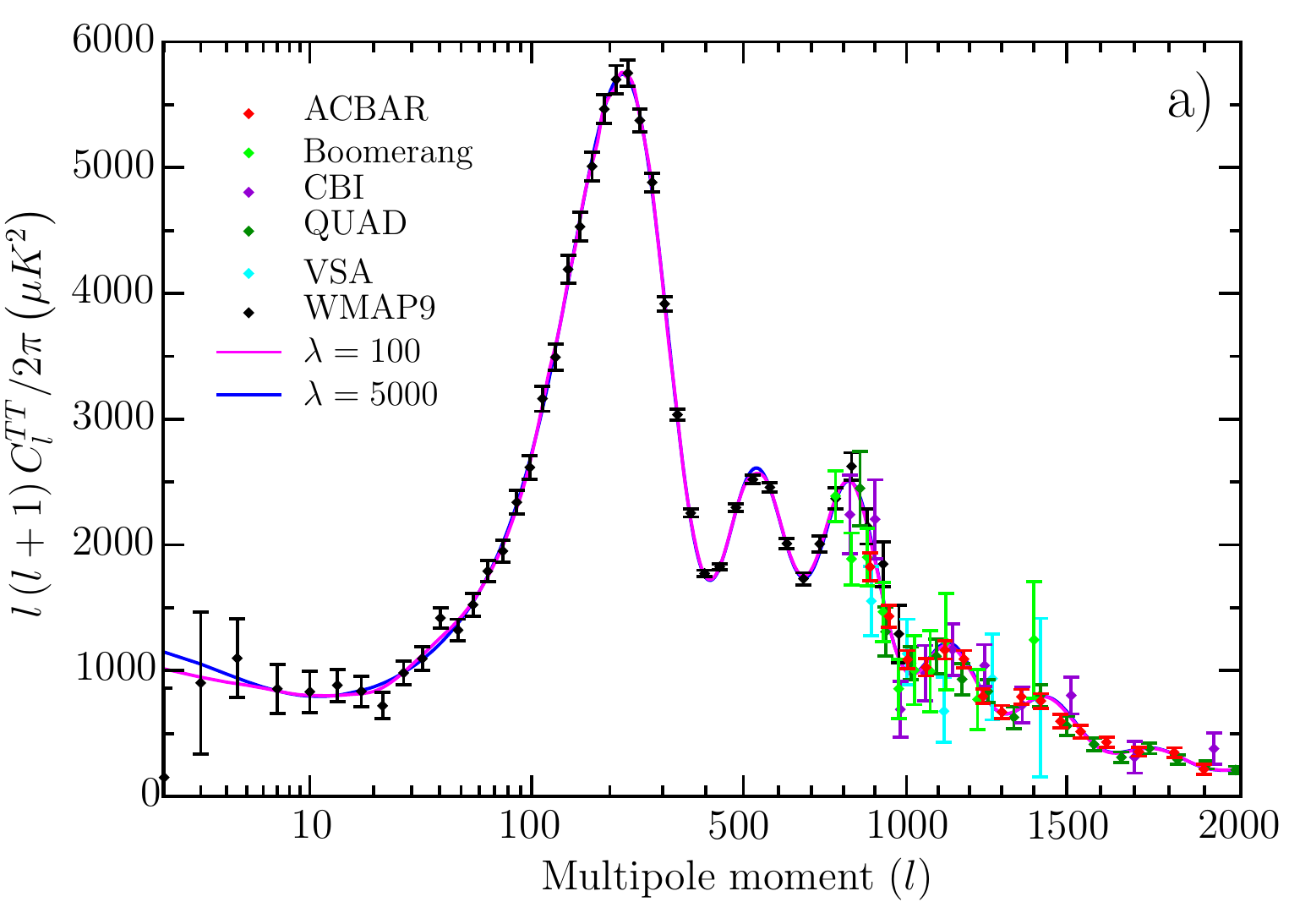}
\includegraphics*[angle=0,width=0.5\columnwidth,trim = 32mm 171mm 23mm
  15mm, clip]{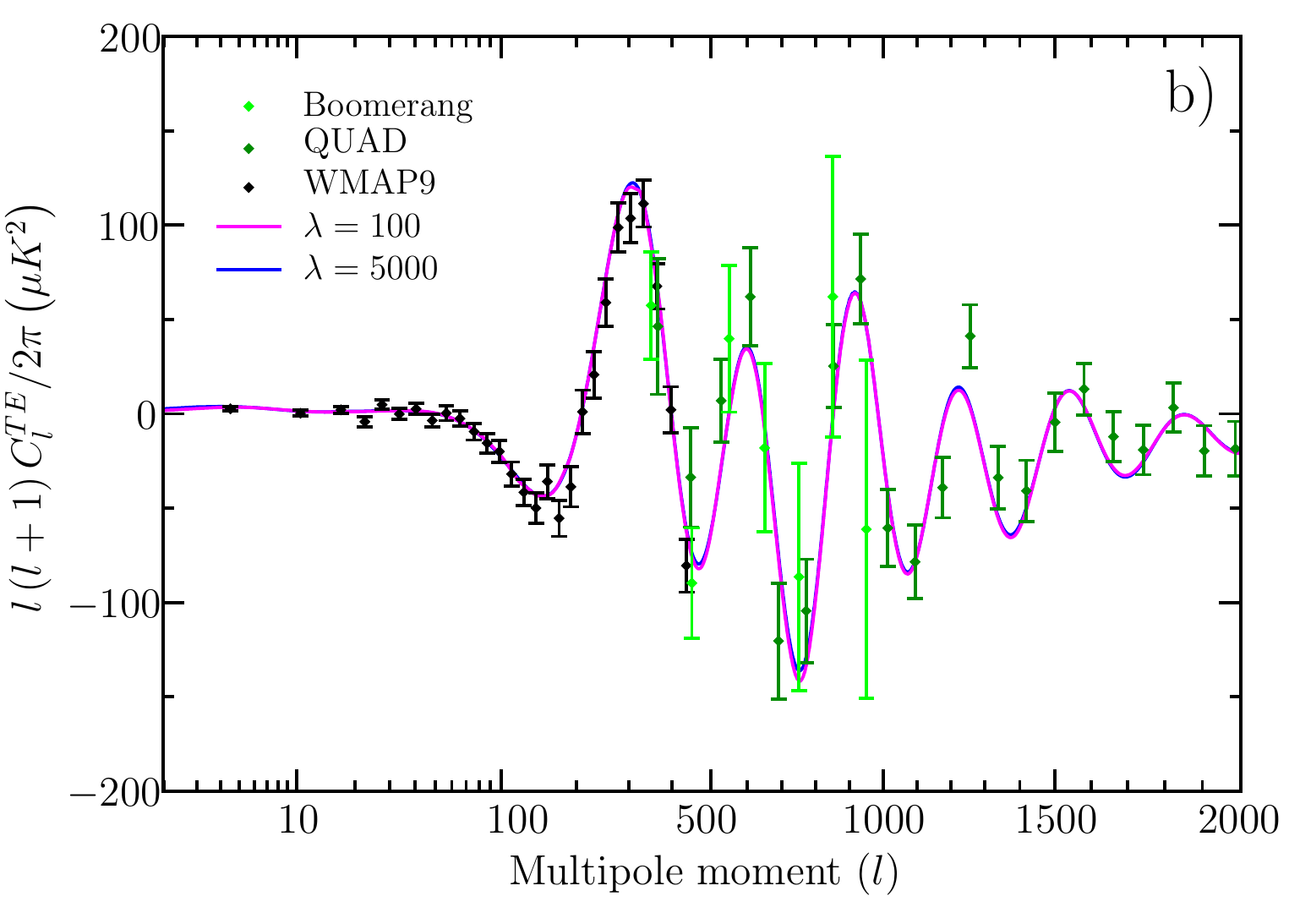}
\includegraphics*[angle=0,width=0.5\columnwidth,trim = 32mm 171mm 23mm
  15mm, clip]{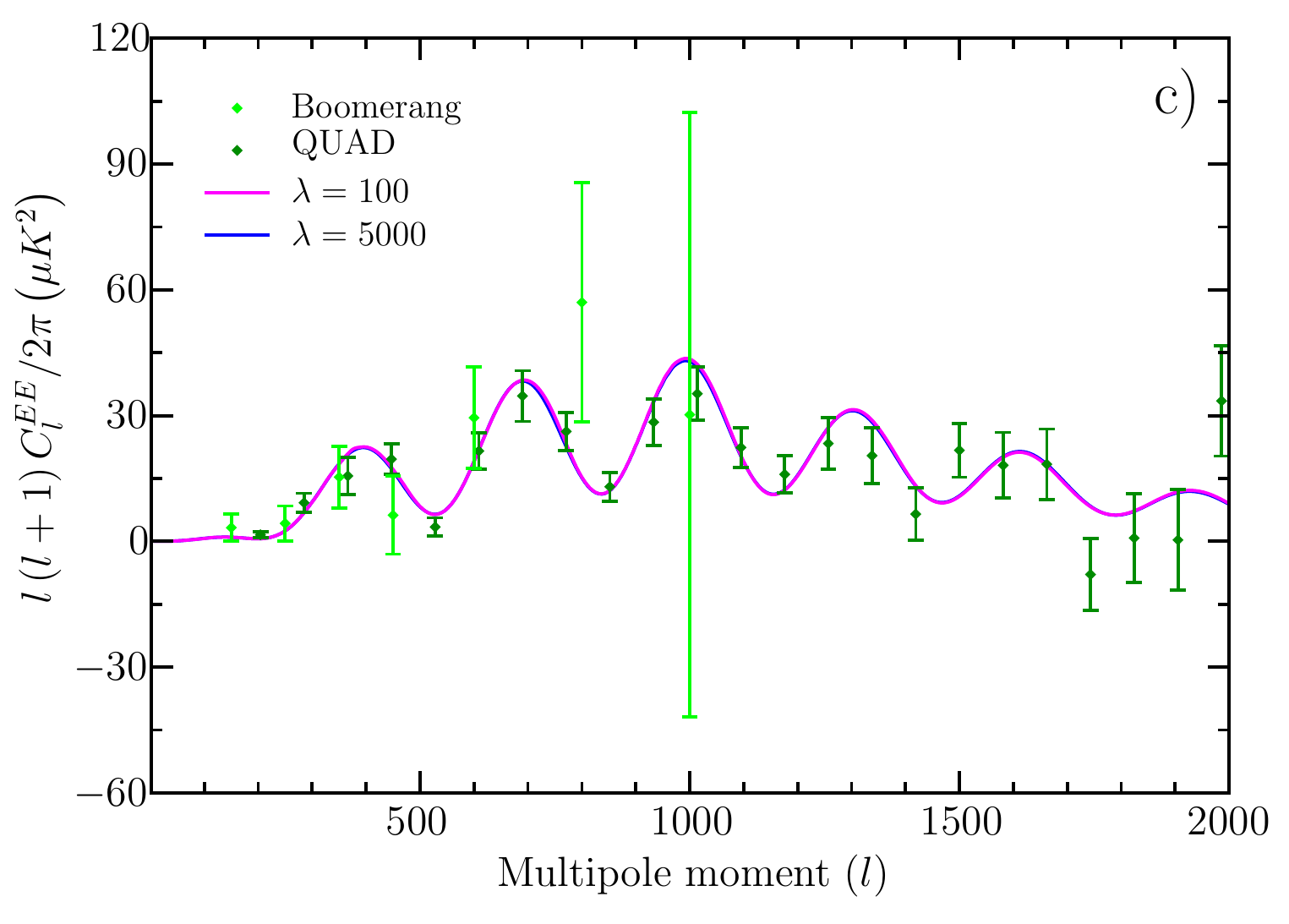}
\includegraphics*[angle=0,width=0.5\columnwidth,trim = 32mm 171mm 23mm
  15mm, clip]{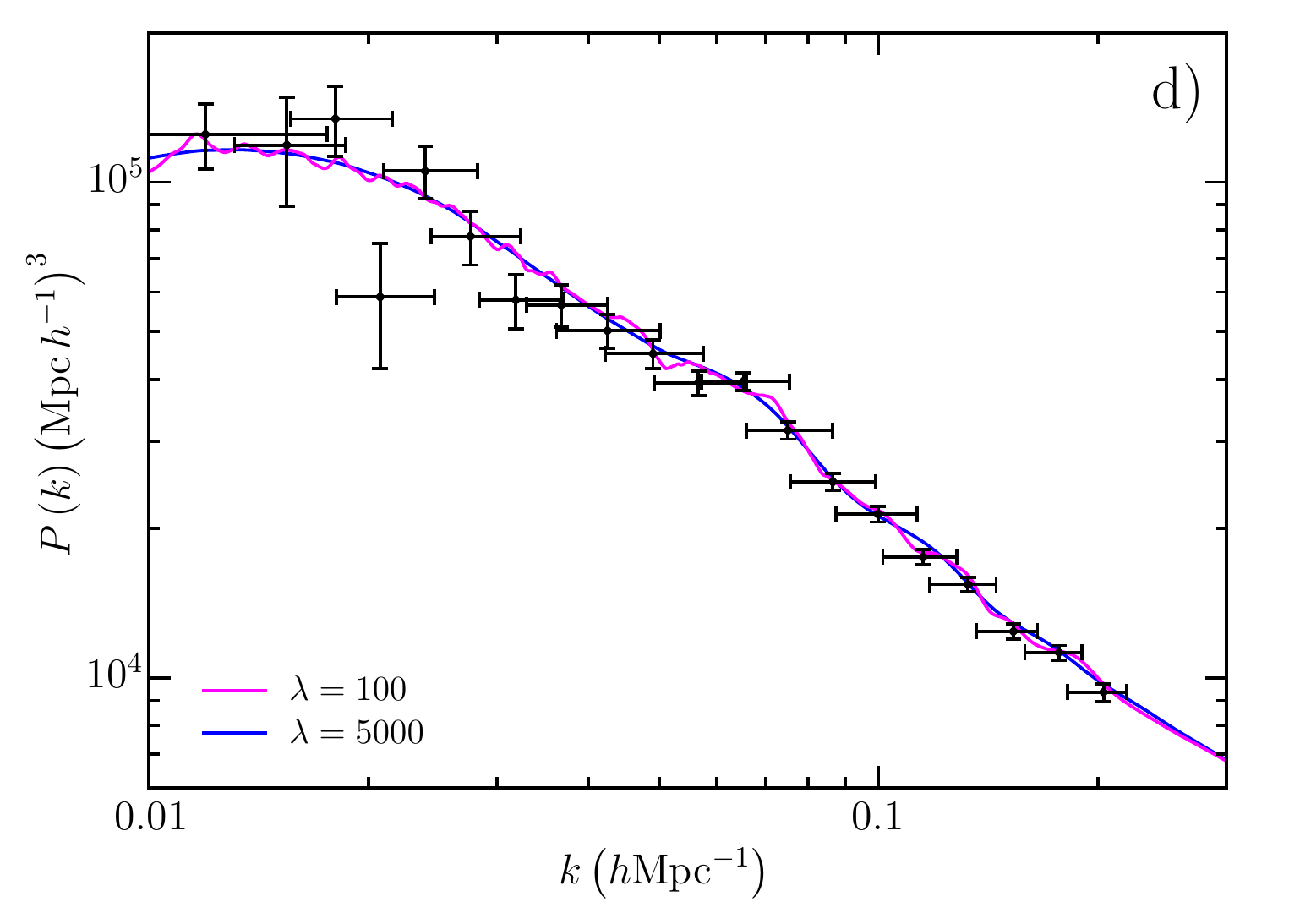}
\caption{\label{datfit} The fit to the data of the a) TT power
  spectrum b) TE power spectrum c) EE power spectrum d) LRG power
  spectrum corresponding to the PPS recovered from the WMAP-9, small
  scale CMB and SDSS-4 LRG data. In each case results are shown for
  both $\lambda=100$ and $\lambda=5000$.}
\end{figure*}

The features in the recovered spectra can be understood with reference
to Fig.\,\ref{datfit}, which shows a comparison to the data of the
angular power spectra and the galaxy power spectrum derived from the
estimated PPS. Since our method maximises the likelihood for a given
roughness of the PPS, the fit to the data is correctly weighted by the
inverse covariance matrices. The recovered spectra all provide good
fits to the observations. The $\lambda=100$ spectrum found from the
WMAP-9, small-scale CMB and SDSS-4 LRG data (1582 data points) has
$\chi^2=1629$ while the $\lambda=5000$ spectrum has $\chi^2=1650$.
These should be compared to $\chi^2=1664$ for the best-fit power-law
model with $n_\mathrm{s}=0.969$.

\subsection{Uncorrelated bandpowers \label{uncorr}}

\begin{figure*}
\begin{minipage}{150mm}
\includegraphics*[angle=0,width=0.5\columnwidth,trim = 32mm 171mm 23mm
  15mm, clip]{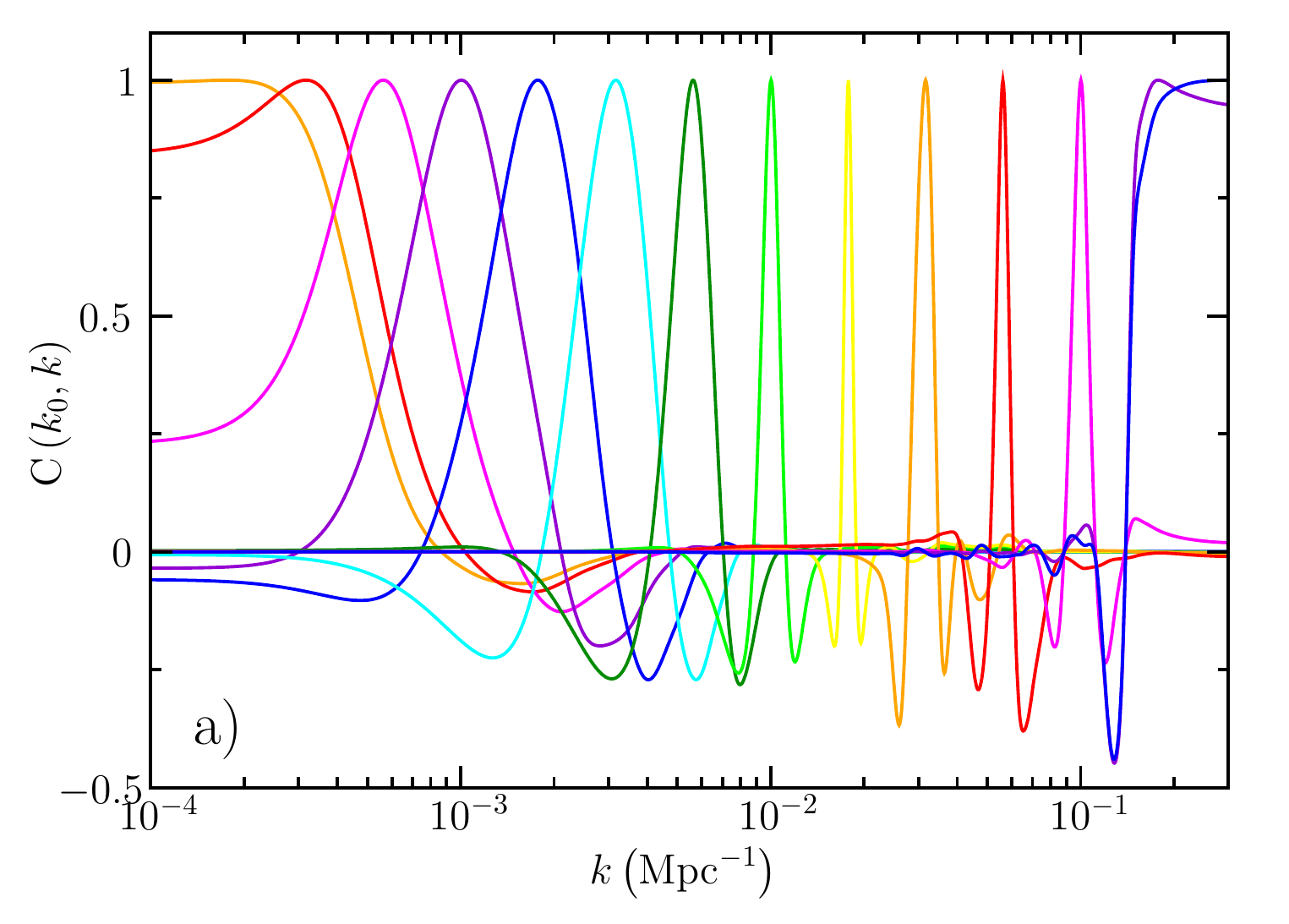}
\includegraphics*[angle=0,width=0.5\columnwidth,trim = 32mm 171mm 23mm
  15mm, clip]{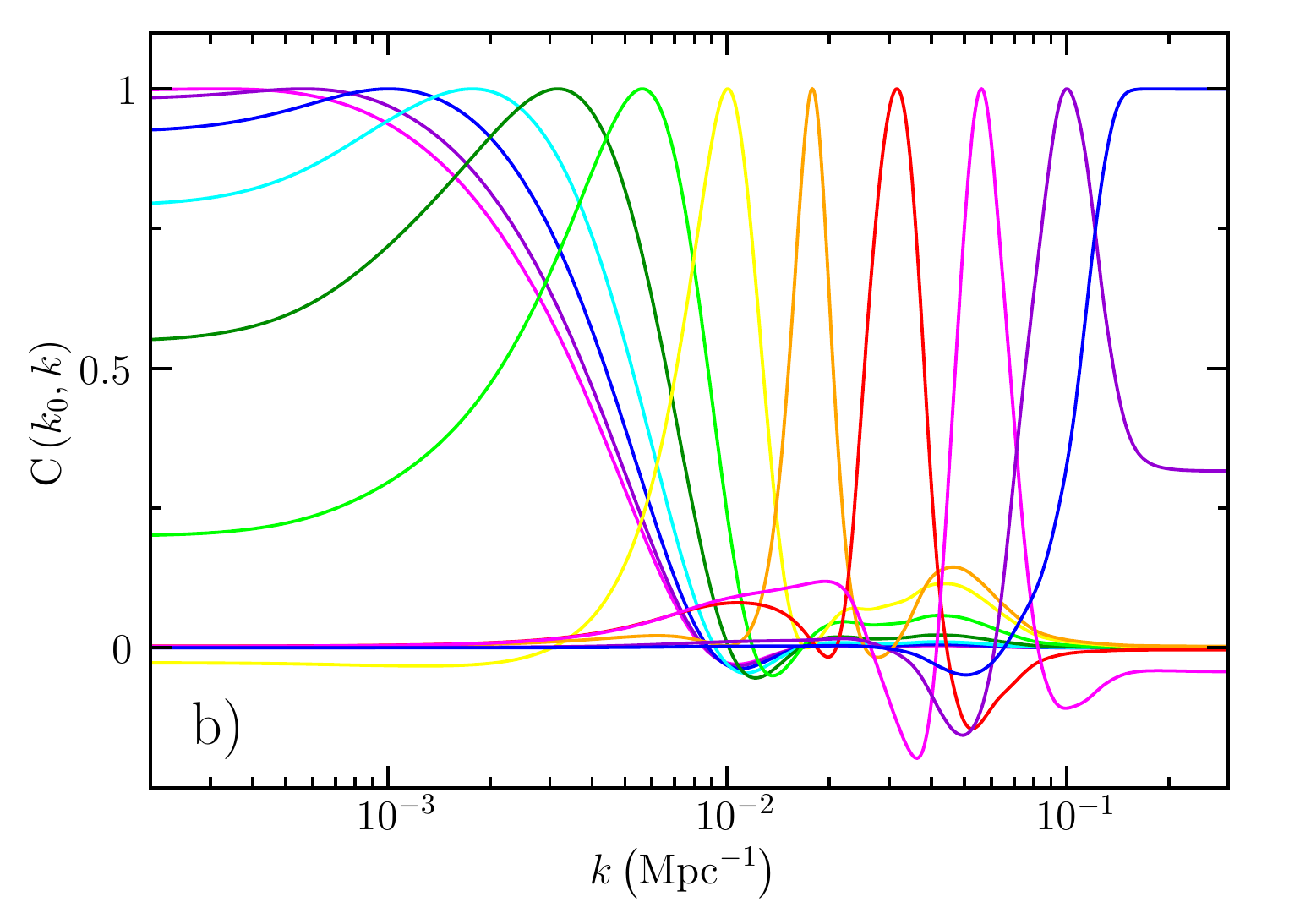}
\caption{\label{corr} The correlation function (eq.\,\ref{equncorr}) of
  the frequentist covariance error matrix $\mathsf{\Sigma}_\mathrm{F}$
  (eq.\,\ref{sigmaf}) for (a) $\lambda=100$ and (b) $\lambda=5000$, for
  the reconstructions using the WMAP-9, small-scale CMB and SDSS-4 LRG
  data.  The lines correspond to logarithmically equally spaced values
  from $k_0=2\times 10^{-4}\;\mathrm{Mpc}^{-1}$ to
  $k_0=0.3\;\mathrm{Mpc}^{-1}$ in panel (a) and from $k_0=3\times
  10^{-4}\;\mathrm{Mpc}^{-1}$ to $k_0=0.2\;\mathrm{Mpc}^{-1}$ in panel
  (b). Note that $\mathrm{C}\left(k_0;k_0\right)=1$ and that
  $\mathrm{C}\left(k_0;k\right)=\mathrm{C}\left(k;k_0\right)$. }
\end{minipage}
\end{figure*}

Since the reconstructed PPS is smooth by design, neighbouring elements
of it are highly \emph{correlated}, particularly on large and small scales.
This can be seen using the correlation function calculated using the
elements of the frequentist covariance matrix (eq.\,\ref{sigmaf}):
\begin{equation}
\label{equncorr}
\mathrm{C}\left(k_0;k\right) \equiv \frac{\sum_{i,j} 
 \Sigma_{\mathrm{F}_{|ij}}\phi_i\left(k_0\right)\phi_j\left(k\right)}
 {\left[\sum_{i,j} \Sigma_{\mathrm{F}_{|ij}}\phi_i\left(k_0\right)\phi_j
 \left(k_0\right)\right]^{1/2}\left[\sum_{i,j} 
 \Sigma_{\mathrm{F}_{|ij}}\phi_i\left(k\right)\phi_j\left(k\right)\right]^{1/2}},
\end{equation}
which is displayed in Fig.\,\ref{corr}. The correlated errors hinder
the interpretation of the significance of any features in the
reconstructed PPS. To display the true information content of the
recovered PPS we follow \cite{Nagata:2008tk,Nagata:2008zj} and
construct statistically \emph{independent} band powers.

Correlated bandpowers $\qB=\mathsf{T}\hat{\pB}$ are defined as the
mean of the PPS over separate neighbouring wavenumber ranges. Thus the
elements of each row of $\mathsf{T}$ are identical when not equal to
zero and sum to unity. The columns only have one nonzero entry. The
frequentist covariance matrix of the bandpowers is
$\mathsf{\Sigma}_\mathrm{N} =
\mathsf{T}\mathsf{\Sigma}_\mathrm{F}\mathsf{T}^\mathrm{T}$. For
$\lambda=100$, the various estimates of the number of effective
parameters (as defined in Appendix~\ref{regularisation}) are
$\nu_1=33.5$, $\nu_2=38.6$ and $\nu_3=28.4$, while for $\lambda=5000$
they are $\nu_1=10.5$, $\nu_2=14.5$ and $\nu_3=6.6$. Hence we choose
35 bandpowers for $\lambda=100$ and 10 bandpowers for $\lambda=5000$.

\begin{figure*}
\begin{minipage}{150mm}
\includegraphics*[angle=0,width=0.5\columnwidth,trim = 32mm 171mm 23mm
  15mm, clip]{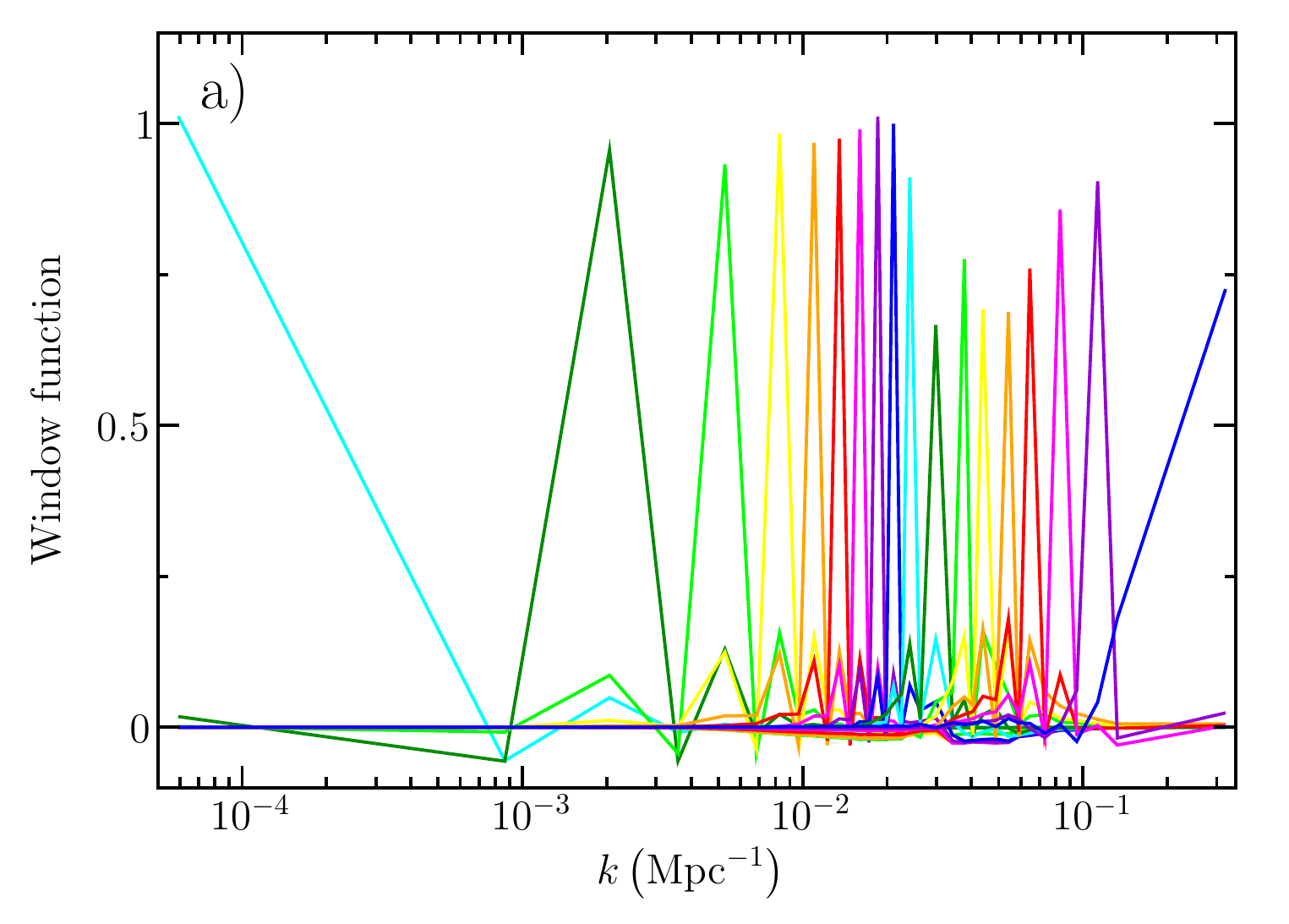}
\includegraphics*[angle=0,width=0.5\columnwidth,trim = 32mm 171mm 23mm
  15mm, clip]{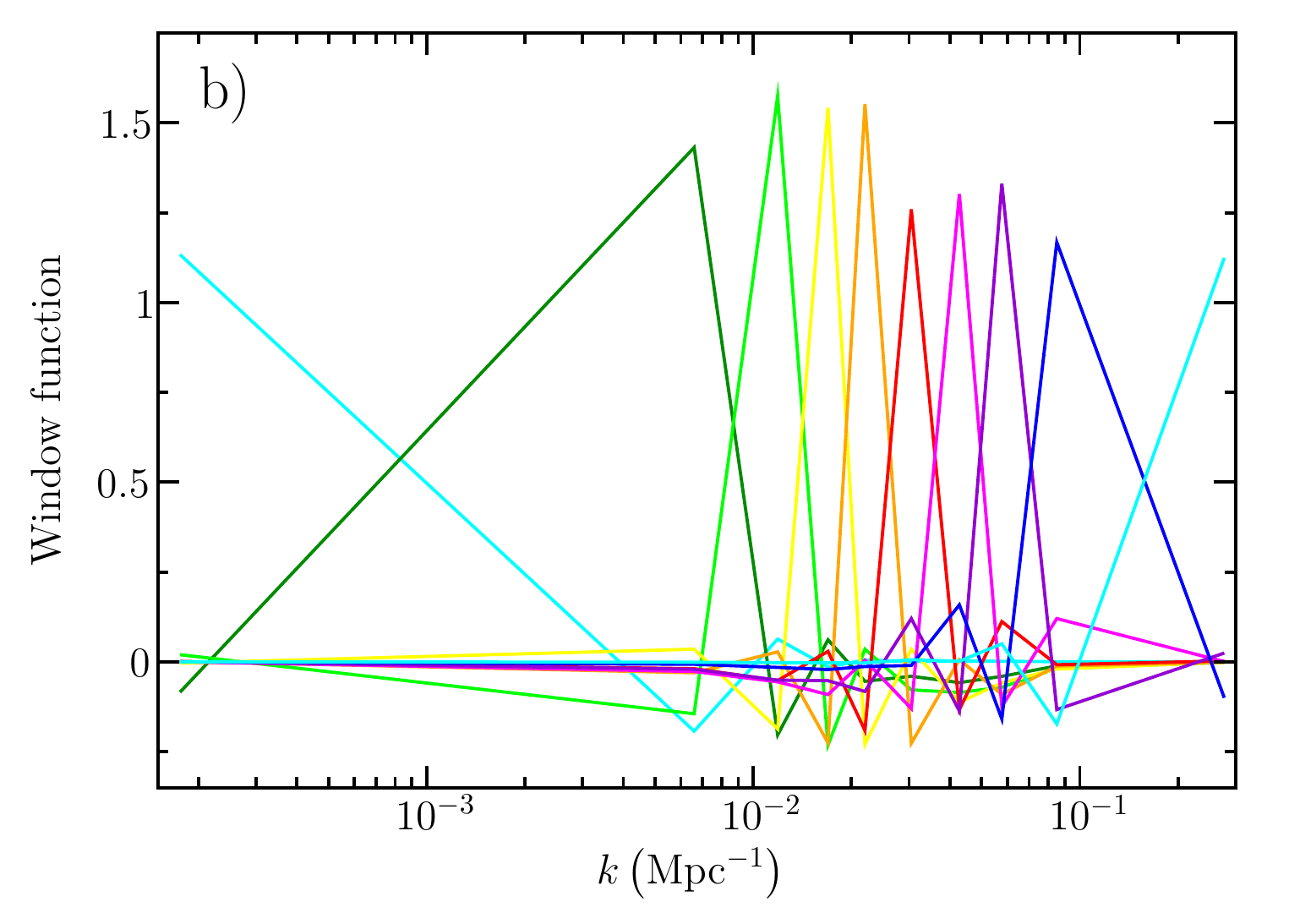}
\caption{\label{win} The bandpower window functions for (a) $\lambda=100$
 and (b) $\lambda=5000$ (right panel) of the reconstructions
  using the WMAP-9, small-scale CMB and SDSS-4 LRG data. Only even
  number window functions are displayed in panel (a) for clarity.}
\end{minipage}
\end{figure*}

\begin{figure*}
\begin{minipage}{150mm}
\includegraphics*[angle=0,width=0.5\columnwidth,trim = 32mm 171mm 23mm
  15mm, clip]{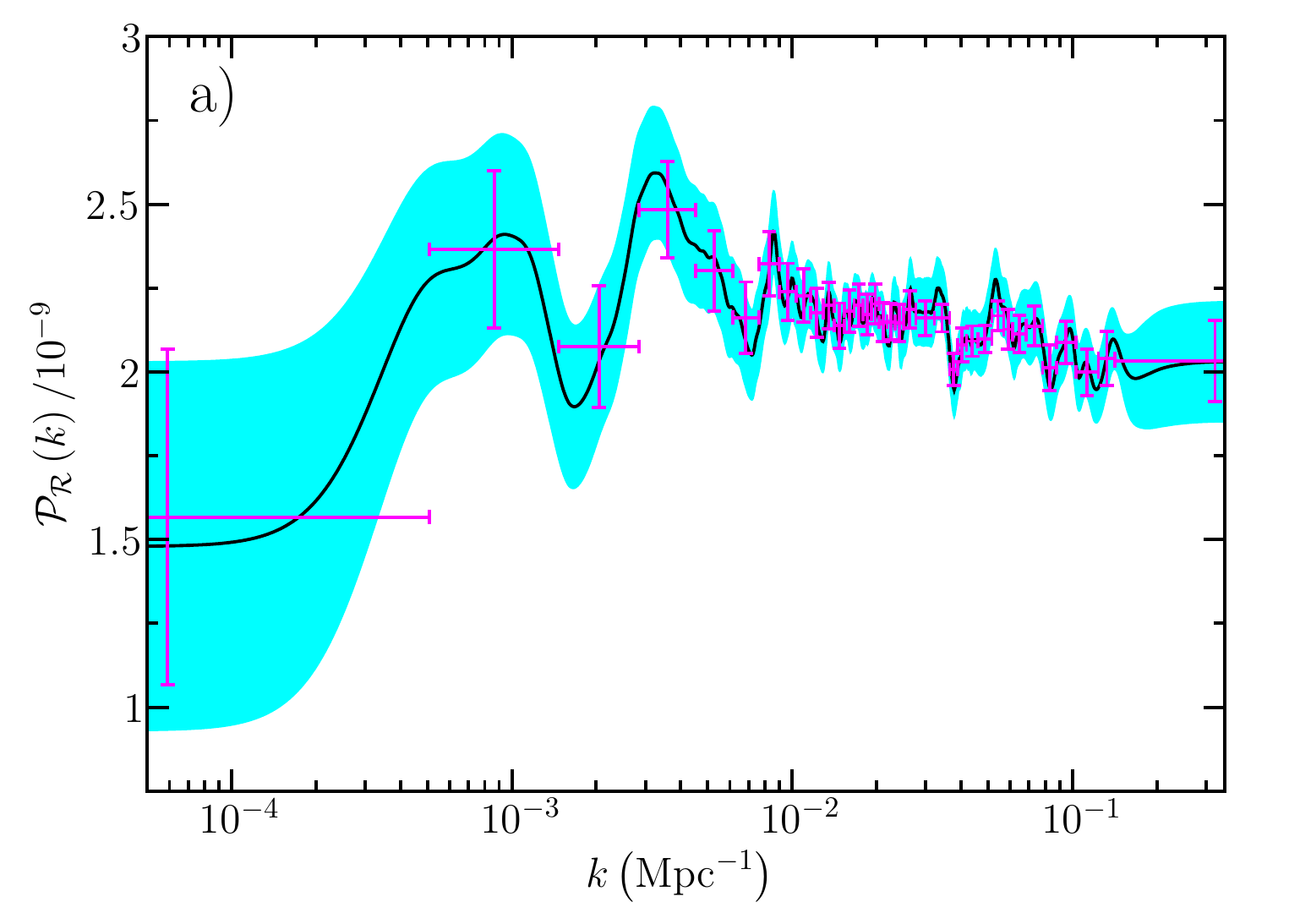}
\includegraphics*[angle=0,width=0.5\columnwidth,trim = 32mm 171mm 23mm
  15mm, clip]{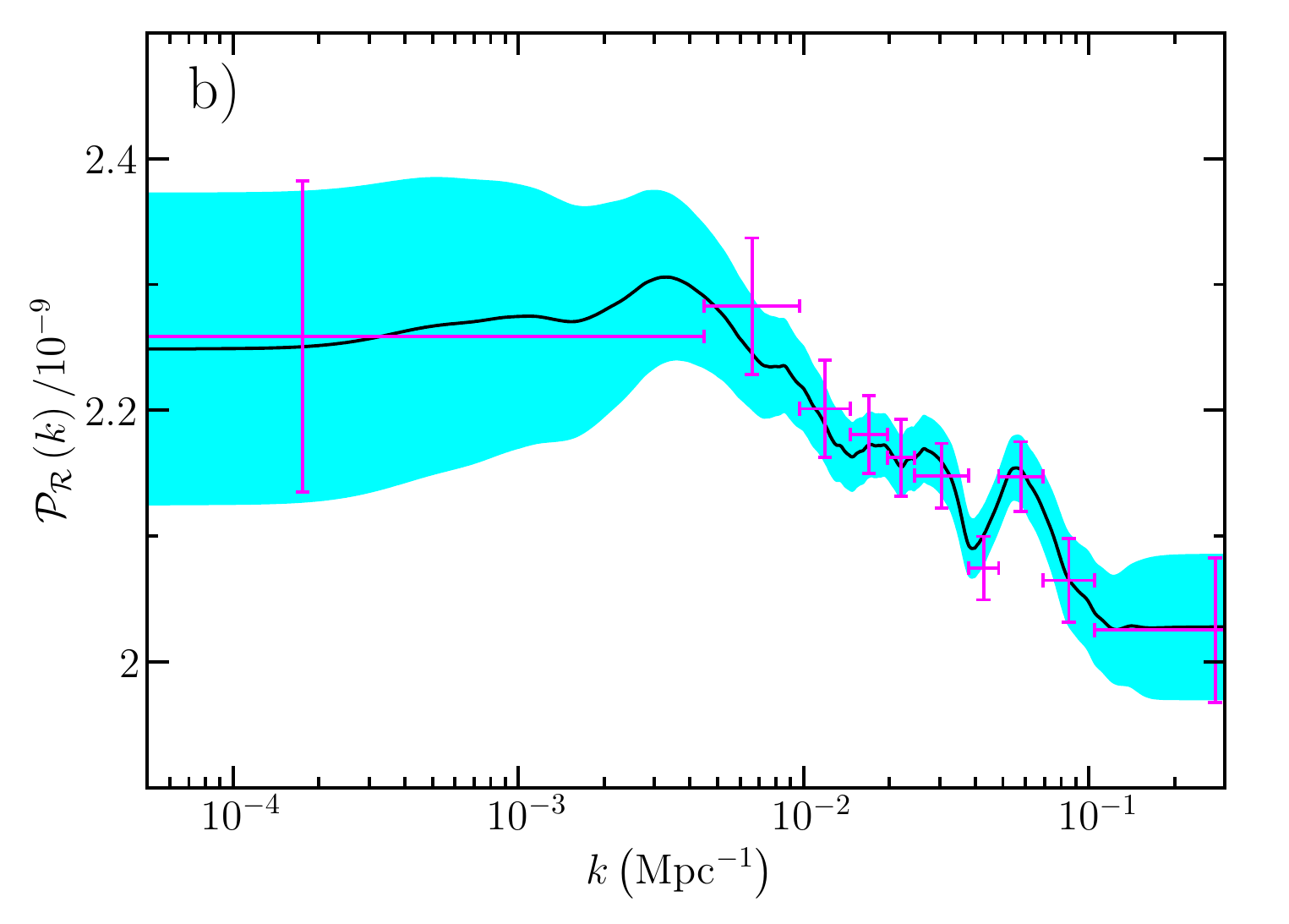}
\caption{\label{decorr} The decorrelated bandpowers for (a)
  $\lambda=100$ and (b) $\lambda=5000$. The black line is the PPS
  recovered from the WMAP-9, small-scale CMB and SDSS-4 LRG data. The
  light band is the $1\sigma$ error obtained from the square root of
  the diagonal elements of the frequentist covariance matrix
  $\mathsf{\Sigma}_\mathrm{F}$ (eq.\,\ref{sigmaf}). The vertical error
  bars are the $1\sigma$ errors $\mathsf{D}_{II}$
  Sec.\,\ref{uncorr}). The horizontal error bars indicate the wavenumber
  range to which the bandpowers are most sensitive.}
\end{minipage}
\end{figure*}

Any transformation of the form $\tilde{\qB}=\mathsf{G}\qB$ where
$\mathsf{G}\equiv\mathsf{D}\mathsf{O}\tilde{\mathsf{G}}$ and
$\mathsf{D}$ is diagonal, $\mathsf{O}$ is orthogonal and
$\tilde{\mathsf{G}}$ satisfies
$\mathsf{\Sigma}_\mathrm{N}^{-1}=\tilde{\mathsf{G}}^\mathrm{T}\tilde{\mathsf{G}}$
will produce uncorrelated bandpowers $\tilde{\qB}$ with the covariance
matrix $\mathsf{D}^2$. As recommended in \cite{Hamilton:1999uw} we
obtain $\tilde{\mathsf{G}}$ by first performing the diagonalisation
$\mathsf{\Sigma}_\mathrm{N}=\mathsf{E}^\mathrm{T}\mathsf{\Pi}\mathsf{E}$,
where $\mathsf{E}$ is the eigenvector matrix and $\mathsf{\Pi}$ is the
diagonal matrix of eigenvalues, and then setting
$\tilde{\mathsf{G}}=\mathsf{E}^\mathrm{T}\mathsf{\Pi}^{-1/2}\mathsf{E}$. After
setting $\mathsf{O}=\mathsf{I}$ and $D_{II}^{-1}=\sum_J\tilde{G}_{IJ}$
the rows of $\mathsf{G}$ form bandpower window functions which are
normalised to unity, $\sum_J G_{IJ}=1$. The bandpowers $\qB$ are
chosen by an iterative algorithm designed to ensure that the window
functions are as well-behaved and non-negative as
possible.\footnote{Note this is different from selecting the
  bandpowers according to the signal-to-noise ratio
  \cite{Paykari:2009ac}.} The window functions and the uncorrelated
bandpowers are shown in Figs.\,\ref{win} and \ref{decorr}
respectively.  The connection between the window functions and the
resolution kernels of Fig.\,\ref{kern1} is clearly apparent; the
window functions are sharper and more densely clustered where the
resolution is higher. Thus the window functions at
$k=0.03\;\mathrm{Mpc}^{-1}$ and $k=0.05\;\mathrm{Mpc}^{-1}$, which
correspond to the first and second troughs in the CMB TT power
spectrum where the resolution is poorer, are lower and broader than
their neighbours. This is particularly interesting given that no
explicit information about the resolution was included in the
algorithm that chose the bandpowers.\footnote{The importance of
  resolution was previously noted in \cite{Hamilton:1999uw} where it
  was reported that the window functions become jittery when the
  natural resolution of the galaxy surveys is exceeded.}

\subsection{Background parameter errors \label{bgpe}}

\begin{figure*}
\includegraphics*[angle=0,width=0.5\columnwidth,trim = 32mm 171mm 23mm
  15mm, clip]{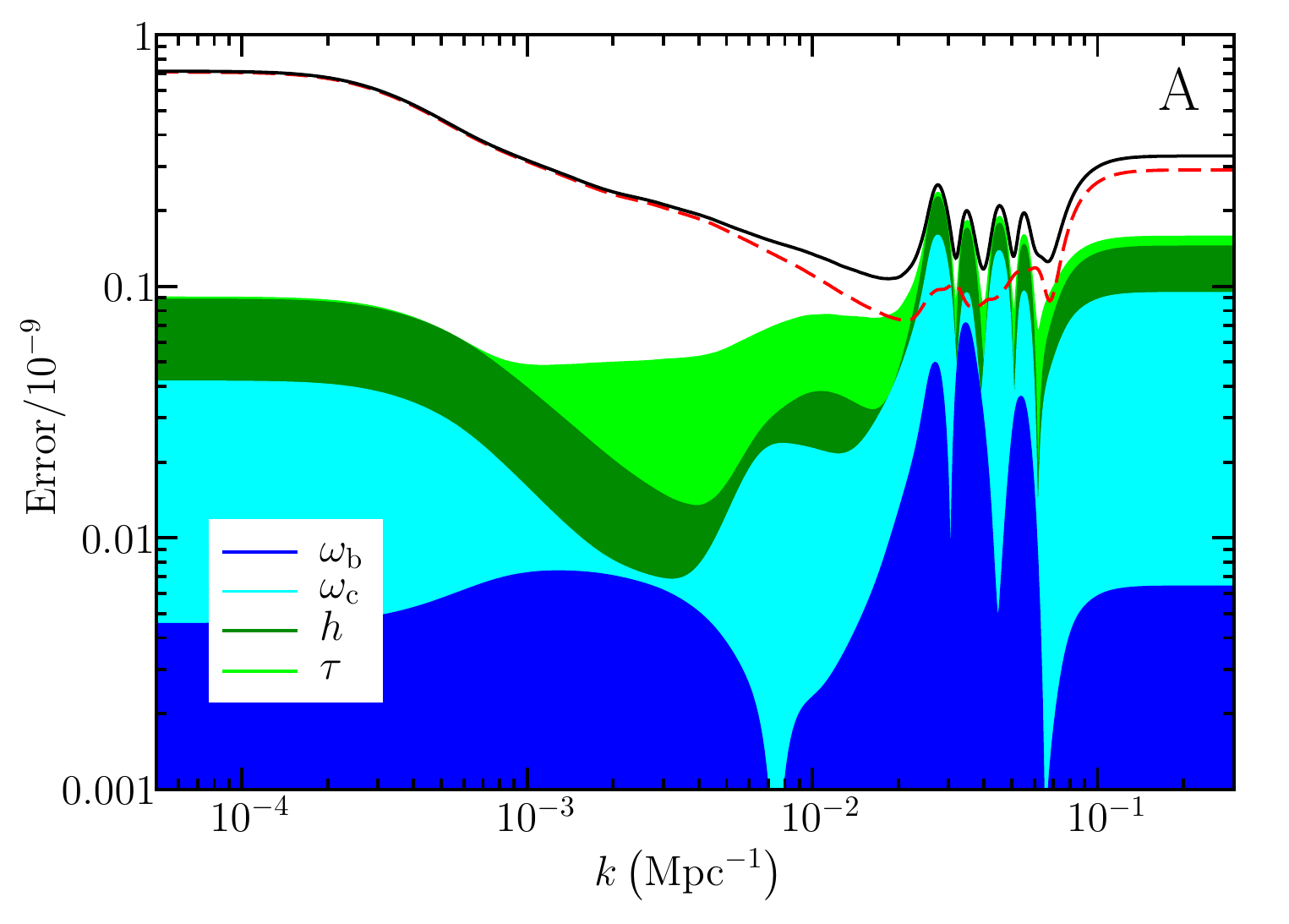}
\includegraphics*[angle=0,width=0.5\columnwidth,trim = 32mm 171mm 23mm
  15mm, clip]{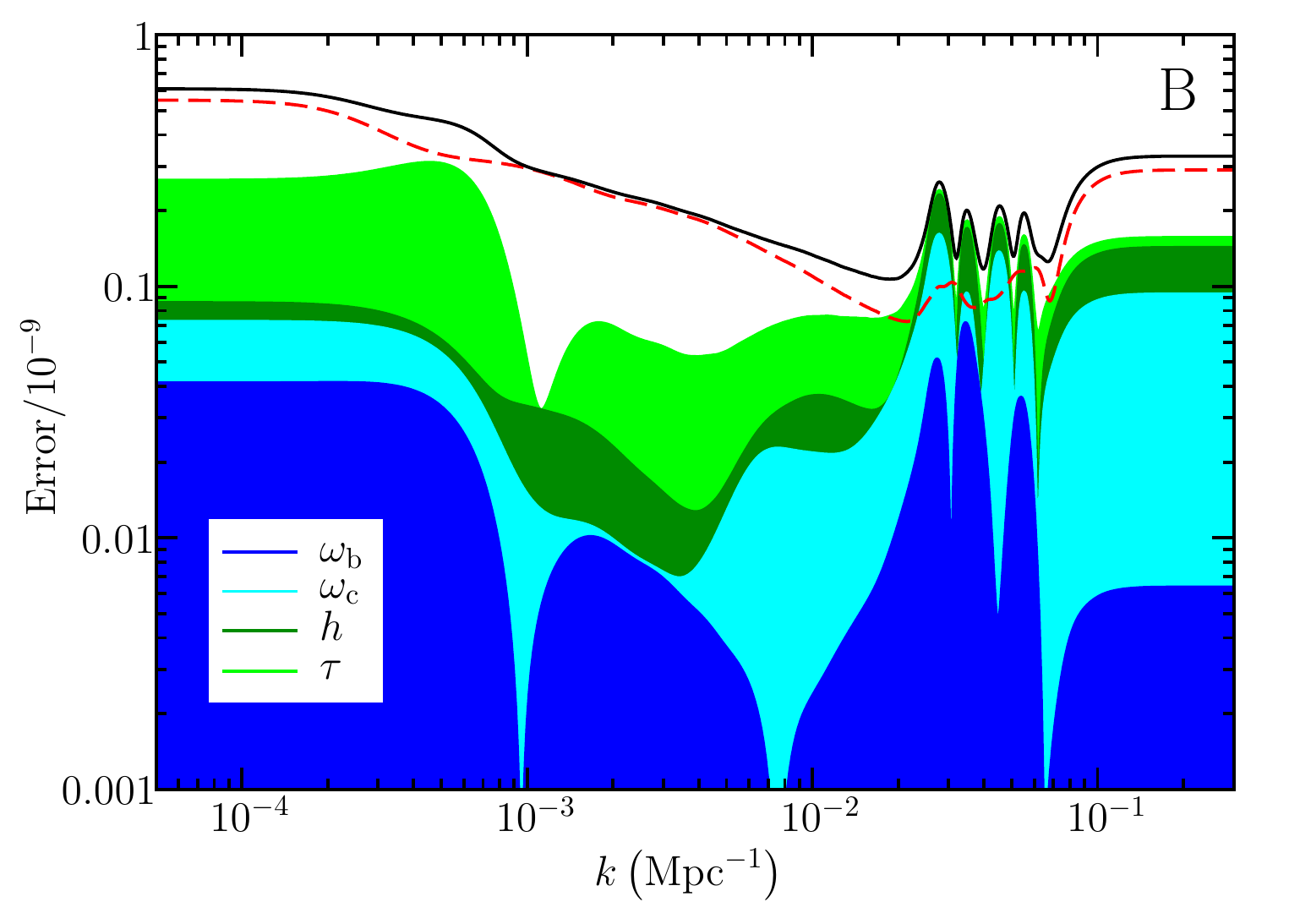}
\includegraphics*[angle=0,width=0.5\columnwidth,trim = 32mm 171mm 23mm
  15mm, clip]{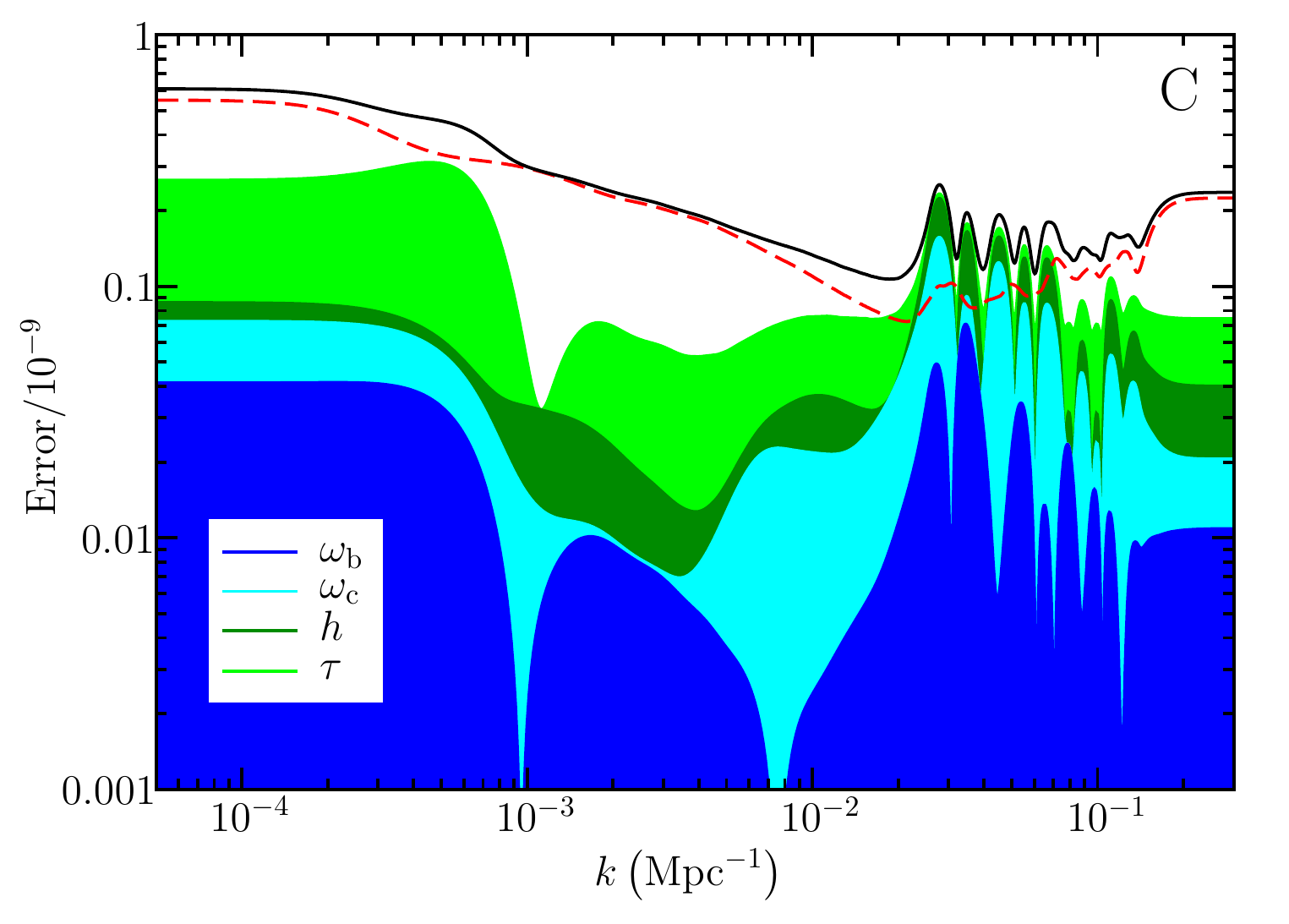}
\includegraphics*[angle=0,width=0.5\columnwidth,trim = 32mm 171mm 23mm
  15mm, clip]{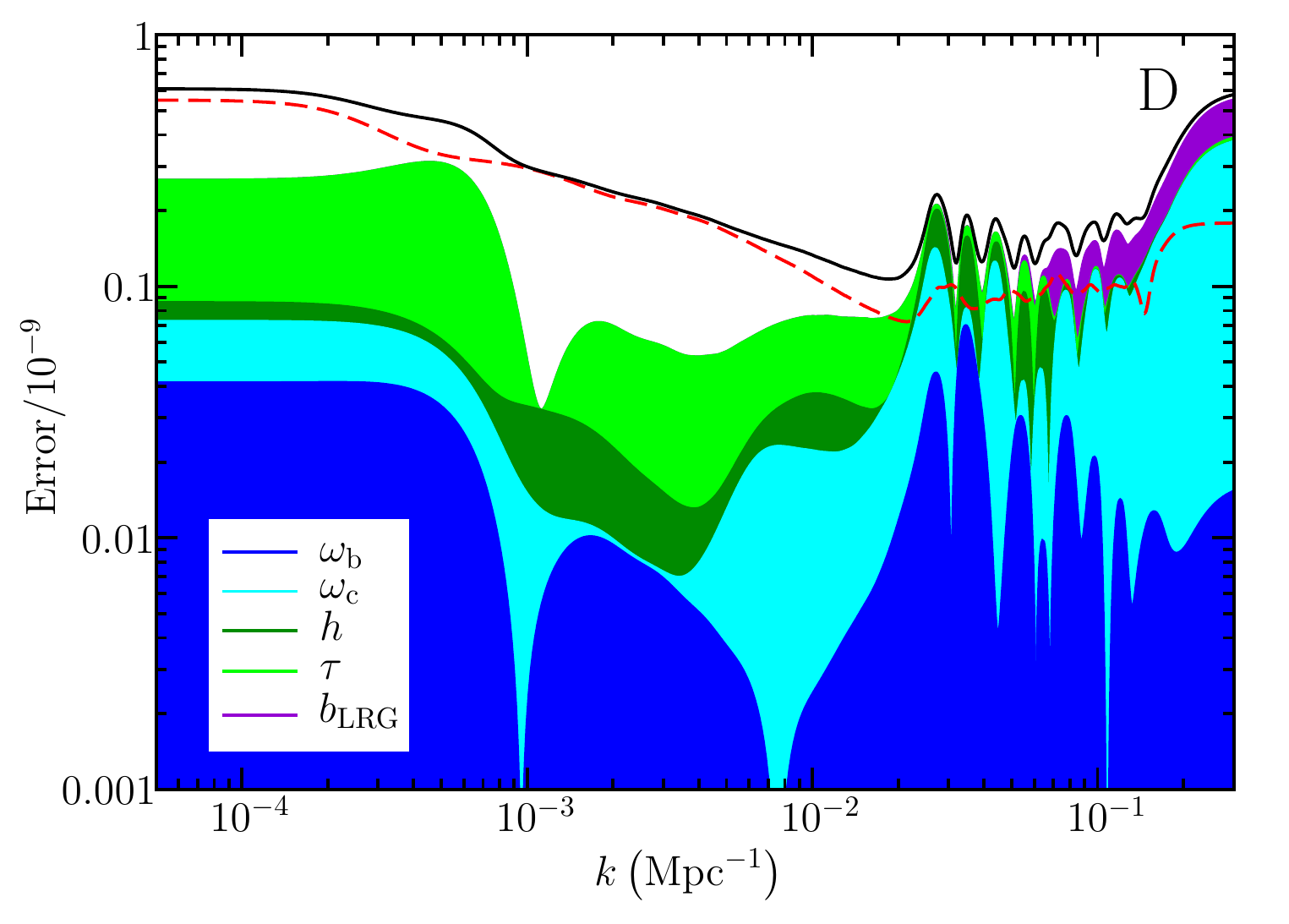}
\caption{\label{covf2a} The contributions of the different background
  parameters to the square root of the diagonal elements of the matrix
  $\mathsf{\Sigma}_\mathrm{P}$ (eq.\,\ref{sigmap}) for the 4 data
  combinations. The error contributions are added in quadrature and
  $\lambda=100$ throughout.  In each panel the dashed line is the
  square root of the diagonal elements of the matrix
  $\mathsf{\Sigma}_\mathrm{F}$ (eq.\,\ref{sigmaf}) and is included for
  comparison. The solid line is the square root of the diagonal
  elements of the matrix $\mathsf{\Sigma}_\mathrm{T}$
  (eq.\,\ref{sigmat}).}
\end{figure*}

As an illustration of how errors in the background parameters affect
the recovered PPS, we calculate the associated covariance matrix
$\mathsf{\Sigma}_\mathrm{P}$ (eq.\,\ref{sigmap}) using the error matrix
\begin{equation}
\label{bkgderror}
\mathsf{U} = \mathrm{diag}\left(\left(0.025\,\omega_\mathrm{b}\right)^2,
\left(0.050\,\omega_\mathrm{c}\right)^2,\left(0.035\,h\right)^2,
\left(0.17\,\tau\right)^2,\left(0.037\,b_\mathrm{LRG}\right)^2\right),
\end{equation}
which corresponds to the uncertainties in the parameters from the WMAP
and SDSS-4 LRG data.\footnote{These uncertainties were obtained under
  the assumption of a power-law PPS.  In principle
  \textit{independent} data sets ought to be used to determine the
  background parameters and the PPS separately, but we are presenting
  an illustrative example here rather than a full analysis.}
Fig.\,\ref{covf2a} shows how these uncertainties contribute to the
diagonal elements of $\mathsf{\Sigma}_\mathrm{P}$ for
$\lambda=100$. The results were obtained by error propagation using
the derivatives $\partial\mathsf{W}_\mathbb{Z}/\partial\bm{\theta}$
evaluated numerically by a modified version of CAMB
\cite{camb}.\footnote{Note that when higher accuracy is required it
  would be better to rely on Monte Carlo simulations rather than
  attempt to calculate higher order derivatives.} Varying the
background parameters causes the reconstructed PPS to alter in such a
way that the predicted data remain almost constant. Hence $\hat{\pB}$
changes in the opposite direction to the change in the predicted data
when the PPS is held fixed. A parameter that, for instance, increases
the height of the integral kernels
$\mathcal{K}_a^{(\mathbb{Z})}\left(\bm{\theta},k\right)$ will reduce
the amplitude of the estimated PPS.

For the reconstruction from the WMAP TT data with $\lambda=100$ the
error is largest on the scales corresponding to the CMB acoustic peaks
where the PPS is most sensitive to the background
parameters. Increasing $\omega_\mathrm{b}$ raises the first acoustic
peak and lowers the second due to greater baryon loading, while
increasing $\omega_\mathrm{c}$ reduces the early ISW effect and lowers
both peaks. The angular diameter distance of the last scattering
surface falls with increasing $h$, which shifts the peaks laterally
towards larger scales. Incorrect values of these three parameters
produce different oscillatory patterns in the recovered PPS.  This
leads to the series of peaks seen in the figures. Most of the error on
large scales originates from $\omega_\mathrm{c}$ and $h$. In a flat
universe increasing $h$ or decreasing $\omega_\mathrm{c}$ with the
other parameters held constant is equivalent to raising
$\Omega_\Lambda$. This enhances the late ISW effect, boosting the
predicted TT angular power spectrum for low multipoles and lowering
$\hat{\pB}$ on large scales.  Increasing $\tau$, the optical depth to
last scattering, suppresses the TT power spectrum for $\ell\gtrsim10$
and raises the reionisation bump at $\ell\simeq4$ in the TE and EE
spectra. Hence increasing $\tau$ raises the reconstructed PPS for
$k\gtrsim 10^{-3}\;\mathrm{Mpc}^{-1}$ but also lowers it on large
scales when the WMAP EE and TE polarisation data is
added. Consequently the contribution to the error on large scales from
the uncertainty in $\tau$ is negligible for the WMAP TT data alone but
greatly increased by the inclusion of the TE data. Adding the
small-scale CMB data introduces features into the error associated
with the higher acoustic peaks. When the SDSS-4 LRG data is used in
the deconvolution, increasing $\omega_\mathrm{c}$ and $b_\mathrm{LRG}$
suppresses $\hat{\pB}$ on small scales.  This is because increasing
$\omega_\mathrm{c}$ delays the epoch of matter-radiation equality,
which moves the matter power spectrum turnover towards smaller scales
and raises the predicted SDSS-4 LRG data. Increasing $b_\mathrm{LRG}$
boosts the overall normalisation of the galaxy power spectrum.  Thus
the $\omega_\mathrm{c}$ and $b_\mathrm{LRG}$ contributions dominate
the error on small scales. These uncertainties are \emph{not} included
in the usual parameter estimates obtained assuming a power-law PPS
(e.g. \cite{Dunkley:2008ie}).

\begin{figure*}
\includegraphics*[angle=0,width=0.5\columnwidth,trim = 32mm 171mm 23mm
  15mm, clip]{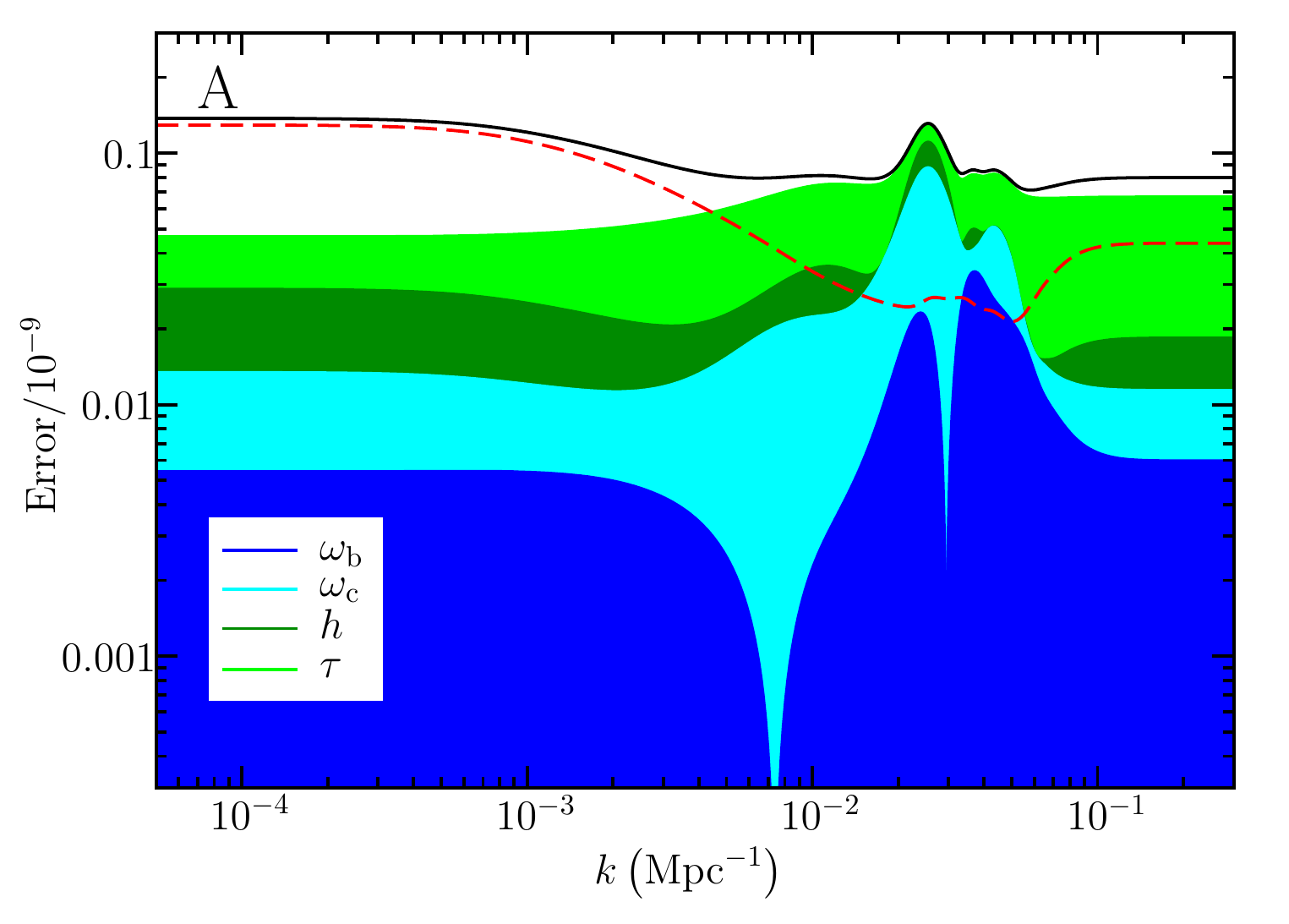}
\includegraphics*[angle=0,width=0.5\columnwidth,trim = 32mm 171mm 23mm
  15mm, clip]{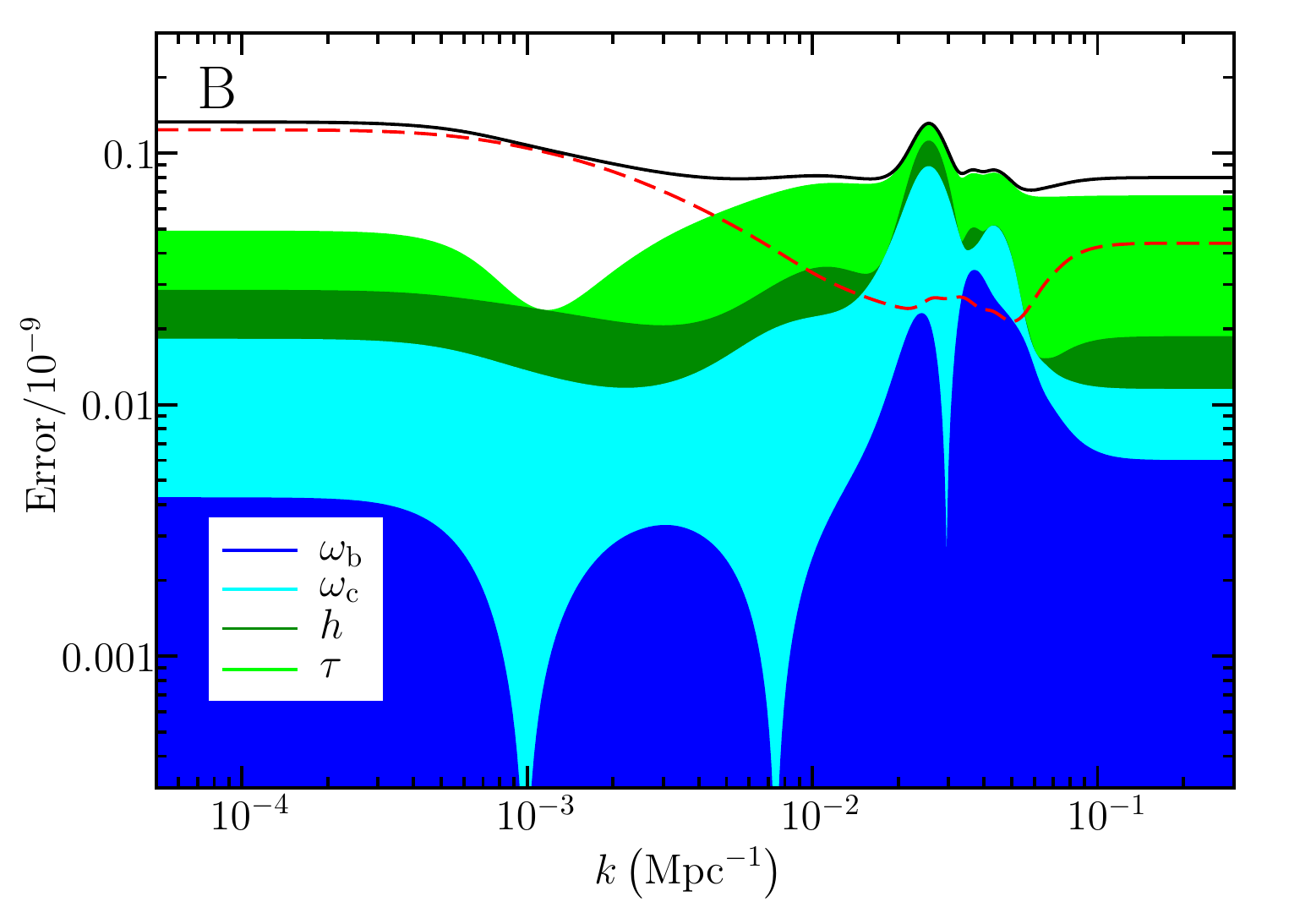}
\includegraphics*[angle=0,width=0.5\columnwidth,trim = 32mm 171mm 23mm
  15mm, clip]{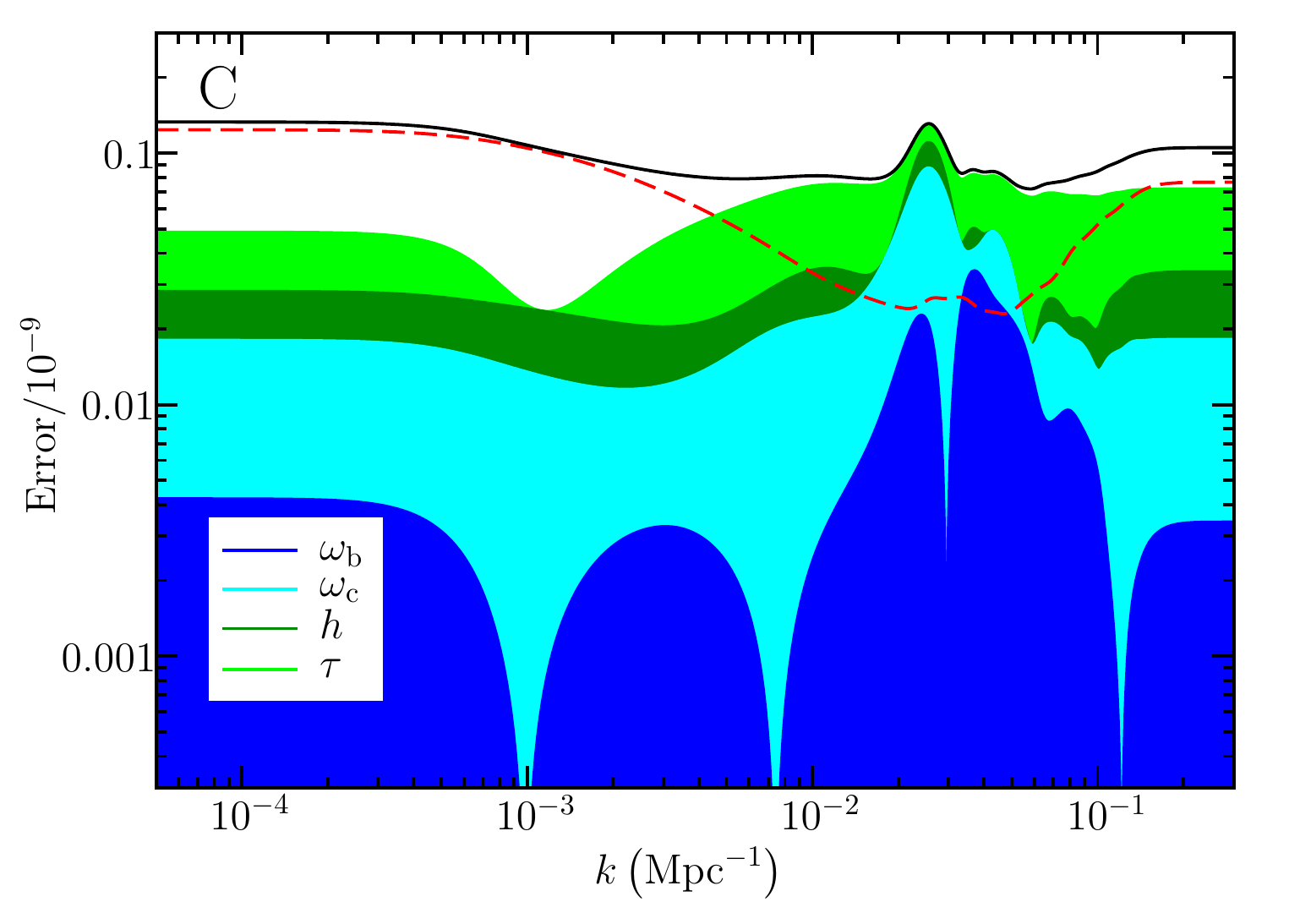}
\includegraphics*[angle=0,width=0.5\columnwidth,trim = 32mm 171mm 23mm
  15mm, clip]{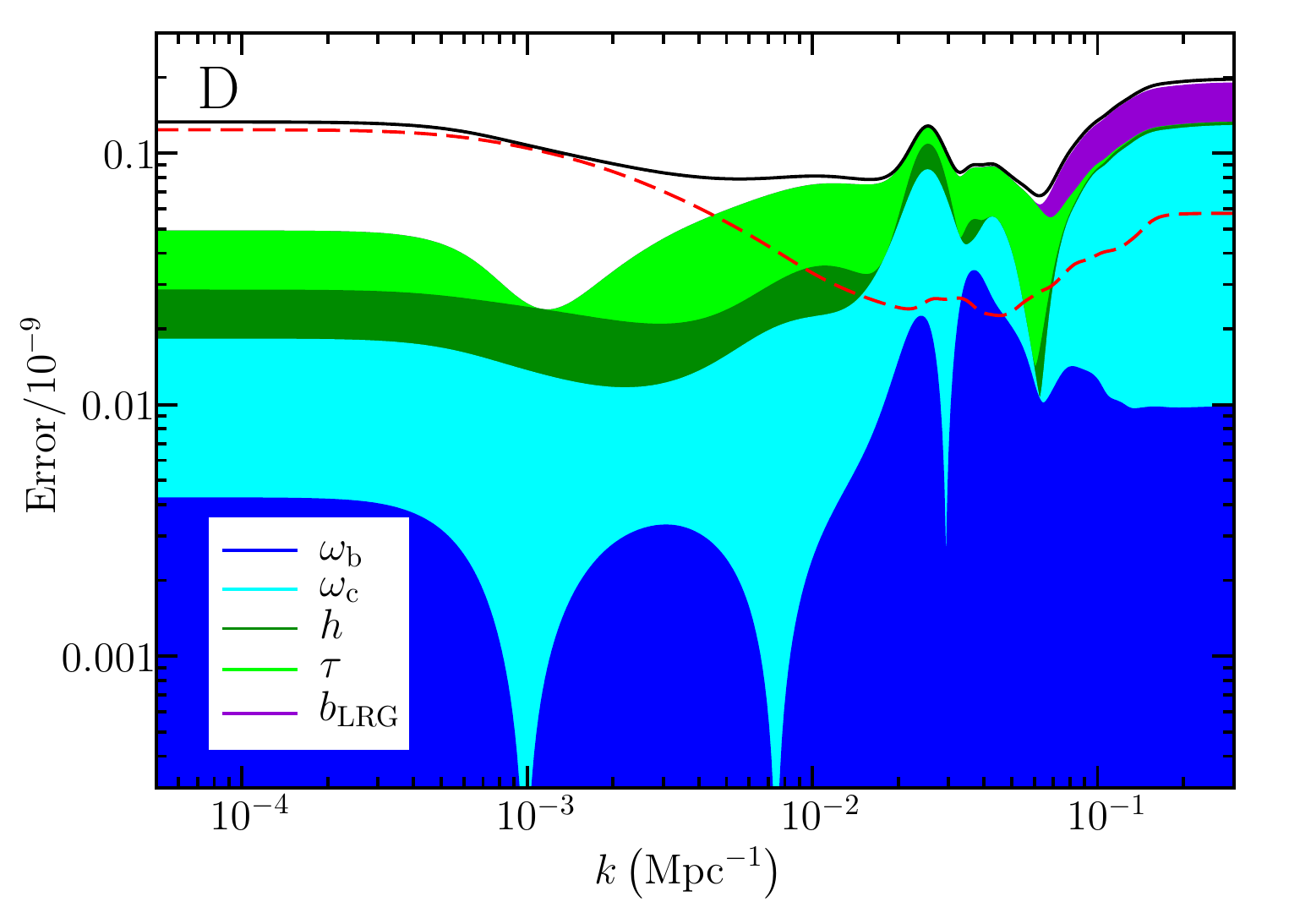}
\caption{\label{covf2b}
Same as Fig.\,\ref{covf2a}, but for $\lambda=5000$.}
\end{figure*}

As shown in Fig.\,\ref{covf2b}, the error in the reconstruction is
lower for $\lambda=5000$.  The lower resolution means that the
reconstructions are more stable and less sensitive to uncertainties in
the background parameters as well as to noise in the data. However the
reconstruction on large scales is particularly sensitive to $\tau$ as
it depends on the TT power spectrum at high multipoles. The error due
to noise in the data dominates the error due to uncertain parameter
values on large scales because of cosmic variance, but on smaller
scales it is subdominant.

\subsection{Statistical significance of the features \label{statsig}}

To investigate whether the features in the recovered spectra are
present in the true PPS or merely noise induced artifacts, we must
determine if the reconstructions are statistically consistent with a
featureless power-law. Following \cite{Hamann:2009bz} and
\cite{Tocchini-Valentini:2005ja} we perform a hypothesis test, with
the null hypothesis being that the true spectrum is a power-law with
$n_s=0.969$.

\begin{figure*}
\begin{minipage}{150mm}
\includegraphics*[angle=0,width=0.5\columnwidth,trim = 32mm 171mm 23mm
  15mm, clip]{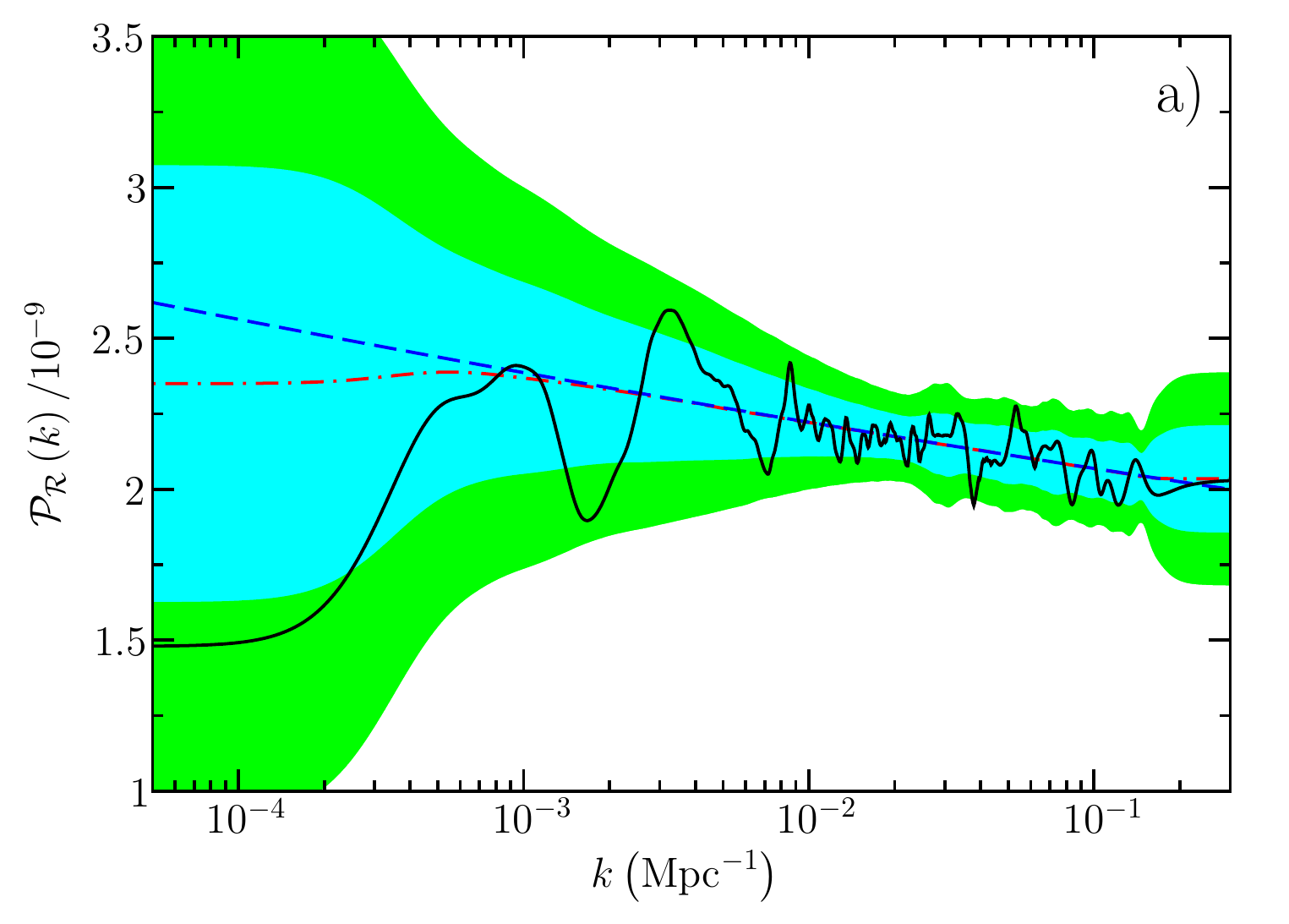}
\includegraphics*[angle=0,width=0.5\columnwidth,trim = 32mm 171mm 23mm
  15mm, clip]{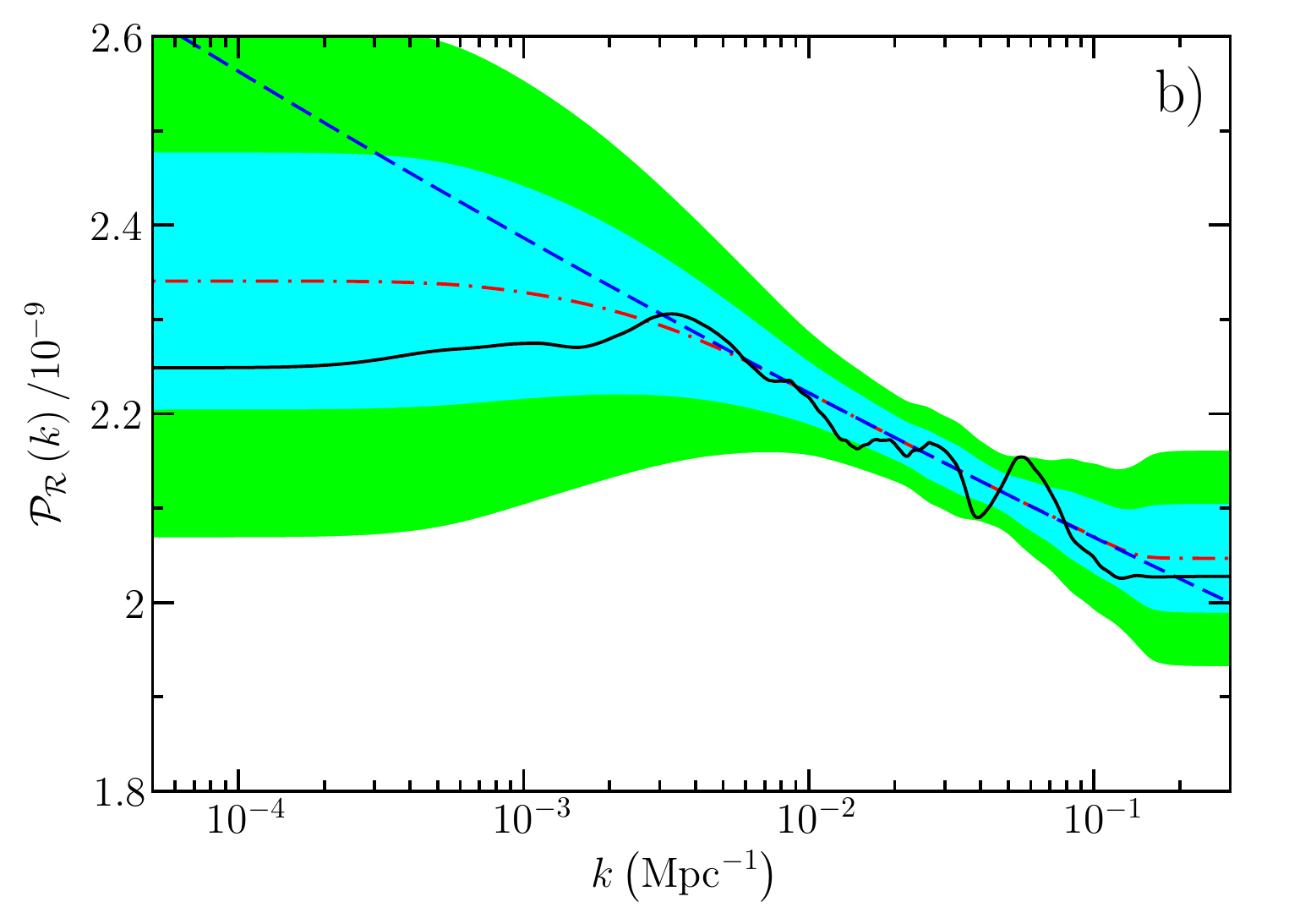}
\caption{\label{hypoth} Comparison of the PPS recovered from the
  WMAP-9, small-scale CMB and SDSS-4 LRG data (full black line) with
  the results of $10^5$ simulated reconstructions for a)
  $\lambda=100$, b) $\lambda=5000$. The simulations were generated
  from a power-law PPS with $n_\mathrm{s}=0.969$ (dashed blue line)
  and the shaded bands indicate the $1\sigma$ and $2\sigma$ error
  estimate from Monte Carlo simulations. The red dot-dashed line is
  the mean of the reconstructions.}
\end{minipage}
\end{figure*}

We generate $10^5$ simulated data realisations under the assumption
that the null hypothesis is true, perform inversions and compare the
distribution of the results with the reconstructions from the actual
data in Fig.\,\ref{hypoth}. From the scatter shown in the plots it is
apparent that the features represent at most $\sim 2\sigma$ deviations
from a power-law.
 
To quantify the significance of the individual features more precisely
we employ two test statistics.  The first is a two-tailed test
statistic \cite{ramsay}
\begin{equation}
\label{t1stat}
  \mathrm{T}_1 \equiv \int^{\infty}_0 \xi\left(k\right)
  \hat{\mathcal{P}}_\mathcal{R}\left(k\right)\,\mathrm{d}k
      = \sum_i r_i \hat{\mathrm{p}}_i,
\end{equation}
where the kernel $\xi\left(k\right)$ is non-zero in the region of the
feature under study, and
\begin{equation}
r_i \equiv \int^{k_{i+1}}_{k_i} \xi\left(k\right)\,\mathrm{d}k.
\end{equation}
The second is a one-tailed statistic \cite{Kogo:2003yb}
\begin{equation}
 \label{t2stat}
  \mathrm{T}_2 \equiv \int^{\infty}_0 \xi\left(k\right)
  \left[\hat{\mathcal{P}}_\mathcal{R}
  \left(k\right)-\mathcal{P}_\mathcal{R}^\mathrm{PL}
  \left(k\right)\right]^2\,\mathrm{d}k
  = \sum_i
  r_i\left(\hat{\mathrm{p}}_i-\hat{\mathrm{p}}_i^\mathrm{PL}\right)^2.
\end{equation}
Here $\mathcal{P}_\mathcal{R}^\mathrm{PL}\left(k\right)\equiv\sum_i
\mathrm{p}_i^\mathrm{PL}\phi_i\left(k\right)$ is the power-law
spectrum associated with the null hypothesis. 

Recall that the $p$-value of a test statistic is the probability under
the null hypothesis that the statistic has a value at least as extreme
as the one actually observed.  For each feature we use a top-hat
kernel $\xi\left(k\right)$ and estimate the $p$-values of
$\mathrm{T}_1$ and $\mathrm{T}_2$ from the distributions of the test
statistics in the simulated PPS reconstructions for two values of the
regularisation parameter $\lambda$. The results are listed in Tables
\ref{tab1} and \ref{tab2}.

A high $\mathrm{T}_2$ value can arise from a peak, dip or an
oscillation about the null hypothesis power-law.  Only the first two
cases will result in an extreme $\mathrm{T}_1$ value. Hence the
$\mathrm{T}_1$ $p$-value of a feature is normally less than half that
of the $\mathrm{T}_2$ $p$-value.\footnote{The exception occurs for the
  infrared cutoff due to the bias in the reconstruction on large
  scales.} Unlike $\mathrm{T}_2$, the test statistic $\mathrm{T}_1$ is
found to closely obey a Gaussian distribution under the null
hypothesis. The dips in the $\lambda=100$ reconstruction at $k\simeq
0.0017$ and $0.038\;\mathrm{Mpc}^{-1}$ and the bump at
$k\simeq0.053\;\mathrm{Mpc}^{-1}$ therefore represent $1.4\sigma$,
$1.9\sigma$ and $1.6\sigma$ deviations, according to their
$\mathrm{T}_1$ $p$-values. Similarly, the dip at $k\simeq
0.038\;\mathrm{Mpc}^{-1}$ and the bump at
$k\simeq0.055\;\mathrm{Mpc}^{-1}$ in the $\lambda=5000$ reconstruction
have $1.6\sigma$ and $1.7\sigma$ statistical significance
respectively.

\begin{table}[t]
\begin{center}
\footnotesize{
\begin{tabular}{c|cccc}
\hline $k$-range /$\mathrm{Mpc}^{-1}$ & $5\times 10^{-5} - 3\times
10^{-4}$ & $0.0013-0.0023$ & $0.0030-0.0040$ & $0.0067-0.0075$
\\ \hline $\mathrm{T}_1$ $p$-value & 0.13 & 0.077 & 0.088 & 0.10
\\ $\mathrm{T}_2$ $p$-value & 0.17 & 0.16 & 0.19 & 0.22 \\ \hline
\end{tabular}
\vspace{0.25cm}
\vspace{0.25cm}
\begin{tabular}{c|cccc}
\hline
$k$-range/$\mathrm{Mpc}^{-1}$ & $0.0084-0.0088$ & $0.0125-0.0131$ 
& $0.0145-0.0150$ & $0.0215-0.0223$  \\
\hline
$\mathrm{T}_1$ $p$-value & 0.089 &  0.14 &  0.13
            & 0.12 \\
$\mathrm{T}_2$ $p$-value & 0.18 &  0.29 & 0.27
            & 0.27  \\
\hline
\end{tabular}
\vspace{0.25cm}
\vspace{0.25cm}
\begin{tabular}{c|cccc}
\hline
$k$-range/$\mathrm{Mpc}^{-1}$ & $0.0326-0.0344$ & $0.0362-0.0402$ 
& $0.0510-0.0560$ & $0.0810-0.0865$  \\
\hline
$\mathrm{T}_1$ $p$-value & 0.13 &  0.026 &  0.055
            & 0.097 \\
$\mathrm{T}_2$ $p$-value & 0.28 &  0.089 & 0.14
            & 0.22  \\
\hline
\end{tabular}}

\caption{\label{tab1} The $p$-values of selected features in the PPS
  recovered from the WMAP-9, small-scale CMB and SDSS-4 LRG data with
  $\lambda$=100, for two-sided (eq.\,\ref{t1stat}) and one-sided
  (eq.\,\ref{t2stat}) statistics.}
\end{center}
\end{table}

\begin{table}[t]
\begin{center}
\footnotesize{
\begin{tabular}{c|ccc}
\hline
$k$-range/$\mathrm{Mpc}^{-1}$ & $0.0125-0.0155$ & $0.036-0.044$ 
& $0.049-0.070$ \\
\hline
$\mathrm{T}_1$ $p$-value & 0.12 &  0.057  &  0.041 \\
$\mathrm{T}_2$ $p$-value & 0.26 &  0.13 & 0.11  \\
\hline
\end{tabular}}
\caption{\label{tab2} Same as Table \ref{tab1}, but for
  $\lambda=5000$.}
\end{center}
\end{table}

\begin{figure*}
\begin{minipage}{150mm}
\includegraphics*[angle=0,width=0.5\columnwidth,trim = 32mm 171mm 23mm
  15mm, clip]{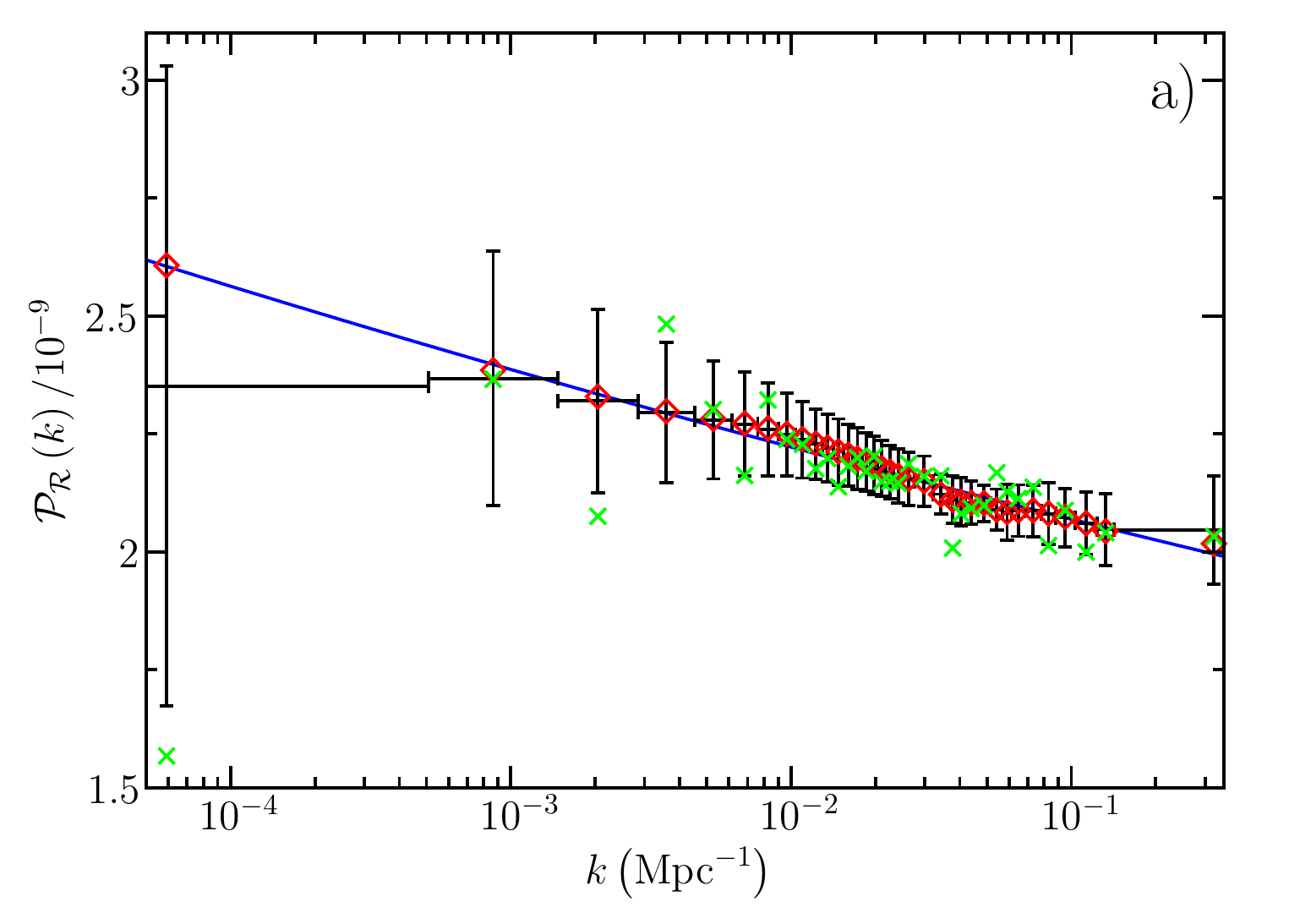}
\includegraphics*[angle=0,width=0.5\columnwidth,trim = 32mm 171mm 23mm
  15mm, clip]{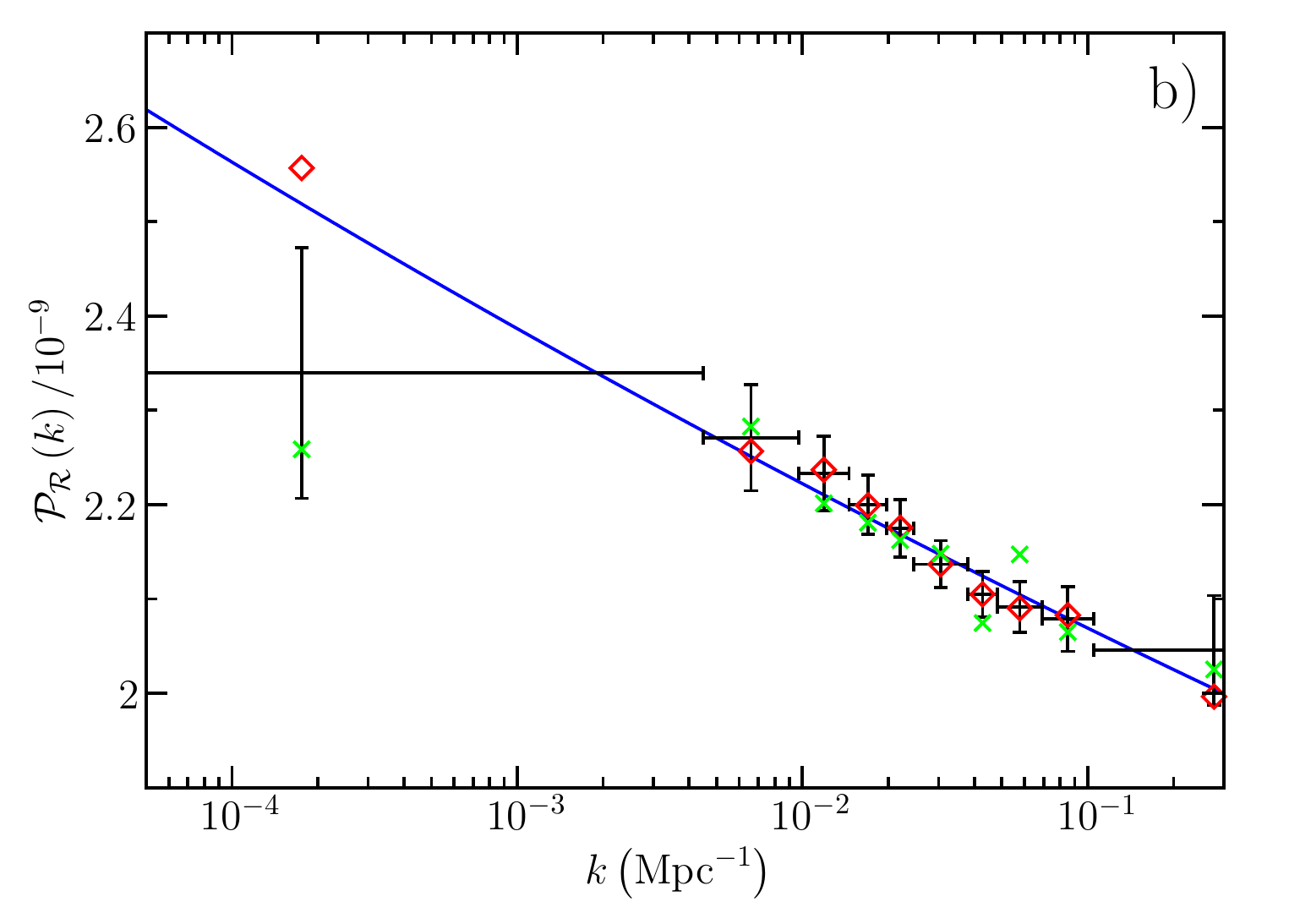}
\caption{\label{binstat} The solid line is the power-law spectrum
  corresponding to the null hypothesis. The red diamonds are the
  bandpowers calculated from the power-law spectrum --- if the
  bandpower window functions were perfectly well-behaved, the diamonds
  would lie on the solid line. The vertical error bars are the
  $1\sigma$ scatter of the bandpowers from $10^5$ simulated
  reconstructions while the horizontal error bars indicate the
  wavenumber range to which the bandpowers are most sensitive. Note
  that these are not centred on the diamonds due to the bias of the
  deconvolution method on very large and very small scales. The green
  crosses are the bandpowers recovered from the actual WMAP-9, small
  scale CMB and SDSS-4 LRG data. Panel a) is for $\lambda=100$ and
  panel b) is for $\lambda=5000$.}
\end{minipage}
\end{figure*}

Turning to the bandpower analysis, Fig.\,\ref{binstat} displays the
distribution of the bandpowers obtained from the $10^5$ simulated
reconstructions together with the bandpowers derived from the observed
data. For $\lambda=100$ the bandpowers at $k=0.0020$, $0.038$ and
$0.054\;\mathrm{Mpc}^{-1}$ correspond to $1.3\sigma$, $2.1\sigma$ and
$1.8\sigma$ fluctuations, while for $\lambda=5000$ the bandpowers at
$k=0.043$ and $0.058\;\mathrm{Mpc}^{-1}$ constitute $1.3\sigma$ and
$2.1\sigma$ fluctuations.

In addition to individual features we ought to consider statistics
associated with the entire PPS in order to avoid \textit{a posteriori}
selection effects. For $\lambda=100$ the $p$-value of the ($\chi^2$ of
the) fit of all $35$ power-law bandpowers to those recovered from the
WMAP-9, small-scale CMB and SDSS-4 LRG data is $0.93$ (equivalent to a
$-1.5\sigma$ deviation). The $p$-value drops to $0.45$ for the $10$
bandpowers of the $\lambda=5000$ reconstruction (a $0.1\sigma$
deviation). Thus the reconstructions with $\lambda=100$ and
$\lambda=5000$ are both statistically \emph{consistent} with a
power-law.

\begin{figure*}
\begin{minipage}{150mm}
\includegraphics*[angle=0,width=0.5\columnwidth,trim = 32mm 171mm 23mm
  15mm, clip]{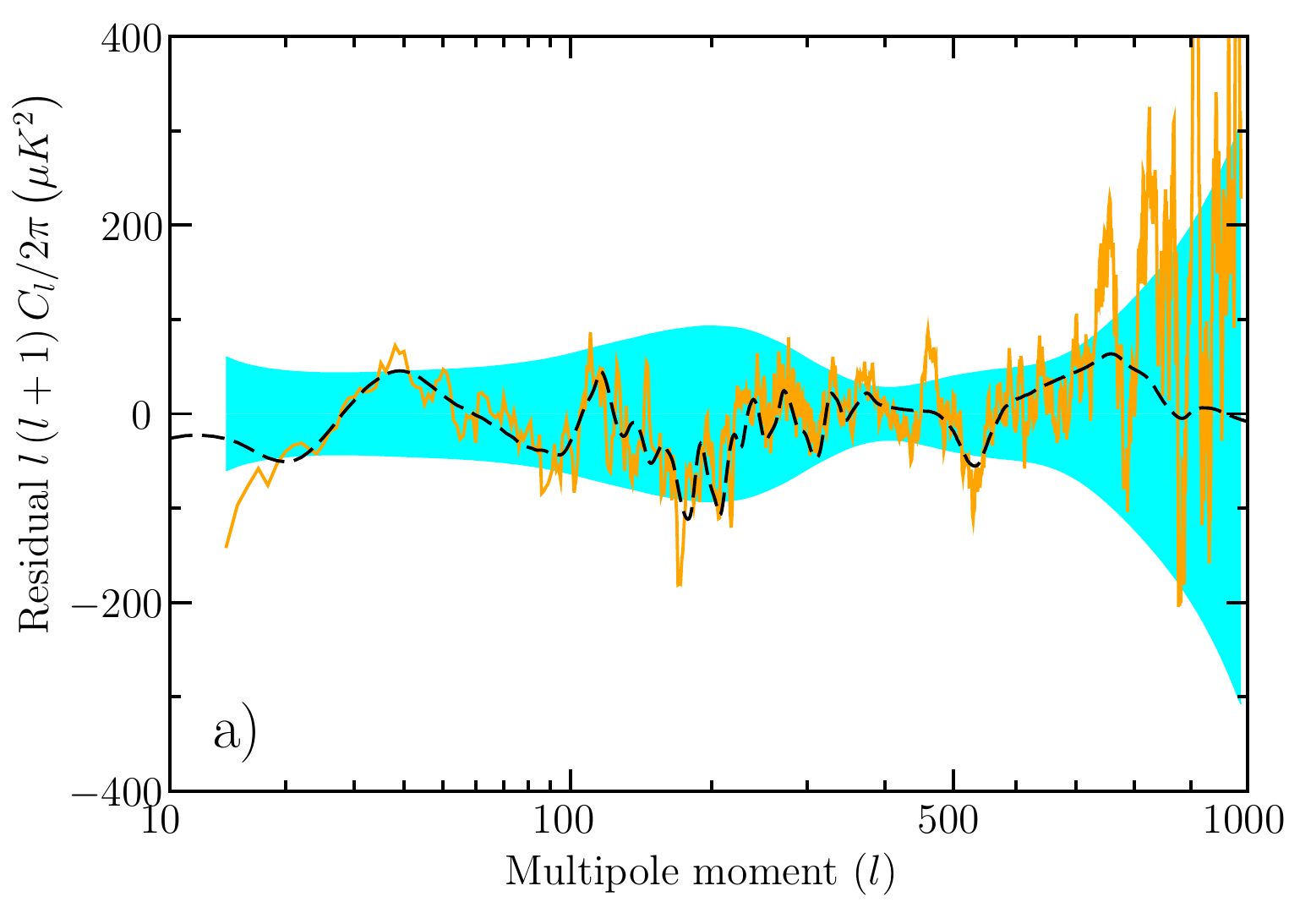}
\includegraphics*[angle=0,width=0.5\columnwidth,trim = 32mm 171mm 23mm
  15mm, clip]{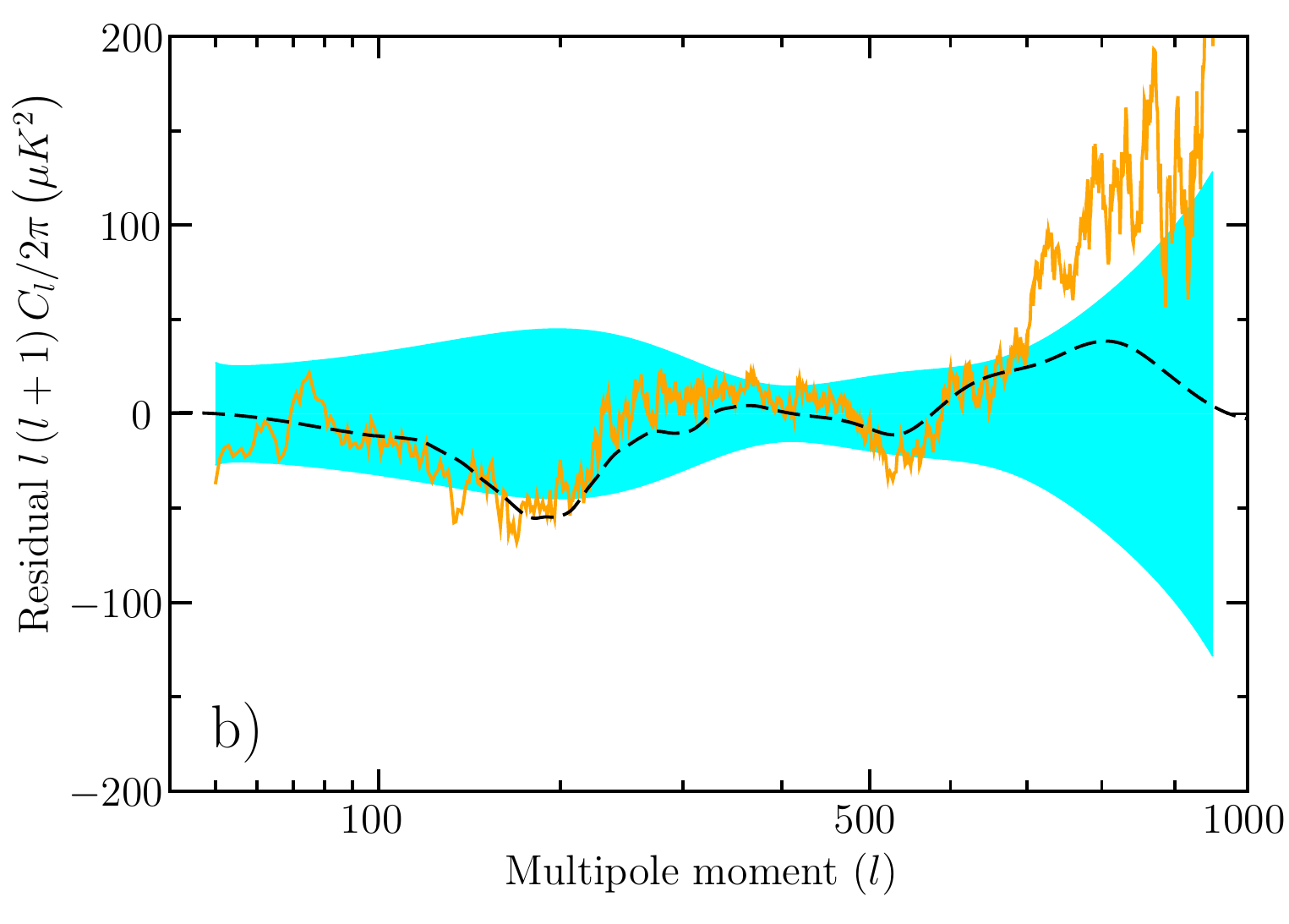}
\caption{\label{ttdev} a) Comparison of residuals for the $\ell=25$
  running average of the WMAP-9 TT data (solid orange line) with the
  residuals corresponding to the $\lambda=100$ reconstruction from the
  WMAP-9, small-scale CMB and SDSS-4 LRG data (dashed black line).
  The residuals are obtained by subtracting off the TT spectrum of the
  null hypothesis power-law model.  The band indicates the $1\sigma$
  scatter of the $\ell=25$ running average data, calculated from the
  WMAP-9 TT covariance matrix. b) Same as panel a) but for the
  $\ell=97$ running average and the $\lambda=5000$ reconstruction.}
\end{minipage}
\end{figure*}

This may seem surprising, given the poor $\chi^2$ of the fit of the
power-law model to the WMAP-9 data. Indeed, of the simulated TT power
spectra generated under the null hypothesis, only $6.6\%$ have as high
a $\chi^2$ value as that of the actual data. However, the
reconstructed PPS is sensitive to the \emph{running average} of the
data. This is illustrated in Fig.\,\ref{ttdev}, which shows how the TT
spectra corresponding to the reconstructions with $\lambda=100$ and
$\lambda=5000$ trace the $\ell=25$ and $\ell=97$ running
averages. From the simulations we find the correlation coefficient
between the bandpower $\chi^2$ (for $\lambda=100$) and the $\chi^2$ of
the $\ell=25$ running average data to be $0.55$, compared to a
correlation coefficient of $0.12$ between the bandpower $\chi^2$ and
the $\chi^2$ of the unaveraged data. For $\lambda=5000$ the equivalent
correlation coefficients are $0.40$ for the $\ell=97$ running average
and $0.077$ for the unaveraged data. The $p$-value of the $\chi^2$ of
the $\ell=25$ running average data is $0.96$ (a $-1.7\sigma$
deviation) while that of the $\ell=97$ running average data is $0.51$
(a $-0.02\sigma$ deviation). Since fluctuations in the running average
are responsible for most of those in the reconstructions, both
reconstructions can be said to be presently consistent with a
power-law for an assumed $\Lambda$CDM cosmology.

\section{Conclusions \label{conclusions}}

The task of inverting noisy data with limited resolution and an
uncertain background cosmology to extract the PPS of curvature
perturbations is a challenging one, with \textit{no} unique
solution. We have studied and validated a robust regularisation
procedure for obtaining a stable estimate of the PPS from multiple
cosmological data sets. A perturbative analysis of the relationship of
the estimate to the true PPS (which can be applied to any inversion
method) has allowed us (see Appendix~\ref{bivar}) to identify the
different components of the bias and variance that characterise the
performance of the reconstruction. The bias is found to depend on
first- and second-order resolution kernels which indicate the
resolving power of the inversion. Moreover, we find that data with a
non-Gaussian likelihood function can increase the bias, even if the
data itself is unbiased.  The variance arises from both noise in the
data and uncertainties in the background cosmological parameters.
 
We have seen how the regularisation parameter governs the trade-off
between the bias and variance, and tested (in
Appendix~\ref{regularisation}) several methods for choosing the
optimum value of the parameter, with mixed results. The recovered
spectra exhibit interesting deviations from the usually assumed
scale-free power-law, with a statistical significance of around
$2\sigma$ as determined from Monte Carlo simulations. However, using a
decorrelated bandpower analysis the spectra are found to be
statistically consistent with a power-law. We emphasise that these
conclusions are dependent on the \emph{assumed} background cosmology
--- here taken to be the standard $\Lambda$CDM model. We do not expect
the statistical significance of the features in the deconvolved
spectrum to change much when the $\Lambda$CDM parameters are varied
within their quoted uncertainties. However if the assumed cosmology is
radically different then the recovered spectrum may indeed look rather
different too (e.g. Spectrum D in Sec.~\ref{spectest} would be
recovered for a flat CHDM model with no dark energy
\cite{Hunt:2007dn,Hunt:2008wp,Nadathur:2010zm}).

Our results are in agreement with previous work by other authors
(e.g. \cite{Tocchini-Valentini:2005ja,
  Nicholson:2009pi,Hamann:2009bz}) who also reported a cutoff in the
PPS on large scales and features at $k\simeq0.0017\;\mathrm{Mpc}^{-1}$
and $k\simeq0.0032\;\mathrm{Mpc}^{-1}$ due to the WMAP glitches. In
particular, \cite{Nagata:2008tk,Nagata:2008zj} found the $\chi^2$ for
$34$ and $46$ bandpowers in the ranges $9\times10^{-3}\lesssim
k\lesssim 0.03\;\mathrm{Mpc}^{-1}$ and $2\times10^{-3}\lesssim
k\lesssim 0.03\;\mathrm{Mpc}^{-1}$ recovered from the WMAP-5 data to
have $-1.5\sigma$ and $-1.1\sigma$ significance respectively, using
the Tikhonov regularisation and cosmic inversion methods. This is
quite consistent with the $-1.5\sigma$ deviation of our $35$
bandpowers. In summary, the statistical significance of these
deviations is not yet compelling, however the potential here for
probing the physical process that generated the primordial curvature
perturbations certainly motivates further investigation.

Many studies have compared regularisation parameter selection methods,
though not in this context (see
\cite{Bauer:2011uw,Hamarik:2012fw,Reichel:2013gw} and references
therein). The consensus in the literature is that the results are
strongly dependent on the application at hand
\cite{vogel,hansen:2010dc}.  Here the selection of the regularisation
parameter is complicated by the fact that the width of the CMB
integral kernels and the noise level of the data vary strongly with
the multipole moment (see Appendix~\ref{regularisation}).\footnote{By
  contrast, other inversion applications (e.g. image restoration) do
  not suffer from an equivalent problem.} Until a suitable
regularisation parameter selection method is found, we recommend
performing inversions using a range of $\lambda$ values informed by
simulations.

There are some subtleties associated with adding data to improve the
inversion. The bias is reduced only if the PPS departs from the flat
H-Z form on the scales covered by the new data. Including a
non-Gaussian distributed data set can increase the bias, particularly
for small $\lambda$. The variance will increase if the additional data
is strongly dependent on background cosmological parameters with large
uncertainties.

The next step would naturally be to apply our method to the recent
data release from the Planck satellite \cite{Planck:2013kta}, as well
as small angular experiments such as ACT \cite{Das:2013zf} and SPT
\cite{Story:2012wx}. Planck has much greater resolution than WMAP on
small scales but the Planck team have applied \cite{Ade:2013uln} only
a parametric method using a smoothing spline
(e.g. \cite{Gauthier:2012aq}), which we consider inadequate
for recovering sharp features in the spectrum, such as we have
uncovered in the WMAP data. Should these turn out to exist in the
Planck data too, it would spell the death of single-field inflationary
models! The ``precision'' estimation of cosmological parameters
\textit{assuming} a power-law spectrum \cite{Ade:2013lta} would also
then need to be revisited.

\section{Acknowledgments}
  We acknowledge use of the {\tt CAMB} and {\tt cosmoMC} codes and
  thank the WMAP and Planck teams for making their data and analysis
  tools publicly available. This work was supported by the EU Marie
  Curie Network `UNILHC' (PITN-GA-2009-237920) and a Niels Bohr
  Professorship awarded by the Danish National Research Foundation.

\begin{appendix}

\section{Error analysis \label{error}}

We analyse Tikhonov regularisation within the frequentist statistical
framework. This approach involves a hypothetical ensemble of PPS
estimates obtained from repeated independent identical measurements of
the data and the background parameters. There exists an unknown true
PPS $\pB_\mathrm{t}$ and a true set of background parameters
$\bm{\theta}_\mathrm{t}$. The estimates constitute samples from the
distribution $P\left(\hat{\pB}|\pB_\mathrm{t}\right)$ which must be
appropriately characterised.

Since
$\dB=\mathsf{W}\left(\bm{\theta}_\mathrm{t}\right)\pB_\mathrm{t}+\nB$
the data depend on the true background parameters
$\bm{\theta}_\mathrm{t}$ but the reconstruction is performed using the
estimate $\hat{\bm{\theta}}$. Thus the recovered PPS can be written as
\begin{equation}
 \hat{\pB}\left(\dB,\hat{\bm{\theta}}\right) =
 \bm{\mathcal{T}}\left(\pB_\mathrm{t},\bm{\theta}_\mathrm{t},
 \hat{\bm{\theta}},\nB\right),
\label{transfer}
\end{equation} 
where the transfer function $\bm{\mathcal{T}}$ gives the relationship
of $\hat{\pB}$ to $\pB_\mathrm{t}$, which is nonlinear in
general. Performing a Taylor expansion of the transfer function about
some PPS $\pB_1$ close to $\pB_\mathrm{t}$ gives
\begin{eqnarray}
  \hat{\mathrm{p}}_i\left(\dB,\hat{\bm{\theta}}\right) & = &
  \mathcal{T}_i\left(\pB_1,\bm{\theta}_\mathrm{t},\bm{\theta}_\mathrm{t},
  \mathbf{0}\right) + \sum_{j}R_{ij}\,\Delta
  \mathrm{p}_j+\frac{1}{2}\sum_{j,k}Y_{ijk}\,\Delta
  \mathrm{p}_j\,\Delta \mathrm{p}_k \nonumber\\ & &
  +\sum_{\mathbb{Z},a}
  M_{ia}^{(\mathbb{Z})}\,\mathrm{n}_a^{(\mathbb{Z})}+\sum_\alpha
  M_{i\alpha}\,\mathrm{u}_\alpha+\sum_{\mathbb{Z},j,a}Z_{ija}^{(\mathbb{Z})}\,\Delta
  \mathrm{p}_j
  \,\mathrm{n}_a^{(\mathbb{Z})}+\sum_{j,\alpha}Z_{ij\alpha}\,\Delta
  \mathrm{p}_j\,\mathrm{u}_\alpha \nonumber\\ & &
  +\frac{1}{2}\sum_{\mathbb{Z},\mathbb{Z'},a,b}X_{iab}^{(\mathbb{Z}\mathbb{Z'})}\,
  \mathrm{n}_a^{(\mathbb{Z})}\,\mathrm{n}_b^{(\mathbb{Z'})} +
  \sum_{\mathbb{Z},a,\alpha}X_{ia\alpha}^{(\mathbb{Z})}\,\mathrm{n}_a^{(\mathbb{Z})}\,
  \mathrm{u}_\alpha +
  \frac{1}{2}\sum_{\alpha,\beta}X_{i\alpha\beta}\,\mathrm{u}_\alpha\,\mathrm{u}_\beta
  +\ldots
\label{recerr}
\end{eqnarray}
Here $\Delta \mathrm{p}_i\equiv \mathrm{p}_{\mathrm{t}i}-\mathrm{p}_{1i}$, and
\begin{eqnarray}
  M_{ia}^{(\mathbb{Z})}\equiv\left.\frac{\partial\hat{\mathrm{p}}_i}{\partial
    \mathrm{d}_{a}^{(\mathbb{Z})}}\right|_{\hat{\dB}_1,\bms{\theta}_\mathrm{t}},
  \qquad
  M_{i\alpha}\equiv\left.\frac{\partial\hat{\mathrm{p}}_i}{\partial
    \theta_\alpha}\right|_{\hat{\dB}_1,\bms{\theta}_\mathrm{t}},
  \qquad\qquad\qquad\qquad \nonumber
  \\ X_{iab}^{(\mathbb{Z}\mathbb{Z'})} \equiv
  \left.\frac{\partial^{2}\hat{\mathrm{p}}_i}{\partial
    \mathrm{d}_{a}^{(\mathbb{Z})}\partial
    \mathrm{d}_{b}^{(\mathbb{Z'})}}\right|_{\hat{\dB}_1,\bms{\theta}_\mathrm{t}},
  \qquad X_{ia\alpha}^{(\mathbb{Z})} \equiv
  \left.\frac{\partial^{2}\hat{\mathrm{p}}_i}{\partial
    \mathrm{d}_{a}^{(\mathbb{Z})}\partial
    \theta_\alpha}\right|_{\hat{\dB}_1,\bms{\theta}_\mathrm{t}},
  \qquad X_{i\alpha\beta} \equiv
  \left.\frac{\partial^{2}\hat{\mathrm{p}}_i}{\partial
    \theta_\alpha\partial
    \theta_\beta}\right|_{\hat{\dB}_1,\bms{\theta}_\mathrm{t}},\nonumber\\ R_{ij}
  \equiv\sum_{\mathbb{Z},a}M_{ia}^{(\mathbb{Z})}W_{aj}^{(\mathbb{Z})} 
  \left(\bm{\theta}_\mathrm{t}\right),
  \qquad Y_{ijk}\equiv\sum_{\mathbb{Z},\mathbb{Z'},a,b}
  X_{iab}^{(\mathbb{Z}\mathbb{Z'})}W_{aj}^{(\mathbb{Z})}
  \left(\bm{\theta}_\mathrm{t}\right)
  W_{bk}^{(\mathbb{Z'})}\left(\bm{\theta}_\mathrm{t}\right),
  \qquad\nonumber \\ 
  Z_{ija}^{(\mathbb{Z})} \equiv 
  \sum_{\mathbb{Z'},b}X_{iab}^{(\mathbb{Z}\mathbb{Z'})}W_{bj}^{(\mathbb{Z'})}
  \left(\bm{\theta}_\mathrm{t}\right), \qquad
  Z_{ij\alpha}\equiv\sum_{\mathbb{Z},a}X_{ia\alpha}^{(\mathbb{Z})}W_{aj}^{(\mathbb{Z})}
  \left(\bm{\theta}_\mathrm{t}\right),\qquad\qquad
\label{matrices}
\end{eqnarray} 
where
$\hat{\dB}_1\equiv\mathsf{W}\left(\bm{\theta}_\mathrm{t}\right)\pB_1$
is the data estimated from $\pB_1$. In eq.(\ref{recerr}),
$\bm{\mathcal{T}}\left(\pB_1,\bm{\theta}_\mathrm{t},\bm{\theta}_\mathrm{t},
\mathbf{0}\right)$ is the reconstruction resulting from the noise-free
data $\hat{\dB}_1$ using the true background parameters and should
ideally equal $\pB_1$, but does not do so in general. The first-order
resolution matrix $\mathsf{R}$ characterises the linear mapping of
$\pB_\mathrm{t}$ to $\hat{\pB}$. Perfect resolution is attained when
$\mathsf{R}$ equals the identity matrix $\mathsf{I}$. However,
multiplication by $\mathsf{R}$ usually acts as a smoothing operation
so that $\hat{\pB}$ is a smoothed version of $\pB_\mathrm{t}$, even
for noise-free data. Components of $\pB_\mathrm{t}$ belonging to the
null space of $\mathsf{R}$ are not reproduced in $\hat{\pB}$. The
second-order resolution matrix $\mathsf{Y}$ describes an undesirable
quadratic mapping from $\pB_\mathrm{t}$ to $\hat{\pB}$ and should
ideally vanish. The remaining contributions to $\hat{\pB}$ represent
artifacts caused by noise in the data and incorrect estimates of the
background parameters.

Using eq.(\ref{recerr}) the leading sources of bias in the
reconstruction can be identified:
\begin{eqnarray}
  \mathrm{Bias}\left(\hat{\mathrm{p}}_i\right)& \equiv &
  \langle\hat{\mathrm{p}}_i-\mathrm{p}_{\mathrm{t}i}\rangle =
  \mathcal{T}_i\left(\pB_1,\bm{\theta}_\mathrm{t},\bm{\theta}_\mathrm{t},\mathbf{0}\right)-\mathrm{p}_{1i}
  +\sum_{j}\left(R_{ij}-I_{ij}\right)\,\Delta \mathrm{p}_j \nonumber\\ &
  & +\frac{1}{2}\sum_{j,k}Y_{ijk}\,\Delta \mathrm{p}_j\,\Delta
  \mathrm{p}_k+\frac{1}{2}\sum_{\mathbb{Z},a,b}
  X_{iab}^{(\mathbb{ZZ})}\, N_{ab}^{(\mathbb{Z})}
  +\frac{1}{2}\sum_{\alpha,\beta} X_{i\alpha\beta}\,
  N_{\alpha\beta}+\ldots.
\end{eqnarray}
Here
$\bm{\mathcal{T}}\left(\pB_1,\bm{\theta}_\mathrm{t},\bm{\theta}_\mathrm{t}, 
\mathbf{0}\right)-\pB_1$
is the error in recovering $\pB_1$. The quantity
$\left(\mathsf{R}-\mathsf{I}\right)\,\Delta\pB$ represents the error
due to the limited resolution of the reconstruction and is labelled
the `null space error' in \cite{Rodgers:1990fa}. The quadratic mapping
associated with the second-order resolution matrix contributes
directly to the bias. It is apparent that a quadratic (or higher order) 
dependence of the estimated PPS on the data or the background parameters 
will also increase the bias.

Rather than work with the resolution matrices it is more convenient to
use their continuous counterparts, the resolution kernels. Resolution
kernels enable the resolving power of different inversion methods to
be compared directly (see \cite{Christensen-Dalsgaard:1990bn} for an
example). First-order resolution kernels were originally introduced as
part of the Backus-Gilbert method \cite{Backus:1968ed} and were
generalised subsequently to higher order in \cite{Sneider:1991rt}. The
continuous analogue of the first-order contribution
$\delta\hat{\mathrm{p}}_i^{(1)}\equiv\sum_{j}R_{ij}\mathrm{p}_{\mathrm{t}j}$
of $\pB_\mathrm{t}$ to $\hat{\pB}$ is the contribution
\begin{equation}
  \delta\hat{\mathcal{P}}_\mathcal{R}^{(1)}\left(k_0\right)\equiv\int^{\infty}_0
  R\left(k_0;k\right) \mathcal{P_R}\left(k\right)\,\mathrm{d}k
\end{equation}
to $\hat{\mathcal{P}}_\mathcal{R}\left(k_0\right)\equiv\sum_i
\hat{\mathrm{p}}_i\phi_i\left(k_0\right).$ Here the first-order resolution
kernel $R\left(k_0;k\right)$ is given by
\begin{equation}
\label{1storderkernel}
R\left(k_0;k\right)\equiv\sum_{\mathbb{Z},i,a}\phi_i\left(k_0\right)
M_{ia}^{(\mathbb{Z})} \mathcal{K}_a^{(\mathbb{Z})}\left(k\right).
\end{equation}
Similarly, the continuous analogue of the second-order contribution
$\delta\hat{\mathrm{p}}_i^{(2)} \equiv 
\frac{1}{2}\sum_{j,k}Y_{ijk}\mathrm{p}_{\mathrm{t}j}\mathrm{p}_{\mathrm{t}k}$
is
\begin{equation}
\delta\hat{\mathcal{P}}_\mathcal{R}^{(2)}\left(k_0\right) \equiv 
\frac{1}{2}\int^{\infty}_0\int^{\infty}_0
Y\left(k_0;k_1,k_2\right)
\mathcal{P_R}\left(k_1\right)\mathcal{P_R} 
\left(k_2\right)\,\mathrm{d}k_1\mathrm{d}k_2
\end{equation}
where the second-order resolution kernel $Y\left(k_0;k_1,k_2\right)$
is given by
\begin{equation}
\label{2ndorderkernel}
Y\left(k_0;k_1,k_2\right) \equiv 
\sum_{\mathbb{Z},\mathbb{Z'},i,a,b}\phi_i \left(k_0\right)
X_{iab}^{(\mathbb{ZZ'})} \mathcal{K}_a^{(\mathbb{Z})}\left(k_1\right) 
K_b^{(\mathbb{Z'})}\left(k_2\right).
\end{equation}
With $k_0$ kept constant the first-order resolution kernel is a
sharply peaked function of $k_1$ that indicates the wavenumber range
of the true PPS to which the estimate
$\hat{\mathcal{P}}_\mathcal{R}\left(k_0\right)$ is sensitive. For the
reconstructed PPS to be correctly scaled, necessary properties of the
first-order kernel are that for each $k_0$ value the kernel must have
a single peak at $k=k_0$ and that the kernel must integrate to
unity, $\int^{\infty}_0
R\left(k_0;k\right)\,\mathrm{d}k=1$. Clearly the closer
$R\left(k_0;k\right)$ is to the Dirac delta function
$\delta\left(k_0-k\right)$ and the closer
$Y\left(k_0;k_1,k_2\right)$ is to zero, the better the resolution of
the recovered PPS.

It follows from eq.(\ref{recerr}) that the total frequentist
covariance matrix $\mathsf{\Sigma}_\mathrm{T}$ which characterises
variations in $\hat{\pB}$ between members of the ensemble due to noise
and errors in the estimated background parameters is given by
\begin{eqnarray}
\label{sigmat}
\mathsf{\Sigma}_\mathrm{T} &\equiv& \langle\left(\hat{\pB} - 
\langle \hat{\pB}\rangle\right) 
\left(\hat{\pB}-\langle \hat{\pB}\rangle\right)^\mathrm{T}\rangle=
\mathsf{\Sigma}_\mathrm{F}+\mathsf{\Sigma}_\mathrm{P}+\ldots,\\ 
\label{sigmaf}
\mathsf{\Sigma}_\mathrm{F} &\equiv& 
\sum_\mathbb{Z}\mathsf{M}_\mathbb{Z}\mathsf{N}_\mathbb{Z}\mathsf{M}_\mathbb{Z}^\mathrm{T}, \\ 
\label{sigmap}
\mathsf{\Sigma}_\mathrm{P} &\equiv& \mathsf{M}\mathsf{U}\mathsf{M}^\mathrm{T},
\end{eqnarray}
where
$\mathsf{\Sigma}_\mathrm{F}$
results from the data noise,
$\mathsf{\Sigma}_\mathrm{P}$ results from
errors in the background parameters, and higher-order terms have been
omitted. 

Thus far the analysis applies to \textit{any} reconstruction method in
which the PPS is parameterised using basis functions as in
eq.(\ref{basis}). In the case of Tikhonov regularisation, analytic
expressions for the matrices of eq.(\ref{matrices}) can be derived. To
this end, consider the estimates
$\hat{\pB}_1=\hat{\pB}\left(\hat{\dB}_1,\bm{\theta}_\mathrm{t}\right)$ and
$\hat{\pB}_2=\hat{\pB}\left(\dB_2,\hat{\bm{\theta}}_2\right)$ where
$\dB_2$ is a realisation of the data and $\hat{\bm{\theta}}_2$ is a
realisation of the estimated background parameters.  Performing a
Taylor expansion of $\left.\partial Q/\partial
\mathrm{p}_{i}\right|_{\hat{\pB}_2,\dB_2,\hat{\bms{\theta}}_2}$ about
$\left.\partial Q/\partial
\mathrm{p}_{i}\right|_{\hat{\pB}_1,\hat{\dB_1},\bms{\theta}_\mathrm{t}}$, we get:
\begin{eqnarray}
  \left.\frac{\partial Q}{\partial
    \mathrm{p}_{i}}\right|_{\hat{\pB}_2,\dB_2,\hat{\bms{\theta}}_2} &
  = & \left.\frac{\partial Q} {\partial
    \mathrm{p}_{i}}\right|_{\hat{\pB}_1,\hat{\dB_1},\bms{\theta}_\mathrm{t}}+\sum_j
  A_{ij}\,\Delta\hat{\mathrm{p}}_j+\sum_{\mathbb{Z},a}
  B_{ia}^{(\mathbb{Z})}\,\Delta \mathrm{d}_a^{(\mathbb{Z})}+\sum_j
  B_{i\alpha}\,\Delta\theta_\alpha \nonumber\\ & &
  +\frac{1}{2}\sum_{j,k}
  C_{ijk}\,\Delta\hat{\mathrm{p}}_j\,\Delta\hat{\mathrm{p}}_k + 
  \frac{1}{2}\sum_{\mathbb{Z},\mathbb{Z'},a,b}
  D_{iab}^{(\mathbb{Z}\mathbb{Z'})}\,\Delta
  \mathrm{d}_a^{(\mathbb{Z})}\,\Delta \mathrm{d}_b^{(\mathbb{Z'})}
  +\sum_{\mathbb{Z},a,\alpha} D_{ia\alpha}^{(\mathbb{Z})}\,\Delta
  \mathrm{d}_a^{(\mathbb{Z})}\,\Delta \theta_\alpha \nonumber\\ & &
  +\frac{1}{2}\sum_{\alpha,\beta}
  D_{i\alpha\beta}\,\Delta\theta_\alpha\,\Delta\theta_\beta
  +\sum_{\mathbb{Z},j,a}
  E_{ija}^{(\mathbb{Z})}\,\Delta\hat{\mathrm{p}}_j\,\Delta
  \mathrm{d}_a^{(\mathbb{Z})} +\sum_{j,\alpha}
  E_{ij\alpha}\,\Delta\hat{\mathrm{p}}_j\,\Delta \theta_\alpha\ldots
\label{exp1}
\end{eqnarray}
Here $\Delta\hat{\mathrm{p}}_i\equiv\hat{\mathrm{p}}_{2i} - 
\hat{\mathrm{p}}_{1i}$, $\Delta
\mathrm{d}_a^{(\mathbb{Z})}\equiv \mathrm{d}_{2a}^{(\mathbb{Z})} - 
\hat{\mathrm{d}}_{1a}^{(\mathbb{Z})}$ ,
$\Delta\theta_\alpha\equiv\hat{\theta}_{2\alpha}-\theta_{\mathrm{t}\alpha}$ and
\begin{equation}
 \begin{array}{ccc}
   A_{ij} \equiv \left.\frac{\displaystyle\partial^{2}Q}
   {\displaystyle\partial \mathrm{p}_i\partial
     \mathrm{p}_j}\right|_{\hat{\pB}_1,\hat{\dB_1},\bms{\theta}_\mathrm{t}}, &
   B_{ia}^{(\mathbb{Z})}\equiv\left.\frac{\displaystyle\partial^{2}Q}
   {\displaystyle\partial \mathrm{p}_i\partial
  \mathrm{d}_a^{(\mathbb{Z})}}\right|_{\hat{\pB}_1,\hat{\dB_1},\bms{\theta}_\mathrm{t}}, &
   B_{i\alpha} \equiv \left.\frac{\displaystyle\partial^{2}Q}
   {\displaystyle\partial \mathrm{p}_i\partial
     \theta_\alpha}\right|_{\hat{\pB}_1,\hat{\dB_1},\bms{\theta}_\mathrm{t}}
   \\ C_{ijk}\equiv\left.\frac{\displaystyle\partial^{3}Q}
      {\displaystyle\partial \mathrm{p}_{i}\partial \mathrm{p}_j\partial
        \mathrm{p}_k}\right|_{\hat{\pB}_1,\hat{\dB_1},\bms{\theta}_T}, &
      D_{iab}^{(\mathbb{Z}\mathbb{Z'})} \equiv \left.\frac{\displaystyle\partial^{3}Q}
      {\displaystyle\partial \mathrm{p}_{i}\partial \mathrm{d}_{a}^{(\mathbb{Z})}\partial
        \mathrm{d}_{b}^{(\mathbb{Z'})}}\right|_{\hat{\pB}_1,\hat{\dB_1},\bms{\theta}_\mathrm{t}},
      & D_{ia\alpha}^{(\mathbb{Z})} \equiv
      \left.\frac{\displaystyle\partial^{3}Q}{\displaystyle\partial
        \mathrm{p}_{i}\partial \mathrm{d}_{a}^{(\mathbb{Z})}\partial \theta_\alpha}
      \right|_{\hat{\pB}_1,\hat{\dB_1},\bms{\theta}_\mathrm{t}}
      \\ D_{i\alpha\beta}\equiv\left.\frac{\displaystyle\partial^{3}Q}
         {\displaystyle\partial \mathrm{p}_{i}\partial \theta_\alpha\partial
           \theta_\beta}\right|_{\hat{\pB}_1,\hat{\dB_1},\bms{\theta}_\mathrm{t}},
         & E_{ija}^{(\mathbb{Z})} \equiv
         \left.\frac{\displaystyle\partial^{3}Q}{\displaystyle\partial
           \mathrm{p}_{i}\partial \mathrm{p}_{j}\partial \mathrm{d}_{a}^{(\mathbb{Z})}}
         \right|_{\hat{\pB}_1,\hat{\dB_1},\bms{\theta}_\mathrm{t}}, &
         E_{ij\alpha} \equiv
         \left.\frac{\displaystyle\partial^{3}Q}{\displaystyle\partial
           \mathrm{p}_{i}\partial \mathrm{p}_{j}\partial \theta_\alpha}
         \right|_{\hat{\pB}_1,\hat{\dB_1},\bms{\theta}_\mathrm{t}}. \\
 \end{array}
\end{equation}
We also have
\begin{eqnarray}
  \hat{\mathrm{p}}_{2i} & = & \hat{\mathrm{p}}_{1i}+\sum_{\mathbb{Z},a} M_{ia}^{(\mathbb{Z})}\,\Delta
  \mathrm{d}_a^{(\mathbb{Z})}+\sum_\alpha M_{i\alpha}\,\Delta \theta_\alpha
  +\frac{1}{2}\sum_{\mathbb{Z},\mathbb{Z'},a,b} X_{iab}^{(\mathbb{Z}\mathbb{Z'})}\,\Delta \mathrm{d}_a^{(\mathbb{Z})}\,\Delta
  \mathrm{d}_b^{(\mathbb{Z'})}\nonumber \\ & & +\sum_{\mathbb{Z},a,\alpha}
  X_{ia\alpha}^{(\mathbb{Z})}\,\Delta \mathrm{d}_a^{(\mathbb{Z})}\,\Delta \theta_\alpha
  +\frac{1}{2}\sum_{\alpha,\beta} X_{i\alpha\beta}\,\Delta
  \theta_\alpha\,\Delta \theta_\beta+\ldots
\label{exp2}
\end{eqnarray}
Both $\left.\partial Q/\partial
\mathrm{p}_i\right|_{\hat{\pB}_1,\hat{\dB_1},\bms{\theta}_\mathrm{t}}$
and $\left.\partial Q/\partial
\mathrm{p}_i\right|_{\hat{\pB}_2,\dB_2,\hat{\bms{\theta}}_2}$ vanish
since $\hat{\pB}_1$ and $\hat{\pB}_2$ minimise
$Q\left(\pB,\hat{\dB}_1,\bm{\theta}_\mathrm{t}\right)$ and
$Q\left(\pB,\dB_2,\hat{\bm{\theta}}_2\right)$ respectively.  Therefore
comparing eqs.(\ref{exp1}) and \eqref{exp2} gives
\begin{equation}
M_{ia}^{(\mathbb{Z})} = -\sum_{j} A_{ij}^{-1} B_{ja}^{(\mathbb{Z})},\qquad M_{i\alpha} =
-\sum_{j} A_{ij}^{-1} B_{j\alpha},
\end{equation}
\begin{equation}
X_{iab}^{(\mathbb{Z}\mathbb{Z'})} = -\sum_{j,k,l} A_{ij}^{-1} C_{jkl} M_{ka}^{(\mathbb{Z})}
M_{lb}^{(\mathbb{Z'})}- \sum_{j,k} A_{ij}^{-1} E_{jka}^{(\mathbb{Z})}
M_{kb}^{(\mathbb{Z'})}-\sum_{j,k} A_{ij}^{-1} E_{jkb}^{(\mathbb{Z'})}
M_{ka}^{(\mathbb{Z})}-\sum_{j} A_{ij}^{-1} D_{jab}^{(\mathbb{Z}\mathbb{Z'})},
\end{equation}
\begin{equation}
X_{ia\alpha}^{(\mathbb{Z})} = -\sum_{j,k,l} A_{ij}^{-1} C_{jkl} M_{ka}^{(\mathbb{Z})}
M_{l\alpha}- \sum_{j,k} A_{ij}^{-1} E_{jka}^{(\mathbb{Z})}
M_{k\alpha}-\sum_{j,k} A_{ij}^{-1} E_{jk\alpha} M_{ka}^{(\mathbb{Z})}-\sum_{j}
A_{ij}^{-1} D_{ja\alpha}^{(\mathbb{Z})}.
\end{equation}
\begin{equation}
X_{i\alpha\beta} = -\sum_{j,k,l} A_{ij}^{-1} C_{jkl} M_{k\alpha}
M_{l\beta}- \sum_{j,k} A_{ij}^{-1} E_{jk\alpha} M_{k\beta}-\sum_{j,k}
A_{ij}^{-1} E_{jk\beta} M_{k\alpha}-\sum_{j} A_{ij}^{-1}
D_{j\alpha\beta},
\end{equation}

If the likelihood function is Gaussian as in eq.(\ref{gauslike}) it
can be shown analytically that
$\hat{\pB}\left(\dB,\hat{\bm{\theta}}\right)=
\sum_\mathbb{Z}\tilde{\mathsf{M}}_\mathbb{Z}\left(\hat{\bm{\theta}}\right)\dB_\mathbb{Z}$
where
$\tilde{\mathsf{M}}_\mathbb{Z}\left(\bm{\theta}\right)\equiv\mathsf{S}
\left(\bm{\theta}\right)
\mathsf{W}_\mathbb{Z}^\mathrm{T}\left(\bm{\theta}\right)\mathsf{N}_\mathbb{Z}^{-1}$
and
\begin{equation}
\mathsf{S}\left(\bm{\theta}\right)\equiv\left[\lambda \mathsf{\Gamma}
  +\sum_\mathbb{Z}\mathsf{W}_\mathbb{Z}^\mathrm{T}\left(\bm{\theta}\right) 
\mathsf{N}_\mathbb{Z}^{-1}\mathsf{W}_\mathbb{Z}\left(\bm{\theta}\right)\right]^{-1}.
\end{equation}
The expression for $\hat{\pB}$ has the form of a sum over the data
sets weighted by the inverse covariance matrices, so that $\hat{\pB}$
preferentially fits more precise data sets. It follows that in this
case the estimated PPS is a linear function of the true PPS, with
\begin{equation}
\bm{\mathcal{T}}\left(\pB_\mathrm{t},\bm{\theta}_\mathrm{t},\hat{\bm{\theta}},\nB\right) =
\sum_\mathbb{Z}\tilde{\mathsf{M}}_\mathbb{Z}\left(\hat{\bm{\theta}}\right) 
\mathsf{W}_\mathbb{Z}\left(\bm{\theta}_\mathrm{t}\right)\pB_\mathrm{t}
+ \sum_\mathbb{Z}\tilde{\mathsf{M}}_\mathbb{Z}\left(\hat{\bm{\theta}}\right)\nB_\mathbb{Z}.
\end{equation}
Eq.(\ref{recerr}) then becomes
\begin{eqnarray}
  \hat{\mathrm{p}}_i\left(\dB,\hat{\bm{\theta}}\right) & = &
  \sum_{j}R_{ij}\,\mathrm{p}_{\mathrm{t}j} +\sum_{\mathbb{Z},a}
  M_{ia}^{(\mathbb{Z})}\,\mathrm{n}_a^{(\mathbb{Z})}+\sum_\alpha
  M_{i\alpha}\,\mathrm{u}_\alpha \nonumber\\ & &
  +\sum_{j,\alpha}Z_{ij\alpha}\,\Delta \mathrm{p}_j\,\mathrm{u}_\alpha
  +
  \sum_{\mathbb{Z},a,\alpha}X_{ia\alpha}^{(\mathbb{Z})}\,\mathrm{n}_a^{(\mathbb{Z})}\,\mathrm{u}_\alpha
  +
  \frac{1}{2}\sum_{\alpha,\beta}X_{i\alpha\beta}\,\mathrm{u}_\alpha\,\mathrm{u}_\beta
  +\ldots,
\end{eqnarray}
where
$\mathsf{M}_\mathbb{Z}=\tilde{\mathsf{M}}_\mathbb{Z}\left(\bm{\theta}_\mathrm{t}\right)$. It
was shown \cite{Ory:1995dc} that the row vectors of the first-order
resolution matrix each sum to unity for Gaussian distributed data with
our choice of roughness function, so the amplitude of the recovered
PPS is correctly scaled. This is a significant advantage of Tikhonov
regularisation over most other inversion methods.  The second-order
resolution matrix vanishes for Gaussian data. The bias is given by
\begin{equation}
\mathrm{Bias}\left(\hat{\mathrm{p}}_i\right) = 
\sum_{j}\left(R_{ij}-I_{ij}\right)\,\mathrm{p}_{\mathrm{t}j}+\frac{1}{2}\sum_{\alpha,\beta}
X_{i\alpha\beta}\, N_{\alpha\beta}+\ldots.
\end{equation}
Thus good resolution and low bias are complementary. The contribution
of the data noise to the frequentist covariance matrix is
\begin{equation}
  \mathsf{\Sigma}_\mathrm{F}=\mathsf{S}\left(\bm{\theta}_\mathrm{t}\right)
  \left[\sum_\mathbb{Z}\mathsf{W}_\mathbb{Z}^\mathrm{T}\left(\bm{\theta}_\mathrm{t}\right)\mathsf{N}_\mathbb{Z}^{-1} 
  \mathsf{W}_\mathbb{Z}\left(\bm{\theta}_\mathrm{t}\right)\right]
  \mathsf{S}^\mathrm{T}\left(\bm{\theta}_\mathrm{t}\right).
\end{equation}
Note that including additional data sets does not necessarily reduce
$\mathsf{\Sigma}_\mathrm{F}$.

\subsection{Bayesian inference \label{bayesian}}

A different approach to inversion uses Bayesian inference, in which
all unknown quantities are treated as random variables with
probability distributions that represent the uncertainty about their
values \cite{tarantola:2004dc}. It is informative to compare Tikhonov
regularisation with a Bayesian inversion method that incorporates our
\textit{a priori} knowledge (or bias) about the smoothness of the
PPS. Consider a two-stage hierarchical Bayes model in which the prior
distribution for the PPS $P\left(\pB|\tilde{\lambda}\right)$ is
conditional on a hyperparameter $\tilde{\lambda}$ that itself has a
hyperprior $P\left(\tilde{\lambda}\right)$. According to Bayes'
theorem, the joint posterior distribution
$P\left(\pB,\bm{\theta},\tilde{\lambda}|\dB\right)$ of the PPS,
background cosmological parameters, and the hyperparameter, given the
data, is:
\begin{equation}
P\left(\pB,\bm{\theta},\tilde{\lambda}|\dB\right) = 
\frac{\mathcal{L}\left(\pB,\bm{\theta}|\dB\right)P\left(\bm{\theta}\right)
  P\left(\pB|\tilde{\lambda}\right)P 
 \left(\tilde{\lambda}\right)}{P\left(\dB\right)}.
\end{equation} 
Here $P\left(\dB\right)$ is the prior distribution of the data and
\begin{equation} 
  P\left(\bm{\theta}\right)\propto\exp\left[-\frac{1}{2} 
  \left(\bm{\theta}-\hat{\bm{\theta}}\right)^\mathrm{T}\mathsf{U}^{-1} 
  \left(\bm{\theta}-\hat{\bm{\theta}}\right)\right]
\end{equation}
is the prior distribution of the background parameters which is
assumed to be Gaussian. A prior distribution which gives smoother
spectra a higher prior probability is
$P\left(\pB|\tilde{\lambda}\right)\propto\exp\left[-\tilde{\lambda}
  \mathrm{R}\left(\pB\right)/2\right]$, in which case larger values of
$\tilde{\lambda}$ penalise roughness more strongly. The marginalised
posterior distribution of the PPS is obtained by integrating over the
background parameters and the hyperparameter,
\begin{equation}
P\left(\pB|\dB\right)=\int
P\left(\pB,\bm{\theta},\tilde{\lambda}|\dB\right)\,
\mathrm{d}\bm{\theta}\mathrm{d}\tilde{\lambda}.
\end{equation}
The maximum \textit{a posteriori} estimate of the PPS,
$\hat{\pB}_\mathrm{MAP}$, is defined as the mode of
$P\left(\pB|\dB\right)$. Assuming that the posterior can be
approximated by a Gaussian, the mean of the distribution is $\langle
\pB \rangle=\hat{\pB}_\mathrm{MAP}$ and the Bayesian covariance matrix
is
\begin{equation}
  \langle\left(\pB-\langle \pB\rangle\right)\left(\pB-\langle
  \pB\rangle\right)^\mathrm{T}\rangle=\left.\mathsf{H}^{-1}\right|_{\hat{\pB}_\mathrm{MAP}},
\end{equation}  
where $H_{ij}\equiv -\partial^{2}\ln P\left(\pB|\dB\right)/\partial
\mathrm{p}_{i}\partial \mathrm{p}_{j}$.

Now, suppose that the value of $\tilde{\lambda}$ is known to be
$\hat{\lambda}$ \textit{a priori}, and that the uncertainty in the
background parameters is negligible compared to the uncertainty in the
data. Then $P\left(\tilde{\lambda}\right) =
\delta\left(\tilde{\lambda}-\hat{\lambda}\right)$ and
$P\left(\bm{\theta}\right)=\delta\left(\bm{\theta}-\hat{\bm{\theta}}\right)$.
With these priors maximising $P\left(\pB|\dB\right)$ is equivalent to
minimising $Q\left(\pB,\dB,\hat{\bm{\theta}},\hat{\lambda}\right)$, so
that $\hat{\pB}_\mathrm{MAP} =
\hat{\mathbf{p}}\left(\dB,\hat{\bm{\theta}},\hat{\lambda}\right)$. Hence
Tikhonov regularisation can be thought of as maximum \textit{a
  posteriori} estimation with $\hat{\lambda}=\lambda$ and negligible
uncertainty in the background parameters. In this case the Bayesian
covariance matrix is denoted by $\mathsf{\Sigma}_\mathrm{B}$. The
elements of the inverse covariance matrix are given by the Hessian of
$Q$ evaluated at
$\pB=\hat{\mathbf{p}}\left(\dB,\hat{\bm{\theta}},\lambda\right)$,
\begin{equation}
\mathsf{\Sigma}^{-1}_{\mathrm{B}_{|ij}}\equiv\left.\frac{1}{2}
\frac{\displaystyle\partial^{2}Q\left(\pB,\dB,\hat{\bm{\theta}},\lambda\right)}
       {\displaystyle\partial \mathrm{p}_i\partial \mathrm{p}_j}\right|_{\hat{\pB}}.
\label{sigb}
\end{equation}
For a given data set (or collection of data sets)
$\mathsf{\Sigma}_\mathrm{B}$ represents a lower bound on the Bayesian
covariance matrix since additional uncertainty in the hyperparameter
and the background parameters will increase the covariance. In general
the Bayesian and frequentist covariance matrices
$\mathsf{\Sigma}_\mathrm{B}$ and $\mathsf{\Sigma}_\mathrm{T}$ are
different due to their fundamentally different statistical
motivations. The elements of $\mathsf{\Sigma}_\mathrm{B}$ and
$\mathsf{\Sigma}_\mathrm{T}$ decrease as $\lambda$ increases,
reflecting the improved stability of the solution.  For data with
Gaussian errors,
$\mathsf{\Sigma}_\mathrm{B}=\mathsf{S}\left(\hat{\bm{\theta}}\right)$;
in this case adding additional data sets reduces
$\mathsf{\Sigma}_\mathrm{B}$, in accordance with our expectation.

\subsection{Performance statistics \label{performance}}

In estimation theory, `loss' functions are used to assess the
performance of an estimator: they quantify the `loss' incurred by
using $\hat{\pB}$ instead of the true $\pB_\mathrm{t}$. The most
common loss function is the squared-error (SE),
\begin{equation}
\label{SE}
\mathrm{SE}\left(\hat{\pB}\right)\equiv\left(\hat{\pB} -
\pB_\mathrm{t}\right)^\mathrm{T}\left(\hat{\pB}-\pB_\mathrm{t}\right).
\end{equation}
In what follows the SE is calculated from the wavenumber $k=10^{-4}\;
\mathrm{Mpc}^{-1}$ up to $k=0.5\; \mathrm{Mpc}^{-1}$. Alternatively,
the predictive error (PE), defined as PE$\left(\hat{\pB}\right)\equiv
L\left(\hat{\pB},\dB_\mathrm{t}\right)$, quantifies the ability of the
estimated spectrum to predict the noise-free data
$\dB_\mathrm{t}\equiv \mathsf{W}\pB_\mathrm{t}$. For multiple data
sets with Gaussian likelihood functions we have,
\begin{equation}
\mathrm{PE}\left(\hat{\pB}\right) =
\sum_\mathbb{Z}\left(\hat{\pB}-\pB_\mathrm{t}\right)^\mathrm{T}\mathsf{W}_\mathbb{Z}^\mathrm{T}
\mathsf{N}_\mathbb{Z}^{-1}\mathsf{W}_\mathbb{Z}\left(\hat{\pB}-\pB_\mathrm{t}\right).
\end{equation}
This loss function emphasises errors on scales where the data more
tightly constrain the PPS.

Taking the ensemble average of a loss function produces the associated
`risk' function. Here this gives the mean squared error (MSE):
\begin{eqnarray}
\label{eqmse}
  \mathrm{MSE}\left(\hat{\pB}\right) & \equiv &
  \langle\left(\hat{\pB}-\pB_\mathrm{t}\right)^\mathrm{T}\left(\hat{\pB}-\pB_\mathrm{t}\right)\rangle = 
  \langle\hat{\pB}-\pB_\mathrm{t}\rangle^\mathrm{T}\langle\hat{\pB}-\pB_\mathrm{t}\rangle + 
  \langle\left(\hat{\pB}-\langle
  \hat{\pB}\rangle\right)^\mathrm{T}\left(\hat{\pB}-\langle
  \hat{\pB}\rangle\right)\rangle,
\end{eqnarray} 
which combines the bias squared and the variance of
$\hat{\pB}$.  The mean predictive error (MPE) is defined as:
\begin{eqnarray}
\label{eqmpe}
\mathrm{MPE}\left(\hat{\pB}\right)
  & \equiv & \langle L\left(\hat{\pB},\dB_\mathrm{t}\right)\rangle.
\end{eqnarray}

\section{Data \label{data}} 

The primordial curvature perturbations generate observable CMB
temperature and electric (E-mode) polarisation anisotropies,
characterised by the TT, TE and EE angular power spectra
$\mathcal{C}_\ell^\mathrm{TT}$, $\mathcal{C}_\ell^\mathrm{TE}$ and
$\mathcal{C}_\ell^\mathrm{EE}$ respectively. These are written in
terms of the spherical harmonic coefficients $a^x_{\ell m}$ as
\begin{equation}
\mathcal{C}_\ell^{X}\equiv\frac{\ell\left(\ell+1\right)}{2\pi}\langle
a^{x*}_{\ell m} a^{x^\prime}_{\ell m} \rangle,
\end{equation}       
where $X= \left(xx^\prime\right)=$ TT, TE and EE. Throughout we consider
only scalar perturbations so that magnetic ($B$-mode) polarisation is
absent (as is expected for generic small-field non-fine-tuned
inflationary potentials
\cite{Hotchkiss:2008sa,Bird:2008cp}). Neglecting nonlinear secondary
effects, each angular power spectrum is related to the PPS by the
integral equation
\begin{equation}
\mathcal{C}_\ell^{X}=4\pi\int_0^\infty\frac{\mathrm{d}k}{k}T^{x*}_\ell 
\left(k\right)T^{x^\prime}_\ell\left(k\right)\mathcal{P_R}\left(k\right).
\label{cmb}
\end{equation}
Here the $T^x_\ell\left(k\right)$ are angular transfer functions
dependent on the background parameters. It follows that
$\mathrm{d}^X_\ell=\mathrm{s}^X_\ell+\mathrm{n}^X_\ell$, where
$\mathrm{d}^X_\ell$ are the measured angular power spectra,
$\mathrm{s}^X_\ell=\sum_i W^X_{\ell i}\mathrm{p}_i$ are the
theoretical power spectra and $\mathrm{n}^X_\ell$ are the noise
vectors for the TT, TE and EE spectra. The matrices $\mathsf{W}^X$
which derive from the discretisation of eq.(\ref{cmb}) are obtained
using a modified version of the CAMB cosmological Boltzmann code
\cite{camb,Lewis:2002ah}.

\subsection{Experimental data sets}

We present the data sets used in this paper. Throughout we treat the
data in the manner recommended in the cited papers.

\subsubsection{WMAP \label{wmap}}

The WMAP team provide a software package which returns the likelihood
of a set of angular power spectra.  The temperature likelihood
function consists of three components,
$L_\mathrm{TT}=L_\mathrm{Gibbs}+L_\mathrm{pTT}+L_\mathrm{bps}$, as
explained below. For multipoles $\ell\leq32$ the likelihood
$L_\mathrm{Gibbs}$ is evaluated using a Blackwell-Rao estimator on
samples drawn by Gibbs sampling from the joint posterior distribution
of the power spectrum and the true sky signal \cite{Dunkley:2008ie}. A
hybrid Gaussian and offset log-Gaussian fitting formula is used for
$33\leq\ell\leq1000$ \cite{Verde:2003ey},
\begin{eqnarray}
L_\mathrm{pTT} & = &\sum_{\ell\ell^\prime}\frac{1}{3}
\left(\mathrm{s}_\ell^\mathrm{TT}-\mathrm{d}_\ell^\mathrm{TT}\right)
\left(N^\mathrm{TT}\right)_{\ell\ell^\prime}^{-1}
\left(\mathrm{s}_\ell^\mathrm{TT}-\mathrm{d}_\ell^\mathrm{TT}\right)\nonumber
\\ & & + \sum_{\ell\ell^\prime}\frac{2}{3}\ln
\left(\frac{\mathcal{S}_\ell^\mathrm{TT}}{\mathcal{D}_\ell^\mathrm{TT}}\right)
\mathcal{S}_\ell^\mathrm{TT}
\left(N^\mathrm{TT}\right)_{\ell\ell^\prime}^{-1}\mathcal{S}_{\ell^\prime}^\mathrm{TT}
\ln\left(\frac{\mathcal{S}_{\ell^\prime}^\mathrm{TT}}{\mathcal{D}_{\ell^\prime}^\mathrm{TT}}\right).
\label{eqptt}
\end{eqnarray}
Here $\mathcal{S}_\ell^\mathrm{TT}\equiv
\mathrm{s}_\ell^\mathrm{TT}+\mathcal{N}_\ell^\mathrm{TT}$ and
$\mathcal{D}_\ell^\mathrm{TT}\equiv
\mathrm{d}_\ell^\mathrm{TT}+\mathcal{N}_\ell^\mathrm{TT}$ where
$\mathcal{N}_\ell^\mathrm{TT}$ is the noise spectrum. The measurements
$\mathrm{d}_\ell^\mathrm{TT}$ were obtained using a pseudo-$C_\ell$
estimator (given that only a fraction $f_\ell$ of the sky is
unmasked). The diagonal elements of the inverse TT covariance matrix
are
\begin{equation}
 \left(N^\mathrm{TT}\right)_{\ell\ell}^{-1}=
\frac{\left(2\ell+1\right)\left(f_\ell^\mathrm{TT}\right)^2}{2
\left(\mathcal{S}_\ell^\mathrm{TT}\right)^2}.
\end{equation}
The WMAP results are also affected by uncertainties in both instrumental
beam reconstruction and extragalactic point source subtraction
\cite{Hinshaw:2006ia}. This leads to the final component $L_\mathrm{bps}$
which we approximate by
\begin{equation}
L_\mathrm{bps}=-\sum_{\ell\ell^\prime}
\left(\mathrm{s}_\ell^\mathrm{TT}-\mathrm{d}_\ell^\mathrm{TT}\right)
\left(N^\mathrm{bps}\right)_{\ell\ell^\prime}^{-1}
\left(\mathrm{s}_\ell^\mathrm{TT}-\mathrm{d}_\ell^\mathrm{TT}\right).
\label{eqbps}
\end{equation} 
The covariance matrix $N^\mathrm{bps}$ is evaluated for a fiducial TT
spectrum (PPS with $n_\mathrm{s}=0.969$).

The polarisation likelihood is given by
$L_\mathrm{pol}=L_\mathrm{pix}+L_\mathrm{pTE}$ where $L_\mathrm{pix}$ is the
likelihood given the TE/EE/BB data for $\ell\leq23$ and
$L_\mathrm{pTE}$ is the TE likelihood for $24\leq\ell\leq450$. The
low-$\ell$ likelihood is evaluated directly from low resolution
pixelised sky maps \cite{Page:2006hz}. The high-$\ell$ likelihood is
approximated by a Gaussian \cite{Verde:2003ey},
\begin{equation}
L_\mathrm{pTE}=\sum_{\ell\ell^\prime}
\left(\mathrm{s}_\ell^\mathrm{TE}-\mathrm{d}_\ell^\mathrm{TE}\right)
\left(N^\mathrm{TE}\right)_{\ell\ell^\prime}^{-1}
\left(\mathrm{s}_\ell^\mathrm{TE}-\mathrm{d}_\ell^\mathrm{TE}\right) - 
\sum_\ell\ln\left(N^\mathrm{TE}\right)_{\ell\ell}^{-1}.
\label{eqpte}
\end{equation} 
The diagonal elements of the inverse TE covariance matrix are
\begin{equation}
 \left(N^\mathrm{TE}\right)_{\ell\ell}^{-1} =
 \frac{\left(2\ell+1\right)
 \left(f_\ell^{TE}\right)^2}{\mathcal{S}_\ell^\mathrm{TT,TE}
\mathcal{S}_\ell^\mathrm{EE} + 
 \left(\mathrm{s}_\ell^\mathrm{TE}\right)^2}\ ,
\end{equation}
where $\mathcal{S}_\ell^\mathrm{TT,TE}\equiv
\mathrm{s}_\ell^\mathrm{TT}+\mathcal{N}_\ell^\mathrm{TT,TE}$, and
$\mathcal{S}_\ell^\mathrm{EE}\equiv
\mathrm{s}_\ell^\mathrm{EE}+\mathcal{N}_\ell^\mathrm{EE}$. 

When first-order derivatives of the likelihood function are calculated
in the BFGS minimisation, the full Gibbs sampling and pixel-based
functions are used, but the pseudo-$C_\ell$ approximations
$L_\mathrm{pTT}$ and $L_\mathrm{pTE}$ are employed to evaluate the
Hessian matrix (see Appendix~\ref{matrix}).

\subsubsection{Small angular scale CMB experiments \label{groundbased}}

We combine the WMAP results with data from a number of ground-based
small angular scale CMB experiments. These include the Very Small
Array (VSA) \cite{Dickinson:2004yr}, the Arcminute Cosmology Bolometer
Array Receiver (ACBAR) \cite{Reichardt:2008ay}, the Cosmic Background
Imager (CBI) \cite{Sievers:2009ah}, BOOMERanG
\cite{Jones:2005yb,Piacentini:2005yq,Montroy:2005yx} and QUaD
\cite{Brown:2009uy}. In particular the `NA pipeline' BOOMERanG results
and the `pipeline $1$' QUaD results are used. All the ground-based
experiments quote bandpower estimates $\mathrm{d}^X_b$ due to their lower sky
coverage. The predicted bandpowers are given by
\begin{equation}
\mathrm{s}^X_b = 
\sum_\ell W^X_{b\ell} \mathrm{s}^X_\ell= \sum_i T^X_{bi} \mathrm{p}_i,
\end{equation}
where the rows of $\mathsf{W}^X$ are bandpower window functions and
$T^X_{bi}\equiv\sum_\ell W^X_{b\ell} T^X_{\ell i}$. To ensure that the
WMAP and small angular scale data sets are statistically independent, we use
only TT bandpowers with mean multipoles of $\ell>750$ and TE
bandpowers with $\ell>350$. In addition, bandpowers with $\ell>2000$
are discarded in order to avoid scales which are affected by 
secondary processes such as gravitational lensing and the
Sunyaev-Zeldovich effect.

The likelihood functions of the small angular scale data sets,
$L_\mathrm{ss}$, are either Gaussian or log-Gaussian in the
bandpowers.  The likelihoods are modified to include beam and
calibration uncertainties and have the form
\begin{equation}
L_\mathrm{ss}=\sum_{X,X^\prime,b,b^\prime}
\mathcal{Z}^X_b\left(V^{XX^\prime}\right)^{-1}_{bb^\prime}\mathcal{Z}^{X^\prime}_{b^\prime}.
\label{ssl}
\end{equation}
Here $\mathcal{Z}^X_b = \mathrm{s}_b^X - \mathrm{d}_b^X$ for a
Gaussian bandpower and
$\mathcal{Z}^X_b=\ln\left(\mathcal{S}_b^X/\mathcal{D}_b^X\right)$ for
a log-Gaussian bandpower, with $\mathcal{S}_b^X\equiv
\mathrm{s}_b^X+\mathcal{N}_b^X$ and $\mathcal{D}_b^X\equiv
\mathrm{d}_b^X+\mathcal{N}_\mathrm{b}^X$. The weight matrices
$\mathsf{V}^{XX^\prime}$ are given by
$\left(V^{XX^\prime}\right)^{-1}_{bb^\prime} =
D_b^X\left(\tilde{N}^{XX^\prime}\right)^{-1}_{bb^\prime}D_{b^\prime}^{X^\prime}$
where for a Gaussian bandpower $D_b^X=1$ and for a log-Gaussian
bandpower $D_b^X=\mathcal{D}_b^X$. For a Gaussian beam of width
$\theta$ the matrices $\tilde{\mathsf{N}}^{XX^\prime}$ are related to
the bandpower covariance matrices $\mathsf{N}^{XX^\prime}$ by
\begin{equation}
\tilde{N}^{XX^\prime}_{bb^\prime}=N^{XX^\prime}_{bb^\prime}+\sigma_\mathrm{cal}^2
\mathrm{d}^X_b
\mathrm{d}^{X^\prime}_{b^\prime}+2\sigma_{\theta}\theta\ell^X_b\ell^{X^\prime}_{b^\prime}
\mathrm{d}^X_b \mathrm{d}^{X^\prime}_{b^\prime},
\end{equation}
where $\sigma_\mathrm{cal}$ is the calibration error,
$\sigma_{\theta}$ is the error in the beam width and $\ell^X_b$ is the
mean multipole of the bandpower labelled by $X$ and $b$. Using
$\tilde{\mathsf{N}}^{XX^\prime}$ instead of $\mathsf{N}^{XX^\prime}$
in eq.(\ref{ssl}) approximates marginalising over nuisance parameters
associated with the calibration and beam uncertainties
\cite{Bridle:2001zv}.

All the bandpowers of the VSA \cite{Dickinson:2004yr}, ACBAR
\cite{Reichardt:2008ay} and CBI \cite{Sievers:2009ah} data sets
consist of TT measurements alone and are taken to be log-Gaussian
distributed. BOOMERanG and QUaD measure both temperature and
polarisation anisotropies. For BOOMERanG the TT bandpowers
\cite{Jones:2005yb} are log-Gaussian while the TE
\cite{Piacentini:2005yq} and EE \cite{Montroy:2005yx} bandpowers are
Gaussian. The QUaD TT and EE bandpowers are log-Gaussian while the TE
bandpowers are Gaussian \cite{Brown:2009uy}. We use the calibration
and beam errors reported by the experimental teams.

\subsubsection{SDSS \label{sdss}}

We use the power spectrum obtained from the luminous red galaxy (LRG)
sample of the Sloan Digital Sky Survey fourth data release (SDSS-4)
\cite{Tegmark:2006az}. The power spectrum measurements extend into the
small-scale nonlinear regime. We use the nonlinear modelling
prescription of \cite{Tegmark:2006az} to calculate the theoretical
galaxy power spectrum $\mathcal{P}_g\left(k\right)$. Baryon acoustic
oscillations in the power spectrum are suppressed on small-scales by
nonlinear structure formation \cite{Eisenstein:2006nj}. This effect is
modelled by the power spectrum
\begin{equation}
\mathcal{P}_\mathrm{suposc}\left(k\right) \equiv 
\left\{e^{-\frac{1}{2}\left(\frac{k}{k_*}\right)^2}
\left[T_\mathrm{osc}^2\left(k\right)-T_\mathrm{noosc}^2
\left(k\right)\right]+T_\mathrm{noosc}^2\left(k\right)\right\}
\mathcal{P_R}\left(k\right),
\end{equation} 
which smoothly interpolates between the linear power spectrum,
$\mathcal{P}_\mathrm{osc}\left(k\right) = T_\mathrm{osc}^2
\left(k\right)\mathcal{P_R}\left(k\right)$,
on large scales and an oscillation-free spectrum,
$\mathcal{P}_\mathrm{noosc}\left(k\right) = 
T_\mathrm{noosc}^2\left(k\right)\mathcal{P_R}\left(k\right)$,
on small scales. Here $T_\mathrm{osc}\left(k\right)$ is the linear
matter transfer function, which we obtain from CAMB, and
$T_\mathrm{noosc}\left(k\right)$ is the analytic transfer function
without oscillations \cite{Eisenstein:1997ik,Eisenstein:1997jh}. The
suppression occurs for wavenumbers $k\gtrsim k_*$ where
\begin{equation}
k_*^{-1}\equiv c D\left(1+f\right)^{1/3}\mathcal{P}_{0.05}^{1/2}.
\end{equation}
Here $D$ is the linear growth factor, $f$ is the linear growth rate
(both evaluated at the mean redshift $z=0.35$ of the LRG sample) and
$c$ is a constant equal to $6.19$ $h^{-1}$ Mpc. For a featureless PPS
$\mathcal{P}_{0.05}$ would be equal to $\mathcal{P_R}$ evaluated at
$k=0.05$ Mpc$^{-1}$ but instead we take it to be the mean amplitude of
the PPS over the interval $0.025<k<0.075$ Mpc$^{-1}$. The Q model
\cite{Cole:2005sx,Hamann:2008we} describes scale-dependent bias and
the enhancement of small-scale power due to nonlinear evolution. It is
used to obtain the galaxy power spectrum from
$\mathcal{P}_\mathrm{suposc}\left(k\right)$:
\begin{equation}
\mathcal{P}_g\left(k\right) = 
b^2_\mathrm{LRG}\frac{1+Q_\mathrm{nl}k^2}{1+A_\mathrm{nl}k}
\mathcal{P}_\mathrm{suposc}\left(k\right),
\end{equation}
where $b_\mathrm{LRG}$ is the scale-independent LRG bias and
$Q_\mathrm{nl}$ and $A_\mathrm{nl}$ are empirical fitting parameters.

The measurement of the LRG power spectrum required a cosmological
model to convert redshift observations into comoving distances. This
was taken to be a fiducial $\Lambda$CDM model. To determine the power
spectrum for a different model the fiducial data points can be
rescaled rather than repeating the entire analysis for the new
model. However it is more convenient to transform instead the
theoretical power spectrum before comparing it with the fiducial
data. The transformed spectrum is
\begin{equation}
\mathcal{P}_g^T\left(k\right)\equiv \frac{1}{\gamma^3}
\mathcal{P}_g\left(\frac{k}{\gamma}\right),
\end{equation}
where the scaling factor $\gamma$ is
\begin{equation}
\gamma \equiv
\left(\frac{D_\mathrm{A}^2 H_\mathrm{fid}}{D_\mathrm{A,fid}^2 H}\right)^{1/3}.
\label{scale}
\end{equation}
Here $D_\mathrm{A}$ is the angular diameter distance and $H$ is the
Hubble parameter, both evaluated at redshift $z=0.35$, and the
subscript `fid' refers to the fiducial model quantities. This
transformation of the spectrum is equivalent to transforming the
window functions $W_a\left(k\right)$ of the data points into
$\mathcal{W}_a\left(k\right)\equiv \gamma^{-2} W_a\left(\gamma
k\right)$. Hence the data points are given by
\begin{equation}
\mathrm{d}_a = \int^{\infty}_0
\mathcal{W}_a\left(k\right)\mathcal{P}_g\left(k\right)\,\mathrm{d}k + 
\mathrm{n}_a.
\end{equation}  
Discretising the above integral gives $s_a=\sum_a
W^\mathrm{LRG}_{ai}\mathrm{p}_i$. The likelihood function is
\begin{equation}
L_\mathrm{LRG}=\sum_{a,a^\prime}
\left(s_a-\mathrm{d}_a\right)\left(\tilde{N}^\mathrm{LRG}\right)^{-1}_{aa^\prime}
\left(s_{a^\prime}-d_{a^\prime}\right)
+\ln\det\tilde{N}^\mathrm{LRG},
\label{eqlrg}
\end{equation}
where
\begin{equation}
\tilde{N}^\mathrm{LRG}_{aa^\prime}=N^\mathrm{LRG}_{aa^\prime}+\sigma_\mathrm{cal}^2
s_a s_{a^\prime}.
\end{equation}

\subsection{Likelihood function derivatives \label{matrix}}

Next we present the likelihood function derivatives used in the calculation 
of $\hat{\pB}$. While the first-order derivatives must be exact, 
only approximations to the second-order derivatives are required by the 
BFGS algorithm used --- the ones listed below were found to be satisfactory.  

\subsubsection{WMAP}

The first-order derivative of the Gibbs sampler component of the WMAP
temperature likelihood function is

\begin{equation}
\frac{\partial L_\mathrm{Gibbs}}{\partial \mathrm{p}_\alpha}=\sum_\ell
W^\mathrm{TT}_{\ell\alpha}\frac{\partial L_\mathrm{Gibbs}}{\partial
  \mathrm{s}_\ell^\mathrm{TT}}.
\end{equation} 
The derivative $\partial L_\mathrm{Gibbs}/\partial \mathrm{s}_\ell^\mathrm{TT}$
is calculated numerically. The first-order derivative of the
high-$\ell$ component is

\begin{eqnarray}
  \frac{\partial L_\mathrm{pTT}}{\partial \mathrm{p}_\alpha} & = &
  \frac{2}{3}\sum_{\ell\ell^\prime}W^\mathrm{TT}_{\ell\alpha}\left(N^\mathrm{TT}\right)^{-1}_{\ell\ell^\prime}\left(\mathrm{s}^\mathrm{TT}_{\ell^\prime}-\mathrm{d}^\mathrm{TT}_{\ell^\prime}\right)+\frac{1}{3}\sum_\ell
  \frac{\partial \left(N^\mathrm{TT}\right)^{-1}_{\ell\ell}}{\partial
    \mathrm{p}_\alpha}\left(\mathrm{s}^\mathrm{TT}_\ell-\mathrm{d}^\mathrm{TT}_\ell\right)^2
  \nonumber\\ & &
  +\frac{4}{3}\sum_{\ell\ell^\prime}\ln\left(\frac{\mathcal{S}_\ell^\mathrm{TT}}{\mathcal{D}_\ell^\mathrm{TT}}\right)
  W_{\ell\alpha}^\mathrm{TT}\left(N^\mathrm{TT}\right)_{\ell\ell^\prime}^{-1}\mathcal{S}_{\ell^\prime}^\mathrm{TT}\ln\left(\frac{\mathcal{S}_{\ell^\prime}^\mathrm{TT}}{\mathcal{D}_{\ell^\prime}^\mathrm{TT}}\right)\nonumber\\ &
  & +\frac{4}{3}\sum_{\ell\ell^\prime}
  W_{\ell\alpha}^\mathrm{TT}\left(N^\mathrm{TT}\right)_{\ell\ell^\prime}^{-1}\mathcal{S}_{\ell^\prime}^\mathrm{TT}\ln\left(\frac{\mathcal{S}_{\ell^\prime}^\mathrm{TT}}{\mathcal{D}_{\ell^\prime}^\mathrm{TT}}\right)
  \nonumber\\ & &
+\frac{2}{3}\sum_\ell
  \frac{\partial \left(N^\mathrm{TT}\right)^{-1}_{\ell\ell}}{\partial
    \mathrm{p}_\alpha}\left(\mathcal{S}^\mathrm{TT}_\ell\right)^2\left[\ln\left(\frac{\mathcal{S}_\ell^\mathrm{TT}}{\mathcal{D}_\ell^\mathrm{TT}}\right)\right]^2,
\end{eqnarray} 
where
\begin{equation}
\frac{\partial \left(N^\mathrm{TT}\right)^{-1}_{\ell\ell}}{\partial
  \mathrm{p}_\alpha}=-\frac{\left(2\ell+1\right)\left(f_\ell^\mathrm{TT}\right)^2
  W^\mathrm{TT}_{\ell\alpha}}{\left(\mathcal{S}_\ell^\mathrm{TT}\right)^3}.
\end{equation}
The first-order derivative of the beam and point source component is
\begin{eqnarray}
\frac{\partial L_\mathrm{bps}}{\partial \mathrm{p}_\alpha} =
-2\sum_{\ell\ell^\prime}W^\mathrm{TT}_{\ell\alpha}\left(N^\mathrm{bps}\right)^{-1}_{\ell\ell^\prime}\left(\mathrm{s}^\mathrm{TT}_{\ell^\prime}- \mathrm{d}^\mathrm{TT}_{\ell^\prime}\right).
\end{eqnarray} 
The second-order derivative of the WMAP TT likelihood is
approximated by
\begin{equation}
\frac{\partial^2 L_\mathrm{TT}}{\partial \mathrm{p}_\alpha \partial \mathrm{p}_\beta}
\simeq
2\sum_{\ell\ell^\prime}W^\mathrm{TT}_{\ell\alpha}\left(N^\mathrm{TT}\right)^{-1}_{\ell\ell^\prime}W^\mathrm{TT}_{\ell\beta}.
\end{equation} 
The first-order derivative of the low-$\ell$ pixel-based component of
the WMAP polarisation likelihood is
\begin{equation}
\frac{\partial L_\mathrm{pix}}{\partial \mathrm{p}_\alpha}=\sum_{\ell X}
W^X_{\ell\alpha}\frac{\partial L_\mathrm{pix}}{\partial \mathrm{s}_\ell^X},
\end{equation} 
where $\partial L_\mathrm{pix}/\partial \mathrm{s}_\ell^X$ is evaluated
numerically. The first-order derivative of the high-$\ell$ component
of the polarisation likelihood is
\begin{eqnarray}
\frac{\partial L_\mathrm{pTE}}{\partial \mathrm{p}_\alpha} & = &
2\sum_{\ell\ell^\prime}W^\mathrm{TE}_{\ell\alpha}\left(N^\mathrm{TE}\right)^{-1}_{\ell\ell^\prime}\left(\mathrm{s}^\mathrm{TE}_{\ell^\prime}-\mathrm{d}^\mathrm{TE}_{\ell^\prime}\right)
+\sum_\ell \frac{\partial \left(N^\mathrm{TE}\right)^{-1}_{\ell\ell}}{\partial \mathrm{p}_\alpha}\left(\mathrm{s}^\mathrm{TE}_\ell-\mathrm{d}^\mathrm{TE}_\ell\right)^2
  \nonumber\\ & &
-\sum_\ell N^\mathrm{TE}_{\ell\ell}\frac{\partial \left(N^\mathrm{TE}\right)^{-1}_{\ell\ell}}{\partial \mathrm{p}_\alpha}, 
\end{eqnarray} 
where
\begin{equation}
\frac{\partial \left(N^\mathrm{TE}\right)^{-1}_{\ell\ell}}{\partial \mathrm{p}_\alpha}  = 
-\frac{\left(2\ell+1\right)\left(f_\ell^\mathrm{TT}\right)^2\left(W^\mathrm{EE}_{\ell\alpha}\mathcal{S}_\ell^\mathrm{TT, TE}+W^\mathrm{TT}_{\ell\alpha}\mathcal{S}_\ell^\mathrm{EE}+
2W^\mathrm{TE}_{\ell\alpha}\mathrm{s}_\ell^\mathrm{TE}\right)}{2\left(\mathcal{S}_\ell^\mathrm{TT}\right)^2\left[\mathcal{S}_\ell^\mathrm{TT, TE}\mathcal{S}_\ell^\mathrm{EE}+\left(\mathrm{s}_\ell^\mathrm{TE}\right)^2\right]}.
\end{equation}
The second-order derivative of the WMAP polarisation likelihood is
approximated by
\begin{equation}
\frac{\partial^2 L_\mathrm{pol}}{\partial \mathrm{p}_\alpha \partial \mathrm{p}_\beta} \simeq
2\sum_{\ell\ell^\prime}W^\mathrm{TE}_{\ell\alpha}\left(N^\mathrm{TE}\right)^{-1}_{\ell\ell^\prime}W^\mathrm{TE}_{\ell\beta}+\sum_\ell
\left(N^\mathrm{TE}_{\ell\ell}\right)^2\frac{\partial
  \left(N^\mathrm{TE}\right)^{-1}_{\ell\ell}}{\partial
  \mathrm{p}_\alpha}\frac{\partial
  \left(N^\mathrm{TE}\right)^{-1}_{\ell\ell}}{\partial \mathrm{p}_\beta}.
\end{equation} 
The first-order derivative of the low-$\ell$ one-dimensional Wishart
likelihood used with mock WMAP TT data alone is
\begin{eqnarray}
\frac{\partial L_\mathrm{W1}}{\partial \mathrm{p}_\alpha} =
2\sum_{\ell}W^\mathrm{TT}_{\ell\alpha}\left(N^\mathrm{TT}\right)^{-1}_{\ell\ell}\left(\mathrm{s}^\mathrm{TT}_\ell-\mathrm{d}^\mathrm{TT}_\ell\right).
\end{eqnarray} 
The first-order derivative of the low-$\ell$ two-dimensional Wishart
likelihood used with mock WMAP temperature and polarisation data is (see eq.\,\ref{wish2})
\begin{eqnarray}
\frac{\partial L_\mathrm{W2}}{\partial \mathrm{p}_\alpha} & = &
\sum_{\ell}n_\mathrm{dof}\left\{\frac{\left(\mathcal{S}_\ell^\mathrm{EE}+\mathcal{D}_\ell^\mathrm{EE}\right)W^\mathrm{TT}_{\ell\alpha}+\left(\mathcal{S}_\ell^\mathrm{TT}+\mathcal{D}_\ell^\mathrm{TT}\right)W^\mathrm{EE}_{\ell\alpha}-2\left(\mathrm{s}_\ell^\mathrm{TE}+\mathrm{d}_\ell^\mathrm{TE}\right)W^\mathrm{TE}_{\ell\alpha}}{\mathcal{S}_\ell^\mathrm{TT}\mathcal{S}_\ell^\mathrm{EE}-\left(\mathrm{s}_\ell^\mathrm{TE}\right)^2}\right. \nonumber
\\ & &
\left.+\frac{\mathcal{S}_\ell^\mathrm{TT}\mathcal{D}_\ell^\mathrm{EE}+\mathcal{D}_\ell^\mathrm{TT}\mathcal{S}_\ell^\mathrm{EE}-2\mathrm{s}_\ell^\mathrm{TE}\mathrm{d}_\ell^\mathrm{TE}}{\left[\mathcal{S}_\ell^\mathrm{TT}\mathcal{S}_\ell^\mathrm{EE}-\left(\mathrm{s}_\ell^\mathrm{TE}\right)^2\right]^2}
\left(2\mathrm{s}_\ell^\mathrm{TE}W^\mathrm{TE}_{\ell\alpha}-\mathcal{S}_\ell^\mathrm{EE}W^\mathrm{TT}_{\ell\alpha}-\mathcal{S}_\ell^\mathrm{TT}W^\mathrm{EE}_{\ell\alpha}\right)\right\}.
\end{eqnarray} 

\subsubsection{Small angular scale CMB experiments}

The first-order derivative of the small angular scale CMB likelihood
is
\begin{eqnarray}
\frac{\partial L_\mathrm{ss}}{\partial \mathrm{p}_\alpha} & =&
2\sum_{X,X^\prime,b,b^\prime}
\mathcal{Z}^X_b\left(V^{XX^\prime}\right)^{-1}_{bb^\prime}\frac{\partial\mathcal{Z}^{X^\prime}_{b^\prime}}{\partial
  \mathrm{p}_\alpha},
\end{eqnarray}
where
\begin{equation}
\frac{\partial\mathcal{Z}^X_b}{\partial \mathrm{p}_\alpha}=\left\{\begin{array}{ll}
T^X_{b\alpha},& \mbox{Gaussian bandpower,}  \\
T^X_{b\alpha}/\mathcal{S}_b^X,&  \mbox{log-Gaussian bandpower.} 
\end{array}\right.
\end{equation}
The second-order derivative of the small angular scale CMB likelihood
is approximated by
\begin{equation}
\frac{\partial^2 L_\mathrm{ss}}{\partial \mathrm{p}_\alpha \partial \mathrm{p}_\beta} \simeq 
2\sum_{X,X^\prime,b,b^\prime} \frac{\partial\mathcal{Z}^X_b}{\partial \mathrm{p}_\alpha}\left(V^{XX^\prime}\right)^{-1}_{bb^\prime}\frac{\partial\mathcal{Z}^{X^\prime}_{b^\prime}}{\partial \mathrm{p}_\alpha}.
\end{equation}

\subsubsection{SDSS}

The SDSS-4 LRG likelihood can be rewritten as \cite{Bridle:2001zv}
\begin{equation}
L_\mathrm{LRG}=\sum_{a,a^\prime} \left(\mathrm{s}_a-\mathrm{d}_a\right)\left(N^\mathrm{LRG}\right)^{-1}_{aa^\prime}\left(\mathrm{s}_{a^\prime}-\mathrm{d}_{a^\prime}\right)
-\frac{A^2}{C}+\ln\left(1+\sigma_\mathrm{cal}^2 B\right)
+\ln\det N^\mathrm{LRG},
\end{equation}
where
\begin{equation}
A \equiv \sum_{a,a^\prime} \mathrm{s}_a\left(N^\mathrm{LRG}\right)^{-1}_{aa^\prime}\left(\mathrm{s}_{a^\prime}-\mathrm{d}_{a^\prime}\right), \qquad
B \equiv \sum_{a,a^\prime}\mathrm{s}_a\left(N^\mathrm{LRG}\right)^{-1}_{aa^\prime}\mathrm{s}_{a^\prime}, \qquad
C \equiv \frac{1+\sigma_\mathrm{cal}^2 B}{\sigma_\mathrm{cal}^2}.
\end{equation}
Hence the first-order derivative of SDSS-4 LRG likelihood is
\begin{eqnarray}
\frac{\partial L_\mathrm{LRG}}{\partial \mathrm{p}_\alpha} & =& 2\sum_{a,a^\prime} W^\mathrm{LRG}_{a\alpha}\left(N^\mathrm{LRG}\right)^{-1}_{aa^\prime}\left(\mathrm{s}_{a^\prime}-\mathrm{d}_{a^\prime}\right)-\frac{2A}{C}\sum_{a,a^\prime} W^\mathrm{LRG}_{a\alpha}\left(N^\mathrm{LRG}\right)^{-1}_{aa^\prime}\left(2\mathrm{s}_{a^\prime}-\mathrm{d}_{a^\prime}\right) \nonumber\\
& & +\frac{A^2+C}{C^2}\sum_{a,a^\prime} W^\mathrm{LRG}_{a\alpha}\left(N^\mathrm{LRG}\right)^{-1}_{aa^\prime}\mathrm{s}_{a^\prime}.
\end{eqnarray}
The second-order derivative of the SDSS-4 LRG likelihood is approximated by
\begin{equation}
\frac{\partial^2 L_\mathrm{LRG}}{\partial \mathrm{p}_\alpha \partial p_\beta} = 2\sum_{a,a^\prime} W^\mathrm{LRG}_{a\alpha}\left(\tilde{N}^{XX^\prime}\right)^{-1}_{aa^\prime}W^\mathrm{LRG}_{a^\prime\alpha}.
\end{equation}

\subsection{Test spectra \label{spectest}}

We test our method on four examples of primordial power spectra,
labelled spectra A to D, to determine how successfully they are
reconstructed from mock data sets. Spectrum A is the best-fit
power-law spectrum found by the WMAP team, with a spectral index of
$n_\mathrm{s}=0.963$ for the WMAP 7-year data \cite{Dunkley:2008ie}.

Spectrum B has the form of a `step' located at $k=k_0$ with
superimposed damped high frequency oscillations. It is generated by a
toy model of inflation with a sharp, discontinuous change in the slope
of the inflaton potential from a value of $V^\prime_-$ to
$V^\prime_+$; The spectrum is obtained analytically as
\cite{Starobinsky:1992ts}:
\begin{eqnarray}
\mathcal{P_R}\left(k\right) & = &
\mathcal{P}_{\mathcal{R}0}\left(k\right)\left\{1-3\left(r-1\right)
\frac{1}{y}\left[\left(1-\frac{1}{y^2}\right)\sin
  2y+\frac{2}{y}\cos2y\right]\right.\nonumber\\ & &
\left.+\frac{9}{2}\left(r-1\right)^2\frac{1}{y^2}
\left(1+\frac{1}{y^2}\right)\left[1+\frac{1}{y^2}+
  \left(1-\frac{1}{y^2}\right)\cos2y-\frac{2}{y}\sin2y\right]\right\}.
\end{eqnarray}
Here $y\equiv k/k_0$, $r\equiv V^\prime_-/V^\prime_+$ and
$\mathcal{P}_{\mathcal{R}0}$ is the underlying power spectrum of the
model, which is taken to be a power-law with $n_\mathrm{s}=0.963$. We set
$k_0=2\times10^{-4}\;\mathrm{Mpc}^{-1}$ and $r=0.5$.

Spectrum C arises from an inflation model in which a scalar field
coupled to the inflaton performs damped oscillations about the origin
during inflation, beginning when the Hubble parameter falls below the
mass of the field. This causes the inflaton mass to decrease from much
greater than, to much less than, the Hubble parameter and leads to an
infrared cutoff in the spectrum. On intermediate scales the spectrum
has a number of irregular features due to a parametric resonance
effect. The parameters of the model \cite{Langlois:2004px} have the
values $q_\mathrm{in}=10$ and $m/H=2\sqrt3$.

Spectrum D is that of the `CHDM bump' model. This Einstein-de Sitter
(E-deS) universe has about 80\% cold, 10\% baryonic and 10\% hot dark
matter (in the form of 3 mass-degenerate neutrinos with mass $\sim
0.5$~eV) and fits both the CMB and large-scale structure data
\textit{without} need for a dominant dark energy component, if there a
suitably located `bump' in the power spectrum
\cite{Hunt:2007dn,Hunt:2008wp}. This can result e.g. from multiple
inflation in $N=1$ supergravity wherein `flat direction' scalar fields
undergo spontaneous symmetry breaking phase transitions, as the
universe cools during inflation \cite{Adams:1997de}. Before each phase
transition the field is confined at the origin by a thermal barrier,
but when the barrier disappears a Hubble-induced mass correction
drives the field rapidly to the global minimum of its potential which
is determined by stabilising higher dimensional operators. The
gravitational coupling between the inflaton and these flat direction
fields causes the inflaton mass to jump at each phase transition and
thus produces characteristic features in the power spectrum
\cite{Hunt:2004vt}. Here the bump has parameters
$k_1=0.03\;\mathrm{Mpc}^{-1}$ and $k_2=0.08\;\mathrm{Mpc}^{-1}$. It
arises from 2 phase transitions which begin about an e-fold of
inflation apart and cause fractional changes in the inflaton
mass-squared of $\Delta m_1^2=0.151$ and $\Delta
m_2^2=0.272$.\footnote{This bump is located at a slightly higher
  wavenumber than that of the best-fit model in \cite{Hunt:2008wp} to
  better illustrate the effects of including small angular scale data
  in the inversion.}

For spectra A to C the background cosmology is taken to be flat
$\Lambda$CDM with $\omega_\mathrm{b}=0.0223$,
$\omega_\mathrm{c}=0.104$, $h=0.73$, $\tau=0.088$ and
$b_\mathrm{LRG}=1.9$. The flat CHDM model of spectrum D has
$\omega_\mathrm{b}=0.0165$, neutrino fraction $f_\nu=0.127$, $h=0.42$,
$\tau=0.074$ and $b_\mathrm{LRG}=2.1$.

\subsection{Mock data \label{mockdata}}

From each test spectrum we generate mock data using the following
method. In the full sky limit the observed CMB multipoles are known to
follow a Wishart distribution, being a sum of squared Gaussian
spherical harmonic coefficients \cite{Percival:2006ss}. When mock WMAP
TT data alone is required, we follow \cite{Tocchini-Valentini:2005ja}
and draw samples from the distribution
\begin{equation}
-2 \ln
P_\mathrm{W1}\left(\left\{\mathrm{d}_\ell^\mathrm{TT}\right\}|\left\{\mathrm{s}_\ell^\mathrm{TT}\right\}\right)
=
\sum_{\ell}n_\mathrm{dof}\left(\ln\mathcal{S}_\ell^\mathrm{TT}-\frac{n_\mathrm{dof}-2}{n_\mathrm{dof}}
\ln\mathcal{D}_\ell^\mathrm{TT} +
\frac{\mathcal{D}_\ell^\mathrm{TT}}{\mathcal{S}_\ell^\mathrm{TT}}\right),
\label{wish1}
\end{equation}
up to a constant, where
$n_\mathrm{dof}=\left(2\ell+1\right)f_\ell^\mathrm{TT}$. This is a
one-dimensional Wishart distribution (i.e. a $\chi^2$
distribution). The samples have the statistical properties:
\begin{equation}
\langle \mathrm{d}_\ell^\mathrm{TT}-\mathrm{s}_\ell^\mathrm{TT}\rangle=0, \qquad \langle
\left(\mathrm{d}_\ell^\mathrm{TT}-\mathrm{s}_\ell^\mathrm{TT}\right)^2\rangle = 
\frac{2\left(\mathcal{S}_\ell^\mathrm{TT}\right)^2}{\left(2\ell+1\right)
\left(f_\ell^\mathrm{TT}\right)^2}\ .
\label{prop1}
\end{equation}
Mock WMAP TT, TE and EE data is created by sampling the
two-dimensional Wishart distribution
\begin{eqnarray}
-2 \ln
P_\mathrm{W2}\left(\left\{\mathrm{d}_\ell^\mathrm{X}\right\}|\left\{\mathrm{s}_\ell^\mathrm{X}\right\}\right)
& = &
\sum_{\ell}n_\mathrm{dof}\left\{\ln\left[\mathcal{S}_\ell^\mathrm{TT}
\mathcal{S}_\ell^\mathrm{EE}-\left(\mathrm{s}_\ell^\mathrm{TE}\right)^2\right] - 
\frac{n_\mathrm{dof}-3}{n_\mathrm{dof}}\ln\left[\mathcal{D}_\ell^\mathrm{TT}\mathcal{D}_\ell^\mathrm{EE} - 
\left(\mathrm{d}_\ell^\mathrm{TE}\right)^2\right]\right.\nonumber\\ &
&
\left. + \frac{\mathcal{S}_\ell^\mathrm{TT}\mathcal{D}_\ell^\mathrm{EE} + 
\mathcal{D}_\ell^\mathrm{TT}\mathcal{S}_\ell^\mathrm{EE} - 
2\mathrm{s}_\ell^\mathrm{TE}\mathrm{d}_\ell^\mathrm{TE}}{\mathcal{S}_\ell^\mathrm{TT}
\mathcal{S}_\ell^\mathrm{EE} - 
\left(\mathrm{s}_\ell^\mathrm{TE}\right)^2}\right\},
\label{wish2}
\end{eqnarray}
where $\mathcal{D}_\ell^\mathrm{EE}\equiv
\mathrm{d}_\ell^\mathrm{EE}+\mathcal{N}_\ell^\mathrm{EE}$. In addition to 
eq.(\ref{prop1}), the data satisfies
\begin{equation}
  \langle \mathrm{d}_\ell^\mathrm{X}-\mathrm{s}_\ell^\mathrm{X}\rangle=0, \quad \langle
  \left(\mathrm{d}_\ell^\mathrm{TE}-\mathrm{s}_\ell^\mathrm{TE}\right)^2\rangle = 
\frac{\mathcal{S}_\ell^\mathrm{TT}\mathcal{S}_\ell^\mathrm{EE} + 
\left(\mathrm{s}_\ell^\mathrm{TE}\right)^2}{\left(2\ell+1\right)\left(f_\ell^\mathrm{TT}\right)^2},
  \quad \langle
  \left(\mathrm{d}_\ell^\mathrm{EE}-\mathrm{s}_\ell^\mathrm{EE}\right)^2\rangle = 
\frac{2\left(\mathcal{S}_\ell^\mathrm{EE}\right)^2}{\left(2\ell + 1\right)
\left(f_\ell^\mathrm{TT}\right)^2},
\end{equation}
\begin{equation}
  \langle\left(\mathrm{d}_\ell^\mathrm{TT}-\mathrm{s}_\ell^\mathrm{TT}\right)\left(\mathrm{d}_\ell^\mathrm{TE}-\mathrm{s}_\ell^\mathrm{TE}\right)\rangle = 
\frac{2\mathcal{S}_\ell^\mathrm{TT}\mathrm{s}_\ell^\mathrm{TE}}{\left(2\ell + 1\right)
\left(f_\ell^\mathrm{TT}\right)^2}, \quad
  \langle\left(\mathrm{d}_\ell^\mathrm{TT}-\mathrm{s}_\ell^\mathrm{TT}\right)\left(\mathrm{d}_\ell^\mathrm{EE}-\mathrm{s}_\ell^\mathrm{EE}\right)\rangle = 
\frac{2\left(\mathrm{s}_\ell^\mathrm{TE}\right)^2}{\left(2\ell + 1\right)
\left(f_\ell^\mathrm{TT}\right)^2},
\end{equation}
\begin{equation}
  \langle\left(\mathrm{d}_\ell^\mathrm{TE}-\mathrm{s}_\ell^\mathrm{TE}\right)\left(\mathrm{d}_\ell^\mathrm{EE}-\mathrm{s}_\ell^\mathrm{EE}\right)\rangle = 
\frac{2\mathrm{s}_\ell^\mathrm{TE}\mathcal{S}_\ell^\mathrm{EE}}{\left(2\ell + 1\right)
\left(f_\ell^\mathrm{TT}\right)^2},
\end{equation}
The variance of the mock WMAP data points thus matches the
diagonal elements of the covariance matrices given in
Sec.\,\ref{wmap}. Although this approach neglects correlations
between different multipoles induced when working on a cut sky, we do
not expect this to significantly affect our results due to the high
sky coverage ($f_\ell \sim 0.8$) of the WMAP experiment.

The quantities $\mathcal{Z}^X_b$ (eq.\,\ref{ssl}) were shown to be
Gaussian distributed \cite{Abroe:2001de}. Thus to simulate
ground-based (small angular scale) CMB experiments, $\mathcal{Z}^X_b$
are drawn from a multivariate Gaussian distribution with vanishing
mean and covariance matrix $\mathsf{V}^{XX^\prime}$, where
$\tilde{\mathsf{N}}^{XX^\prime}$ is now calculated using theoretical
bandpowers instead of data bandpowers. Similarly mock SDSS-4 LRG data
is produced by contaminating $\sB$ from the test spectrum with
Gaussian noise with covariance matrix
$\tilde{\mathsf{N}}^\mathrm{LRG}$.

Rather than use the full WMAP likelihood functions, we adopt less
computationally expensive approximations when inverting the mock
data.  These are $L_\mathrm{TT}\simeq L_\mathrm{W1}+L_\mathrm{pTT}$
for the WMAP temperature data alone and
$L_\mathrm{TT}+L_\mathrm{pol}\simeq L_\mathrm{W2}+
L_\mathrm{pTT}+L_\mathrm{pTE}$ for the WMAP temperature and
polarisation data.  Here $L_\mathrm{W1,2}\equiv -2\ln P_\mathrm{W1,2}$
are the likelihood functions corresponding to the Wishart
distributions of eq.(\ref{wish1}) and eq.(\ref{wish2}).  They are used
for $\ell\leq23$ while $L_\mathrm{pTT}$ and $L_\mathrm{pTE}$ are
employed for $24\leq\ell\leq1000$ and $24\leq\ell\leq450$
respectively. Furthermore following \cite{Tocchini-Valentini:2005ja}
we neglect the off-diagonal elements of the matrices
$\left(\mathsf{N}^\mathrm{TT}\right)^{-1}$.

\section{Validation of the inversion method \label{validation}}

\subsection{The integral and resolution kernels \label{kernels}}

\begin{figure*}
\begin{center}
\includegraphics*[angle=0,width=0.7\columnwidth]{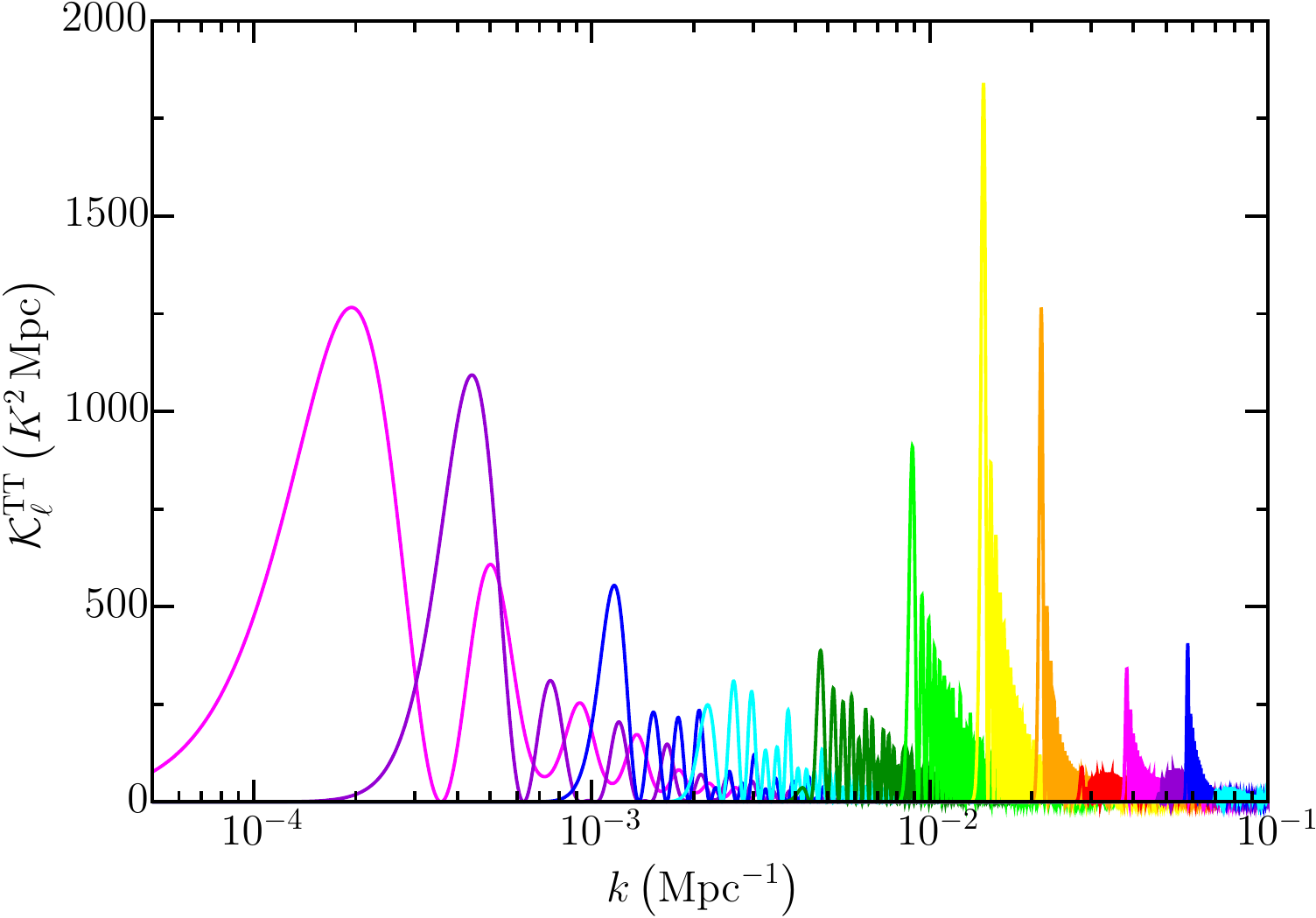}
\end{center}
\caption{\label{kern1} Integral kernels
  $\mathcal{K}^\mathrm{TT}_\ell\left(k\right)$ (eq.\,\ref{int1}) for
  $\ell=2$, 5, 15, 30, 60 (corresponding to the Sachs-Wolfe plateau),
  $\ell=120$, 200, 300 (first acoustic peak), $\ell=400$ (first
  trough), $\ell=540$ (second peak), $\ell=680$ (second trough),
  $\ell=820$ (third peak) and $\ell=999$ (third trough), from left to
  right.}
\end{figure*}

The integral kernels $\mathcal{K}^\mathrm{TT}_\ell\left(k\right)$ for the CMB
temperature angular power spectrum are fundamental to the properties
of the inversion. Some examples of the kernels are presented in
Fig.\,\ref{kern1}. In general the kernels become narrower for higher
multipoles, so that convolution with them involves less smoothing and
the relation between $k$-space and $\ell$-space becomes more direct on
small scales. However, the kernels which correspond to troughs in the
angular power spectrum are broader than those associated with
neighbouring acoustic peaks.

\begin{figure*}
\includegraphics*[angle=0,width=0.5\columnwidth,trim = 32mm 171mm 23mm
  15mm, clip]{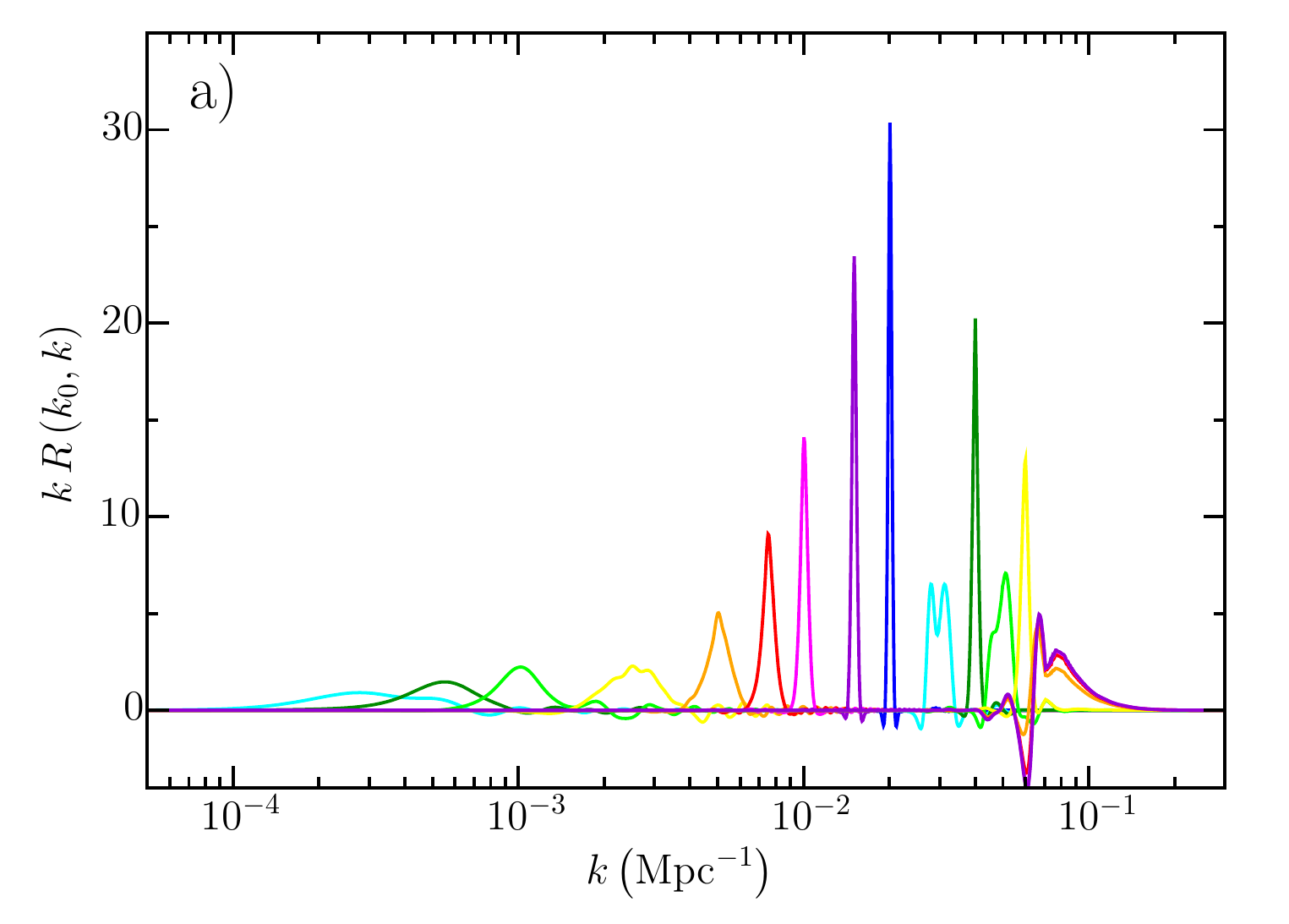}
\includegraphics*[angle=0,width=0.5\columnwidth,trim = 32mm 171mm 23mm
  15mm, clip]{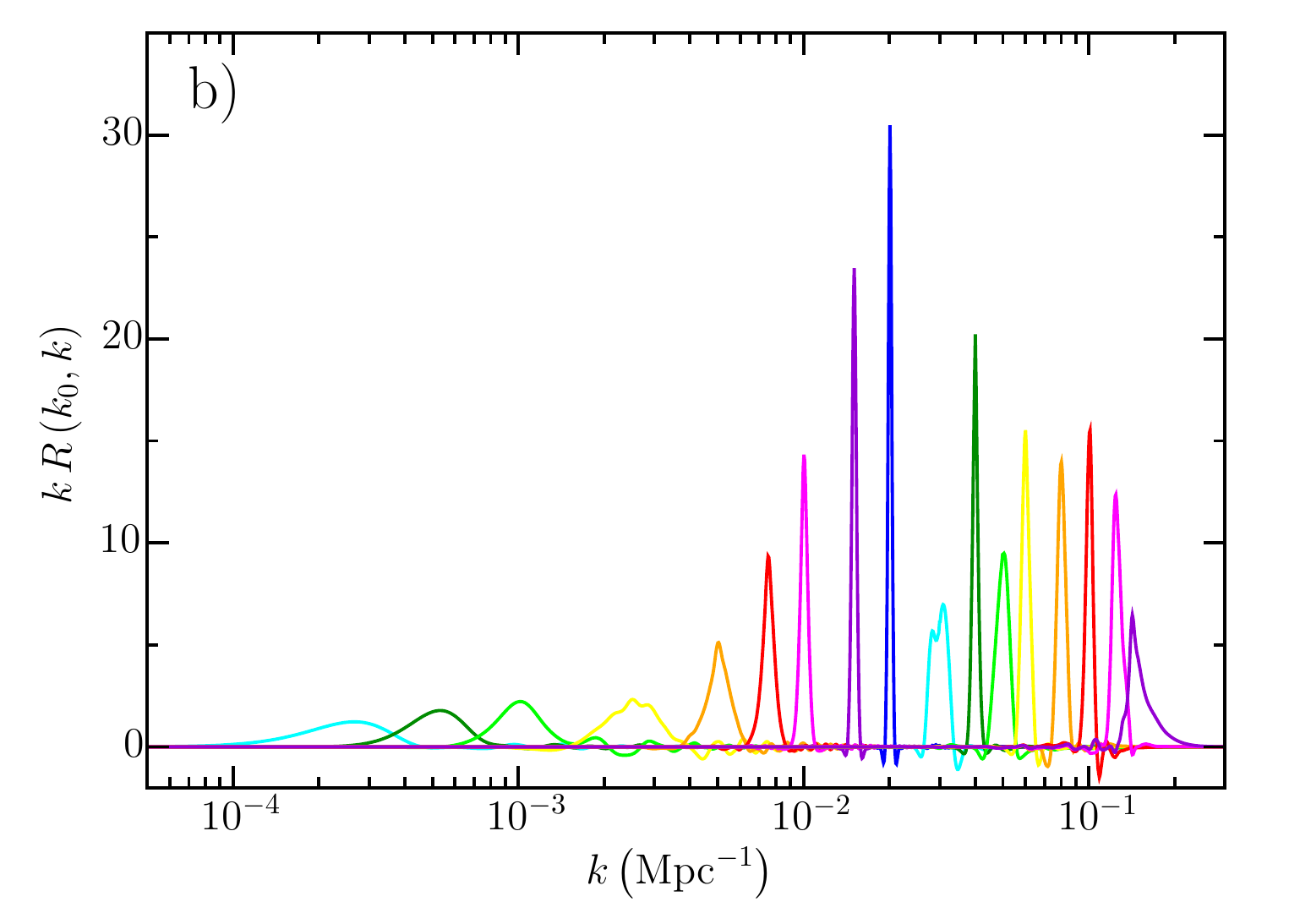}
\includegraphics*[angle=0,width=0.5\columnwidth,trim = 32mm 171mm 23mm
  15mm, clip]{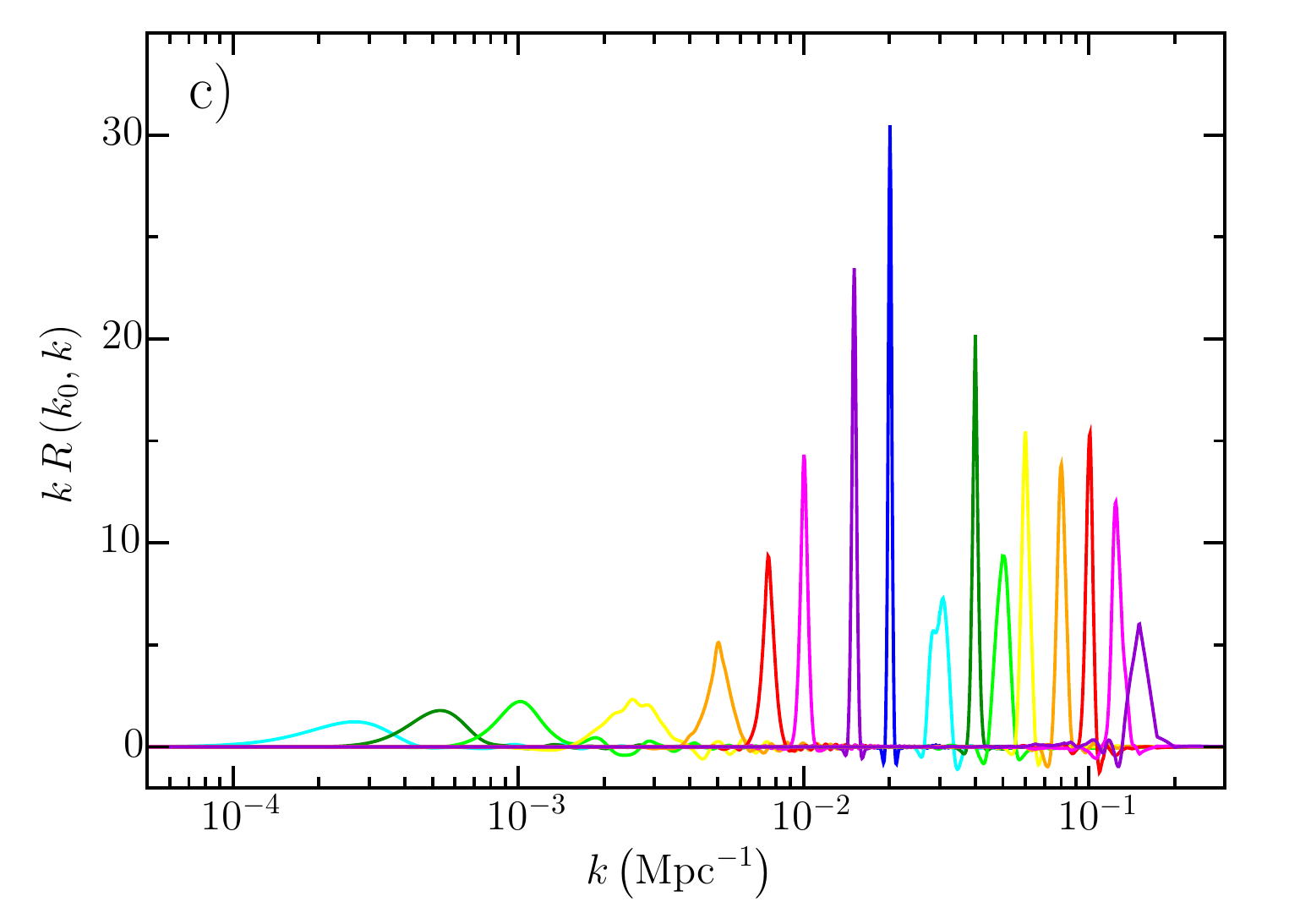}
\includegraphics*[angle=0,width=0.5\columnwidth,trim = 32mm 171mm 23mm
  15mm, clip]{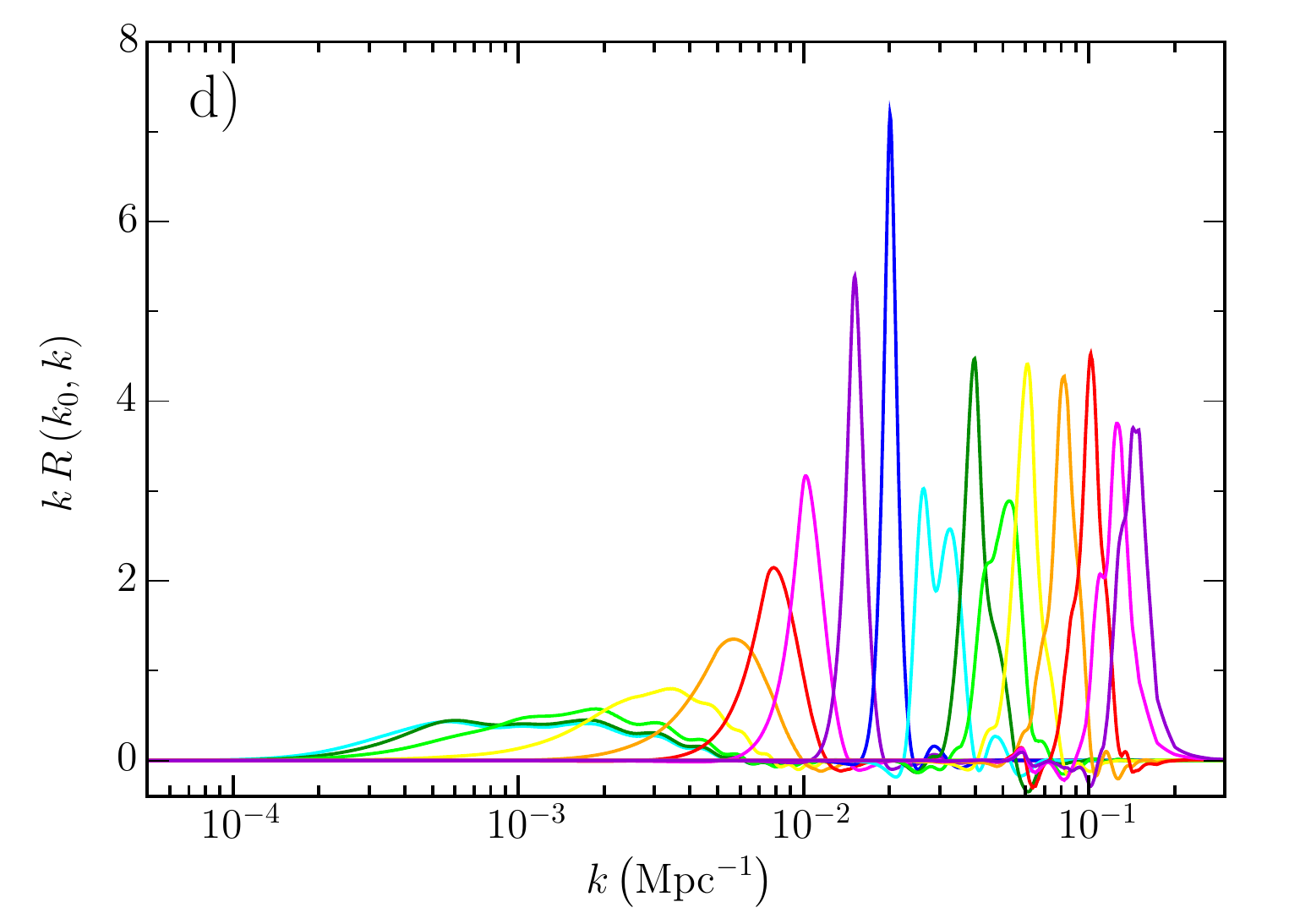}
\caption{\label{resol} First-order resolution kernels $R(k_0,k)$
  (eq.\,\ref{1storderkernel}) for $k_0=10^{-4}$, $5\times10^{-4}$,
  $10^{-3}$, $2.5\times10^{-3}$, $5\times10^{-3}$, $7.5\times10^{-3}$,
  0.01, 0.015, 0.02, 0.03, 0.04, 0.05, 0.06, 0.08, 0.1, 0.125,
  $0.15\;\mathrm{Mpc}^{-1}$. The top left hand plot a) shows the
  kernels for the WMAP-5 TT data alone. The top right hand plot b)
  shows the kernels for the WMAP-5 TT, TE and EE data together with
  the small-scale CMB data.  The bottom left hand plot c) shows the
  kernels for the WMAP-5, small-scale CMB and SDSS-4 LRG data. The
  bottom right hand plot d) is the same as the bottom left but with
  $\lambda=1000$ instead of $\lambda=10$.}
\end{figure*}

\begin{figure*}
\includegraphics*[angle=0,width=0.5\columnwidth,trim = 32mm 171mm 23mm
  15mm, clip]{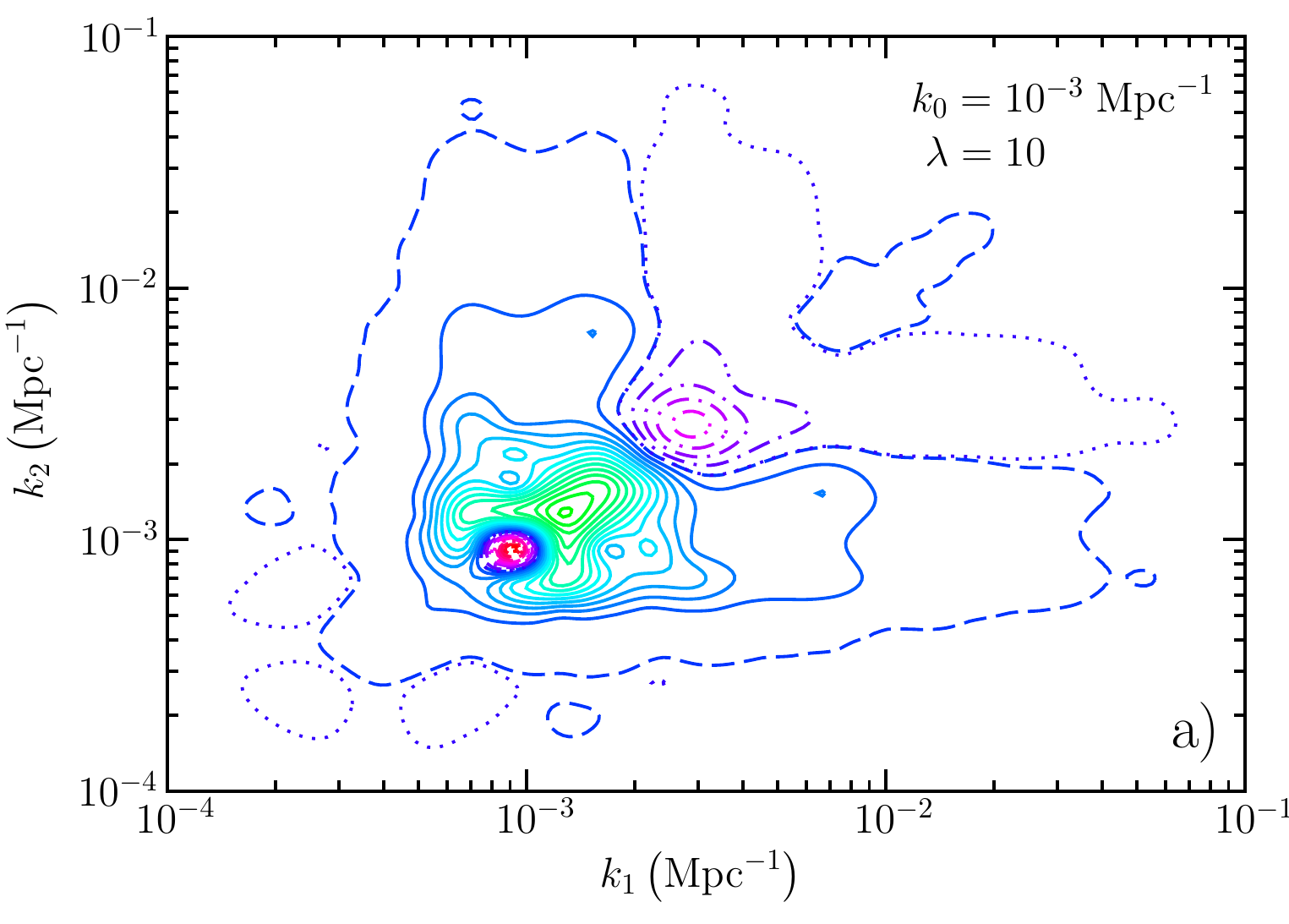}
\includegraphics*[angle=0,width=0.5\columnwidth,trim = 32mm 171mm 23mm
  15mm, clip]{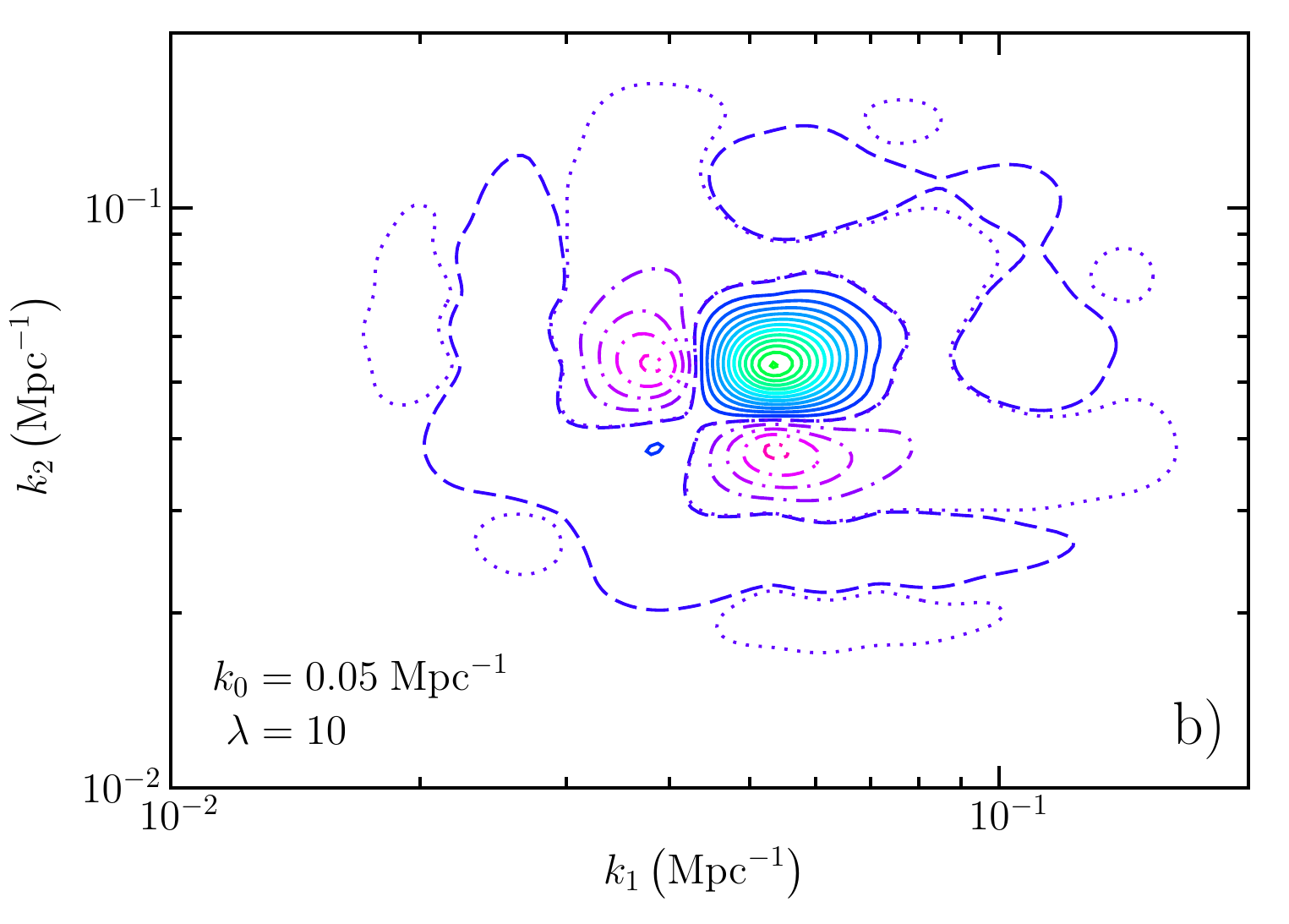}
\includegraphics*[angle=0,width=0.5\columnwidth,trim = 32mm 171mm 23mm
  15mm, clip]{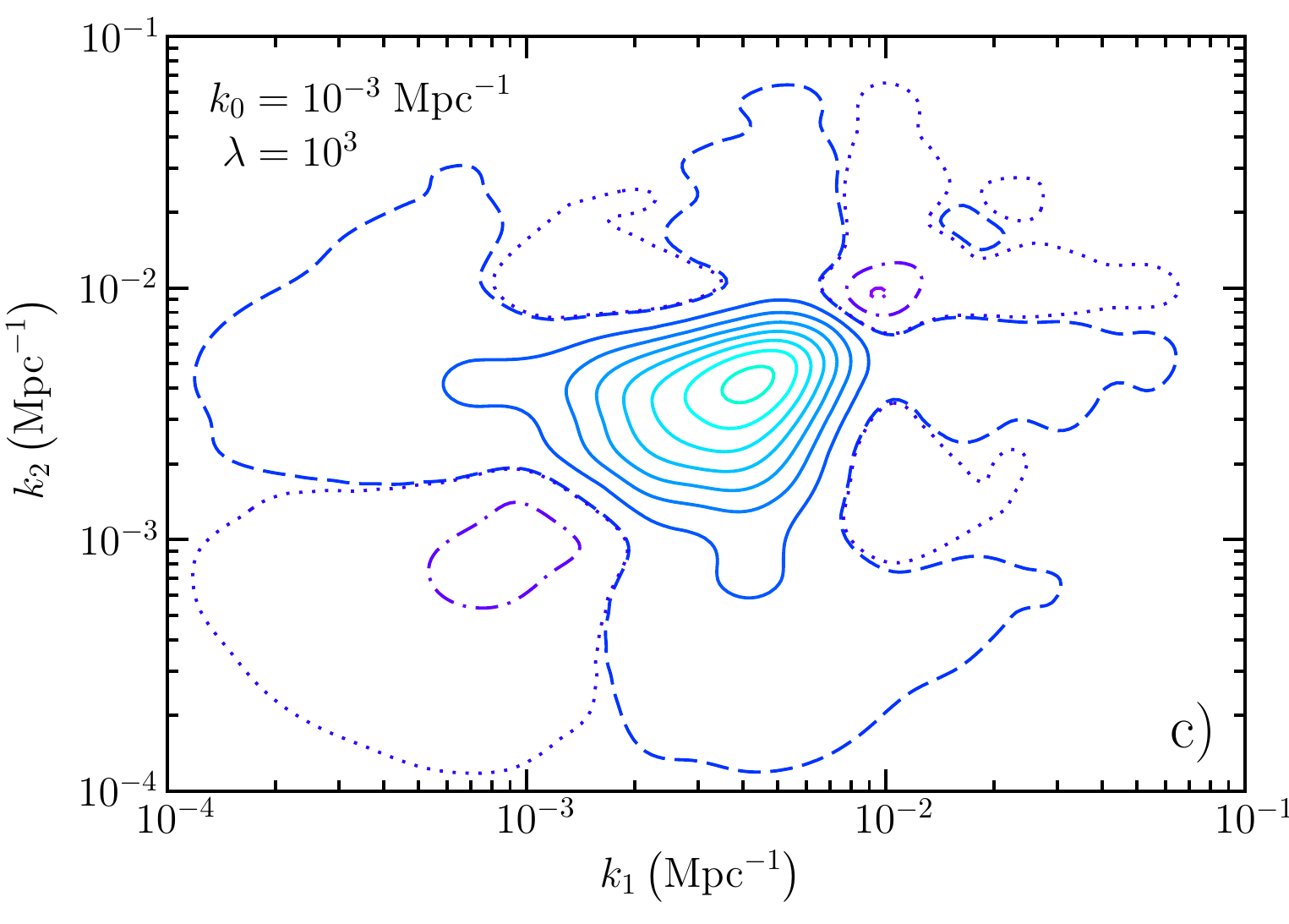}
\includegraphics*[angle=0,width=0.5\columnwidth,trim = 32mm 171mm 23mm
  15mm, clip]{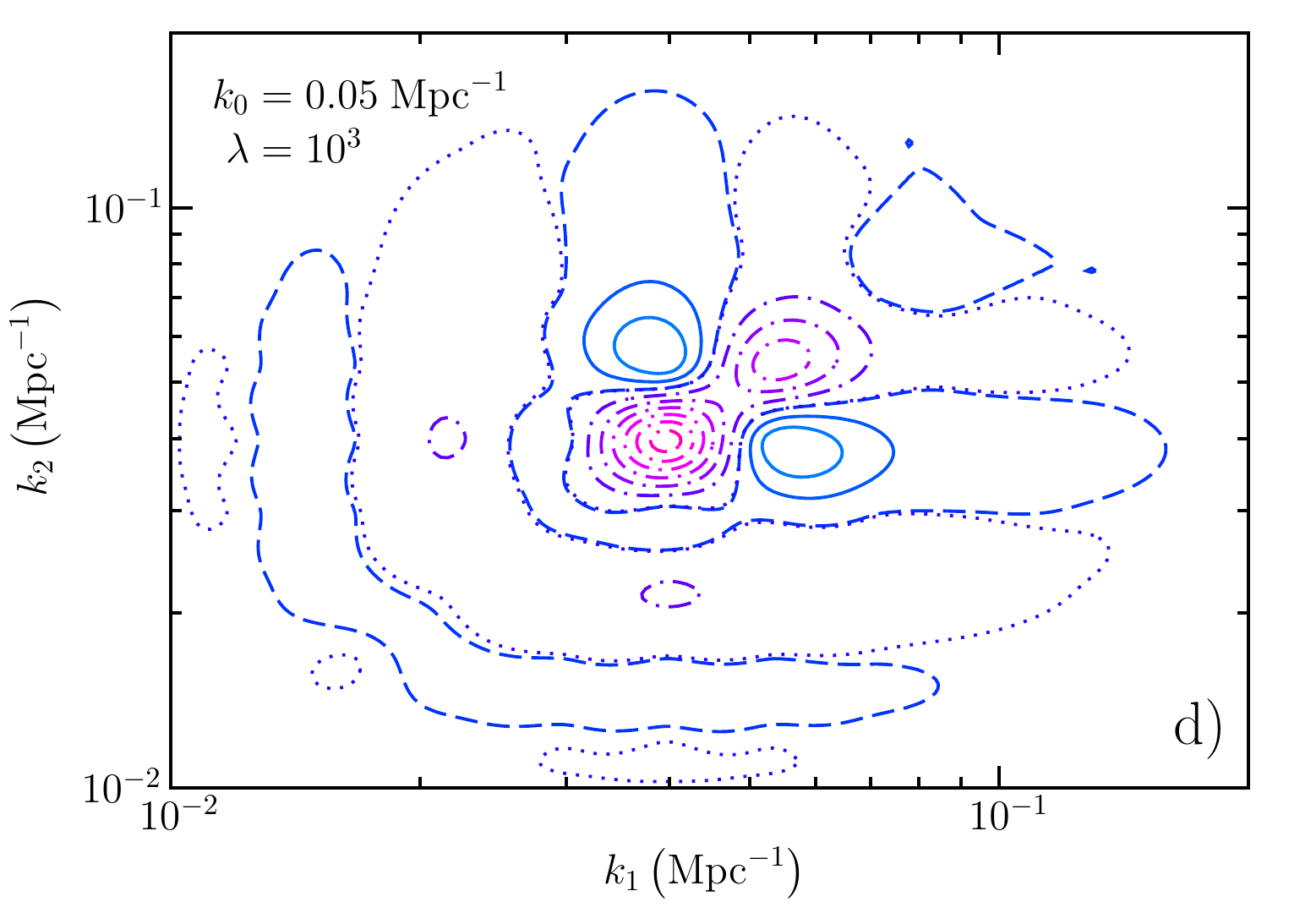}
\caption{\label{resol2} Second-order resolution kernels
  $Y\left(k_0;k_1,k_2\right)$ (eq.\,\ref{2ndorderkernel}) for the
  WMAP-5 TT data. The two left hand plots are for
  $k_0=10^{-3}\;\mathrm{Mpc}^{-1}$ and the right hand plots are for
  $k_0=0.05\;\mathrm{Mpc}^{-1}$. The upper plots are for $\lambda=10$
  and the lower plots are for $\lambda=1000$.  The dashed contours are
  at $Y=10^{-3}$ while the dotted contours are at $Y=-10^{-3}$.  The
  solid contours are at $Y=0.05,0.1,0.15,\ldots$ and the dot-dashed
  contours are at $Y=-0.05,-0.1,-0.15,\ldots$.}
\end{figure*}

The estimated PPS $\hat{\mathcal{P}}_\mathcal{R}\left(k_0\right)$
cannot resolve features in the true PPS
$\mathcal{P}_\mathcal{R}\left(k\right)$ which are narrower than the
resolution kernel $R\left(k_0,k\right)$. Fig.\,\ref{resol} shows the
first-order kernels (eq.\,\ref{1storderkernel}) for some selected
values of the target wavenumber $k_0$. Ideally the resolution kernel
$R\left(k_0,k\right)$ would be sharply peaked at $k=k_0$ and
negligible everywhere else. The resolution kernels depend on both the
integral kernels and on the noise in the data. Consider first the
resolution kernels for the WMAP-5 TT data alone. The resolution
kernels are broadest on large scales due to the wide integral kernels
at low $\ell$, as well as the uncertainty in the data caused by cosmic
variance. The resolution kernels narrow with higher wavenumber until
the greatest resolution is attained at
$k\simeq0.02\;\mathrm{Mpc}^{-1}$. This corresponds to $\ell\simeq280$
on the right hand side of the first acoustic peak where the integral
kernels are sharply peaked and the WMAP error is smallest. The
resolution kernels at $k_0=0.03$ and $0.05\;\mathrm{Mpc}^{-1}$ are
misshapen, reflecting the broad integral kernels of the first and
second troughs in the TT angular power spectrum.  The more strongly
defined kernels at $k_0=0.04$ and $0.06\;\mathrm{Mpc}^{-1}$ correspond
to the second and third acoustic peaks. The large WMAP measurement
errors at high multipoles decrease the resolution on small scales. The
resolution kernels are mostly centred closely on the target
wavenumbers, showing that the reconstructed PPS is
meaningful. However, the $k_0=10^{-4}\;\mathrm{Mpc}^{-1}$ kernel is
centred at $k\simeq 3\times10^{-4}\;\mathrm{Mpc}^{-1}$ instead. This
is a consequence of the fact that at the very lowest wavenumbers, the
estimated PPS is an extrapolation from higher wavenumbers. A similar
phenomenon occurs at the very highest wavenumbers.

Adding the WMAP-5 polarisation data slightly improves the resolution
on large scales.  The $k_0=10^{-4}\;\mathrm{Mpc}^{-1}$ and
$k_0=5\times10^{-4}\;\mathrm{Mpc}^{-1}$ kernels become slightly better
localised. The resolution at high wavenumbers is greatly increased by
the small angular scale CMB data. The kernels for $k_0=0.08$, 0.1 and
$0.125\;\mathrm{Mpc}^{-1}$ which were close together with the WMAP-5
TT data become well separated and strongly peaked. The SDSS-4 LRG data
improves the $k_0=0.03\;\mathrm{Mpc}^{-1}$ and
$k_0=0.05\;\mathrm{Mpc}^{-1}$ kernels, causing them to become more
peaked.  Finally, Fig.\,\ref{resol} shows that increasing the
regularisation parameter broadens the resolution kernels and decreases
the resolution. The kernels on large scales become almost identical,
which indicates that increased extrapolation occurs in the inversion.

The second-order resolution kernel (eq.\,\ref{2ndorderkernel}) is shown
in Fig.\,\ref{resol2}. For $\lambda=10$ the kernel is centred
approximately at $k_1=k_2=k_0$. The kernel is broader for low $k_0$
and more localised for higher $k_0$ due to the narrower integral
kernels at higher multipoles. For $\lambda=1000$ the
$k_0=10^{-3}\;\mathrm{Mpc}^{-1}$ kernel is centred at $k_1=k_2\simeq
4\times10^{-3}\;\mathrm{Mpc}^{-1}$.  This is similar to the way in
which the peak of $R\left(k_0;k_1\right)$, located at $k_1=k_0$ for
low $\lambda$, shifts as $\lambda$ increases towards higher $k_1$ for
low $k_0$ and lower $k_1$ for high $k_0$.  The second-order kernel is
broader and lower in magnitude for higher $\lambda$.

\subsection{Test spectra results \label{testres}}

\begin{figure*}
\includegraphics*[angle=0,width=0.5\columnwidth,trim = 32mm 171mm 23mm
  15mm, clip]{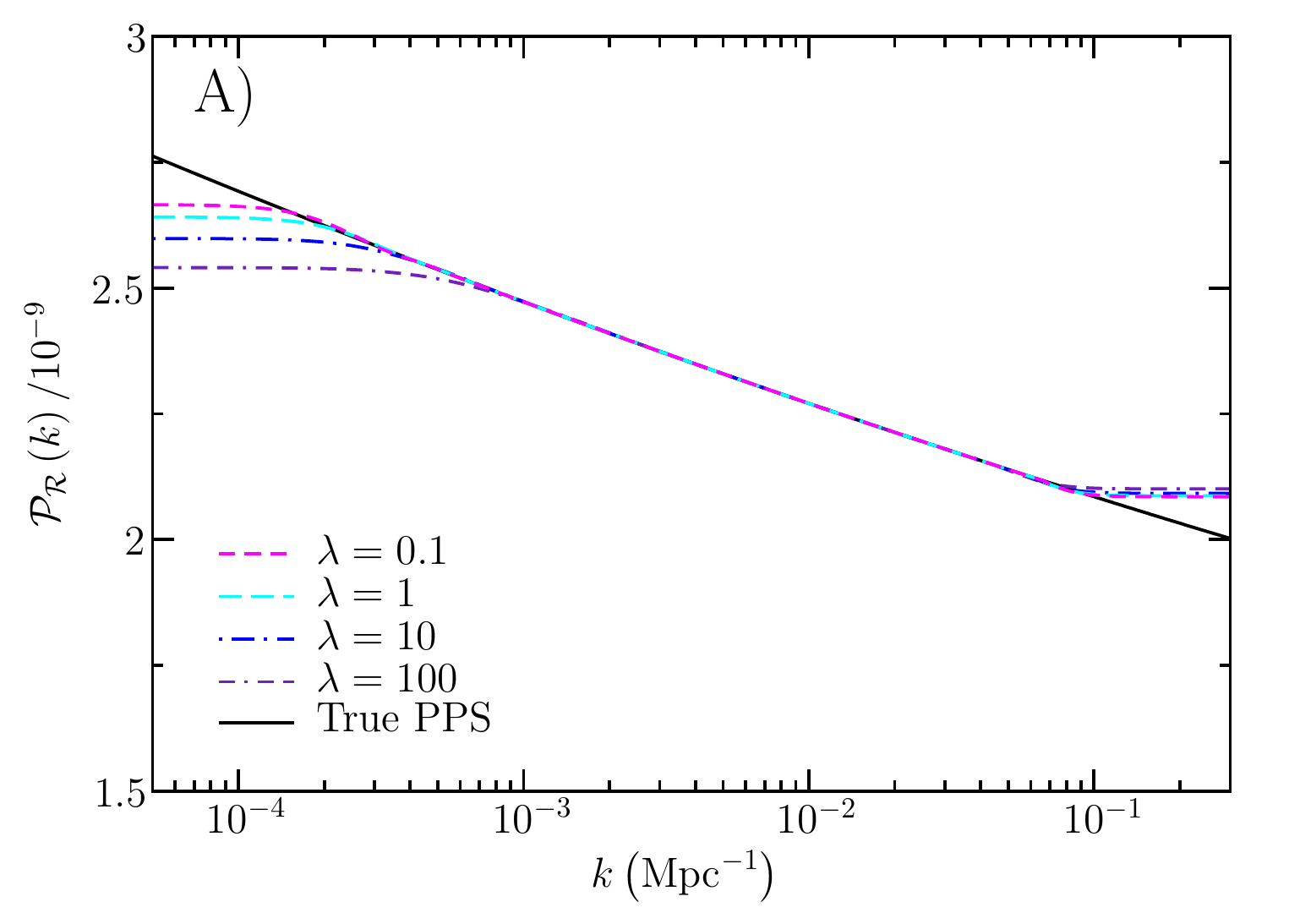}
\includegraphics*[angle=0,width=0.5\columnwidth,trim = 32mm 171mm 23mm
  15mm, clip]{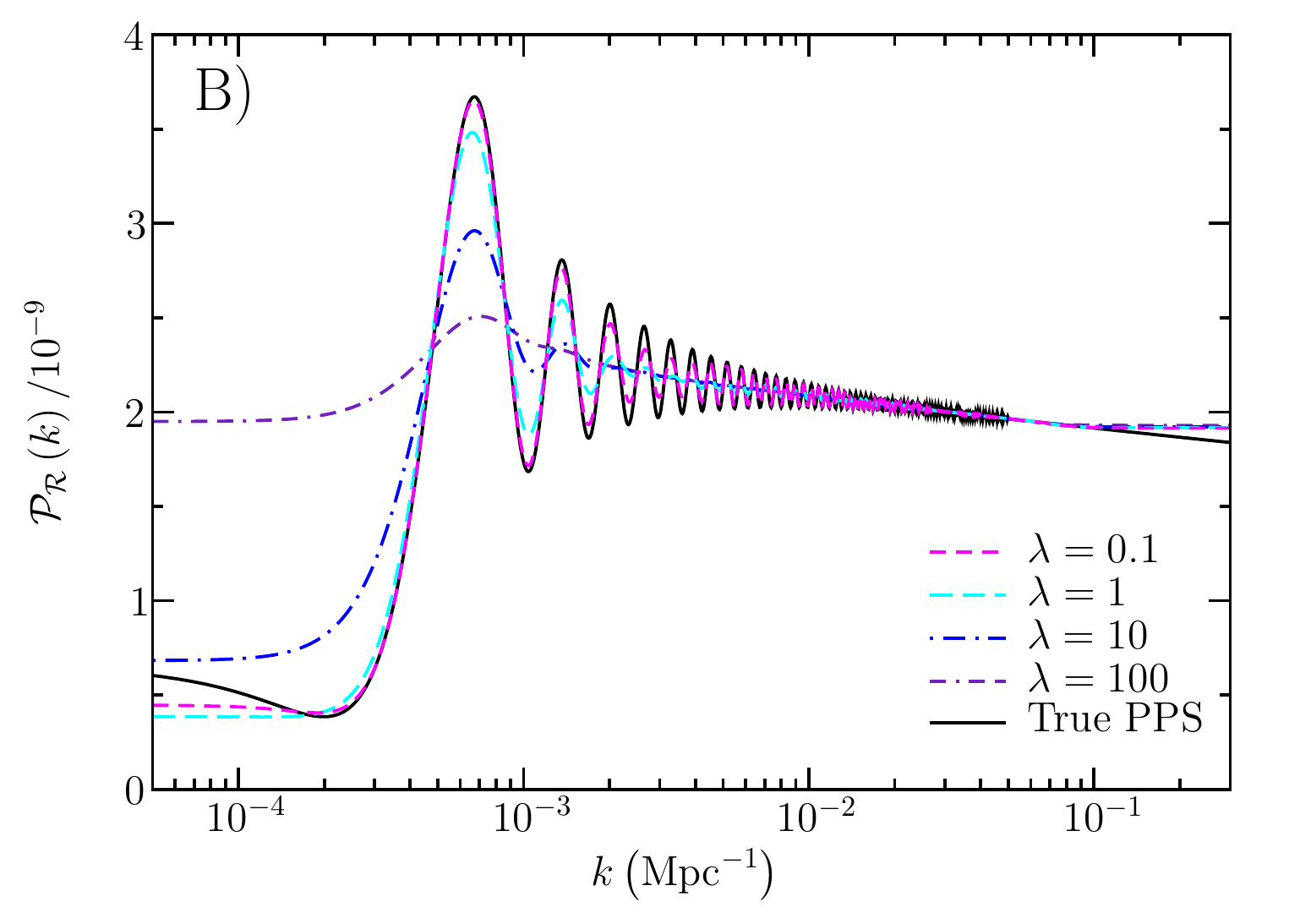}
\includegraphics*[angle=0,width=0.5\columnwidth,trim = 32mm 171mm 23mm
  15mm, clip]{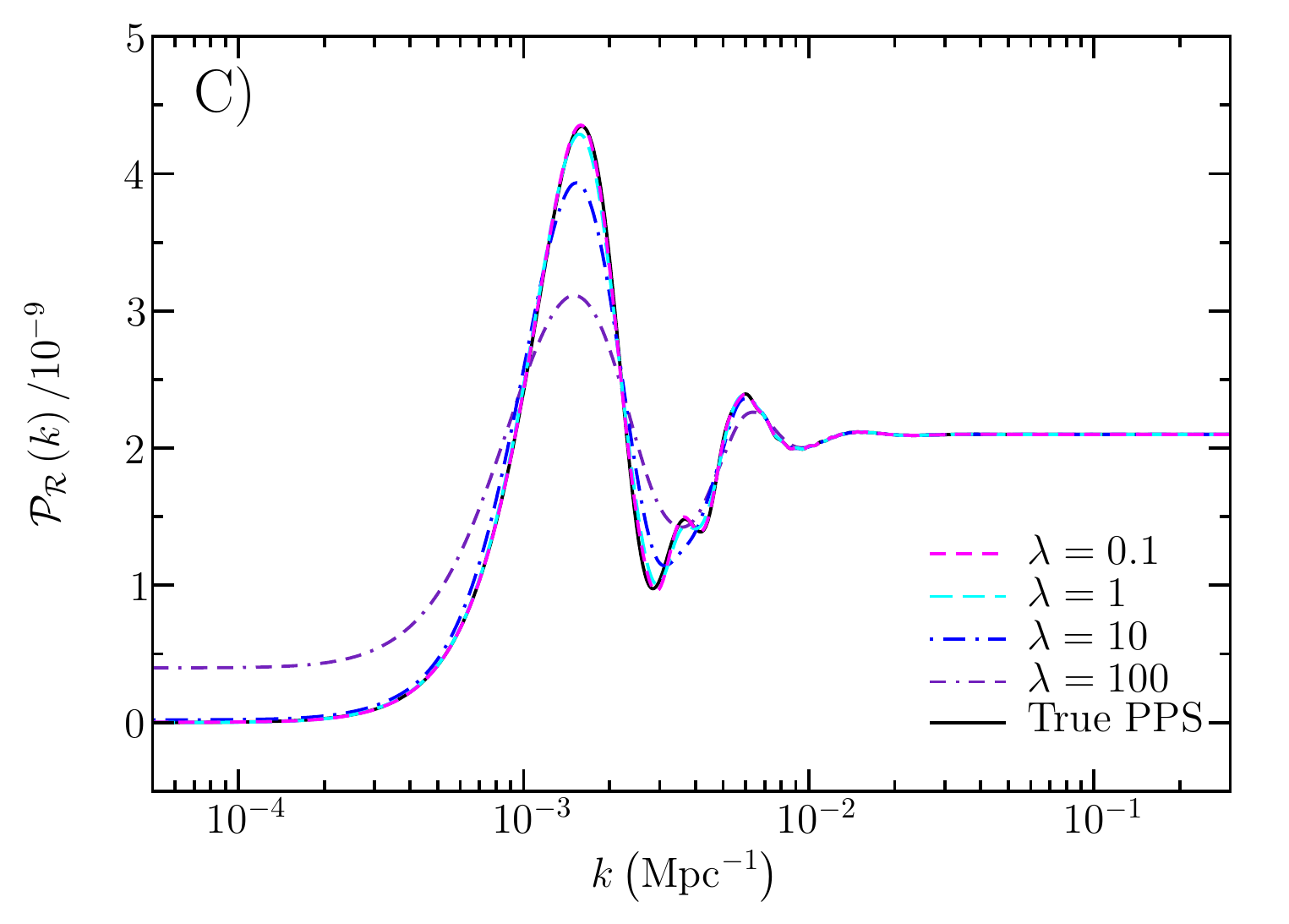}
\includegraphics*[angle=0,width=0.5\columnwidth,trim = 32mm 171mm 23mm
  15mm, clip]{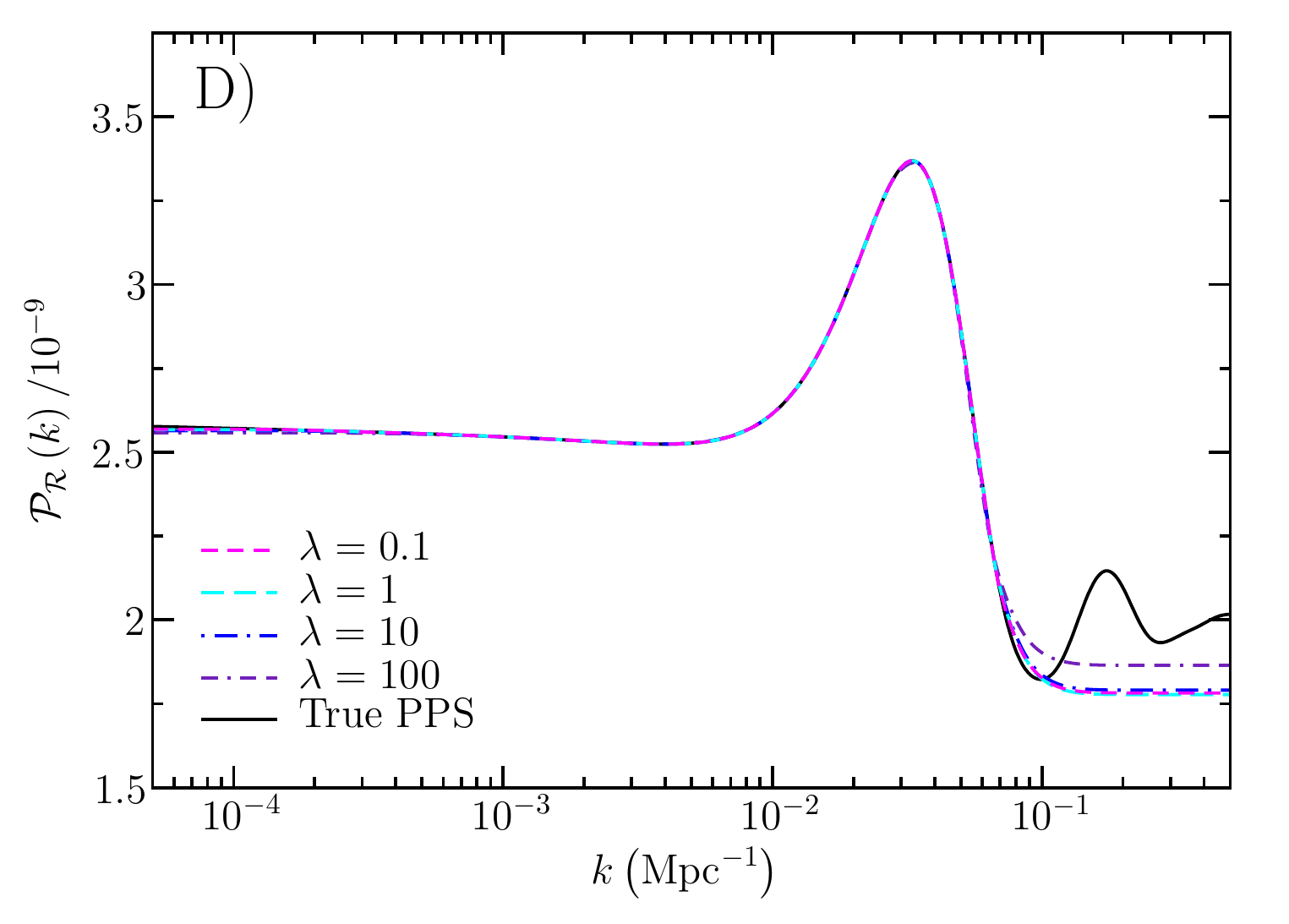}
\caption{\label{nonoise} Spectra recovered from noise-free mock 
  WMAP-5 TT ``data'' with regularisation parameter $\lambda=0.1$, $1$,
  $10$ and $100$.  Results are shown for spectra A to D (see
  Sec.\,\ref{spectest}).}
\end{figure*}

We start by inverting \emph{mock} WMAP-5 TT ``data'' without
noise. The reconstructed spectra resemble versions of the test spectra
smoothed towards flatness, as seen in Fig.\,\ref{nonoise}. The degree
of smoothing increases with $\lambda$, in accordance with
expectation. Since the resolution kernels integrate to unity, there
are no spurious vertical scalings of the reconstructed spectra. For
small values of $\lambda$ the test spectra are recovered with
impressive accuracy. (However since we are using first-order Tikhonov
regularisation based on a H-Z spectrum, the \textit{tilted} spectrum A
is not fully recovered both at high and low $k$.) The loss of
resolution with increased regularisation is clearly apparent for
spectrum B, where the oscillations become increasingly smoothed with
larger values of $\lambda$. For $\lambda=0.1$ the oscillations are
recovered least successfully at
$k\simeq3\times10^{-3}\;\mathrm{Mpc}^{-1}$ because the width of the
oscillations there is comparable to that of the resolution kernels; at
higher and lower $k$ the oscillations are broader than the kernels. At
very high and low $k$ where the data is insensitive to the PPS (since
the integral kernels vanish), the estimated PPS is extrapolated
horizontally from lower and higher $k$ respectively. The amount of
extrapolation increases with $\lambda$, as can be seen most clearly
for spectrum A on large scales.  Note that the WMAP data do not extend
to small scales beyond $k \sim 0.1\;\mathrm{Mpc}^{-1}$, hence the
wiggle in spectrum D cannot be recovered there for any $\lambda$.

\begin{figure*}
\includegraphics*[angle=0,width=0.5\columnwidth,trim = 32mm 171mm 23mm
  15mm, clip]{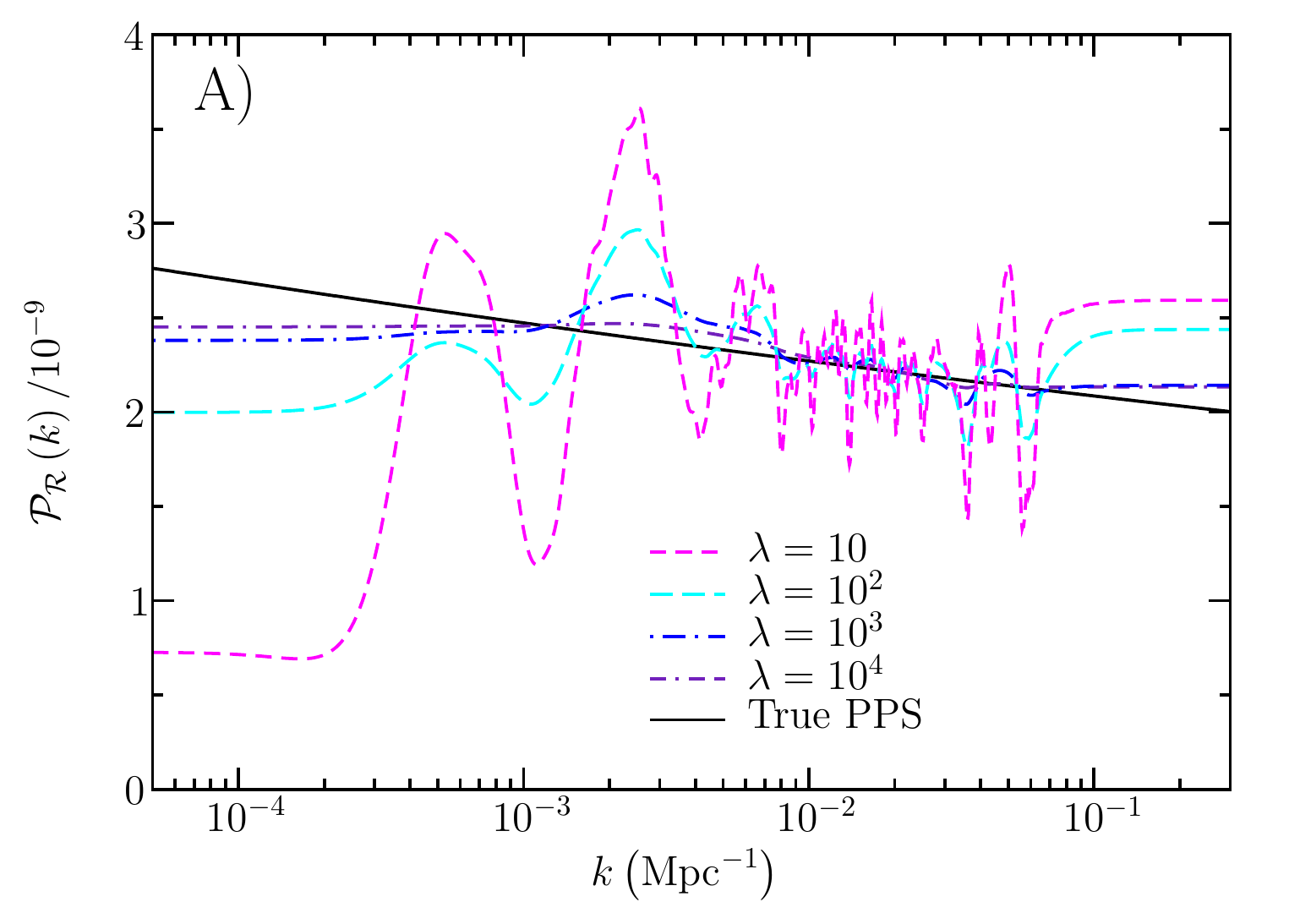}
\includegraphics*[angle=0,width=0.5\columnwidth,trim = 32mm 171mm 23mm
  15mm, clip]{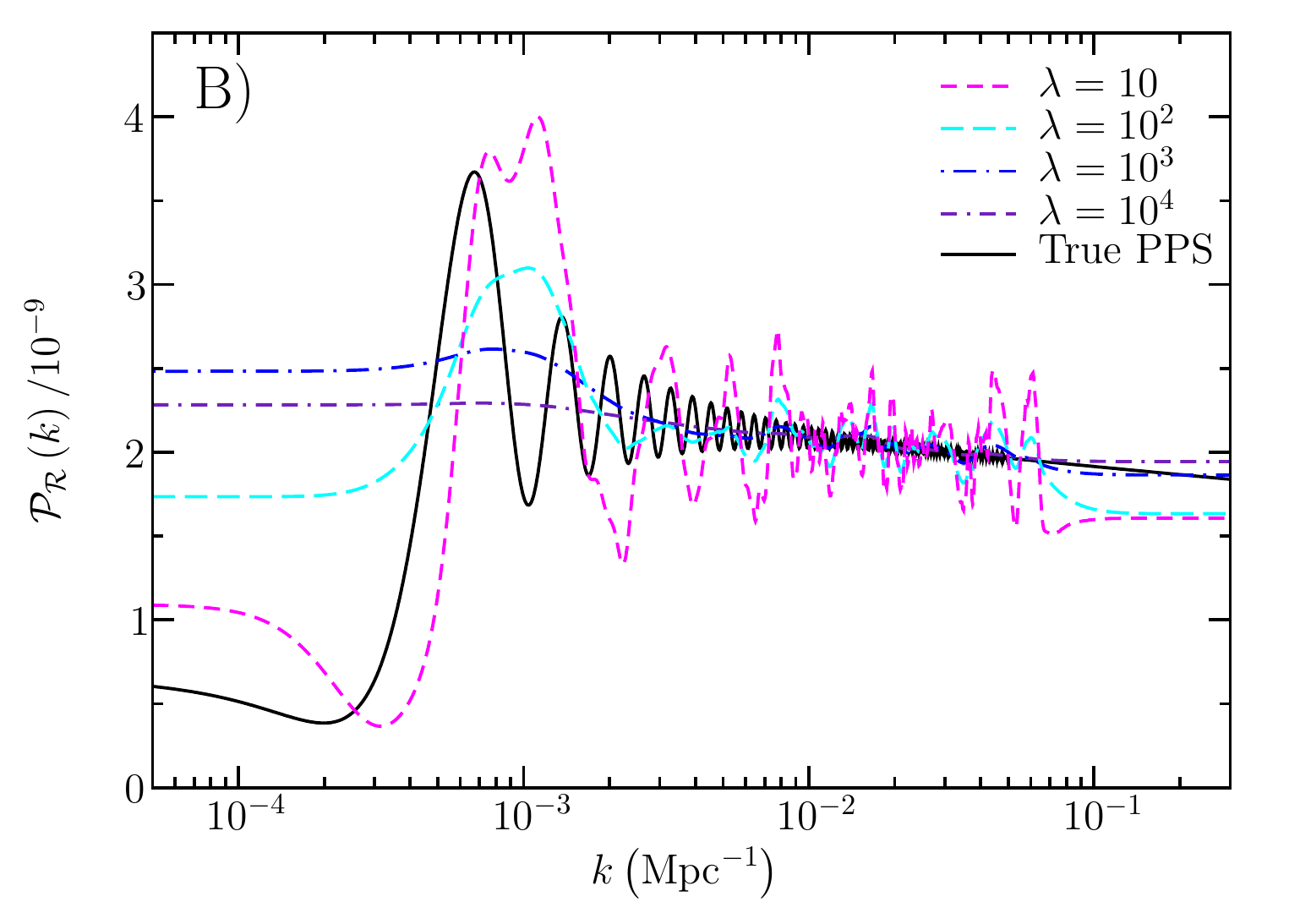}
\includegraphics*[angle=0,width=0.5\columnwidth,trim = 32mm 171mm 23mm
  15mm, clip]{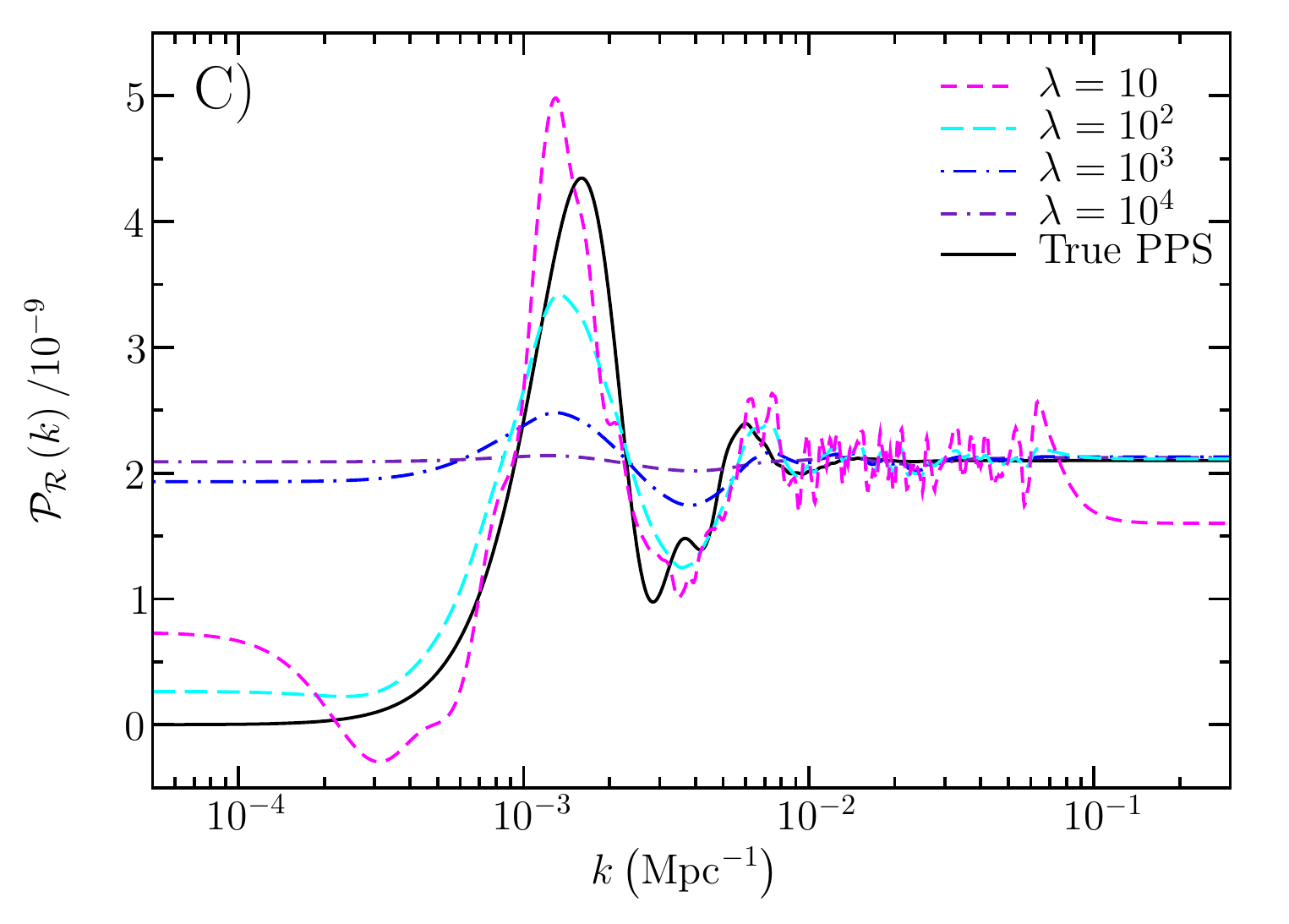}
\includegraphics*[angle=0,width=0.5\columnwidth,trim = 32mm 171mm 23mm
  15mm, clip]{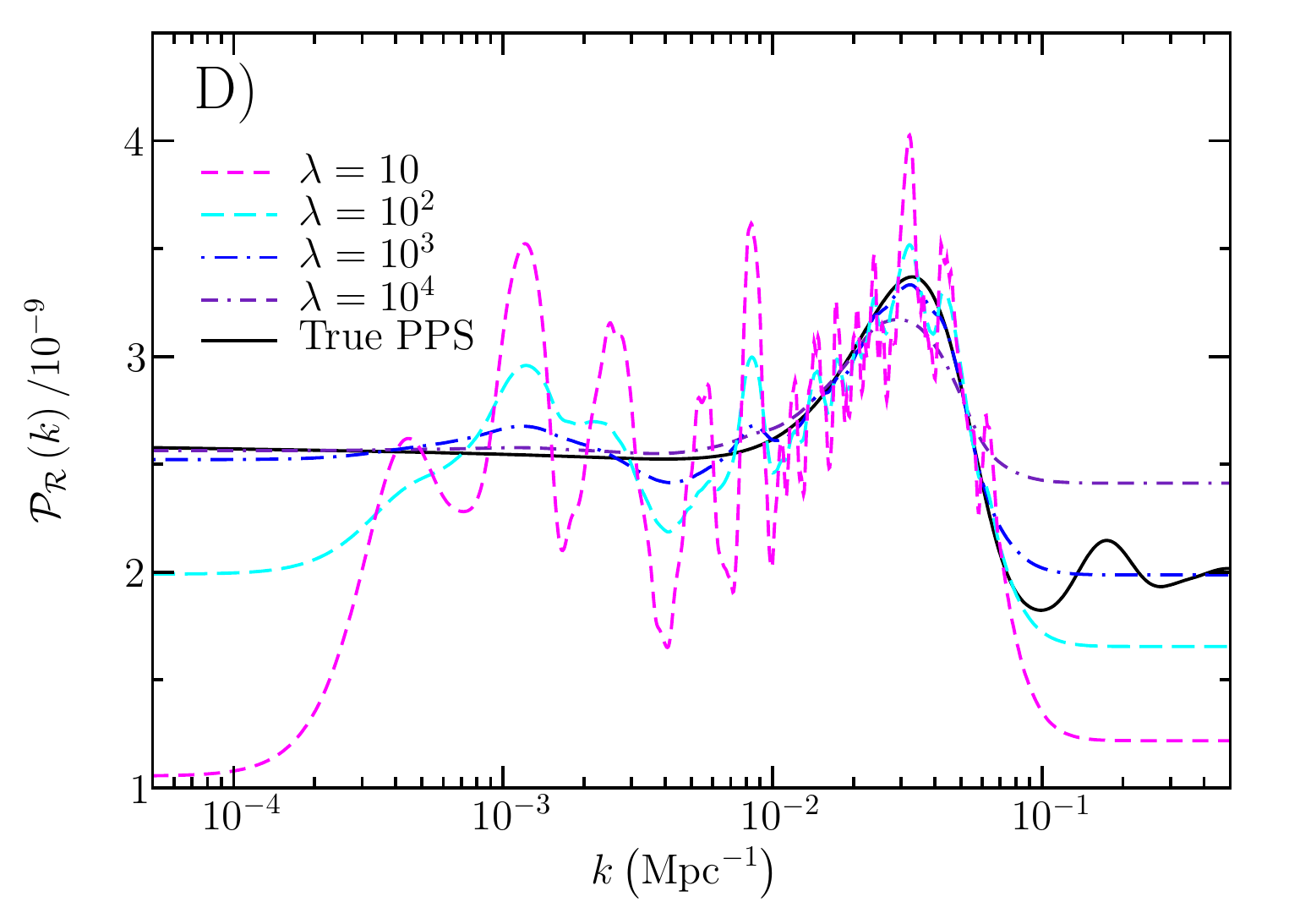}
\caption{\label{noise} Spectra recovered from noisy mock WMAP-5
  TT data with regularisation parameter $\lambda=10$, $10^2$, $10^3$
  and $10^4$. Results are shown for spectra A to D (see
  Sec.\,\ref{spectest}).}
\end{figure*}

Fig.\,\ref{noise} shows typical estimated spectra from mock WMAP-5
TT data, but now including noise. The presence of noise greatly
reduces the accuracy with which the PPS can be found.  The
noise-induced oscillatory features are broadest on large scales due to
the wide integral kernels at low $\ell$, and have high amplitude due
to cosmic variance. The increased suppression of the noise-induced
features can clearly be seen as $\lambda$ is increased. Note that the
best reconstructions are obtained using \textit{different} values of
$\lambda$ for different test spectra.

\begin{figure*}
\includegraphics*[angle=0,width=0.5\columnwidth,trim = 32mm 171mm 23mm
  15mm, clip]{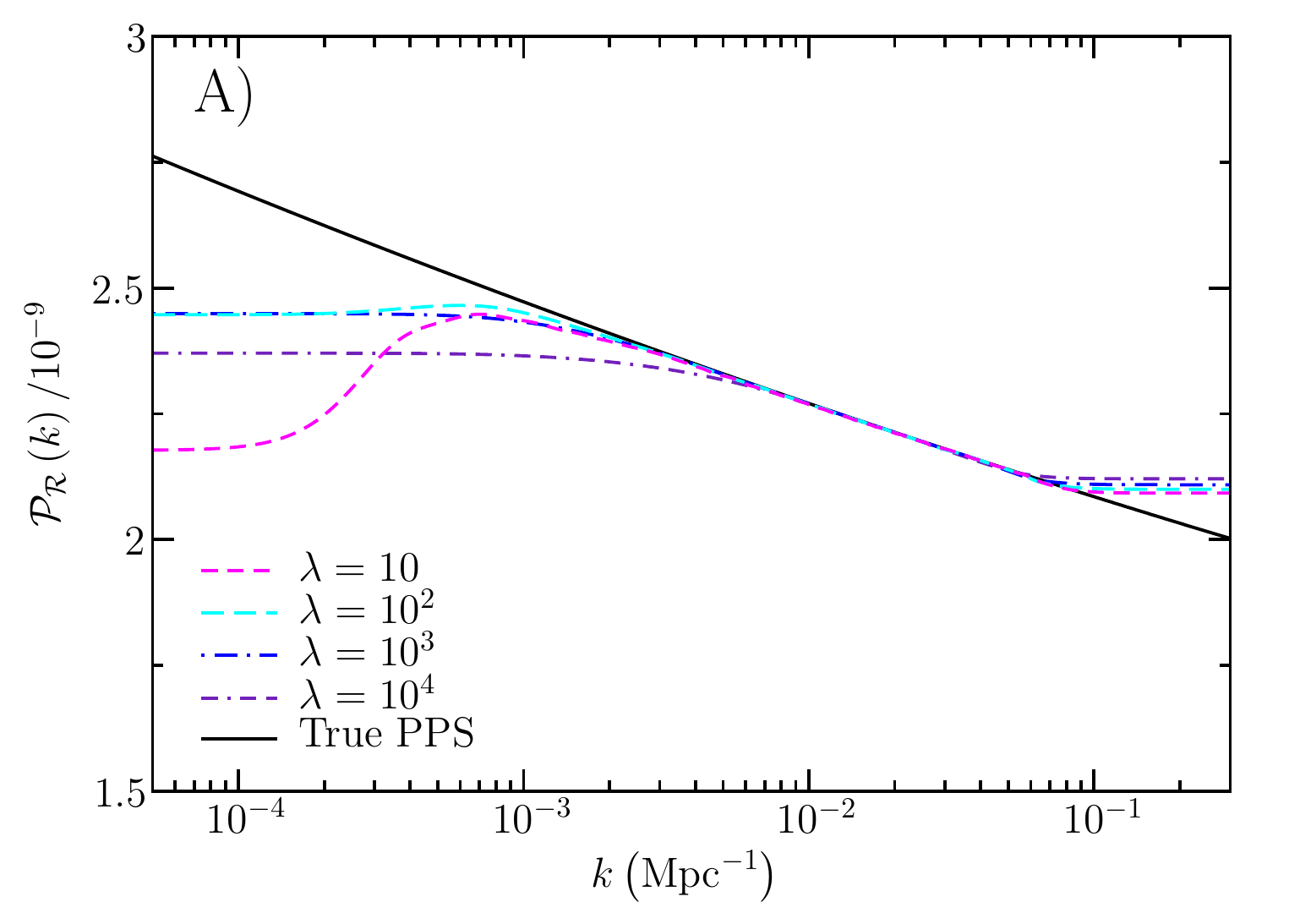}
\includegraphics*[angle=0,width=0.5\columnwidth,trim = 32mm 171mm 23mm
  15mm, clip]{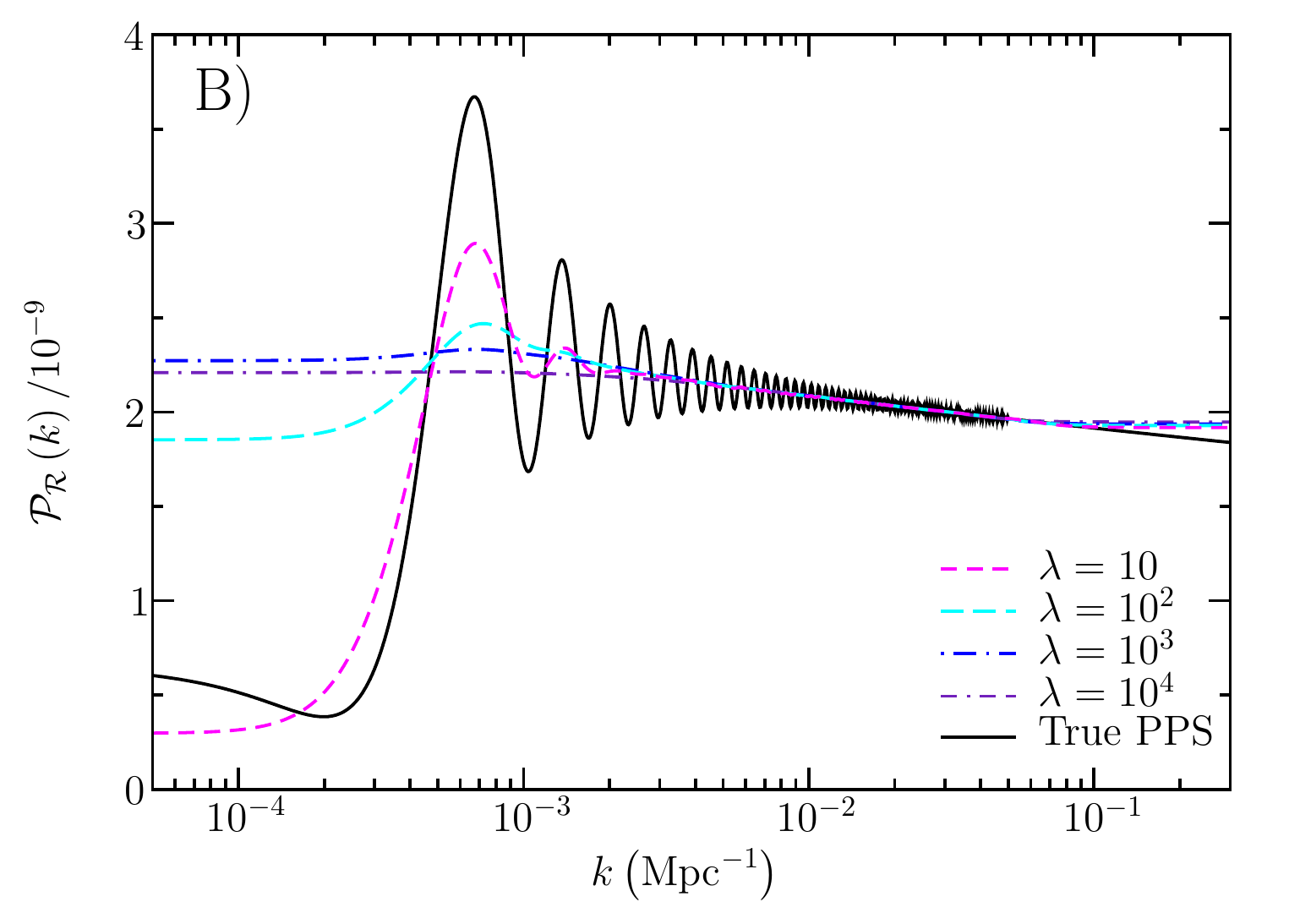}
\includegraphics*[angle=0,width=0.5\columnwidth,trim = 32mm 171mm 23mm
  15mm, clip]{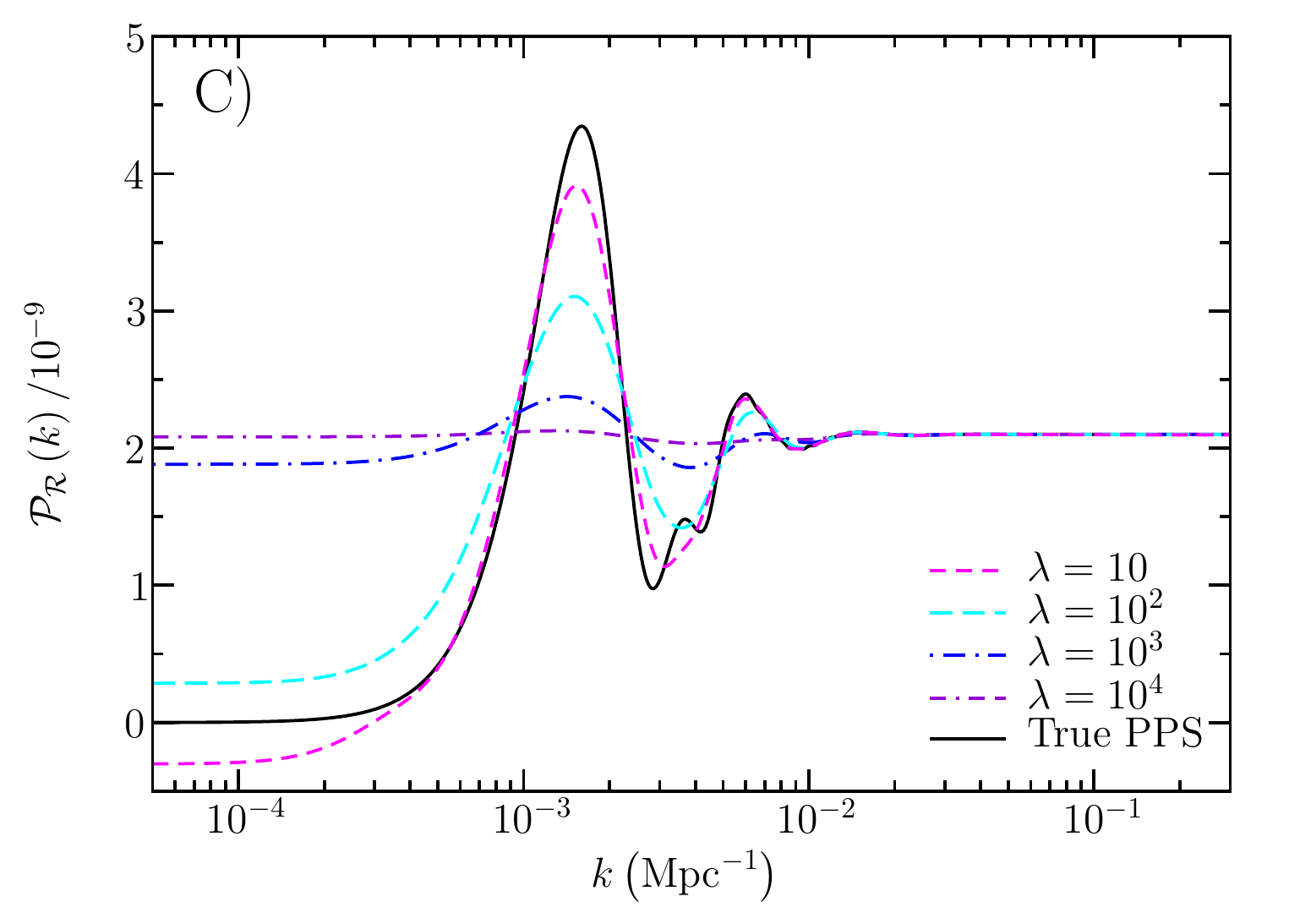}
\includegraphics*[angle=0,width=0.5\columnwidth,trim = 32mm 171mm 23mm
  15mm, clip]{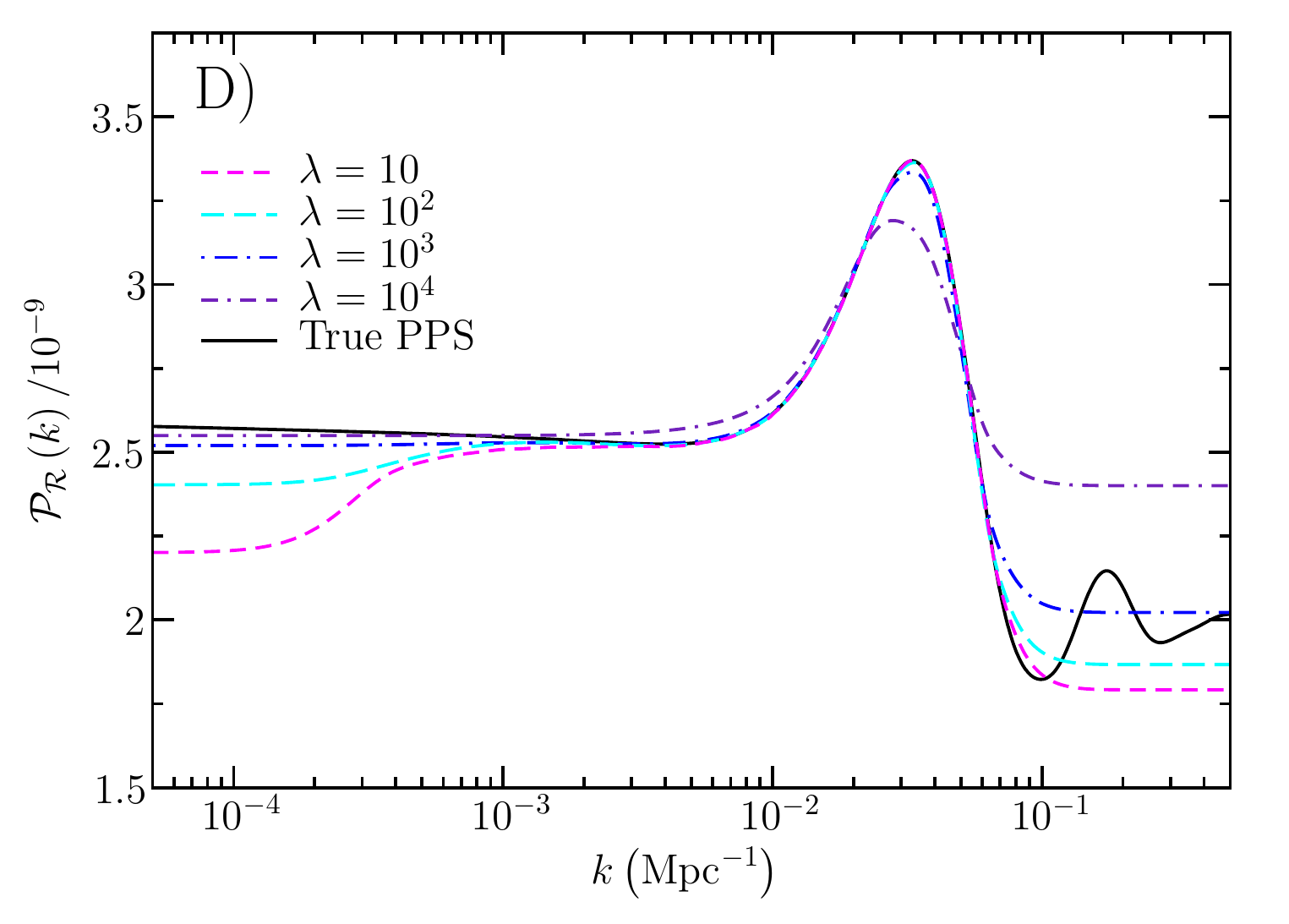}
\caption{\label{mean} Mean reconstructed spectra from $10^5$ mock 
  WMAP-5 TT data realisations with regularisation parameter
  $\lambda=10$, $10^2$, $10^3$ and $10^4$. Results are shown for
  spectra A to D (see Sec.\,\ref{spectest}).}
\end{figure*}

We applied Tikhonov regularisation to $10^5$ mock WMAP-5 TT data
realisations and took the mean of the estimated spectra, with the
results displayed in Fig.\,\ref{mean}. On small and intermediate scales
the mean reconstructions are equivalent to reconstructions obtained
from noise-free data. However, on large scales this is not the case
for small values of $\lambda$: the mean reconstructions have less
power and are clearly biased. 

\begin{figure*}
\includegraphics*[angle=0,width=0.5\columnwidth,trim = 32mm 171mm 23mm
  15mm, clip]{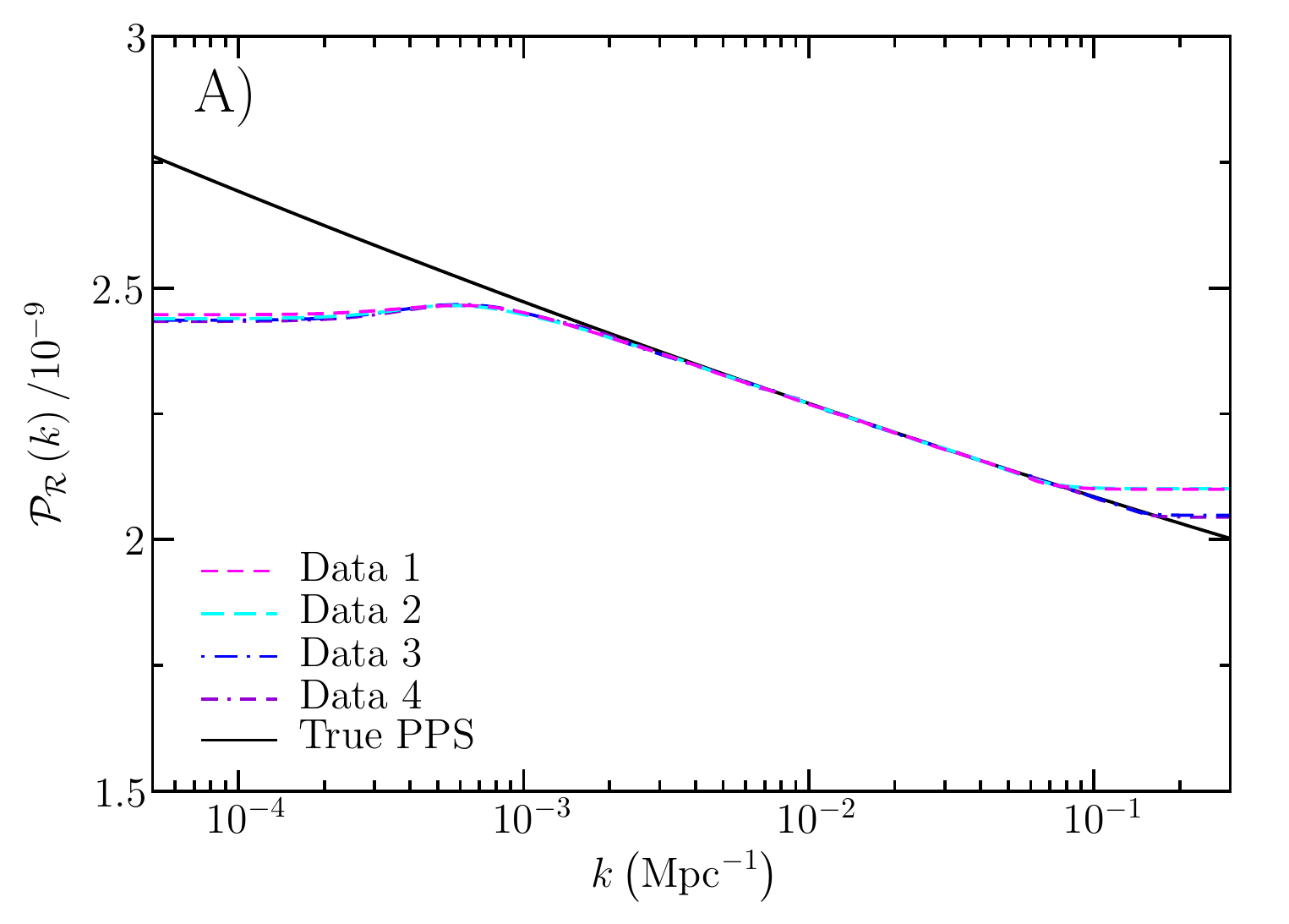}
\includegraphics*[angle=0,width=0.5\columnwidth,trim = 32mm 171mm 23mm
  15mm, clip]{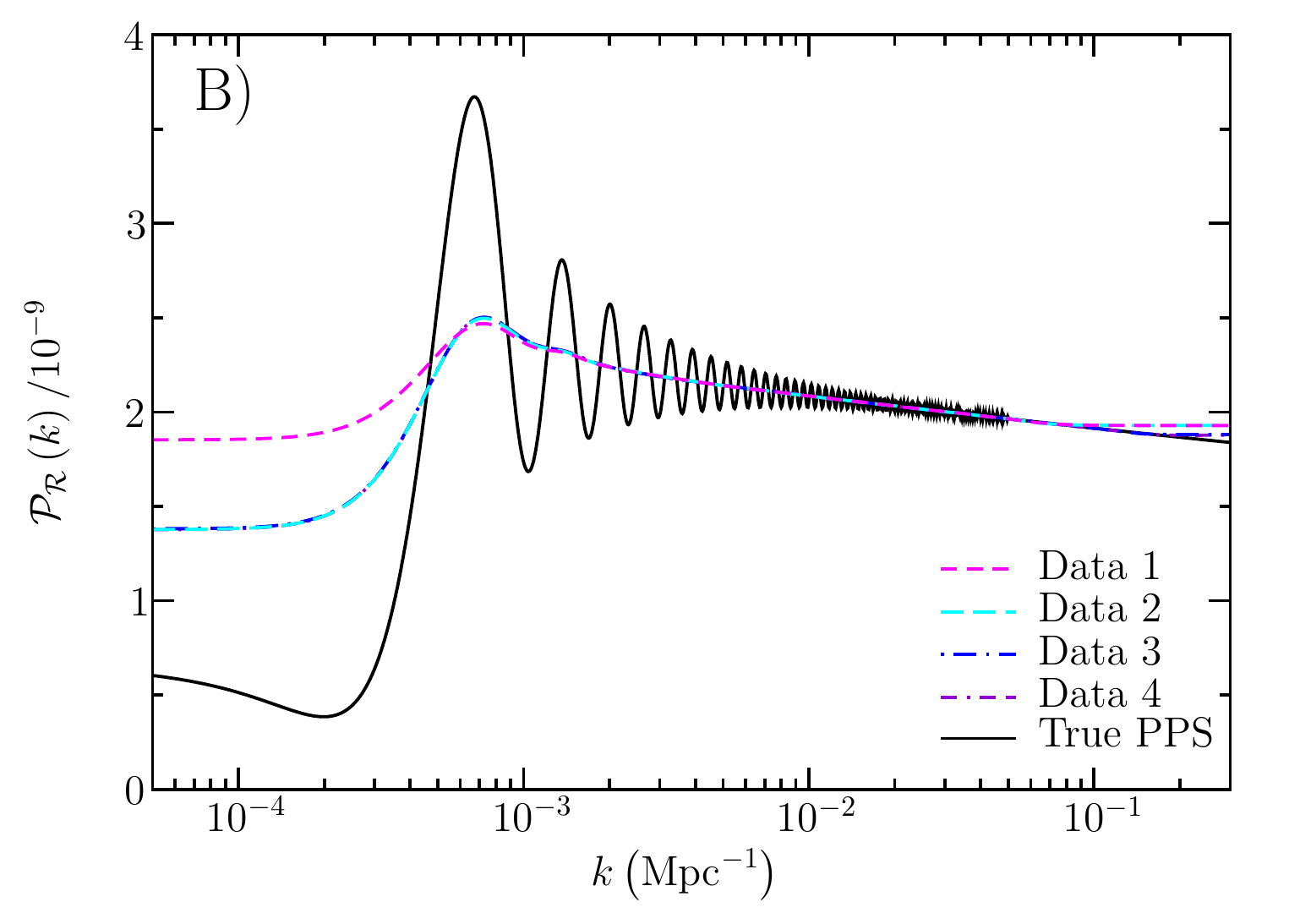}
\includegraphics*[angle=0,width=0.5\columnwidth,trim = 32mm 171mm 23mm
  15mm, clip]{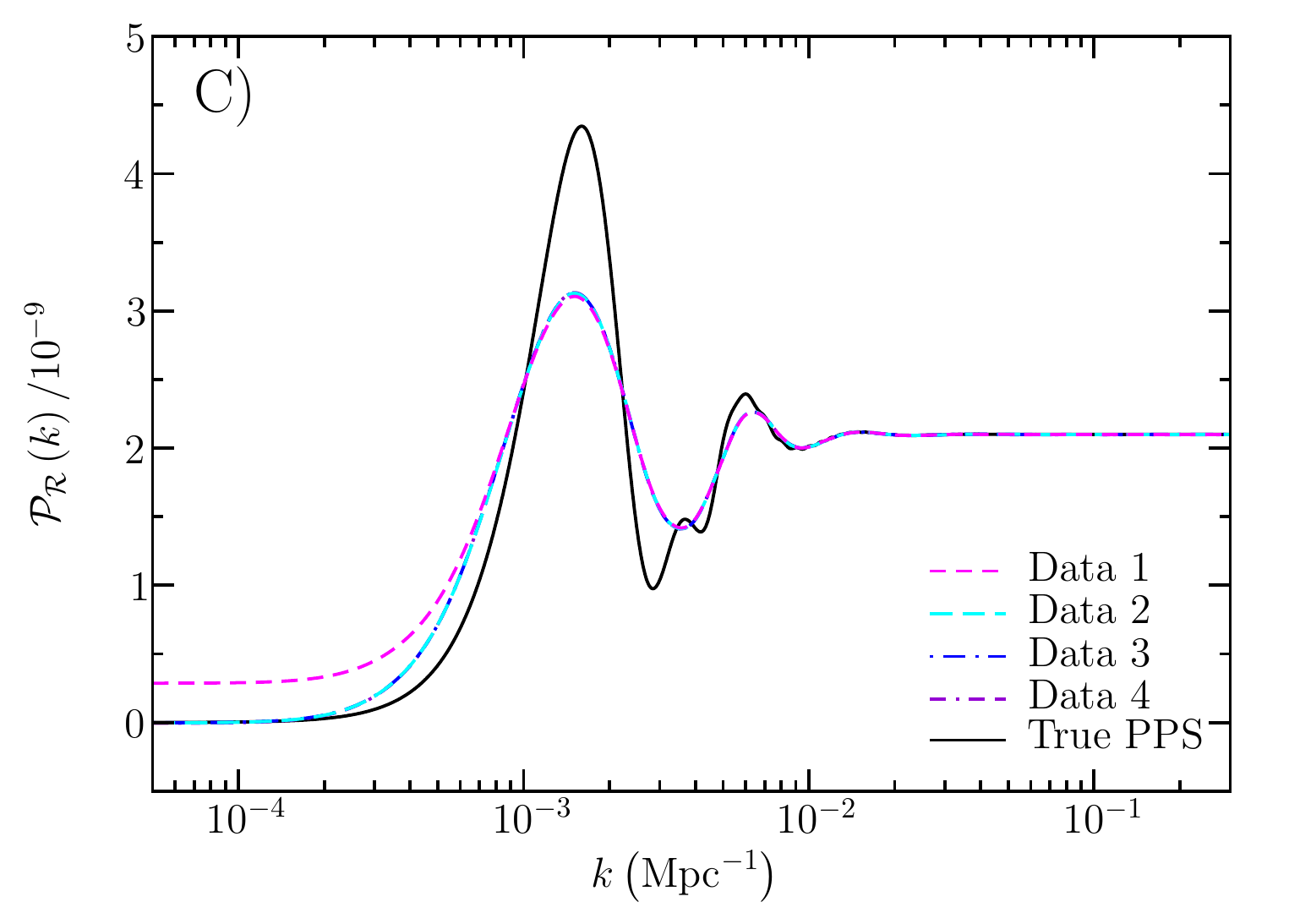}
\includegraphics*[angle=0,width=0.5\columnwidth,trim = 32mm 171mm 23mm
  15mm, clip]{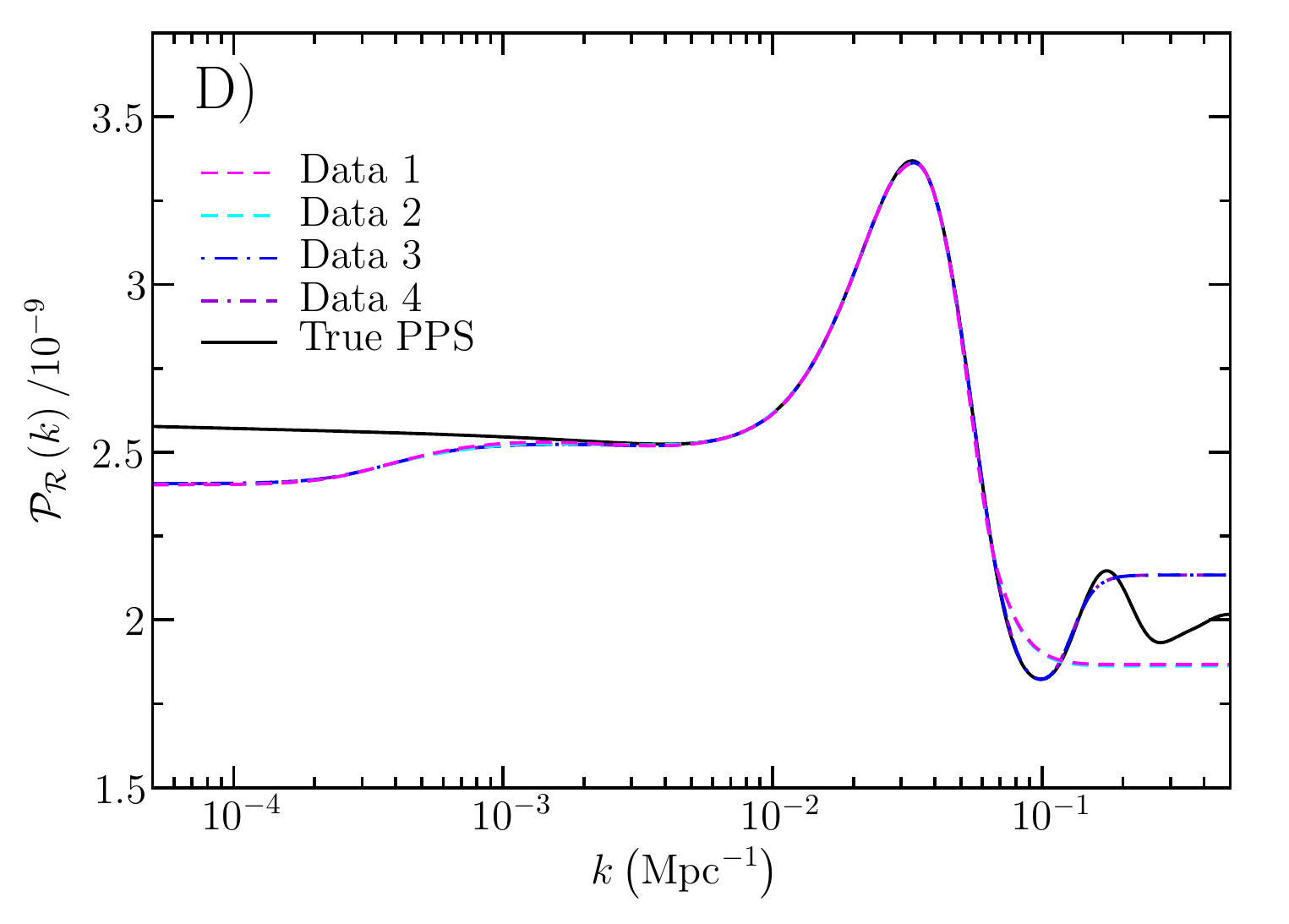}
\caption{\label{mean2} Mean reconstructed test spectra A--D (see
  Sec.\,\ref{spectest}) from $10^5$ mock realisations of the four
  different data set combinations (see Sec.\,\ref{spectra}), with
  $\lambda=100$.}
\end{figure*}

Fig.\,\ref{mean2} shows the mean reconstructed spectra taking
$\lambda=100$, but using all four data combinations to illustrate the
gain when non-WMAP data is available, especially at high $k$.

\subsection{The bias and variance \label{bivar}}

In order to understand the bias on large scales, the Taylor expansion
eq.(\ref{recerr}) is useful since Monte Carlo simulations do not
provide immediate insights into the different sources of bias.
Setting $\hat{\bm{\theta}}=\bm{\theta}_\mathrm{t}$ we define zeroth-,
first- and second-order approximations to $\hat{\pB}$,
\begin{eqnarray}
\label{pz}
\hat{\mathrm{p}}^\mathrm{(z)}_i & \equiv &
\mathcal{T}_i\left(\pB_1,\bm{\theta}_\mathrm{t},\bm{\theta}_\mathrm{t},
\mathbf{0}\right),\\
\label{pf}
\hat{\mathrm{p}}^\mathrm{(f)}_i & \equiv &
\hat{\mathrm{p}}^\mathrm{(z)}_i+\delta\hat{\mathrm{p}}^\mathrm{(f)}_i,\\
\label{ps}
\hat{\mathrm{p}}^\mathrm{(s)}_i & \equiv &
\hat{\mathrm{p}}^\mathrm{(z)}_i+\delta\hat{\mathrm{p}}^\mathrm{(f)}_i+\delta\hat{\mathrm{p}}^\mathrm{(s)}_i,
\end{eqnarray} 
where
\begin{eqnarray}
\label{deltapf}
\delta\hat{\mathrm{p}}^\mathrm{(f)}_i & \equiv &
\sum_{j}R_{ij}\,\Delta \mathrm{p}_j+\sum_{\mathbb{Z},a}
M_{ia}^{(\mathbb{Z})}\,\mathrm{n}_a^{(\mathbb{Z})},\\
\label{deltaps}
\delta\hat{\mathrm{p}}^\mathrm{(s)}_i & \equiv &
\frac{1}{2}\sum_{j,k}Y_{ijk}\, \Delta \mathrm{p}_j\,\Delta
\mathrm{p}_k+ \sum_{\mathbb{Z},j,a}Z_{ija}^{(\mathbb{Z})}\,\Delta
\mathrm{p}_j\,\mathrm{n}_a^{(\mathbb{Z})}+
\frac{1}{2}\sum_{\mathbb{Z},\mathbb{Z'},a,b}X_{iab}^{(\mathbb{Z}\mathbb{Z'})}\,\mathrm{n}_a^{(\mathbb{Z})}\,\mathrm{n}_b^{(\mathbb{Z'})}.
\end{eqnarray} 

\begin{figure*}
\includegraphics*[angle=0,width=0.5\columnwidth,trim = 32mm 171mm 23mm
  15mm, clip]{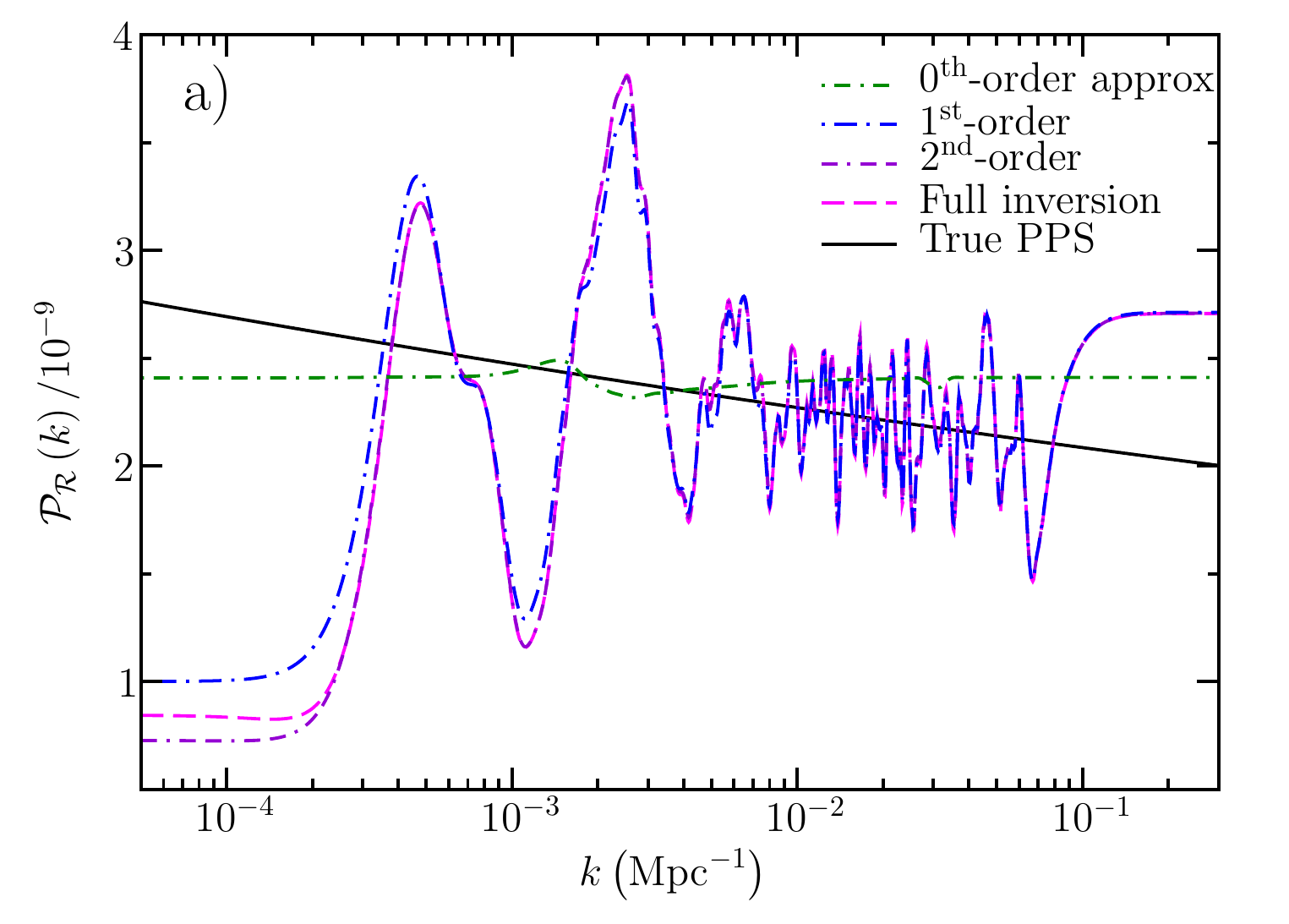}
\includegraphics*[angle=0,width=0.5\columnwidth,trim = 32mm 171mm 23mm
  15mm, clip]{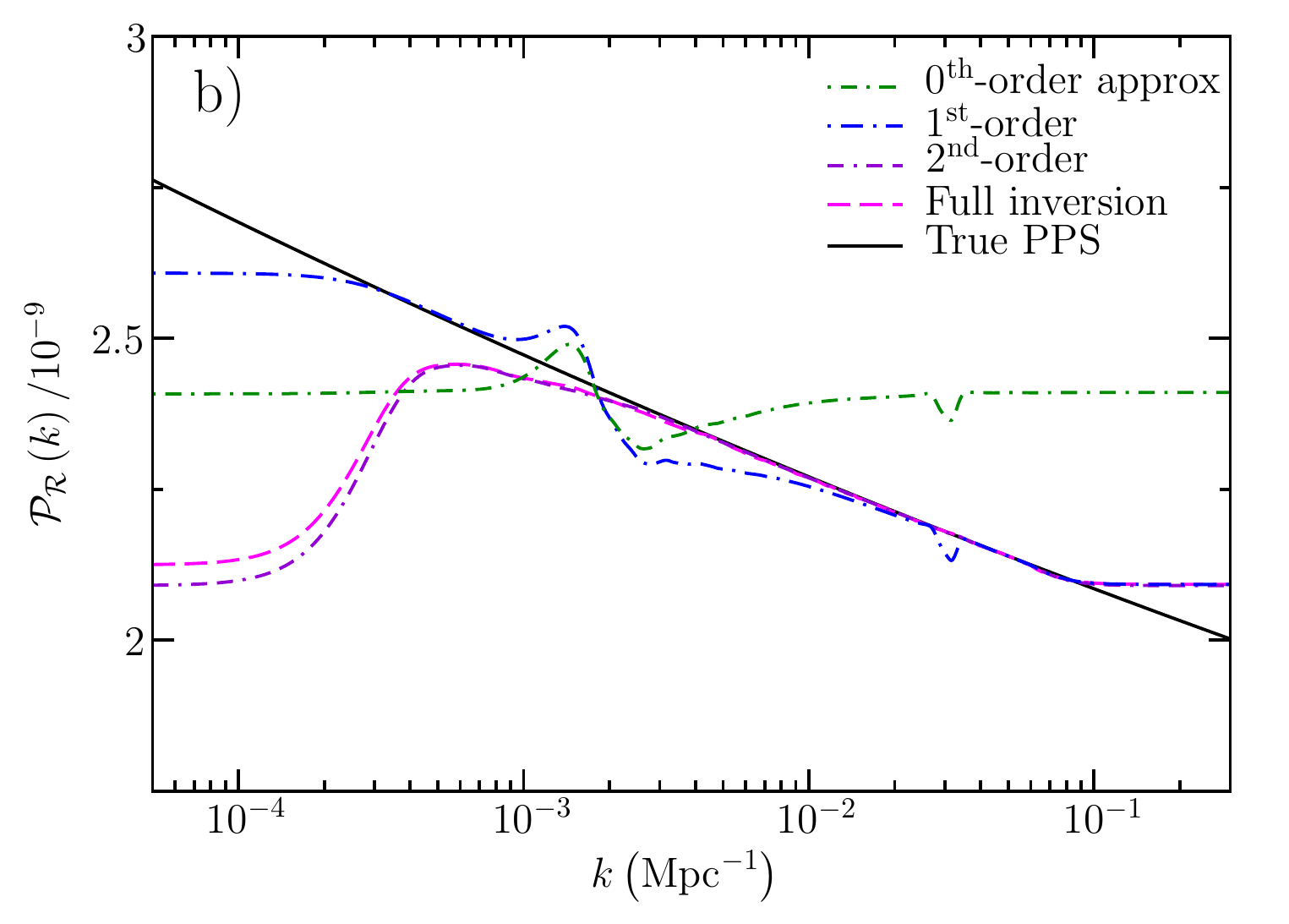}
\caption{\label{approx} Panel (a) shows test spectrum A, the full
  reconstruction $\hat{\pB}\left(\dB\right)$ and the zeroth-, first-
  and second-order approximations $\hat{\pB}^\mathrm{(z)}$
  (eq.\,\ref{pz}), $\hat{\pB}^\mathrm{(f)}$ (eq.\,\ref{pf}) and
  $\hat{\pB}^\mathrm{(s)}$ (eq.\,\ref{ps}) using WMAP-5 temperature and
  polarisation data and $\lambda=10$. Panel (b) again shows test
  spectrum A, and the \emph{mean} full reconstruction
  $\langle\hat{\pB}\left(\dB\right)\rangle$ as well as the mean
  zeroth-, first- and second-order approximations
  $\hat{\pB}^\mathrm{(z)}$, $\langle\hat{\pB}^\mathrm{(f)}\rangle$
  (eq.\,\ref{mpf}) and $\langle\hat{\pB}^\mathrm{(s)}\rangle$
  (eq.\,\ref{mps}).}
\end{figure*}

In the left panel of Fig.\,\ref{approx} the approximations are
compared to the full reconstruction for a simulated WMAP temperature
and polarisation data set, with $\pB_1$ equal to a H-Z spectrum of
amplitude $2.41\times10^{-9}$. It can be seen that $\pB^\mathrm{(z)}$ exhibits
significant departures from $\pB_1$, for the following reason. The
likelihood function for Gaussian distributed data $\dB^\mathrm{G}$
with a covariance matrix $\mathbf{N}^\mathrm{G}$ dependent on the PPS
is
\begin{equation}
L_\mathrm{G}=\left(\sB^\mathrm{G} - 
\dB^\mathrm{G}\right)^\mathrm{T}\mathsf{N}^\mathrm{G}\left(\sB^\mathrm{G} - 
\dB^\mathrm{G}\right)+\ln\det N^\mathrm{G},
\end{equation}   
where $\sB^\mathrm{G}\equiv\mathsf{W}^\mathrm{G}\pB$. It has the derivative:
\begin{equation}
\frac{\partial L_\mathrm{G}}{\partial \mathrm{p}_\alpha} = 
2\sum_{a,a^\prime} W^\mathrm{G}_{a\alpha}
\left(N^\mathrm{G}\right)^{-1}_{aa^\prime}
\left(\mathrm{s}_{a^\prime}^\mathrm{G}-\mathrm{d}_{a^\prime}^\mathrm{G}\right)+
\sum_{a,a^\prime}\left[\left(\mathrm{s}_a^\mathrm{G}-\mathrm{d}_a^\mathrm{G}\right)
\left(\mathrm{s}_{a^\prime}^\mathrm{G}-\mathrm{d}_{a^\prime}^\mathrm{G}\right)
-N_{aa^\prime}^\mathrm{G}\right]
\frac{\partial \left(N^\mathrm{G}\right)^{-1}_{aa^\prime}}{\partial \mathrm{p}_\alpha}.
\end{equation}
The WMAP TE data for $\ell\geq24$ has this type of likelihood
function. The approximation $\pB^\mathrm{(z)}$ differs from $\pB_1$
because the derivative does not vanish when $\pB=\pB_\mathrm{t}$ for noise free
data (recall that $\hat{\pB}$ is defined by $\left.\partial
Q\left(\pB,\dB,\hat{\bm{\theta}},\lambda\right)/\partial
\mathrm{p}_\alpha\right|_{\pB=\hat{\pB}}=0$).

Due to the non-Gaussianity of the WMAP likelihood function at low
multipoles, $\hat{\pB}^\mathrm{(s)}$ is closer to $\hat{\pB}$ than
$\hat{\pB}^\mathrm{(f)}$ for $k\lesssim
5\times10^{-3}\;\mathrm{Mpc}^{-1}$. The means of the first- and
second-order approximations are
\begin{eqnarray}
\label{mpf}
\langle\hat{\mathrm{p}}^\mathrm{(f)}_i\rangle & = &
\hat{\mathrm{p}}^\mathrm{(z)}_i+\sum_{j}R_{ij}\,\Delta \mathrm{p}_j,\\
\label{mps}
\langle\hat{\mathrm{p}}^\mathrm{(s)}_i\rangle & \equiv &
\langle\hat{\mathrm{p}}^\mathrm{(f)}_i\rangle +
\frac{1}{2}\sum_{j,k}Y_{ijk}\,\Delta \mathrm{p}_j\,\Delta p_k
+\frac{1}{2}\sum_{\mathbb{Z},a,b} X_{iab}^{(\mathbb{ZZ})}\,
N_{ab}^{(\mathbb{Z})}.
\end{eqnarray} 
These are shown in the right panel of Fig.\,\ref{approx} together with
the mean full reconstruction from the simulated data
realisations. Since $\langle\partial L_\mathrm{G}/\partial
\mathrm{p}_\alpha\rangle=0$ for $\pB=\pB_\mathrm{t}$ the approximation
$\hat{\pB}^\mathrm{(s)}$ (and $\hat{\pB}$) is unbiased for
$5\times10^{-3}\lesssim k\lesssim 0.1\;\mathrm{Mpc}^{-1}$, unlike
$\hat{\pB}^\mathrm{(f)}$. As
$\langle\hat{\mathrm{p}}^\mathrm{(s)}_i\rangle$ is a good
approximation to the mean reconstruction, we conclude that $\hat{\pB}$
is biased low on large scales by the non-Gaussian likelihood function.

\begin{figure*}
\includegraphics*[angle=0,width=0.5\columnwidth,trim = 32mm 171mm 23mm
  15mm, clip]{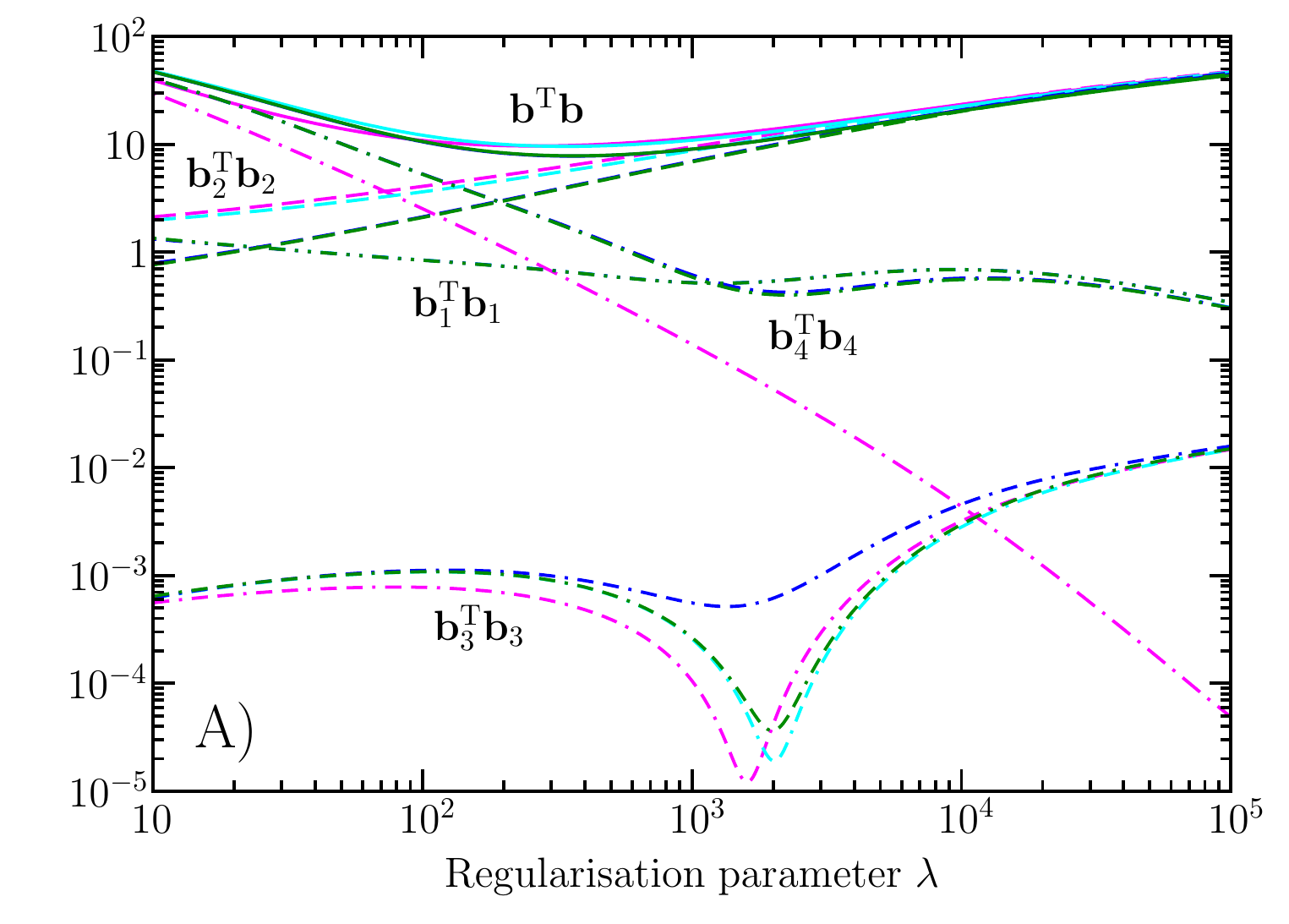}
\includegraphics*[angle=0,width=0.5\columnwidth,trim = 32mm 171mm 23mm
  15mm, clip]{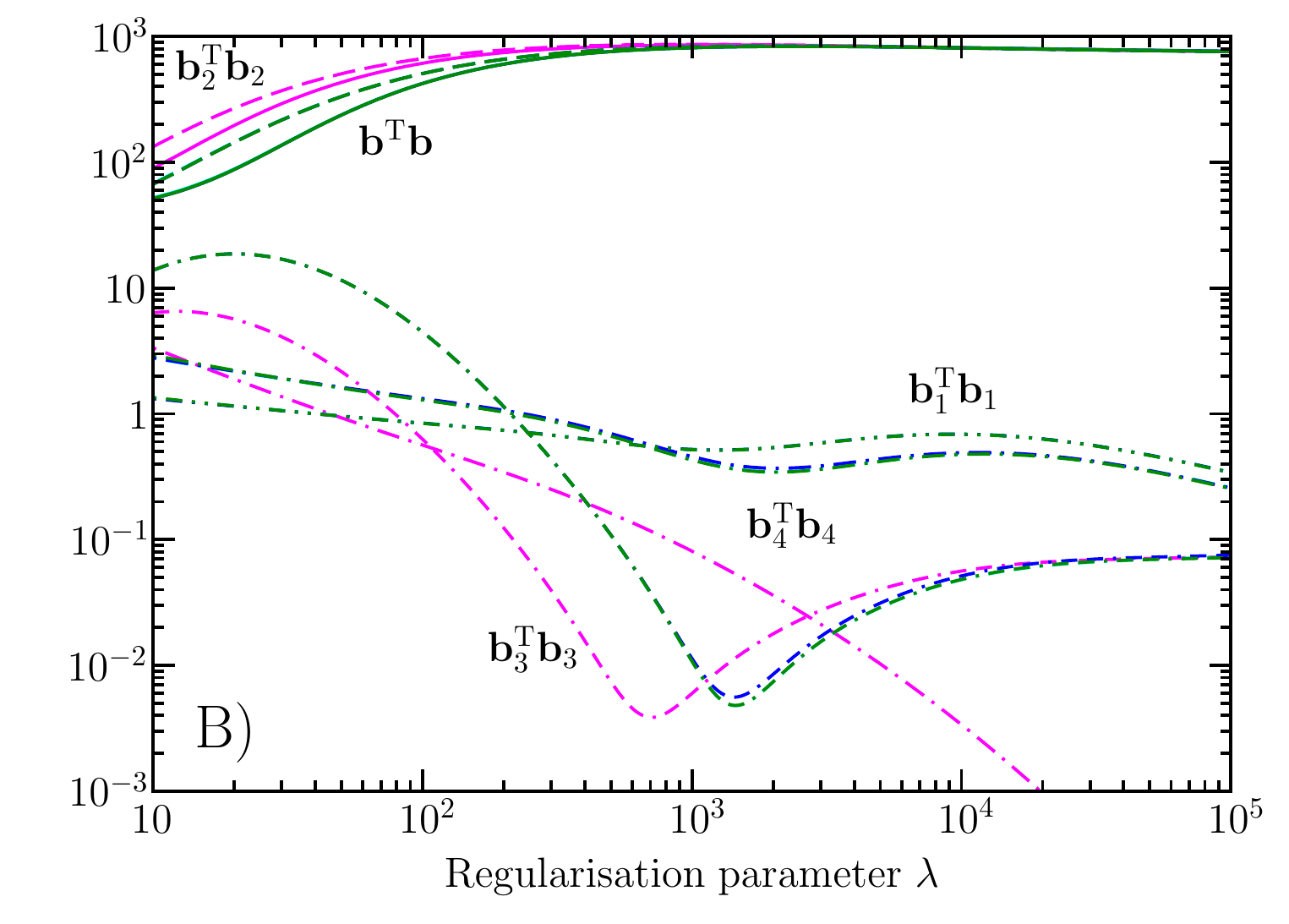}
\includegraphics*[angle=0,width=0.5\columnwidth,trim = 32mm 171mm 23mm
  15mm, clip]{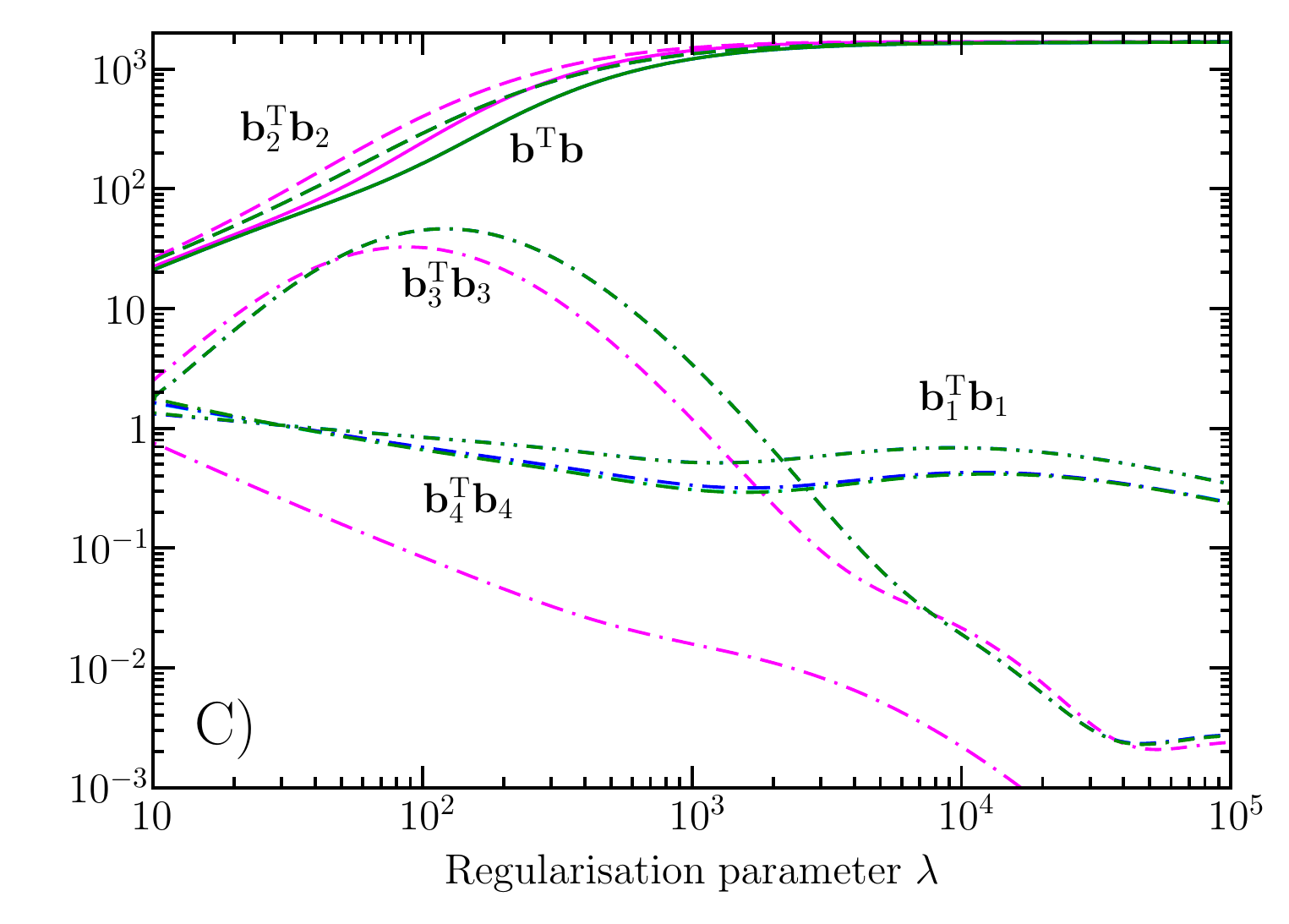}
\includegraphics*[angle=0,width=0.5\columnwidth,trim = 32mm 171mm 23mm
  15mm, clip]{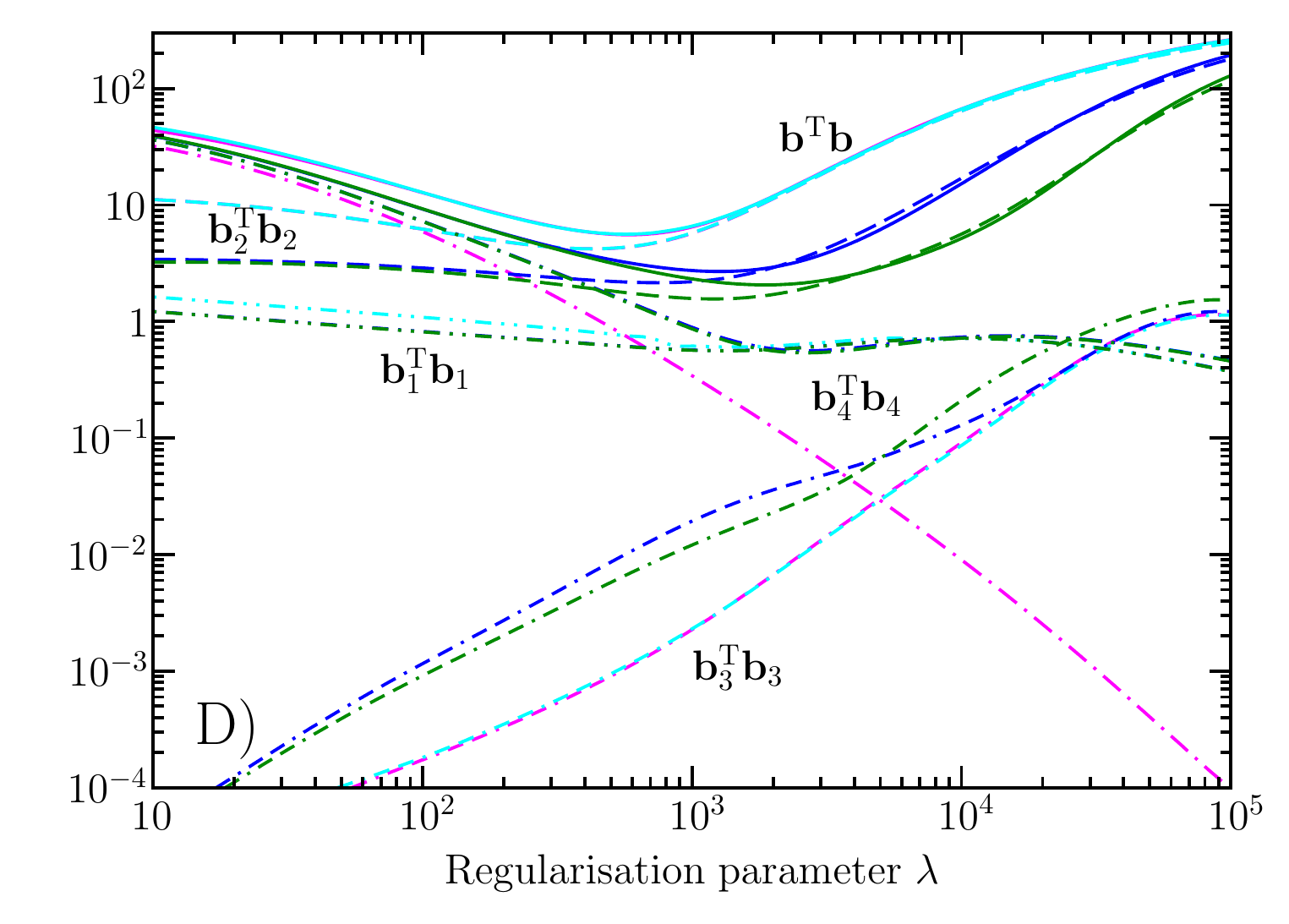}
\caption{\label{bias} The norm of the bias vectors as labelled versus
  the regularisation parameter $\lambda$, for spectra A to D (see
  Sec.\,\ref{spectest}), and various data combinations as defined in
  Sec.\,\ref{spectra} (Combination 1: magenta lines, Combination 2: cyan
  lines, Combination 3: blue lines, Combination 4: green lines). The
  dot-dot-dashed lines show $\bB_1^\mathrm{T}\bB_1$ (eq.\,\ref{bias1}),
  the dashed lines $\bB_2^\mathrm{T}\bB_2$ (eq.\,\ref{bias2}), the
  dot-dash-dashed lines show $\bB_3^\mathrm{T}\bB_3$
  (eq.\,\ref{bias3}), the dot-dashed lines $\bB_4^\mathrm{T}\bB_4$
  (eq.\,\ref{bias4}), and the solid lines show $\bB^\mathrm{T}\bB$.}
\end{figure*}

To further study the bias we introduce the four vectors
\begin{eqnarray}
\label{bias1}
b_{1i} & \equiv & \mathcal{T}_i\left(\pB_1,\bm{\theta}_\mathrm{t},\bm{\theta}_\mathrm{t},\mathbf{0}\right) - 
\mathrm{p}_{1i},\\
\label{bias2}
b_{2i} & \equiv & \sum_{j}\left(R_{ij}-I_{ij}\right)\,\Delta \mathrm{p}_j,\\
\label{bias3}
b_{3i} & \equiv & \frac{1}{2}\sum_{j,k}Y_{ijk}\,\Delta \mathrm{p}_j\,\Delta \mathrm{p}_k,\\
\label{bias4}
b_{4i} & \equiv & \frac{1}{2}\sum_{\mathbb{Z},a,b} X_{iab}^{(\mathbb{ZZ})}\, N_{ab}^{(\mathbb{Z})},
\end{eqnarray}  
and their sum $\bB$, the second-order approximation to the bias. In
Fig.\,\ref{bias} the norm of these vectors is plotted as a function of
$\lambda$ for the various test spectra and combinations of data sets,
again with $\pB_1$ set to a H-Z spectrum of amplitude
$2.41\times10^{-9}$. For the WMAP TT data alone $\bB_1=\mathbf{0}$
because the H-Z spectrum $\pB_1$ is successfully recovered by the
inversion. However, once the WMAP TE data is included $\bB_1$ no
longer vanishes as $\pB_1$ is imperfectly recovered due to the
determinant term in the WMAP TE likelihood function, as discussed
above. The vector $\bB_1$ is the least dependent on $\lambda$ of the
four we study.

The norm of $\bB_2$ generally increases with $\lambda$ as the
resolution of the reconstruction is reduced. The exception occurs for
spectrum D, where for $10 \lesssim\lambda\lesssim 1000$ the norm falls
as $\lambda$ increases. This is because, as can be seen in
Fig.\,\ref{mean}, increasing the regularisation parameter from
$\lambda=10$ to $\lambda=1000$ brings the mean reconstruction closer
to true spectrum on the smallest scales, decreasing the bias.  Adding
WMAP-5 polarisation data to the temperature data reduces the norm of
$\bB_2$ if there are features in the PPS on large scales, such as an
infrared cutoff. Hence the reduction is largest for spectra B and C,
less so for spectrum A, and smallest for spectrum D. Similarly, adding
the small-scale CMB data reduces the extrapolation required in the
estimated PPS at high wavenumbers, improving the mean reconstruction
if the PPS departs from scale invariance there. Thus
$\bB_2^\mathrm{T}\bB_2$ decreases the most for spectrum D when the
small-scale CMB data is used. The SDSS-4 LRG data does not cover
higher wavenumbers than the small-scale CMB data. The latter
constrains the PPS sufficiently well that adding the SDSS-4 LRG data
does not further reduce the bias for small values of $\lambda$, as can
be seen for spectrum D in Fig.\,\ref{mean2}.  For higher $\lambda$ the
inversions are less sensitive to the data and the mean reconstruction
using the small scale CMB data is smoothed away from the true PPS.
Adding the SDSS-4 LRG data then reduces the smoothing and the bias,
provided the true PPS is not scale invariant at high wavenumbers.

The vector $\bB_3$ represents the bias component originating in the
quadratic mapping from the true PPS to the reconstruction.  It is a
subdominant contribution and depends strongly on the true PPS due to
the complicated structure of the second-order resolution kernel. The
vector $\bB_4$ arises from the non-Gaussianity of the WMAP likelihood
function at low multipoles and corresponds to the resulting
suppression of the recovered PPS on large scales. For our choice of
roughness function $Q\left(\pB,\dB\right)$ becomes more quadratic in
$\pB$ as the regularisation parameter increases. Thus
$\bB_4^\mathrm{T}\bB_4$ is less for larger values of $\lambda$.
Adding the WMAP TE data makes $Q\left(\pB,\dB\right)$ more
non-Gaussian and leads to a greater suppression on large scales. When
the polarisation data is included $\bB_4$ contains a component that
cancels with $\bB_1$ to leave the reconstruction unbiased on
intermediate scales. Hence $\bB_1^\mathrm{T}\bB_1$ and
$\bB_4^\mathrm{T}\bB_4$ are similar for high $\lambda$. For low
$\lambda$ values $\bB_4^\mathrm{T}\bB_4$ is greater for spectra A and
D than for B and C. This is because the low multipole diagonal
elements of the covariance matrix $\mathsf{N}^\mathrm{TT}$ are larger
for spectra without an infrared cutoff.

The frequentist covariance matrix of $\hat{\pB}^\mathrm{(f)}$ is given
by $\mathsf{\Sigma}_\mathrm{F}$ (eq.\,\ref{sigmaf}), the elements of
which are given by:
\begin{eqnarray}
\label{sigf}
\mathsf{\Sigma}_{\mathrm{F}_{|ij}} =
\mathsf{\Sigma}^{\mathrm{(f)}}_{ij} & \equiv &
\langle\left(\delta\hat{\mathrm{p}}^\mathrm{(f)}_i-\langle
\delta\hat{\mathrm{p}}^\mathrm{(f)}_i\rangle\right)\left(\delta\hat{\mathrm{p}}^\mathrm{(f)}_j-\langle
\delta\hat{\mathrm{p}}^\mathrm{(f)}_j\rangle\right)\rangle,\\ & = &
\sum_{\mathbb{Z},a,b} M_{ia}^{(\mathbb{Z})}\,
N_{ab}^{(\mathbb{Z})}\,M_{jb}^{(\mathbb{Z})} ,
\end{eqnarray}  
in terms of the first-order differences
$\delta\hat{\mathrm{p}}^\mathrm{(f)}_i$ (eq.\,\ref{deltapf}).  The
frequentist covariance matrix of $\hat{\pB}^\mathrm{(s)}$ is
similarly:
\begin{eqnarray}
\label{sigs}
\mathsf{\Sigma}^{(\mathrm{s})} & \equiv &
\langle\left(\hat{\pB}^\mathrm{(s)}-\langle
\hat{\pB}^\mathrm{(s)}\rangle\right)\left(\hat{\pB}^\mathrm{(s)}-\langle
\hat{\pB}^\mathrm{(s)}\rangle\right)^\mathrm{T}\rangle,\\ & = &
\mathsf{\Sigma}^{(\mathrm{f})}+\mathsf{\Sigma}^\mathrm{(ss)}+2\mathsf{\Sigma}^{(\mathrm{fs})},
\end{eqnarray}  
where
\begin{eqnarray}
\label{sigss}
\mathsf{\Sigma}^{\mathrm{(ss)}}_{ij} & \equiv & \langle\left(\delta\hat{\mathrm{p}}^\mathrm{(s)}_i-\langle
\delta\hat{\mathrm{p}}^\mathrm{(s)}_i\rangle\right)\left(\delta\hat{\mathrm{p}}^\mathrm{(s)}_j-\langle
\delta\hat{\mathrm{p}}^\mathrm{(s)}_j\rangle\right)\rangle,\\ & = &
\frac{1}{4}\sum_{\mathbb{Z},a,b,c,d}
X_{iab}^{(\mathbb{ZZ})}\left(N_{abcd}^{(\mathbb{Z})4}-N_{ab}^{(\mathbb{Z})}\,N_{cd}^{(\mathbb{Z})}\right) 
X_{jcd}^{(\mathbb{ZZ})}\nonumber\\ &
& +\sum_{\mathbb{Z},k,l,a,b} Z_{ika}^{(\mathbb{Z})}\,\Delta \mathrm{p}_k\,N_{ab}^{(\mathbb{Z})}\,\Delta
\mathrm{p}_l\, Z_{jlb}^{(\mathbb{Z})} +\sum_{\mathbb{Z},k,a,b,c}
X_{iab}^{(\mathbb{Z})}\,N_{abc}^{(\mathbb{Z})3}\,\Delta \mathrm{p}_k\,
Z_{jkc}^{(\mathbb{Z})},\\ 
\label{sigfs}
\mathsf{\Sigma}^{\mathrm{(fs)}}_{ij} & \equiv & \langle\left(\delta\hat{\mathrm{p}}^\mathrm{(f)}_i-\langle
\delta\hat{\mathrm{p}}^\mathrm{(f)}_i\rangle\right)\left(\delta\hat{\mathrm{p}}^\mathrm{(s)}_j-\langle
\delta\hat{\mathrm{p}}^\mathrm{(s)}_j\rangle\right)\rangle,\\ & = &
\frac{1}{2}\sum_{\mathbb{Z},a,b,c} M_{ia}^{(\mathbb{Z})}\,
N_{abc}^{(\mathbb{Z})3}\,X_{jbc}^{(\mathbb{ZZ})} +\sum_{\mathbb{Z},k,a,b}
M_{ia}^{(\mathbb{Z})}\,N_{ab}^{(\mathbb{Z})}\,\Delta \mathrm{p}_k\, Z_{jkb}^{(\mathbb{Z})}.
\end{eqnarray}  
Here $N_{abc}^{(\mathbb{Z})3}\equiv\langle \mathrm{n}_a^{(\mathbb{Z})} \mathrm{n}_b^{(\mathbb{Z})}
\mathrm{n}_c^{(\mathbb{Z})}\rangle$ and $N_{abc}^{(\mathbb{Z})4}\equiv\langle \mathrm{n}_a^{(\mathbb{Z})} \mathrm{n}_b^{(\mathbb{Z})}
\mathrm{n}_c^{(\mathbb{Z})} \mathrm{n}_d^{(\mathbb{Z})}\rangle$.

\begin{figure*}
\includegraphics*[angle=0,width=0.5\columnwidth,trim = 32mm 171mm 23mm
  15mm, clip]{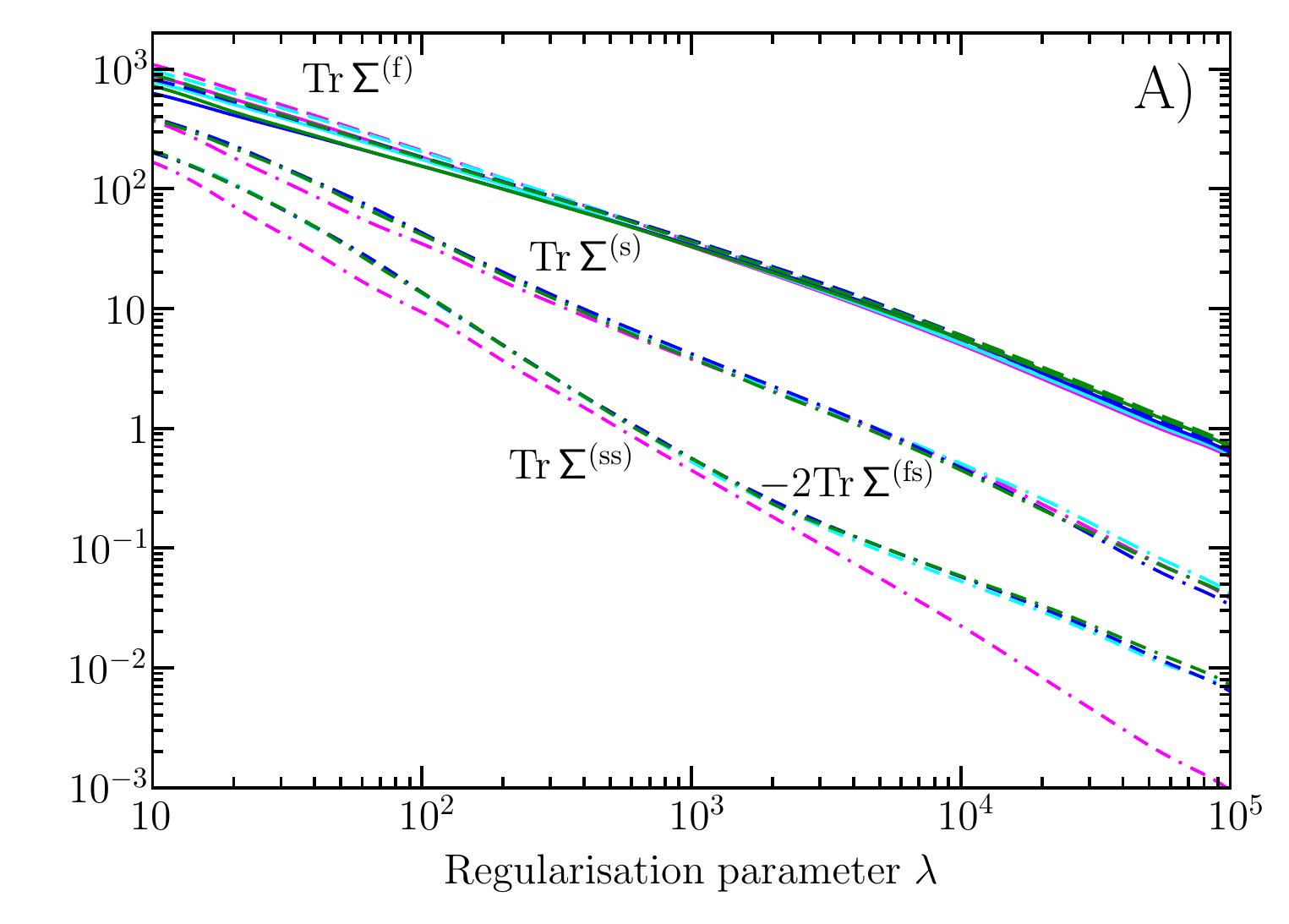}
\includegraphics*[angle=0,width=0.5\columnwidth,trim = 32mm 171mm 23mm
  15mm, clip]{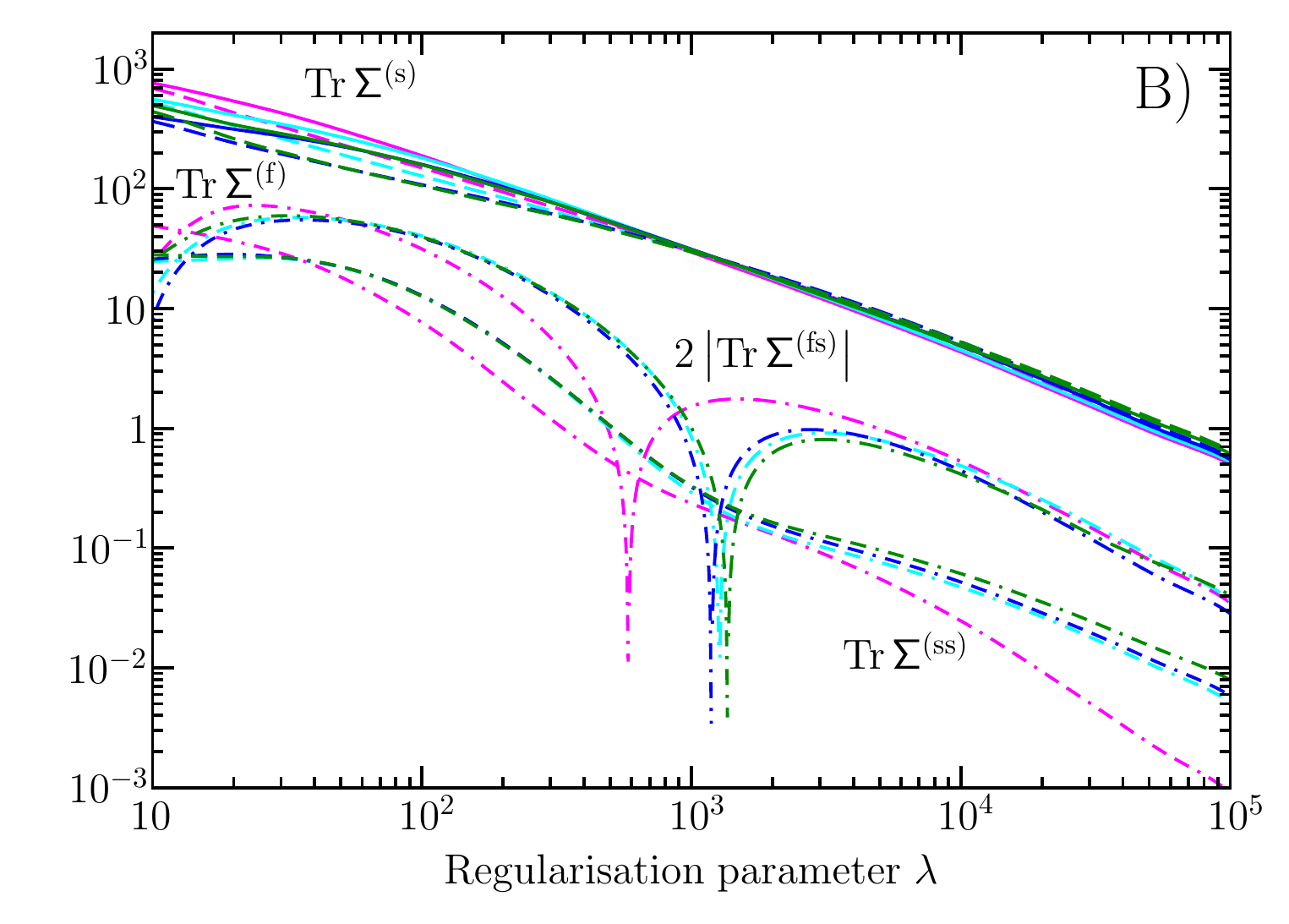}
\includegraphics*[angle=0,width=0.5\columnwidth,trim = 32mm 171mm 23mm
  15mm, clip]{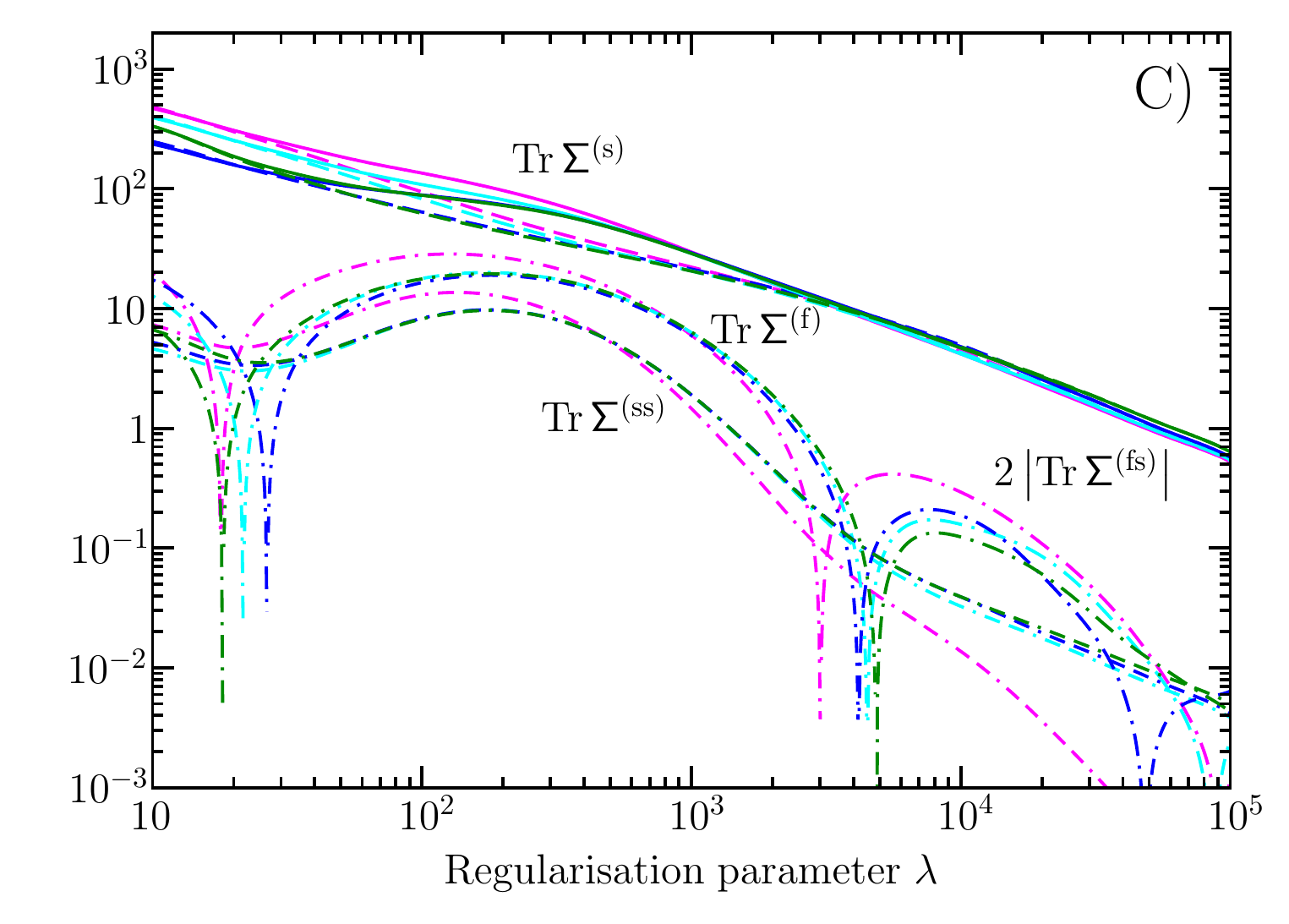}
\includegraphics*[angle=0,width=0.5\columnwidth,trim = 32mm 171mm 23mm
  15mm, clip]{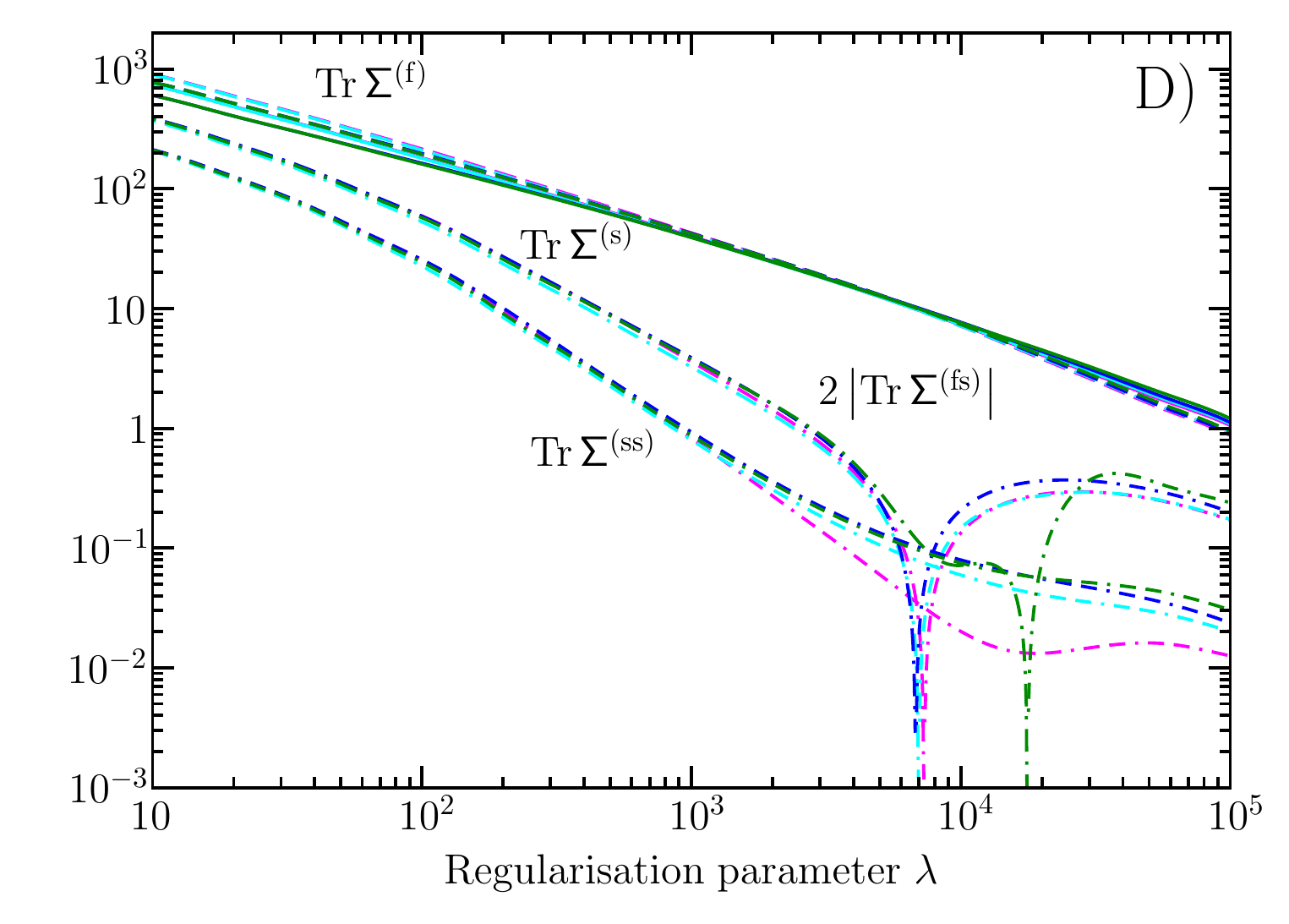}
\caption{\label{variance} The trace of the frequentist covariance
  matrices versus the regularisation parameter $\lambda$ for test
  spectra A to D (see Sec.\,\ref{spectest}) and various data combinations
  as defined in Sec.\,\ref{spectra} (Combination 1: magenta lines,
  Combination 2: cyan lines, Combination 3: blue lines, Combination 4:
  green lines). The dashed lines show
  $\mathrm{Tr}\,\mathsf{\Sigma}^{(\mathrm{f})}$ (eq.\,\ref{sigf}), the
  dot-dash-dashed lines show
  $\mathrm{Tr}\,\mathsf{\Sigma}^\mathrm{(ss)}$ (eq.\,\ref{sigss}), the
  dot-dashed lines show
  $2\left|\mathrm{Tr}\,\mathsf{\Sigma}^{(\mathrm{fs})}\right|$
  (eq.\,\ref{sigfs}), and the solid lines show
  $\mathrm{Tr}\,\mathsf{\Sigma}^{(\mathrm{s})}$ (eq.\,\ref{sigs}).}
\end{figure*}

We repeated the following procedure for many values of $\lambda$:
$\delta \hat{\pB}^{(\mathrm{f})}$ and $\delta
\hat{\pB}^{(\mathrm{s})}$ were calculated for $\Re=10^5$ data
realisations and the quantities $\mathrm{Tr}\,
\mathsf{\Sigma}^{(\mathrm{f})}$, $\mathrm{Tr}\,
\mathsf{\Sigma}^\mathrm{(ss)}$ and $\mathrm{Tr}\,
\mathsf{\Sigma}^\mathrm{(fs)}$ were computed using the estimators
\begin{eqnarray}
\sum_{j=1}^{\Re}
\left(\delta\hat{\pB}_j^\mathrm{(f)} - \delta\bar{\pB}^\mathrm{(f)}\right)^\mathrm{T}
\left(\delta\hat{\pB}_j^\mathrm{(f)} - \delta\bar{\pB}^\mathrm{(f)}\right)/\Re, \\ \nonumber
\sum_{j=1}^{\Re}
\left(\delta\hat{\pB}_j^\mathrm{(s)} - \delta\bar{\pB}^\mathrm{(s)}\right)^\mathrm{T}
\left(\delta\hat{\pB}_j^\mathrm{(s)} - \delta\bar{\pB}^\mathrm{(s)}\right)/\Re, \\ \nonumber
\sum_{j=1}^{\Re}
\left(\delta\hat{\pB}_j^\mathrm{(f)} - \delta\bar{\pB}^\mathrm{(f)}\right)^\mathrm{T}
\left(\delta\hat{\pB}_j^\mathrm{(s)} - \delta\bar{\pB}^\mathrm{(s)}\right)/\Re,
\end{eqnarray}
respectively. Here $\delta\hat{\pB}_j^\mathrm{(f)}$ and
$\delta\hat{\pB}_j^\mathrm{(s)}$ are the vectors $\delta
\hat{\pB}^{(\mathrm{f})}$ and $\delta \hat{\pB}^{(\mathrm{s})}$ for
the $j$th data realisation and $\delta\bar{\pB}^\mathrm{(f)} \equiv
\sum_{j=1}^{\Re} \delta\hat{\pB}_j^\mathrm{(f)}/{\Re}$ and
$\delta\bar{\pB}^\mathrm{(s)} \equiv \sum_{j=1}^{\Re}
\delta\hat{\pB}_j^\mathrm{(s)}/{\Re}$ are the means. This allowed us
to plot $\mathrm{Tr}\, \mathsf{\Sigma}^\mathrm{(f)}$,
$\mathrm{Tr}\,\mathsf{\Sigma}^\mathrm{(ss)}$,
$2\mathrm{Tr}\,\mathsf{\Sigma}^\mathrm{(fs)}$ and their sum,
$\mathrm{Tr}\, \mathsf{\Sigma}^\mathrm{(s)}$, as a function of the
regularisation parameter for the various test spectra and combinations
of data sets. The results are shown in Fig.\,\ref{variance}, again for
$\pB_1=2.41\times10^{-9}$.

Noise-induced artifacts in a reconstruction are more suppressed for
high values of $\lambda$.  Since the artifacts vary with different
data realisations, this means that $\mathrm{Tr}\,
\mathsf{\Sigma}^{(\mathrm{f})}$ and $\mathrm{Tr}\,
\mathsf{\Sigma}^{(\mathrm{s})}$ decrease as $\lambda$ increases. At
higher $\lambda$ values second-order effects are smaller and
$\mathrm{Tr}\, \mathsf{\Sigma}^{(\mathrm{s})}$ is closer to
$\mathrm{Tr}\, \mathsf{\Sigma}^{(\mathrm{f})}$ because
$Q\left(\pB,\dB\right)$ becomes more quadratic for larger
$\lambda$. Generally $2\mathrm{Tr}\, \mathsf{\Sigma}^{(\mathrm{fs})}$
dominates $\mathrm{Tr}\, \mathsf{\Sigma}^\mathrm{(ss)}$ and determines
how $\mathrm{Tr}\, \mathsf{\Sigma}^{(\mathrm{s})}$ differs from
$\mathrm{Tr}\, \mathsf{\Sigma}^{(\mathrm{f})}$. For spectra A and D,
$\mathrm{Tr}\, \mathsf{\Sigma}^{(\mathrm{fs})}$ is predominantly
negative so $\mathrm{Tr}\, \mathsf{\Sigma}^{(\mathrm{s})}$ is mostly
smaller than $\mathrm{Tr}\, \mathsf{\Sigma}^{(\mathrm{f})}$. A slight
bump in $\mathrm{Tr}\, \mathsf{\Sigma}^{(\mathrm{s})}$ for spectra B
and C corresponds to the maximum of $\mathrm{Tr}\,
\mathsf{\Sigma}^{(\mathrm{fs})}$.  For small $\lambda$, adding the
WMAP polarisation and the small-scale CMB data both reduce
$\mathrm{Tr}\, \mathsf{\Sigma}^{(\mathrm{f})}$ and $\mathrm{Tr}\,
\mathsf{\Sigma}^{(\mathrm{s})}$, but including the SDSS-4 LRG data
increases them. For large $\lambda$, $\mathrm{Tr}\,
\mathsf{\Sigma}^{(\mathrm{f})}$ and $\mathrm{Tr}\,
\mathsf{\Sigma}^{(\mathrm{s})}$ increase with each additional data
set.

\subsection{The MSE and MPE \label{msempe}}

Next, inversions were performed on $10^5$ data realisations and the
squared bias
$\langle\hat{\pB}-\pB_\mathrm{t}\rangle^\mathrm{T}\langle\hat{\pB}-\pB_\mathrm{t}\rangle$,
variance $\langle\left(\hat{\pB}-\langle
\hat{\pB}\rangle\right)^\mathrm{T}\left(\hat{\pB}-\langle
\hat{\pB}\rangle\right)\rangle$ and their sum, the MSE were computed
using the estimators
$\left(\bar{\pB}-\pB_\mathrm{t}\right)^\mathrm{T}\left(\bar{\pB}-\pB_\mathrm{t}\right)$,
$\sum_{j=1}^\Re\left(\hat{\pB}_j -
\bar{\pB}\right)^\mathrm{T}\left(\hat{\pB}_j-\bar{\pB}\right)/\Re$,
and $\sum_{j=1}^R\left(\hat{\pB}_j -
\pB_\mathrm{t}\right)^\mathrm{T}\left(\hat{\pB}_j-\pB_\mathrm{t}\right)/\Re$,
respectively. Here $\hat{\pB}_j$ is the $j$th of the $\Re=10^5$
reconstructions and $\bar{\pB}\equiv\sum_{j=1}^\Re\hat{\pB}_j/\Re$ is
the mean reconstruction. The MPE of the reconstructions was also
calculated.  This was repeated for many values of the regularisation
parameter.

\begin{figure*}
\includegraphics*[angle=0,width=0.5\columnwidth,trim = 32mm 171mm 23mm
  15mm, clip]{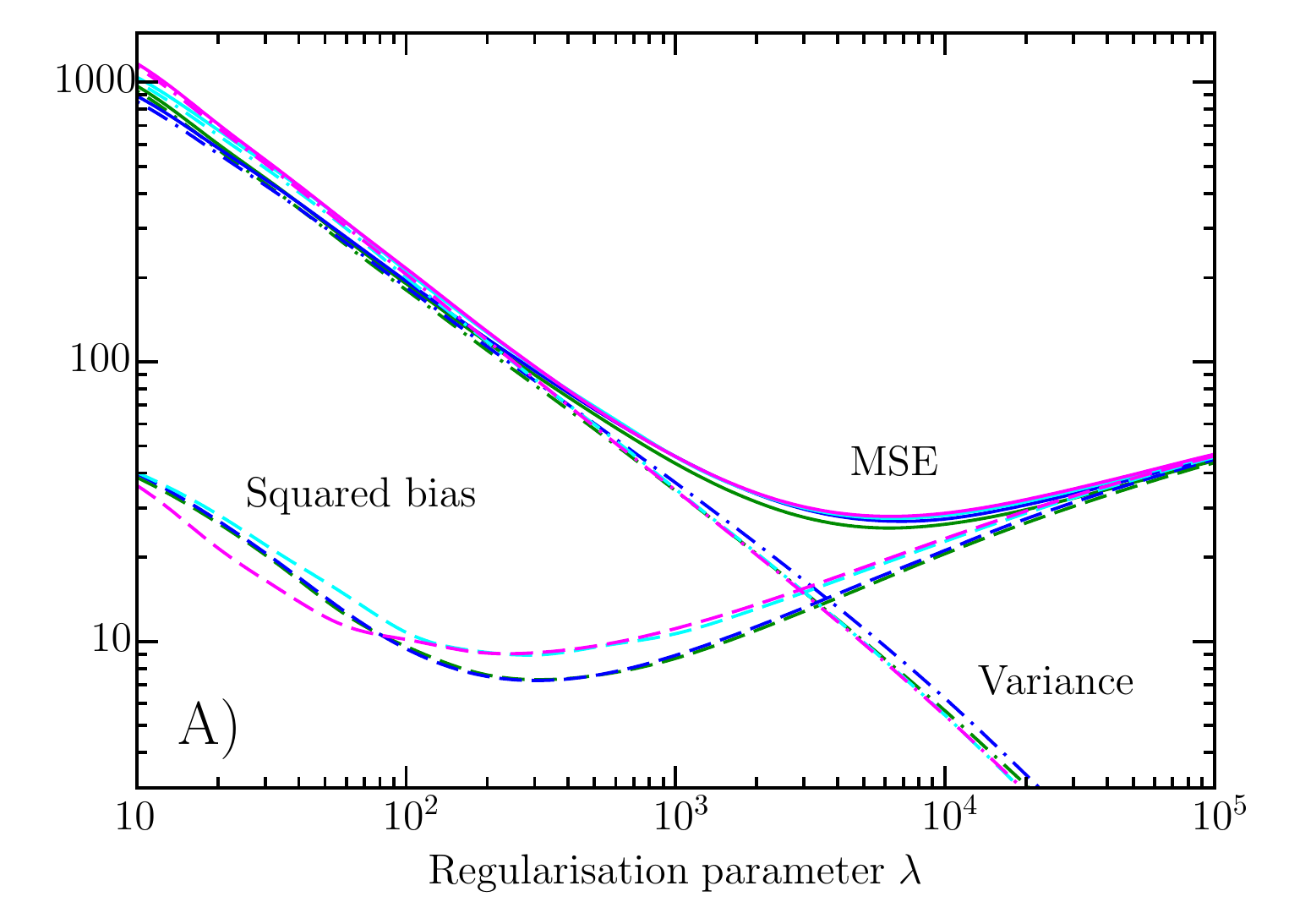}
\includegraphics*[angle=0,width=0.5\columnwidth,trim = 32mm 171mm 23mm
  15mm, clip]{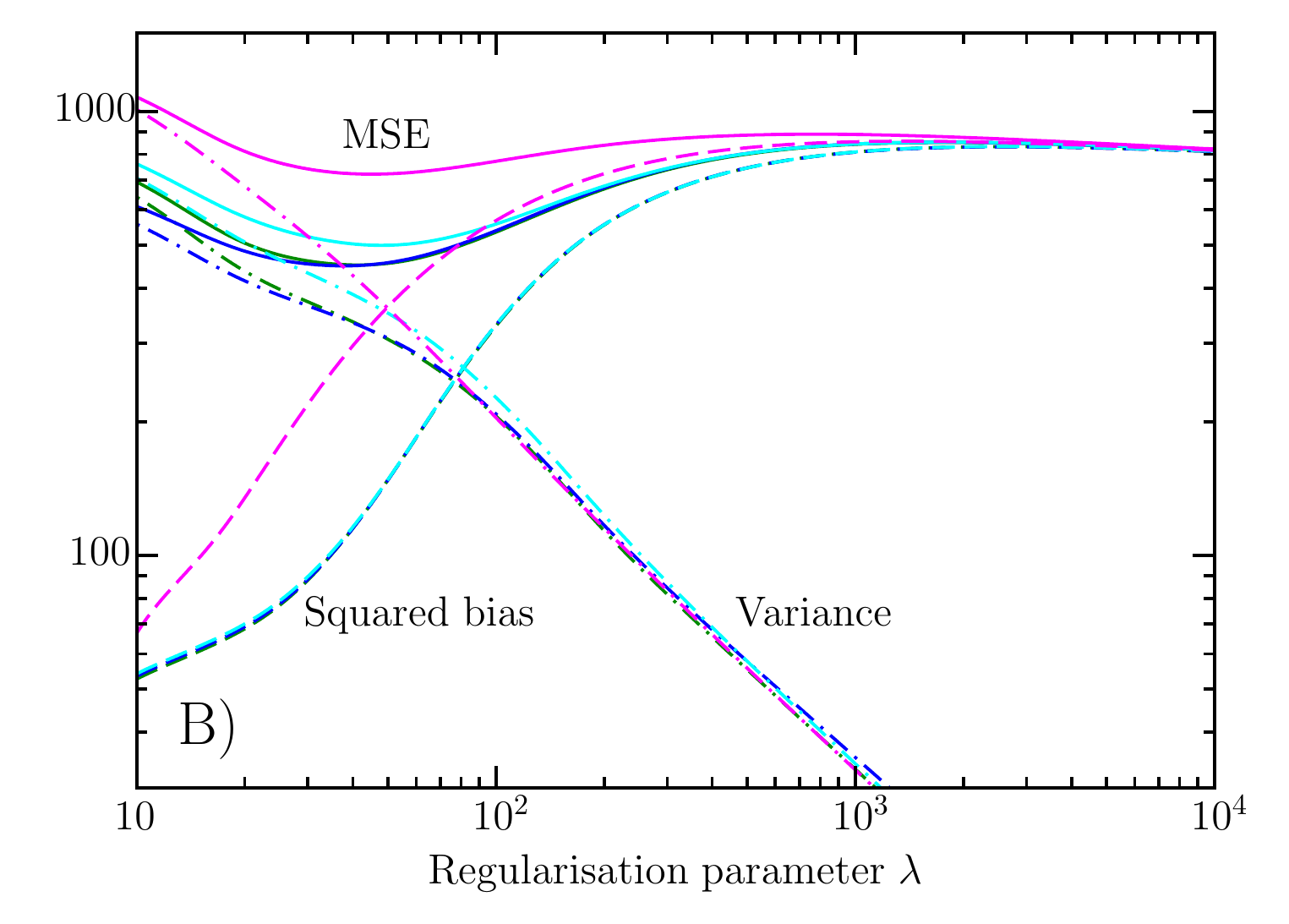}
\includegraphics*[angle=0,width=0.5\columnwidth,trim = 32mm 171mm 23mm
  15mm, clip]{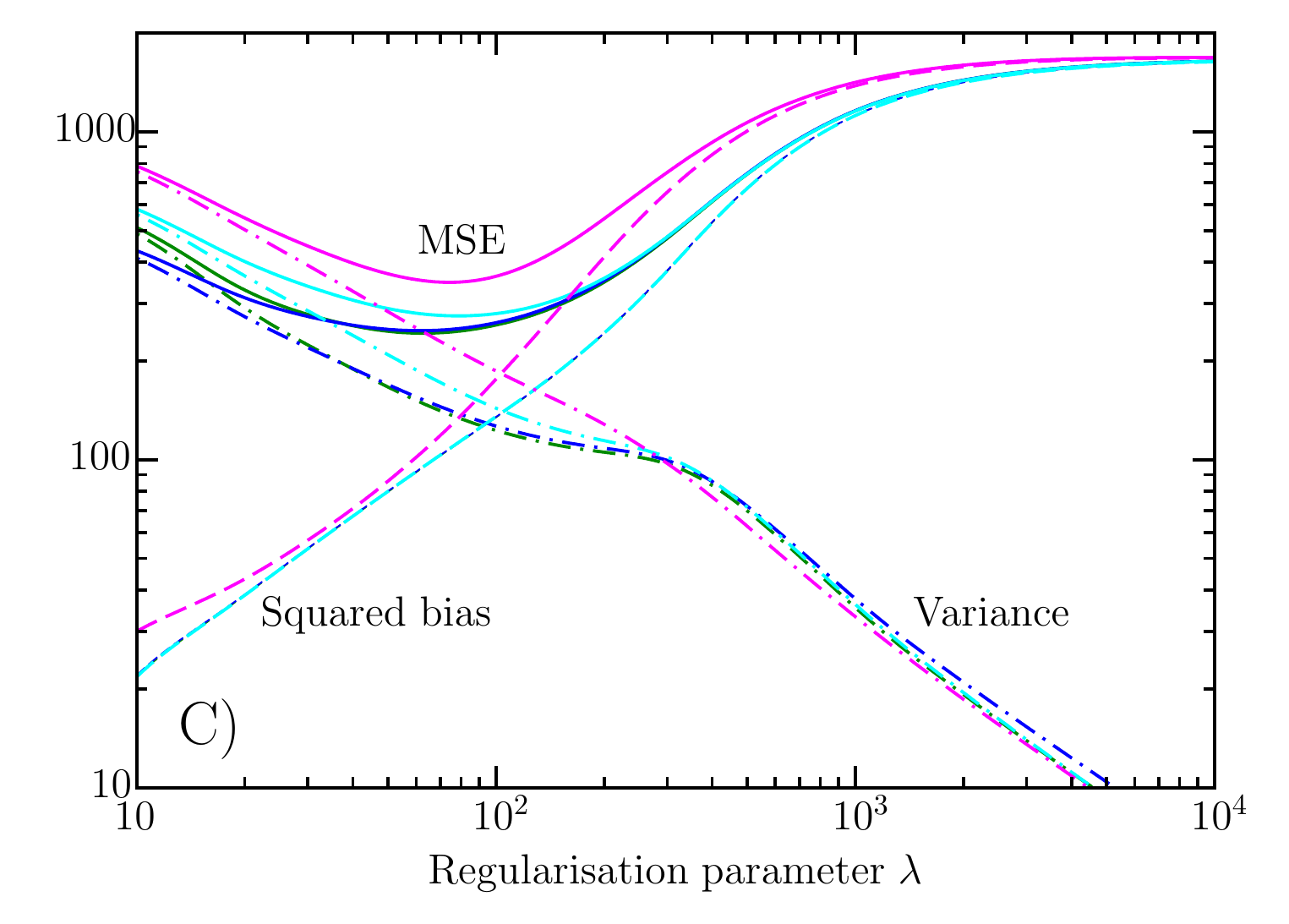}
\includegraphics*[angle=0,width=0.5\columnwidth,trim = 32mm 171mm 23mm
  15mm, clip]{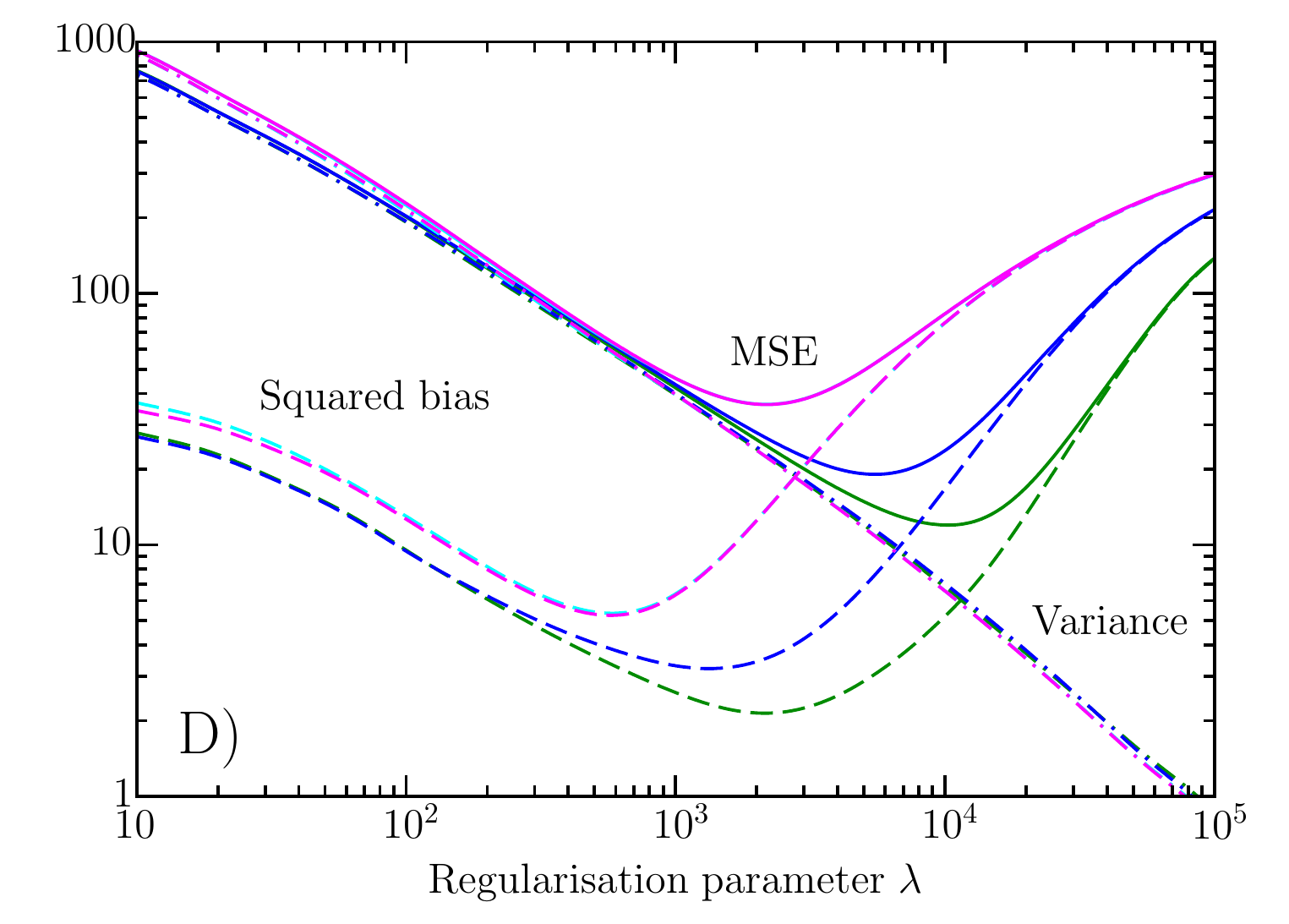}
\caption{\label{mse} The squared bias, variance and (their sum) the
  mean squared error (eq.\,\ref{eqmse}) versus the regularisation
  parameter $\lambda$ for spectra A--D (see Sec.\,\ref{spectest}) and
  various data combinations as defined in Sec.\,\ref{spectra}
  (Combination 1: magenta lines, Combination 2: cyan lines,
  Combination 3: blue lines, Combination 4: green lines). In each case
  the solid lines show the MSE, the dashed lines show the squared bias
  and the dot-dashed lines show the variance.}
\end{figure*}

The squared bias, variance and MSE are shown in Fig.\,\ref{mse} as a
function of $\lambda$ for the various test spectra and combinations of
data sets. The squared bias and variance found from the $10^5$ data
realisations exhibit similar behaviour to their second-order
approximations $\bB^\mathrm{T}\bB$ and $\mathrm{Tr}\,
\mathsf{\Sigma}^{(\mathrm{s})}$. The contributions of the squared bias
and the variance to the MSE vary with $\lambda$: for low values of
$\lambda$ the variance dominates, whereas for high values the bias
dominates. Clearly there is a trade-off between bias and
variance.\footnote{The bias-variance trade-off is a general feature of
  model selection and is well known in estimation theory.}
Regularisation solves the inversion problem because for an appropriate
value of $\lambda$ a large reduction in the variance can be achieved
by accepting a small bias, with the result that the MSE is greatly
improved.

The value of $\lambda$ which minimises the MSE depends on the test
spectrum.  It is smaller for spectra B and C since the bias increases
more rapidly with $\lambda$ for these spectra due to their large
deviations from scale invariance. The minimum MSE is greater for
spectra B and C due to the large bias and larger variance at low
$\lambda$.

\begin{figure*}
\includegraphics*[angle=0,width=0.5\columnwidth,trim = 32mm 171mm 23mm
  15mm, clip]{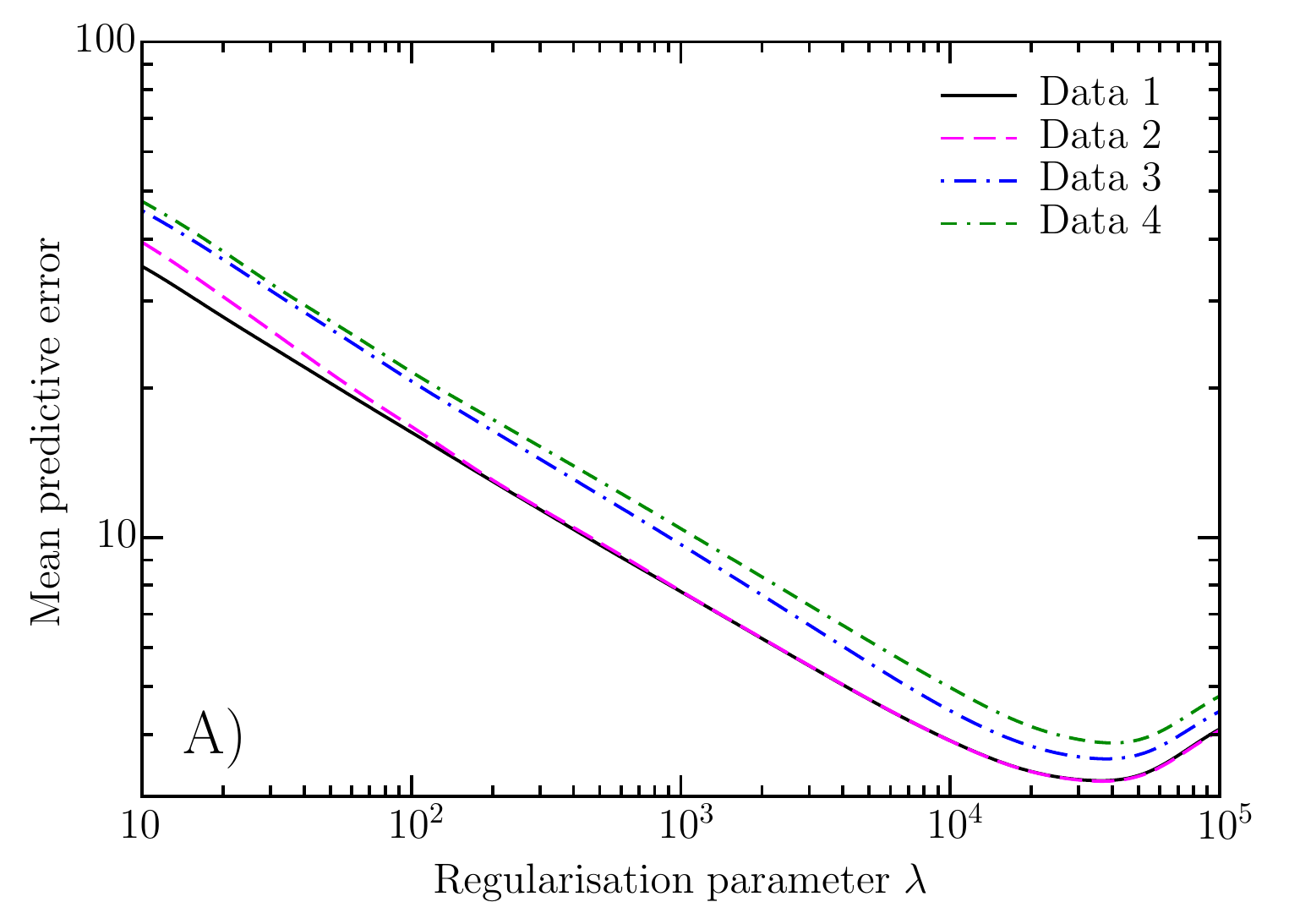}
\includegraphics*[angle=0,width=0.5\columnwidth,trim = 32mm 171mm 23mm
  15mm, clip]{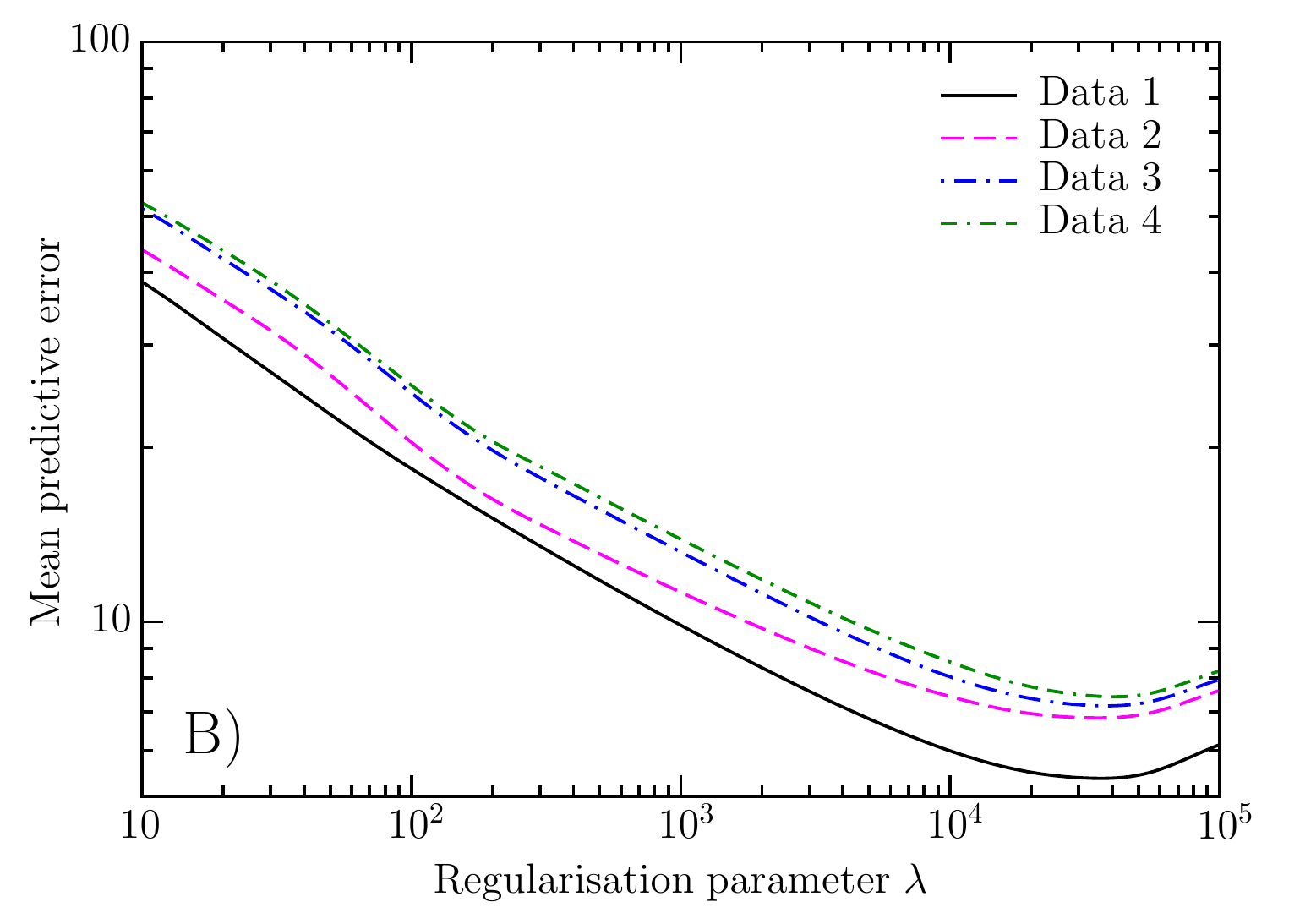}
\includegraphics*[angle=0,width=0.5\columnwidth,trim = 32mm 171mm 23mm
  15mm, clip]{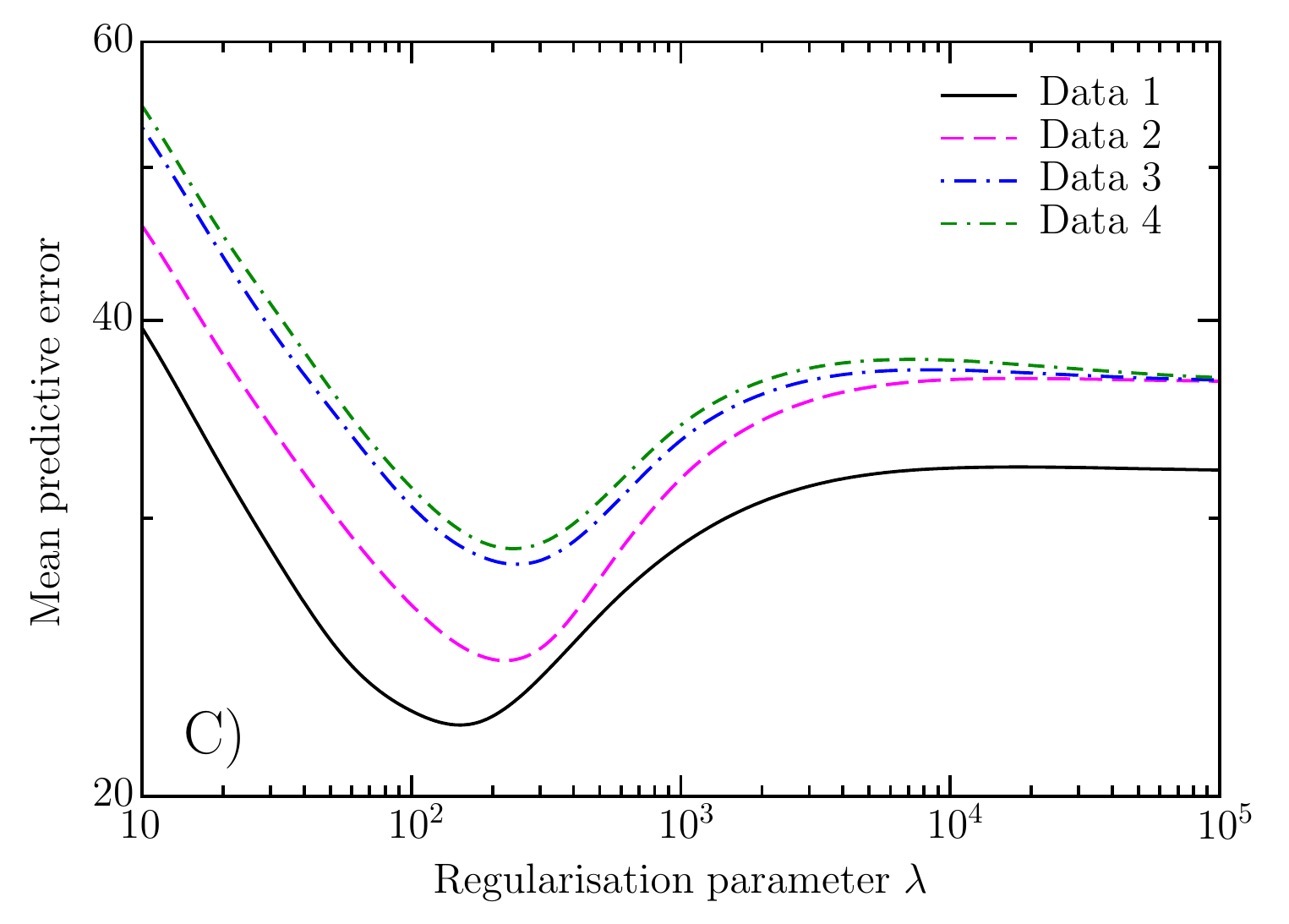}
\includegraphics*[angle=0,width=0.5\columnwidth,trim = 32mm 171mm 23mm
  15mm, clip]{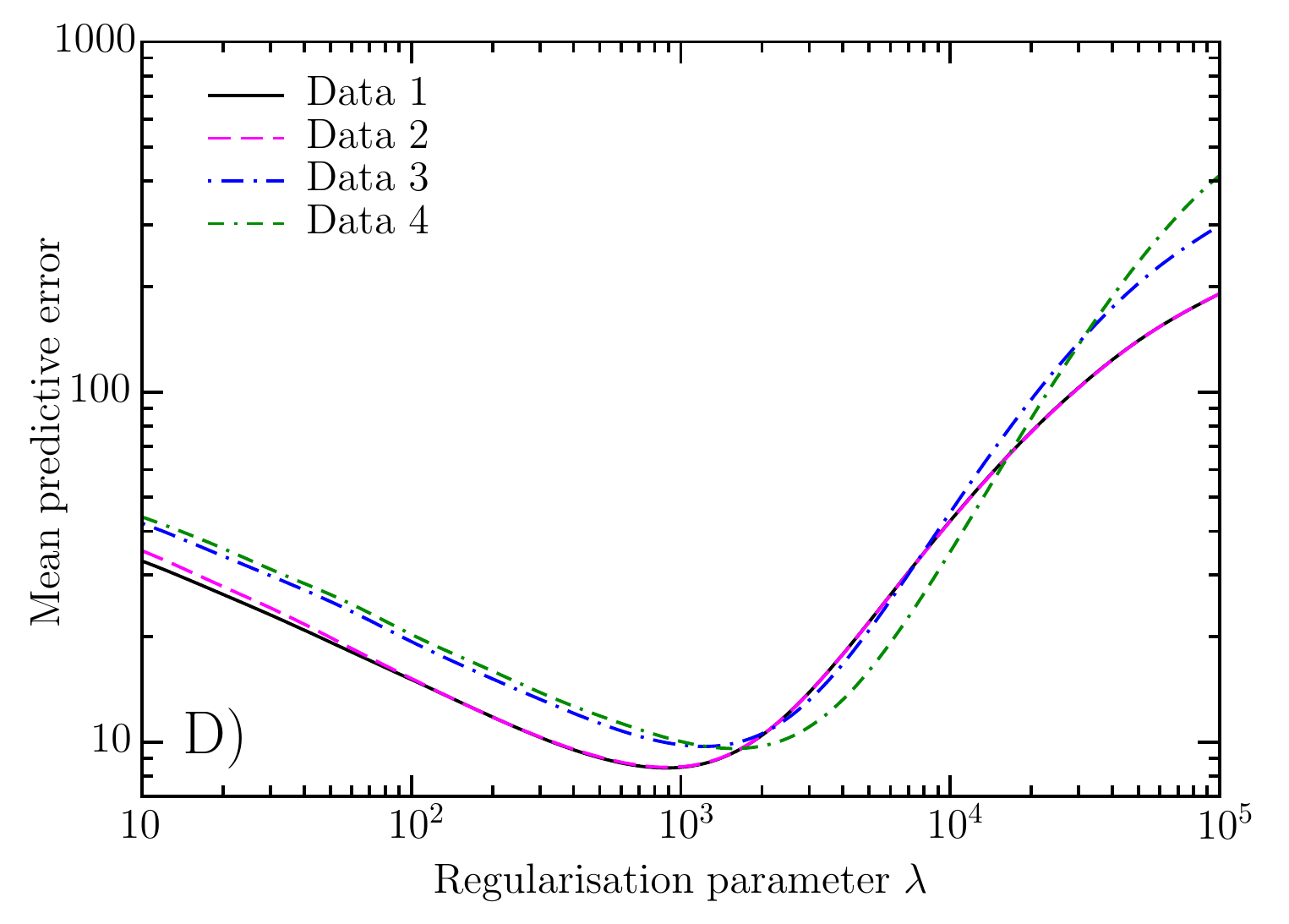}
\caption{\label{mpe} The mean predicted error (eq.\,\ref{eqmpe}) versus
  the regularisation parameter $\lambda$. The plots are for spectrum A
  to D (see Sec.\,\ref{spectest}).}
\end{figure*}

Fig.\,\ref{mpe} shows the estimated MPE as a function of
$\lambda$. For small $\lambda$ the MPE is large as the predicted
reconstructions are close to the noisy data rather than the noise-free
data. For larger $\lambda$ the predicted data resemble a smoothed
version of the noisy data.  Since this is closer to the noise-free
data the MPE is reduced.  The MPE increases again for very large
$\lambda$ because the predicted data are oversmoothed and approach
that of the best-fit H-Z spectrum.

The MPE exhibits similar behaviour to the MSE for spectra A, C and D
(e.g. the minimum of the functions are at the same $\lambda$ value to
within a factor of 2--3). However the MPE of spectrum B is very
different to the corresponding MSE and instead resembles the MSE of
spectrum A.

The MPE can be understood as a version of the MSE weighted towards
wavenumbers where the PPS is more strongly constrained by the data.
The MPE for the WMAP-5 TT data is most sensitive to the reconstructed
PPS on intermediate scales, between about
$k=3\times10^{-3}\;\mathrm{Mpc}^{-1}$ and $k=0.1\;\mathrm{Mpc}^{-1}$.
 
Spectrum C has the largest features in this wavenumber range and so
the value of $\lambda$ which minimises the MPE is smallest for this
spectrum.  The $\lambda$ value which minimises the MPE is higher for
spectrum D since the `bump' in this spectrum is less pronounced. The
MPE is insensitive to the large scale cutoff in spectrum B. On
intermediate scales where it exhibits low amplitude oscillations
spectrum B is comparable to spectrum A and has the same tilt. This
accounts for the similarity of the MPE of the two spectra.

The estimated PPS between $k=3\times10^{-3}\;\mathrm{Mpc}^{-1}$ and
$k=0.04\;\mathrm{Mpc}^{-1}$ largely determines the contribution to the
MPE of the WMAP-5 polarisation data. Since spectra B and C possess
features at the lower end of this wavenumber interval which are
imperfectly recovered, the polarisation contribution is larger for
these spectra than the two others.

The small-scale CMB and SDSS-4 LRG components of the MPE depend on the
reconstructed PPS over the ranges $9\times10^{-3}\lesssim k\lesssim
0.2\;\mathrm{Mpc}^{-1}$ and $0.02\lesssim k\lesssim
0.2\;\mathrm{Mpc}^{-1}$ respectively.  For spectrum D adding the small
scale CMB data significantly improves the inversion, due to the bump
in the spectrum. This leads to a reduced WMAP-5 TT contribution to the
MPE sufficient for intermediate values of $\lambda$ to offset the
additional small-scale CMB component, so that the MPE falls.
Similarly including the SDSS-4 LRG data further decreases the WMAP-5 TT
and small-scale CMB components for intermediate $\lambda$, and the MPE
is again reduced.

\section{Choosing the regularisation parameter \label{regularisation}}

For a successful reconstruction of the PPS the value of the
regularisation parameter must be chosen correctly. If $\lambda$ is too
small $\hat{\pB}$ will be dominated by artifacts caused by fitting the
noise. On the other hand, if $\lambda$ is too large $\hat{\pB}$ will
be oversmoothed and could miss features in the true PPS. Choosing the
regularisation parameter is akin to an exercise in model
selection. The complexity of $\hat{\pB}$ clearly decreases with
increasing $\lambda$. For $\lambda=0$ all the elements of $\hat{\pB}$
are independent whereas for $\lambda=\infty$ the elements are totally
correlated, so that there is effectively only one free parameter, the
amplitude of the H-Z spectrum. One measure of the effective number of
free parameters is the quantity
$\nu_1\equiv\sum_\mathbb{Z}\mathrm{Tr}\left(\mathsf{S}_\mathbb{Z}\right)$
where
$\mathsf{S}_\mathbb{Z}\equiv\mathsf{W}_\mathbb{Z}\mathsf{M}_\mathbb{Z}$
\cite{hastie}. Its value decreases monotonically from
$\sum_\mathbb{Z}\mathrm{Tr}\left(\mathsf{S}_\mathbb{Z}\right)=N_j$ for
$\lambda=0$ to
$\sum_\mathbb{Z}\mathrm{Tr}\left(\mathsf{S}_\mathbb{Z}\right)=1$ for
$\lambda=\infty$.  Other measures are
$\nu_2\equiv\sum_\mathbb{Z}\mathrm{Tr}\left(\mathsf{S}_\mathbb{Z}^\mathrm{T}\mathsf{S}_\mathbb{Z}\right)$
and $\nu_3\equiv\sum_\mathbb{Z}\mathrm{Tr}\left(2\mathsf{S}_\mathbb{Z}
- \mathsf{S}_\mathbb{Z}^\mathrm{T}\mathsf{S}_\mathbb{Z}\right).$

\subsection{Parameter selection methods\label{pselm}}

In general, the optimum value of the regularisation parameter depends
on the noise level, the regularisation scheme and the object to be
recovered. Unfortunately no universal strategy for selecting the
regularisation parameter is known which works in all
situations. Instead numerous methods have been proposed in the
literature, based upon different principles. Here we consider five
different methods: the discrepancy principle (DP), equivalent degrees
of freedom (EDF), normalised cumulative periodogram (NCP), Mallow's
$C_p$ and generalised cross-validation (GCV). The aim of the first
three techniques is to ensure that the estimated PPS has a
statistically reasonable fit to the data, while the final two seek to
minimise the PE. Although most of the methods were originally devised
for single data sets with white noise, it is simple to recast them in
forms suitable for multiple data sets with correlated noise. For
comparison purposes we also include a method in which the SE is
minimised, even though it requires knowledge of $\pB_\mathrm{t}$.

The discrepancy principle method \cite{morozov} is also known as the
constrained least squares method \cite{Hunt:1973ay} and chi-squared
method \cite{Thompson:1991gf}. According to the discrepancy principle
the estimated PPS should fit the data only to within the noise, which
for multiple data sets with Gaussian noise is equivalent to requiring
that
\begin{equation}
\chi^2\left(\hat{\pB}\right)\equiv\sum_\mathbb{Z}\left(\mathsf{W}_{\mathbb{Z}}\hat{\pB}
- \dB_{\mathbb{Z}}\right)^\mathrm{T}\mathsf{N}_{\mathbb{Z}}^{-1}
\left(\mathsf{W}_{\mathbb{Z}}\hat{\pB}-\dB_{\mathbb{Z}}\right)=N_d.
\label{chi}
\end{equation}
The value of $\lambda$ for which eq.(\ref{chi}) holds is denoted by
$\lambda_\mathrm{DP}$. This method is popular due to its simplicity
and was previously used in
\cite{Tocchini-Valentini:2004ht,Tocchini-Valentini:2005ja}.  However,
it implicitly assumes that $\chi^2\left(\hat{\pB}\right)$, like
$\chi^2\left(\pB_\mathrm{t}\right)$, has a $\chi^2$ distribution with $N_d$
degrees of freedom. In this case $\langle
\chi^2\left(\hat{\pB}\right)\rangle$ would equal $N_d$ and
$\chi^2\left(\hat{\pB}\right)=N_d$ would signify an acceptable fit to
the data. In fact this is untrue: it can be shown that
\begin{equation}
\langle\chi^2\left(\hat{\pB}\right)\rangle =
\sum_\mathbb{Z}\pB_\mathrm{t}^\mathrm{T}\mathsf{W}_\mathbb{Z}^\mathrm{T}\left(\mathsf{S}_\mathbb{Z}^\mathrm{T}-\mathsf{I}\right)^\mathrm{T}
\mathsf{N}_\mathbb{Z}^{-1}
\left(\mathsf{S}_\mathbb{Z}-\mathsf{I}\right)\mathsf{W}_\mathbb{Z}\pB_\mathrm{t}
+
\sum_{\mathbb{Z}}\mathrm{Tr}\left(\mathsf{S}_\mathbb{Z}^\mathrm{T}\mathsf{N}_\mathbb{Z}^{-1}
\mathsf{S}_\mathbb{Z}\mathsf{N}_\mathbb{Z}\right)-2\sum_\mathbb{Z}\mathrm{Tr}\left(\mathsf{S}_\mathbb{Z}\right)+N_d.
\label{chi2}
\end{equation} 
We approximate the sum of the first two terms on the right of
eq.(\ref{chi2}) by $\sum_\mathbb{Z}\mathrm{Tr}\left(\mathsf{S}_\mathbb{Z}\right)$, to
which it reduces as $\lambda$ tends to zero, and require that
\begin{equation}
\chi^2\left(\hat{\pB}\right)=N_\mathrm{eff},
\label{chi3}
\end{equation} 
where $N_\mathrm{eff}\equiv
N_d-\sum_\mathbb{Z}\mathrm{Tr}\left(\mathsf{S}_\mathbb{Z}\right)$. The
value of $\lambda$ which satisfies eq.(\ref{chi3}) is denoted by
$\lambda_\mathrm{EDF}$. With the interpretation of
$\sum_\mathbb{Z}\mathrm{Tr}\left(\mathsf{S}_\mathbb{Z}\right)$ as the
effective number of parameters, $N_\mathrm{eff}$ represents the number
of degrees of freedom of the reconstruction. This is known as the
equivalent degrees of freedom method \cite{Thompson:1991gf} and also
the compensated discrepancy principle method \cite{hansen}. A
different Bayesian motivation for it can be found in
\cite{Thompson:1992qq}.

The rationale for the NCP method is that the residual $\hat{\dB}-\dB$
should have the same statistical behaviour as the noise in the data
when the PPS is correctly recovered
\cite{Hansen:2006eq,Rust:2008ix}. This suggests that the optimal value
of $\lambda$ is the one for which the statistical properties of the
residual are closest to those of the noise. To apply the NCP method we
construct random variables $y^{(\mathbb{Z})}_a$ from the predicted data
intended to behave like white noise when $\hat{\pB}=\pB_\mathrm{t}$. For the
WMAP data these are
\begin{equation}
y_\ell^X=\sum_{\ell^\prime}\left(J^X\right)_{\ell\ell^\prime}^{-1}
\left(\hat{\mathrm{s}}_{\ell^\prime}^X-\mathrm{d}_{\ell^\prime}^X\right),
\end{equation}
where $\hat{\mathrm{s}}^X_\ell\equiv\sum_i W^X_{\ell i}\hat{\mathrm{p}}_i$ and
$\mathsf{J}^X$ is given by the Cholesky decomposition
$\mathsf{N}^X=\mathsf{J}^X\mathsf{J}^{X\mathrm{T}}$.  The random variables for
the small-scale CMB data are
\begin{equation}
y_b^X=\sum_{X^\prime,b^\prime}\left(J^{XX^\prime}\right)_{bb^\prime}^{-1}\hat{\mathcal{Z}}_{b^\prime}^{X^\prime}.
\end{equation}
Here $\hat{\mathcal{Z}}^X_b=\hat{\mathrm{s}}_b^X-\mathrm{d}_b^X$ for a
Gaussian bandpower and
$\hat{\mathcal{Z}}^X_b=\ln\left(\hat{\mathcal{S}}_b^X/\mathcal{D}_b^X\right)$
for a log-Gaussian bandpower, with $\hat{\mathrm{s}}^X_b\equiv \sum_i
T^X_{bi} \hat{\mathrm{p}}_i$ and $\hat{\mathcal{S}}_b^X\equiv
\hat{\mathrm{s}}_b^X+\mathcal{N}_b^X$. The matrices
$\mathsf{J}^{XX^\prime}$ are related to the matrices
$\mathsf{V}^{XX^\prime}$ by a Cholesky decomposition. For the SDSS-4
LRG data the variables are
\begin{equation}
y_a=\sum_{a^\prime}\left(J^\mathrm{LRG}\right)_{aa^\prime}^{-1}
\left(\hat{\mathrm{s}}_{a^\prime}-\mathrm{d}_{a^\prime}\right),
\end{equation}
where $\hat{\mathrm{s}}_a=\sum_a T^\mathrm{LRG}_{ai}\mathrm{p}_i$ and
$\mathsf{J}^\mathrm{LRG}$ is given by the Cholesky decomposition of
$\tilde{\mathsf{N}}^\mathrm{LRG}$.

The quantities $y^{(\mathbb{Z})}_a$ of the different data sets are assembled
into a single vector $\yB$. If $\lambda$ is too high or low
$\yB$ will be dominated by low or high frequency components,
respectively. The discrete Fourier transform (DFT) of $\yB$ is
calculated using the fast Fourier transform algorithm after
zero-padding the vector out to $N_y$, a convenient power of two.  The
sine and cosine coefficients of the DFT are
\begin{eqnarray}
c_k & = & \sum_{a=1}^{N_y} y_a \cos\;\frac{2\pi
  \left(k-1\right)\left(a-1\right)}{N_y},\\ s_k & = &
-\sum_{a=1}^{N_y} y_a \sin\;\frac{2\pi
  \left(k-1\right)\left(a-1\right)}{N_y},
\end{eqnarray} 
for $k=1,\ldots,N_y/2+1$. The NCP of $\yB$ is the vector $\hB$
defined as
\begin{equation}
h_j=\frac{\sum_{k=1}^{j} c_k^2+s_k^2}{\sum_{k=1}^{N_y/2+1} c_k^2+s_k^2}.
\label{ncph}
\end{equation}
The NCP is a test for white noise, for which it lies close to the
vector $v_j=2\left(j-1\right)/N_y$. In the NCP method the value of the
regularisation parameter is taken to be the one which minimises the
function
$\gamma_\mathrm{NCP}\left(\lambda\right)\equiv\left[\hB\left(\lambda\right)-\vB\right]^\mathrm{T}
\left[\hB\left(\lambda\right)-\vB\right]$. It is labelled
$\lambda_\mathrm{NCP}$.

An alternative approach to the selection of the regularisation parameter
involves attempting to minimise a loss function. It can be shown that
\begin{eqnarray}
\mathrm{MPE}\left(\hat{\pB}\right) & = &
\sum_\mathbb{Z}\pB_\mathrm{t}^\mathrm{T}\mathsf{W}_\mathbb{Z}^\mathrm{T}
\left(\mathsf{S}_\mathbb{Z}^\mathrm{T}-\mathsf{I}\right)^\mathrm{T}\mathsf{N}_\mathrm{Z}^{-1}
\left(\mathsf{S}_\mathbb{Z}-\mathsf{I}\right)\mathsf{W}_\mathbb{Z}\pB_\mathrm{t}
+
\sum_\mathbb{Z}\mathrm{Tr}\left(\mathsf{S}_\mathbb{Z}^\mathrm{T}\mathsf{N}_\mathbb{Z}^{-1}
\mathsf{S}_\mathbb{Z}\mathsf{N}_\mathbb{Z}\right),\\ & = &
\langle\chi^2\left(\hat{\pB}\right)\rangle +
2\sum_\mathbb{Z}\mathrm{Tr}\left(\mathsf{S}_\mathbb{Z}\right)-N_d.
\end{eqnarray} 
The second equality follows from eq.(\ref{chi2}). Thus Mallow's $C_p$
statistic
\begin{equation}
C_p\left(\lambda\right)\equiv\chi^2\left(\hat{\pB}\right) + 
2\sum_\mathbb{Z}\mathrm{Tr}\left(\mathsf{S}_\mathbb{Z}\right)-N_d
\label{cp}
\end{equation} 
is an unbiased estimator of the MPE. Mallow's $C_p$ method
\cite{Mallows:1973qw} chooses the value of $\lambda$ which minimises
$C_p\left(\lambda\right)$, denoted by $\lambda_\mathrm{CP}$, since on
average $\lambda_\mathrm{CP}$ minimises the PE. This is also referred
to as the unbiased predicative risk estimator method \cite{vogel} and
as Stein's unbiased risk estimator method \cite{Stein:1981qw}. Note
that the $C_p$ statistic is essentially a special case of the Akaike
information criterion used in model selection \cite{Akaike:1973ip}.

The idea behind cross-validation is that the optimum estimate of the
PPS is the one best at predicting new or unused data. In leave-one-out
cross-validation a statistic is employed which quantifies the accuracy
with which $\hat{\pB}$ predicts individual data points using all the
other data. Let $\hat{\pB}^{(\mathbb{Z})}_{-a}$ be the estimate of the
PPS obtained when the data point $\mathrm{d}^{(\mathbb{Z})}_a$ is
unused. Then $\hat{\mathrm{d}}^{(\mathbb{Z})}_{-a}\equiv
\left(\mathsf{W}_\mathbb{Z}\hat{\pB}^{(\mathbb{Z})}_{-a}\right)_a$ is
the estimate of $\mathrm{d}^{(\mathbb{Z})}_a$ found using the other
data points. It can be argued that optimum value of $\lambda$ is the
one which minimises the function
\begin{equation}
V_\mathrm{CV}\left(\lambda\right)\equiv\sum_{\mathbb{Z},a,b}
\left(\hat{\mathrm{d}}^{(\mathbb{Z})}_{-a}-\mathrm{d}^{(\mathbb{Z})}_a\right)\mathsf{N}^{(\mathbb{Z})-1}_{ab}
\left(\hat{\mathrm{d}}^{(\mathbb{Z})}_{-b}-\mathrm{d}^{(\mathbb{Z})}_b\right).
\end{equation}  
Using a rank one update formula yields
\begin{equation}
\hat{\mathrm{d}}^{(\mathbb{Z})}_{-a}-\mathrm{d}^{(\mathbb{Z})}_a = 
\frac{\left(\mathsf{W}_\mathbb{Z}\hat{\pB}\right)_a-\mathrm{d}^{(\mathbb{Z})}_a}{1-S^{(\mathbb{Z})}_{aa}}.
\end{equation}  
In generalised cross-validation \cite{Golub:1979qw} the elements
$S^{(\mathbb{Z})}_{aa}$ are replaced by their mean value
$\sum_\mathbb{Z}\mathrm{Tr}\left(\mathsf{S}_\mathbb{Z}\right)/N_d$
which leads to the more convenient statistic
\begin{equation}
V_\mathrm{GCV}\left(\lambda\right)\equiv\frac{\chi^2\left(\hat{\pB}\right)}
{\left[1-N_d^{-1}\sum_\mathbb{Z}\mathrm{Tr}\left(\mathsf{S}_\mathbb{Z}\right)\right]^2}.
\label{gcv}
\end{equation}  
The value of $\lambda$ which minimises eq.(\ref{gcv}) is written
$\lambda_\mathrm{GCV}$. The mean value of $\lambda_\mathrm{GCV}$
satisfies $\mathrm{d}\langle
V_\mathrm{GCV}\left(\lambda\right)\rangle/\mathrm{d}\lambda=0$ which is
equivalent to
\begin{equation}
\frac{\mathrm{d}}{\mathrm{d}\lambda}\left[\mathrm{MPE}(\hat{\pB})\right]-2\left[1-
  \frac{\langle\chi^2\left(\hat{\pB}\right)\rangle}{N_d-\sum_\mathbb{Z}\mathrm{Tr}
  \left(\mathsf{S}_\mathbb{Z}\right)}\right]
\frac{\mathrm{d}}{\mathrm{d}\lambda}\left[\sum_\mathbb{Z}\mathrm{Tr}\left(\mathsf{S}_\mathbb{Z}\right)\right]=0.
\label{gcv2}
\end{equation}
Provided the second term on the left above is small, which typically
the case in practice \cite{vogel}, it can be seen that
$\lambda_\mathrm{GCV}$ will on average equal $\lambda_\mathrm{CP}$ and
minimise the PE.

Finally, $\lambda_\mathrm{LSE}$ is defined as the minimiser of the
SE (eq.\,\ref{SE}). We refer to this as the least squared error (LSE) method and
identify $\lambda_\mathrm{LSE}$ as the optimum value of the
regularisation parameter. Although calculation of
$\lambda_\mathrm{LSE}$ requires $\pB_\mathrm{t}$ to be known, we include it as
a benchmark for the other methods.

\subsection{Application to test spectra\label{testspec}}

\begin{figure*}
\includegraphics*[angle=0,width=0.5\columnwidth,trim = 32mm 171mm 23mm
  15mm, clip]{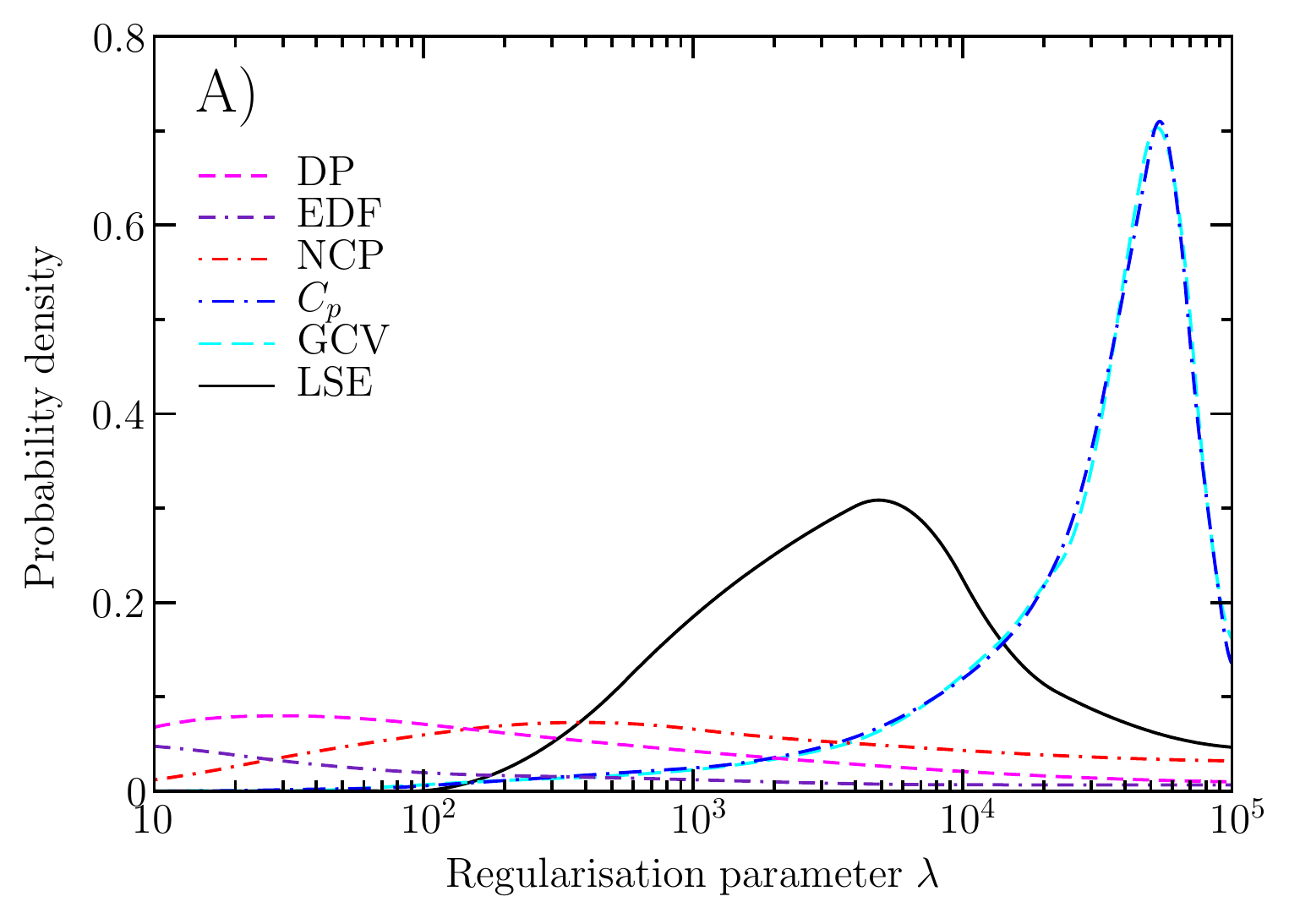}
\includegraphics*[angle=0,width=0.5\columnwidth,trim = 32mm 171mm 23mm
  15mm, clip]{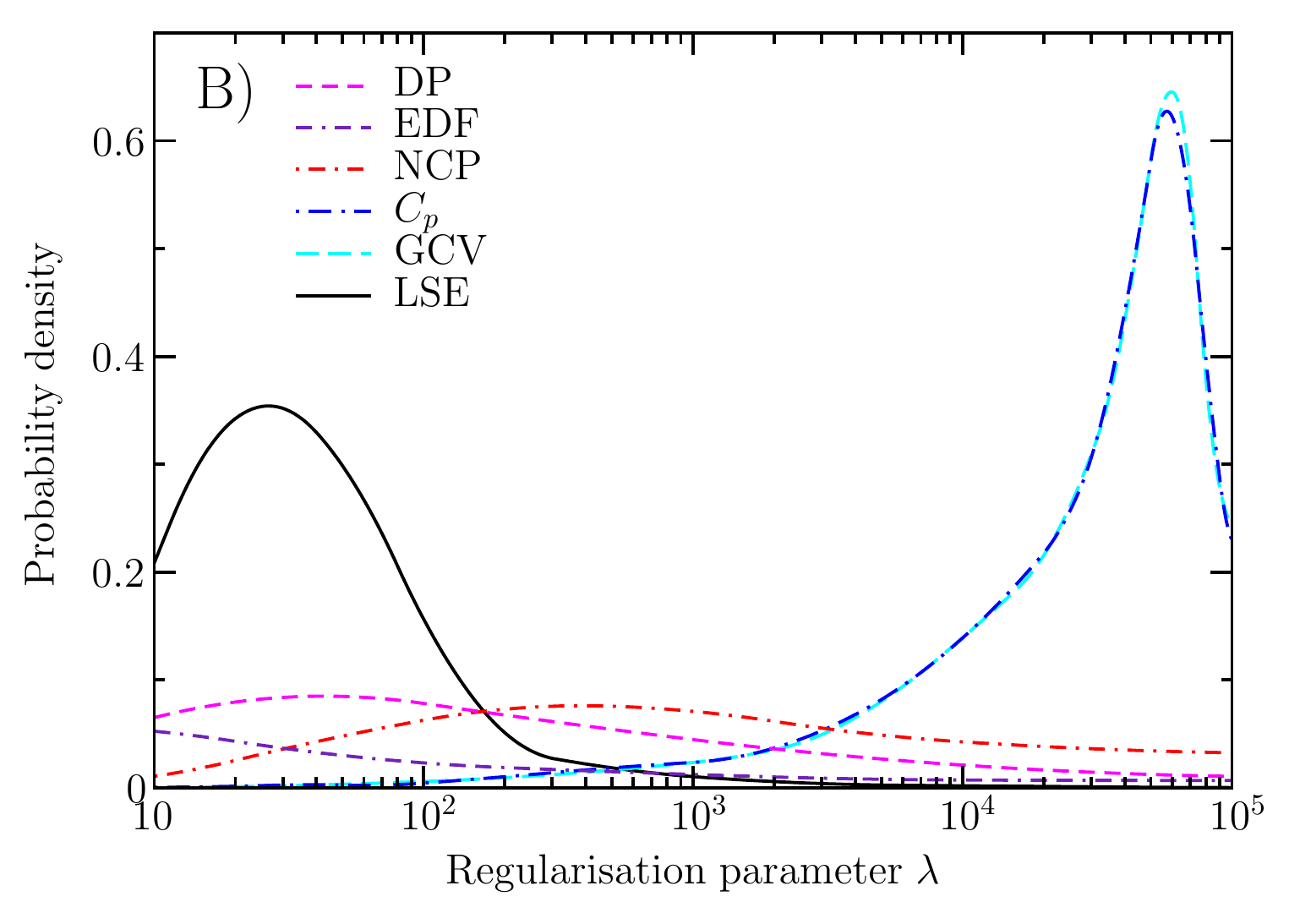}
\includegraphics*[angle=0,width=0.5\columnwidth,trim = 32mm 171mm 23mm
  15mm, clip]{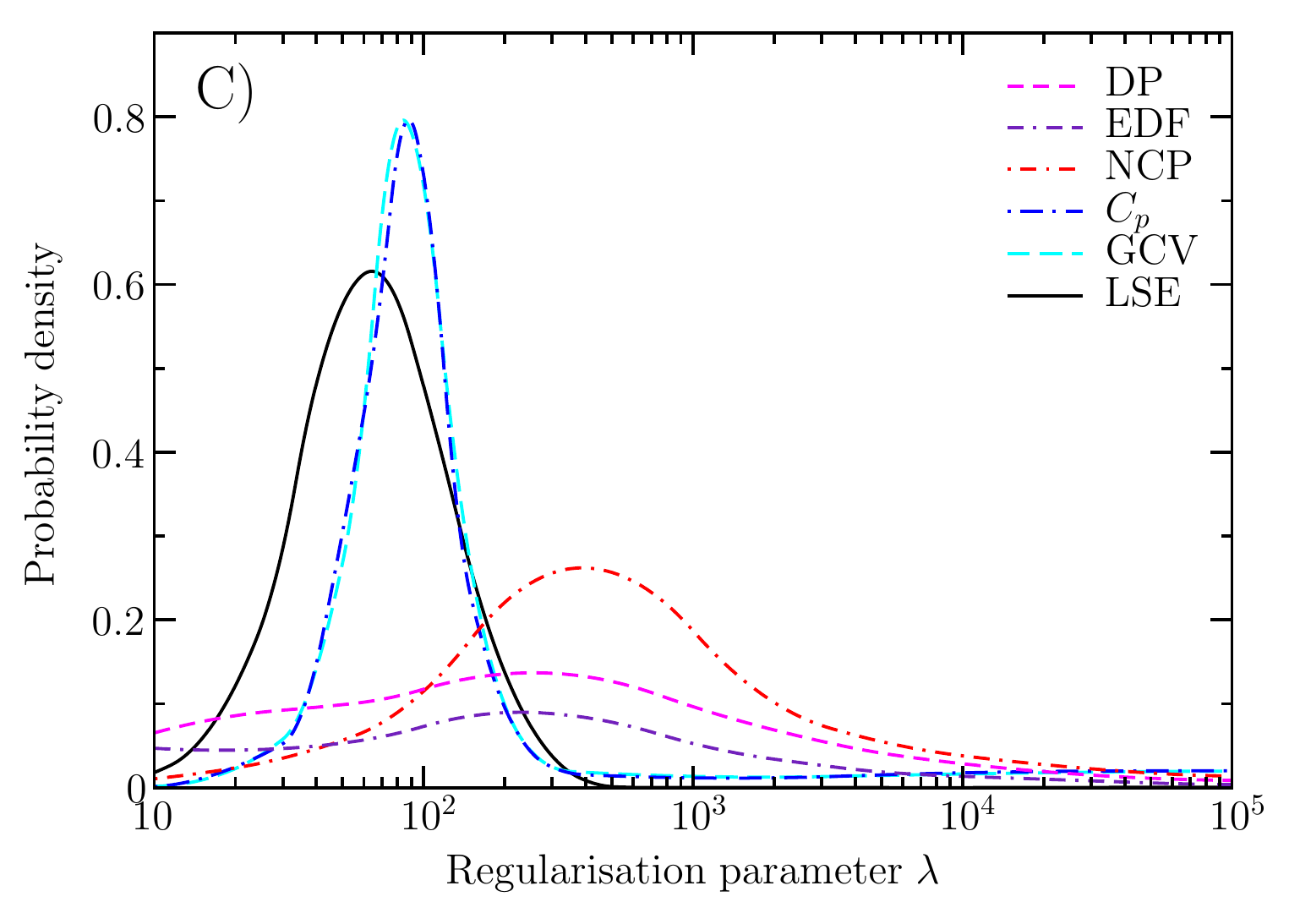}
\includegraphics*[angle=0,width=0.5\columnwidth,trim = 32mm 171mm 23mm
  15mm, clip]{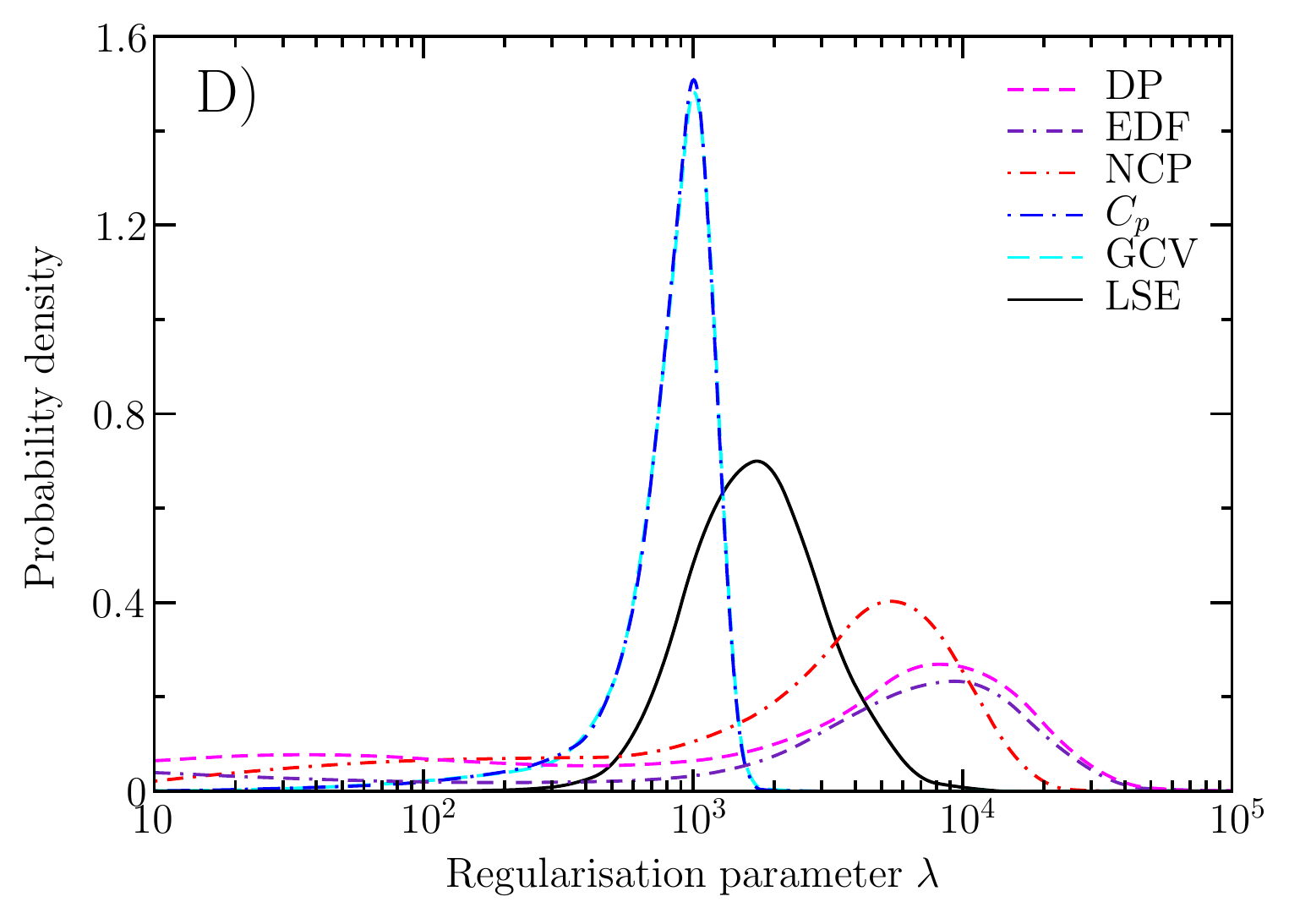}
\caption{\label{dist1} Distributions of the regularisation parameter
  chosen by the different methods for inversions using the WMAP-5
  temperature data (combination
  1, Sec.\,\ref{spectra}) and for spectra A--D
  Sec.\,\ref{spectest}).}
\end{figure*}

\begin{figure*}
\includegraphics*[angle=0,width=0.5\columnwidth,trim = 32mm 171mm 23mm
  15mm, clip]{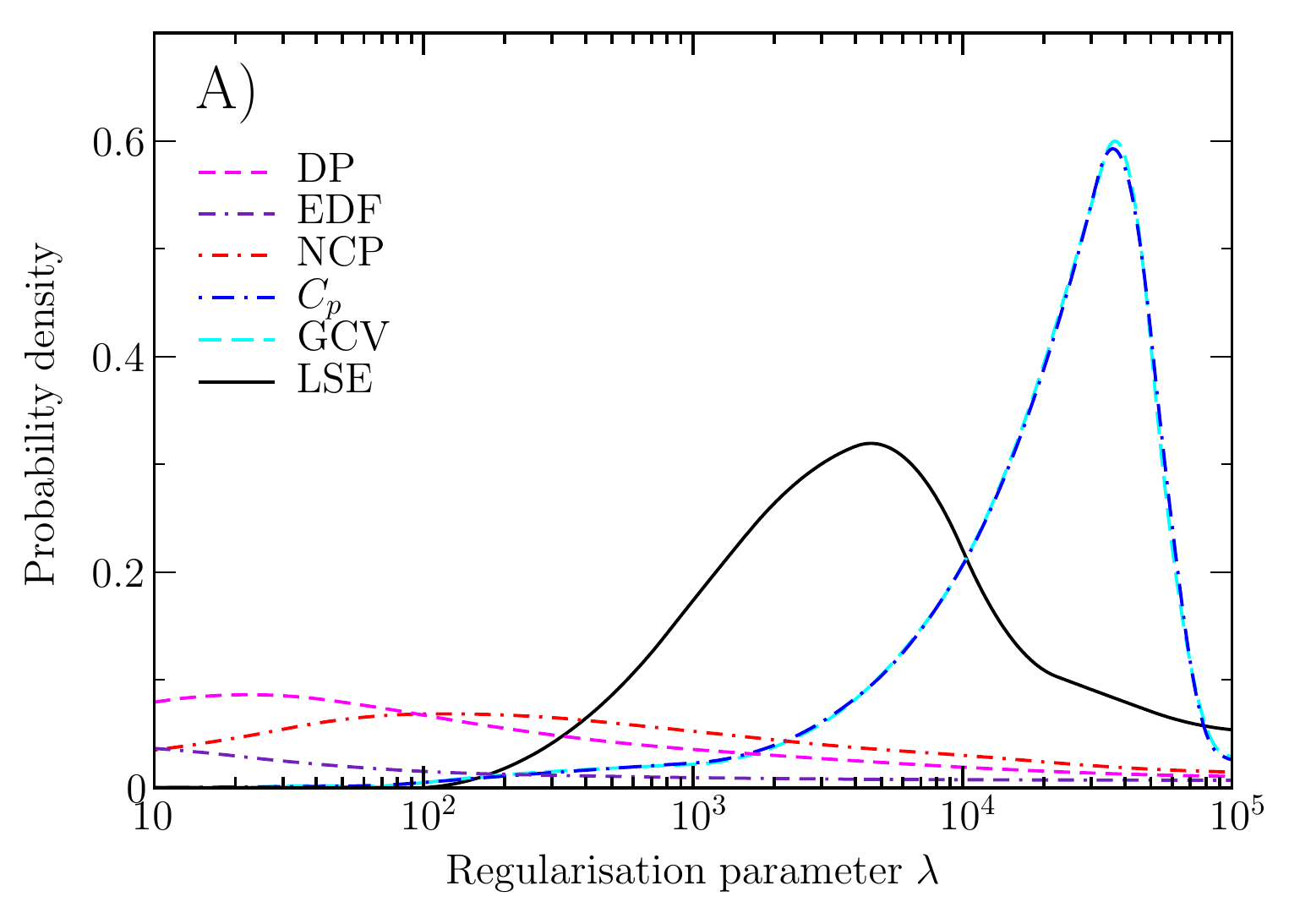}
\includegraphics*[angle=0,width=0.5\columnwidth,trim = 32mm 171mm 23mm
  15mm, clip]{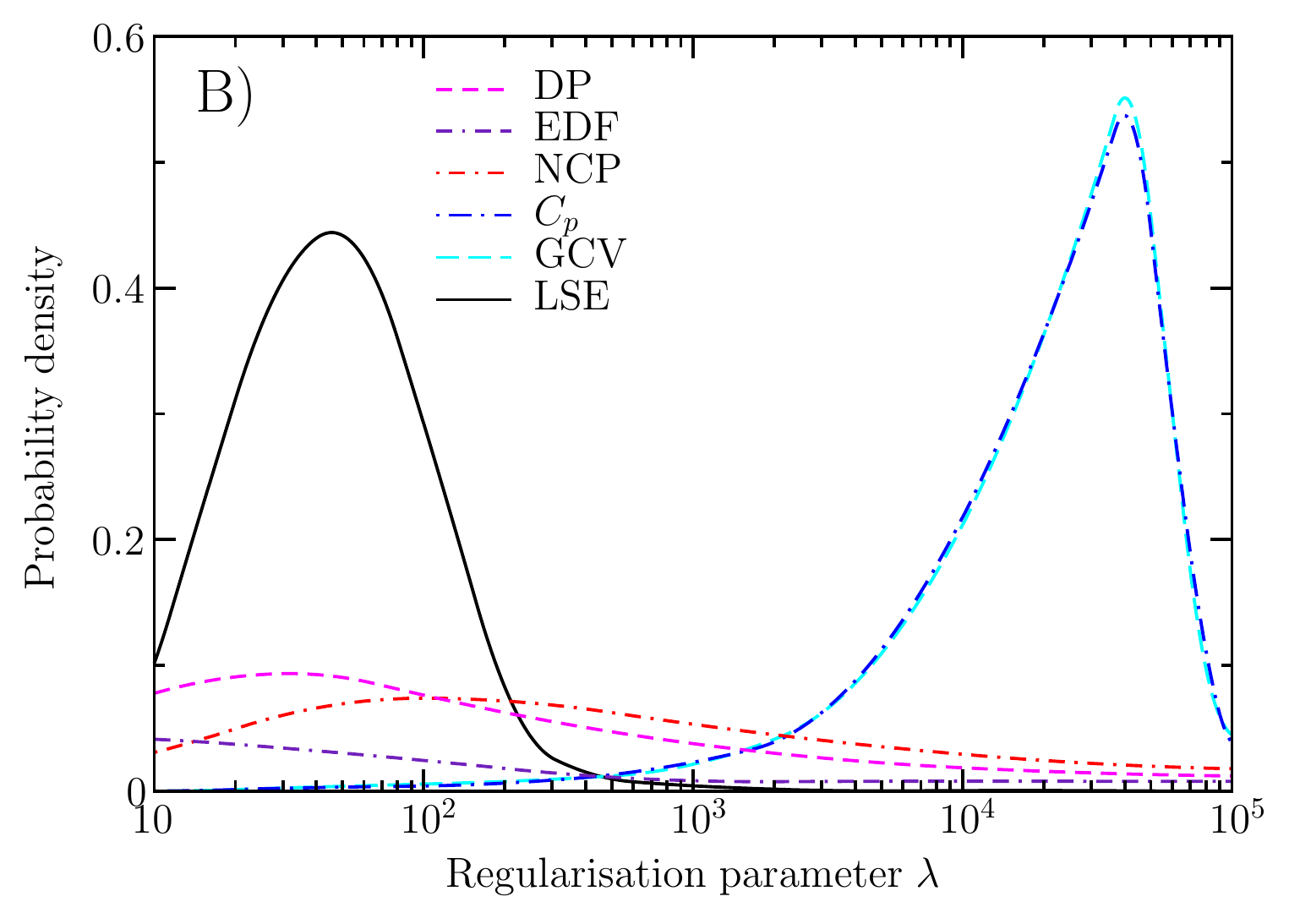}
\includegraphics*[angle=0,width=0.5\columnwidth,trim = 32mm 171mm 23mm
  15mm, clip]{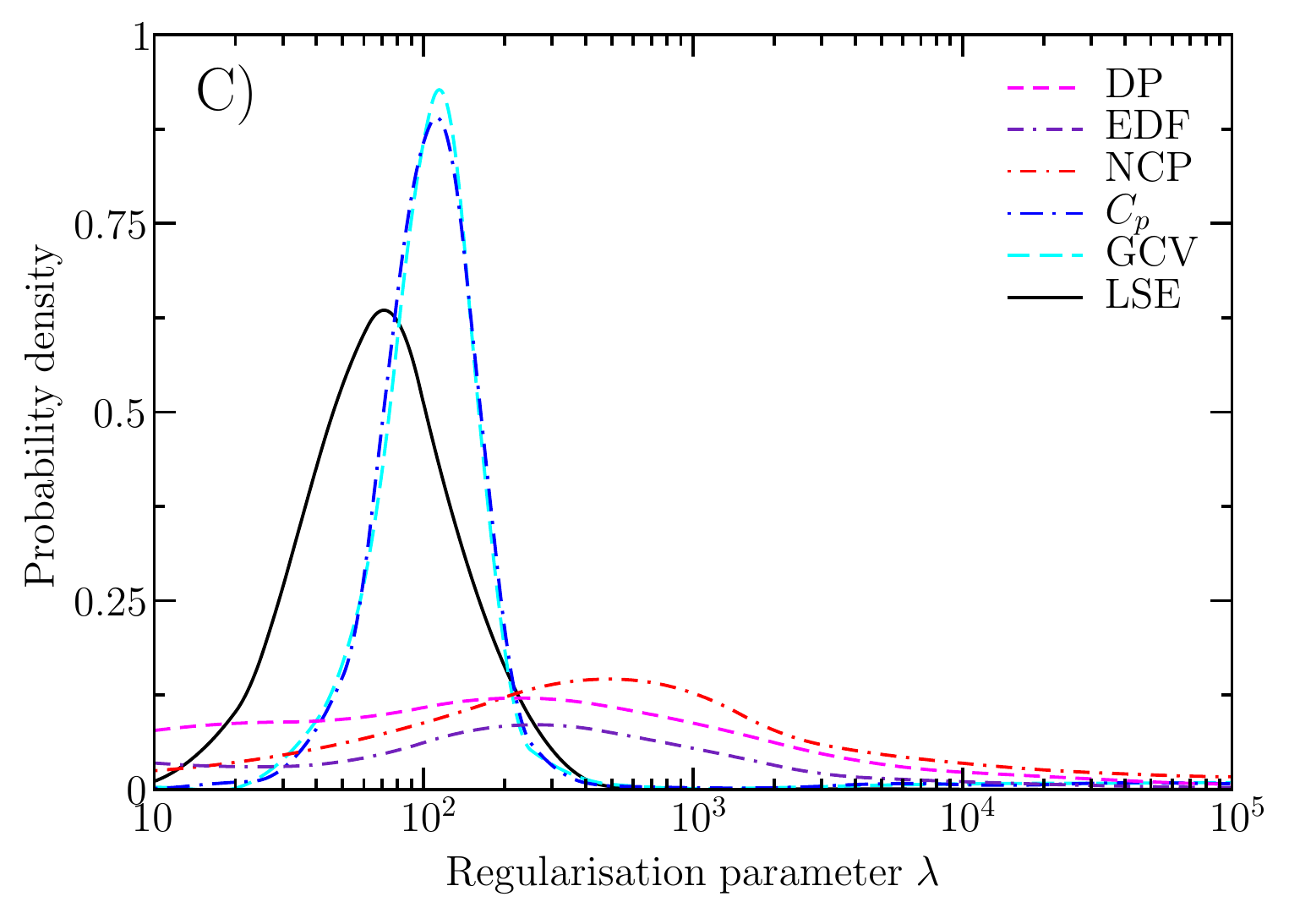}
\includegraphics*[angle=0,width=0.5\columnwidth,trim = 32mm 171mm 23mm
  15mm, clip]{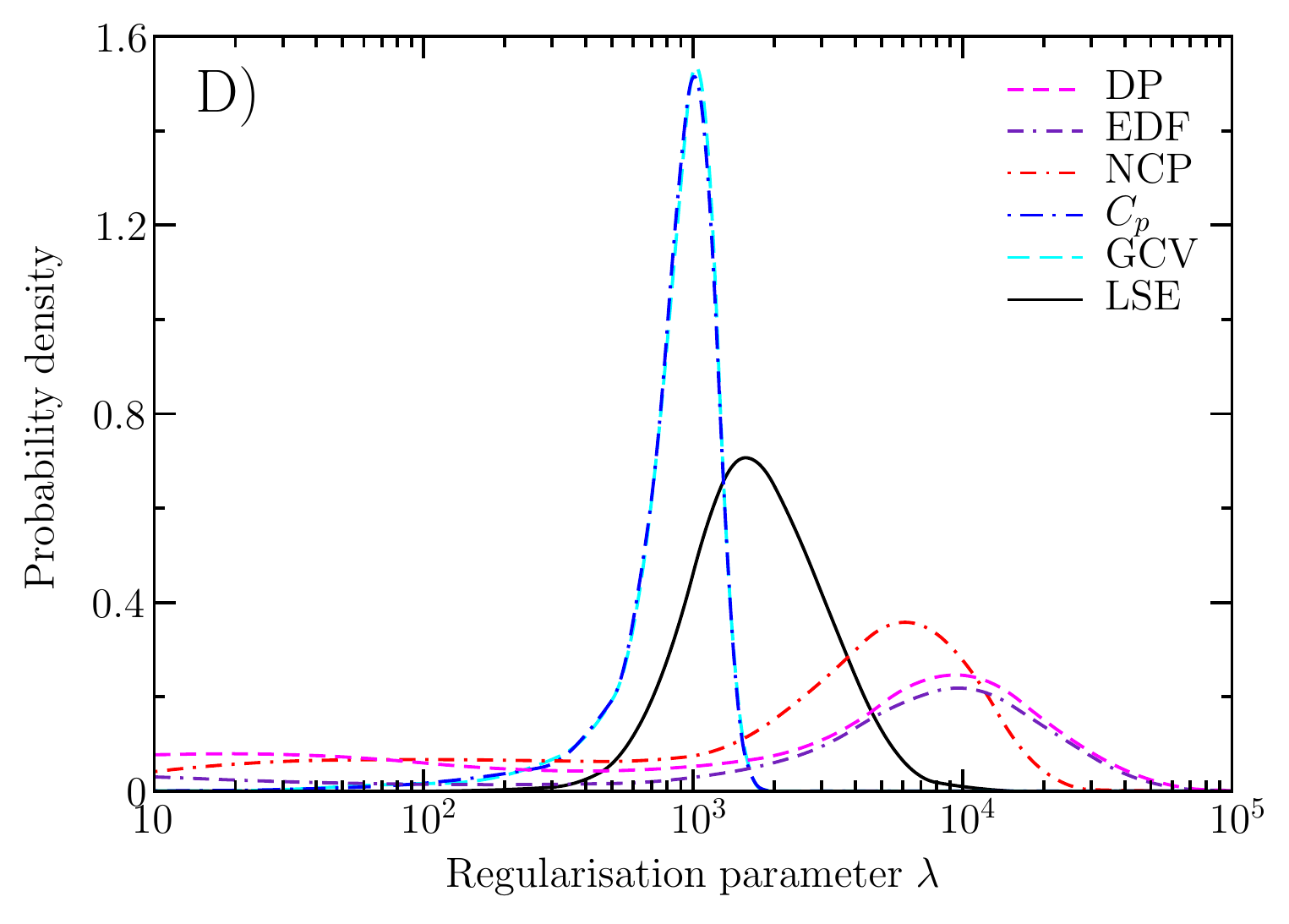}
\caption{\label{dist2} Same as Fig.\,\ref{dist1} for inversions
  using the WMAP-5 temperature and polarisation data (combination
  2, Sec.\,\ref{spectra}), and for spectra A--D Sec.\,\ref{spectest}).}
\end{figure*}

\begin{figure*}
\includegraphics*[angle=0,width=0.5\columnwidth,trim = 32mm 171mm 23mm
  15mm, clip]{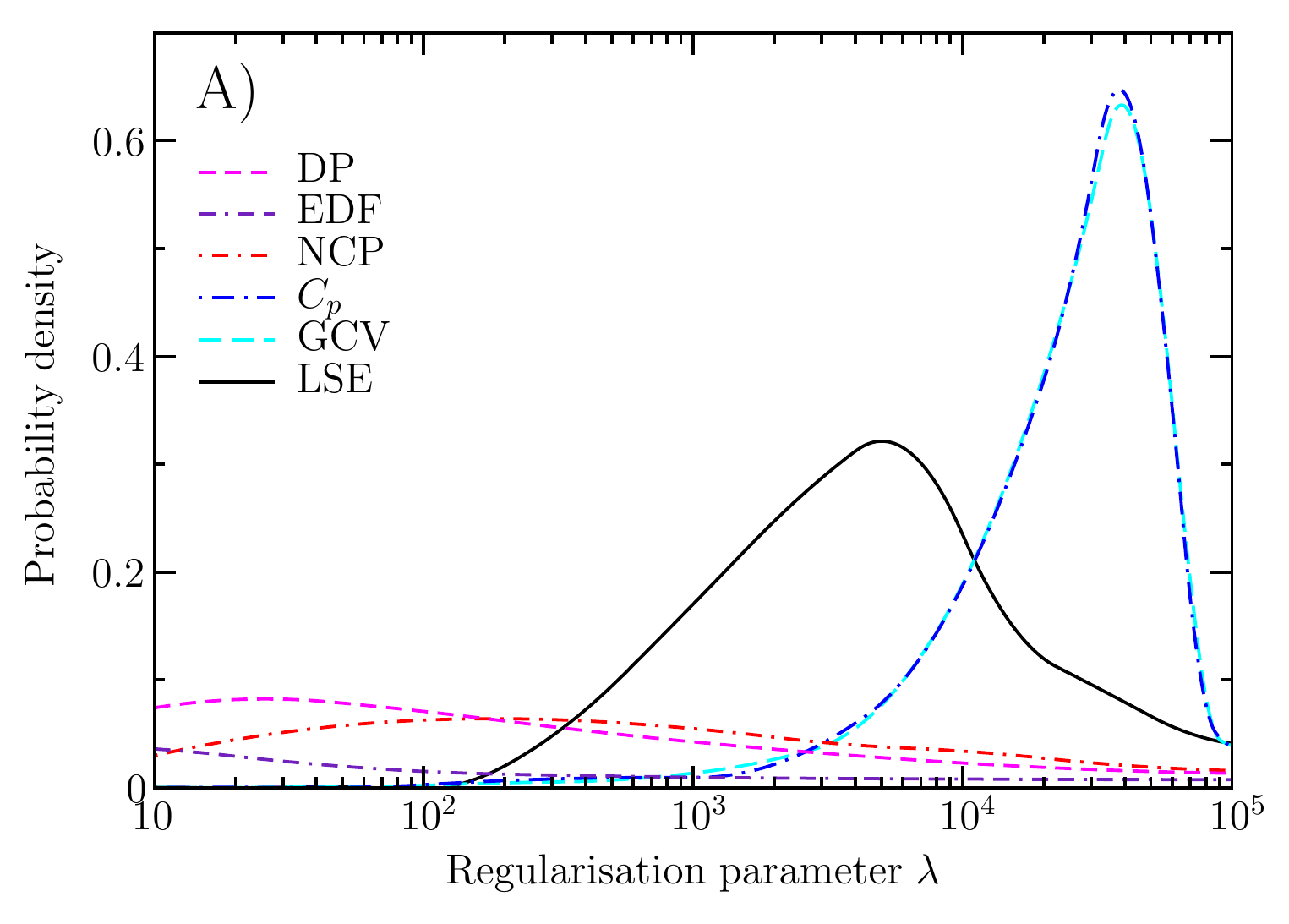}
\includegraphics*[angle=0,width=0.5\columnwidth,trim = 32mm 171mm 23mm
  15mm, clip]{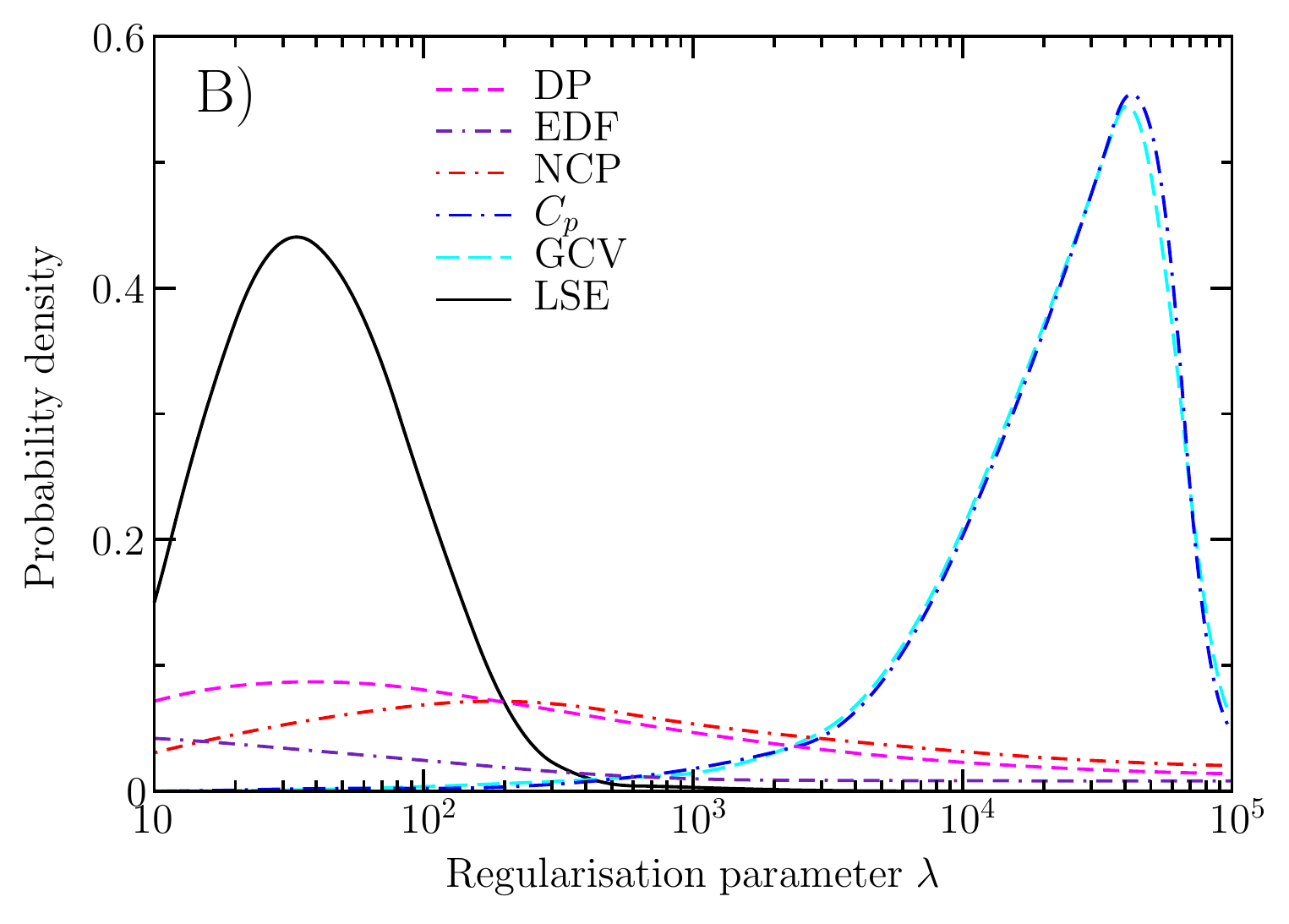}
\includegraphics*[angle=0,width=0.5\columnwidth,trim = 32mm 171mm 23mm
  15mm, clip]{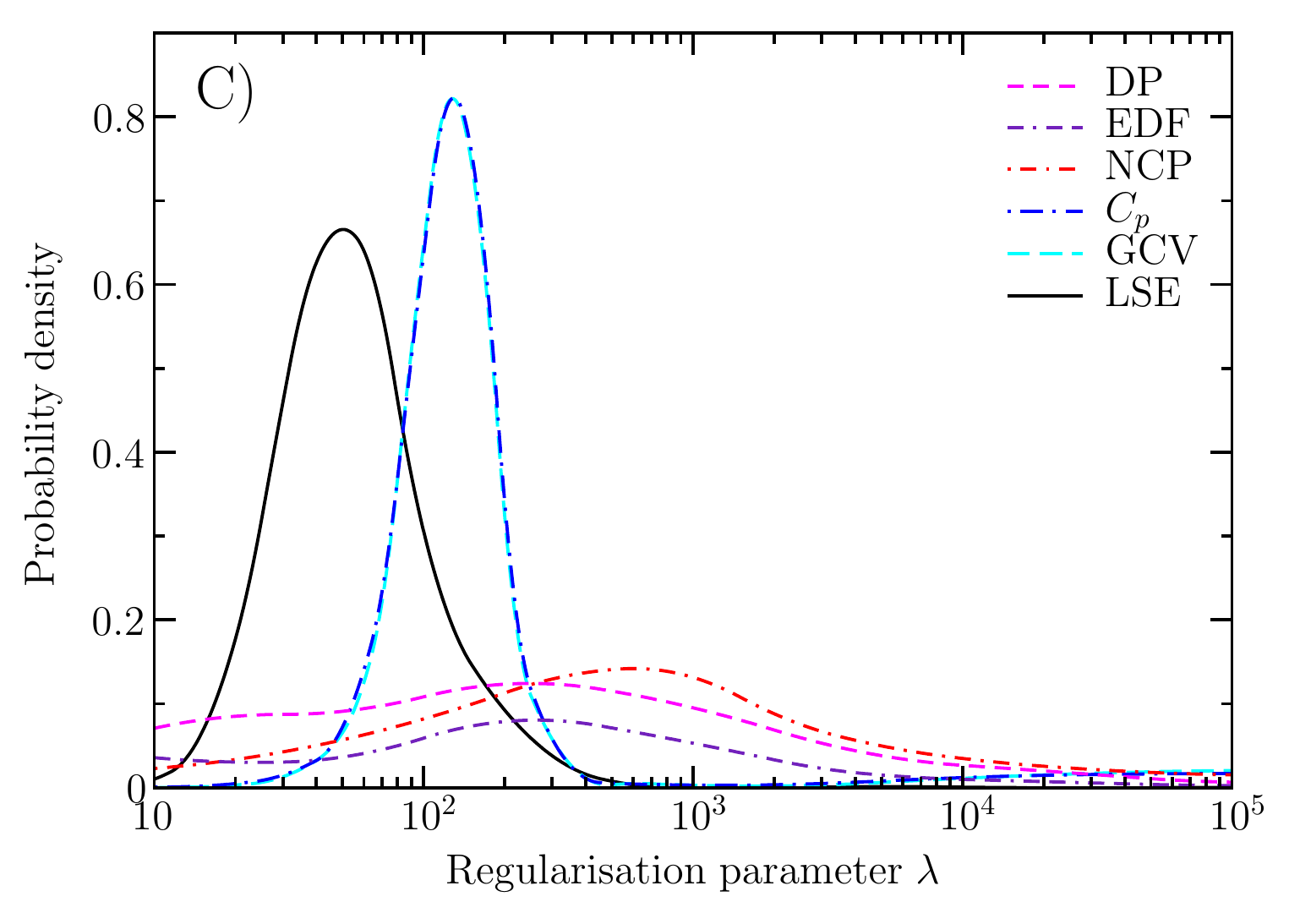}
\includegraphics*[angle=0,width=0.5\columnwidth,trim = 32mm 171mm 23mm
  15mm, clip]{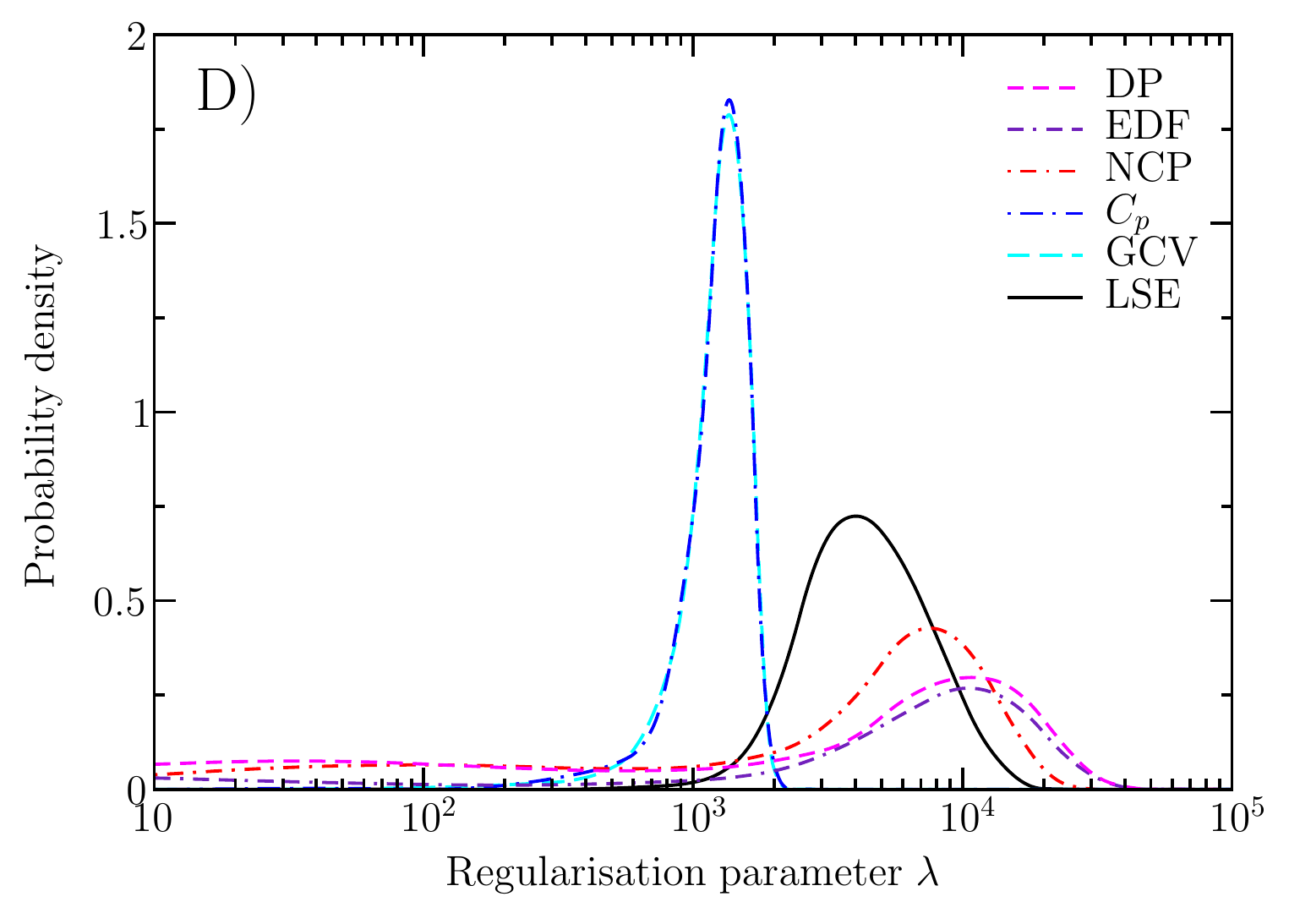}
\caption{\label{dist3} Same as Fig.\,\ref{dist1} for inversions
  using the WMAP-5 and small-scale CMB data (combination 3,
  Sec.\,\ref{spectra}), and for spectra A--D
  Sec.\,\ref{spectest}).}
\end{figure*}

\begin{figure*}
\includegraphics*[angle=0,width=0.5\columnwidth,trim = 32mm 171mm 23mm
  15mm, clip]{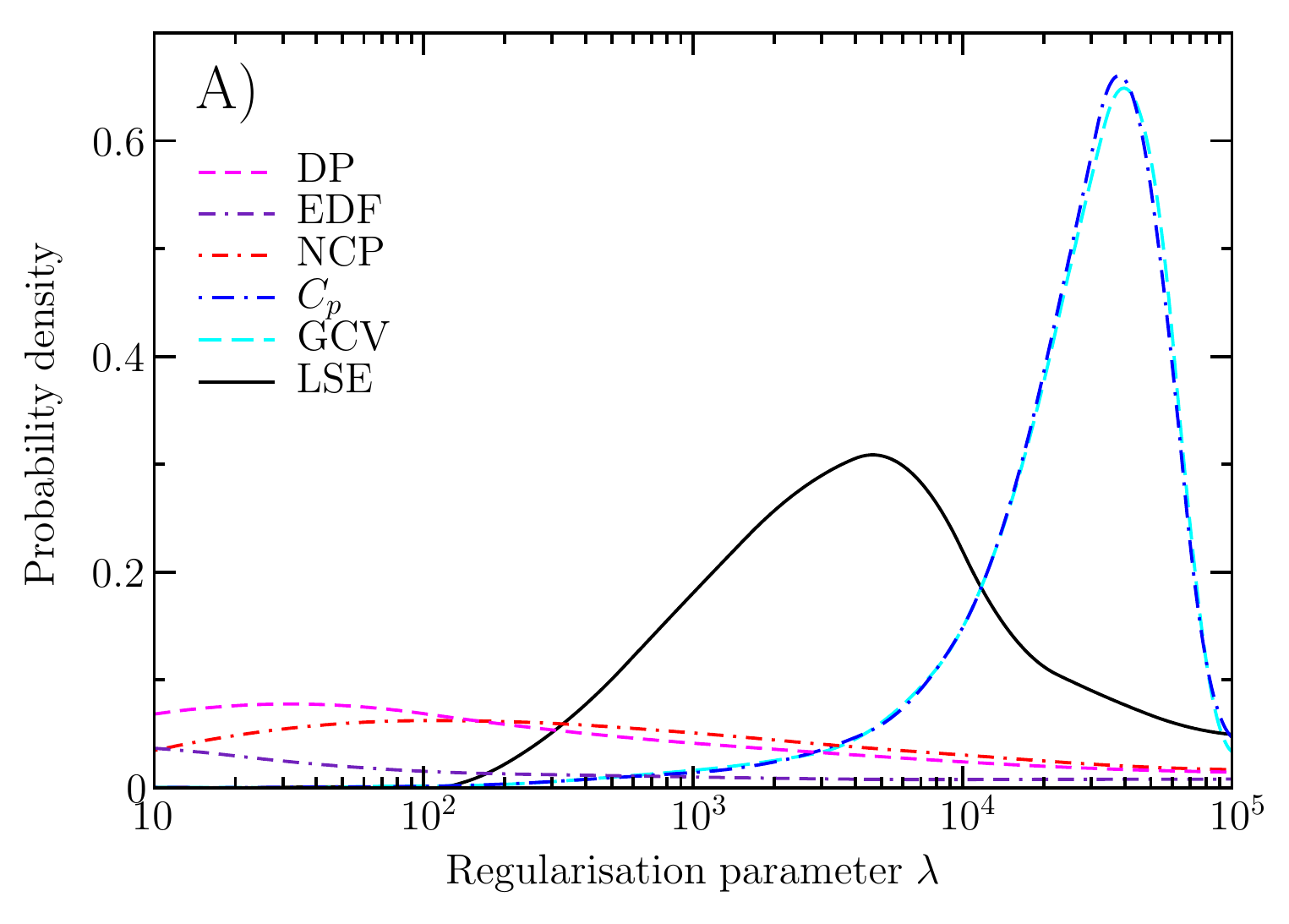}
\includegraphics*[angle=0,width=0.5\columnwidth,trim = 32mm 171mm 23mm
  15mm, clip]{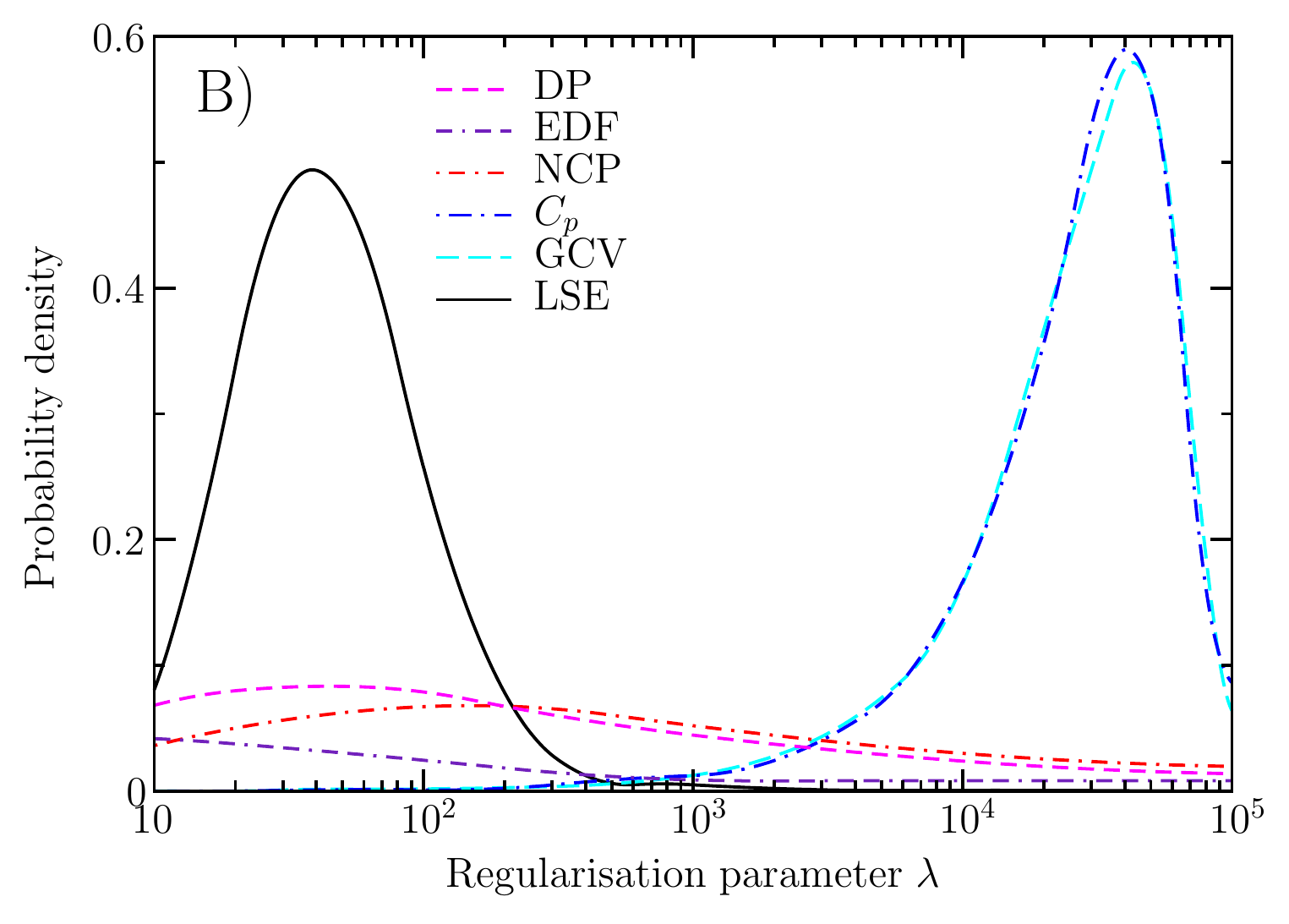}
\includegraphics*[angle=0,width=0.5\columnwidth,trim = 32mm 171mm 23mm
  15mm, clip]{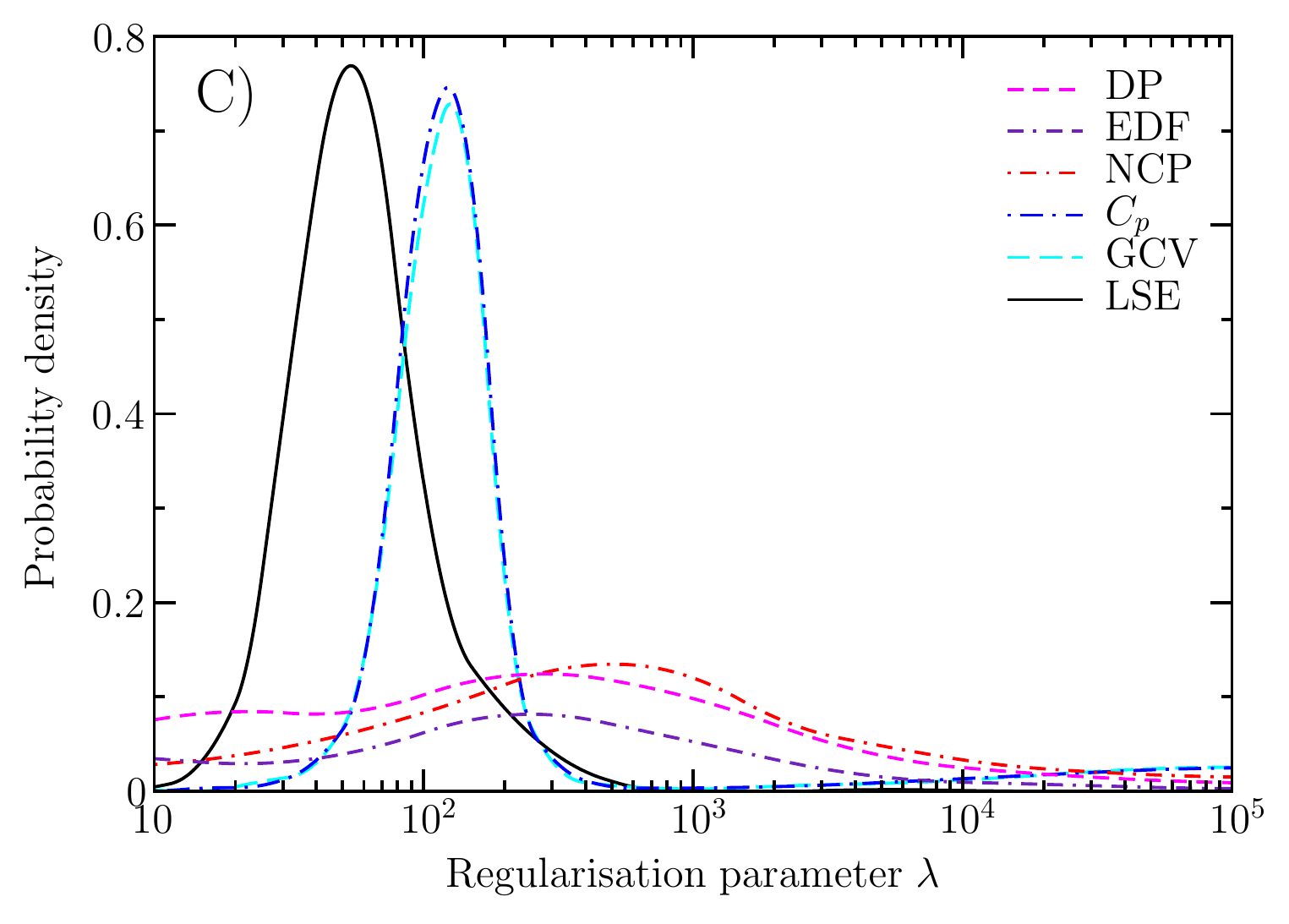}
\includegraphics*[angle=0,width=0.5\columnwidth,trim = 32mm 171mm 23mm
  15mm, clip]{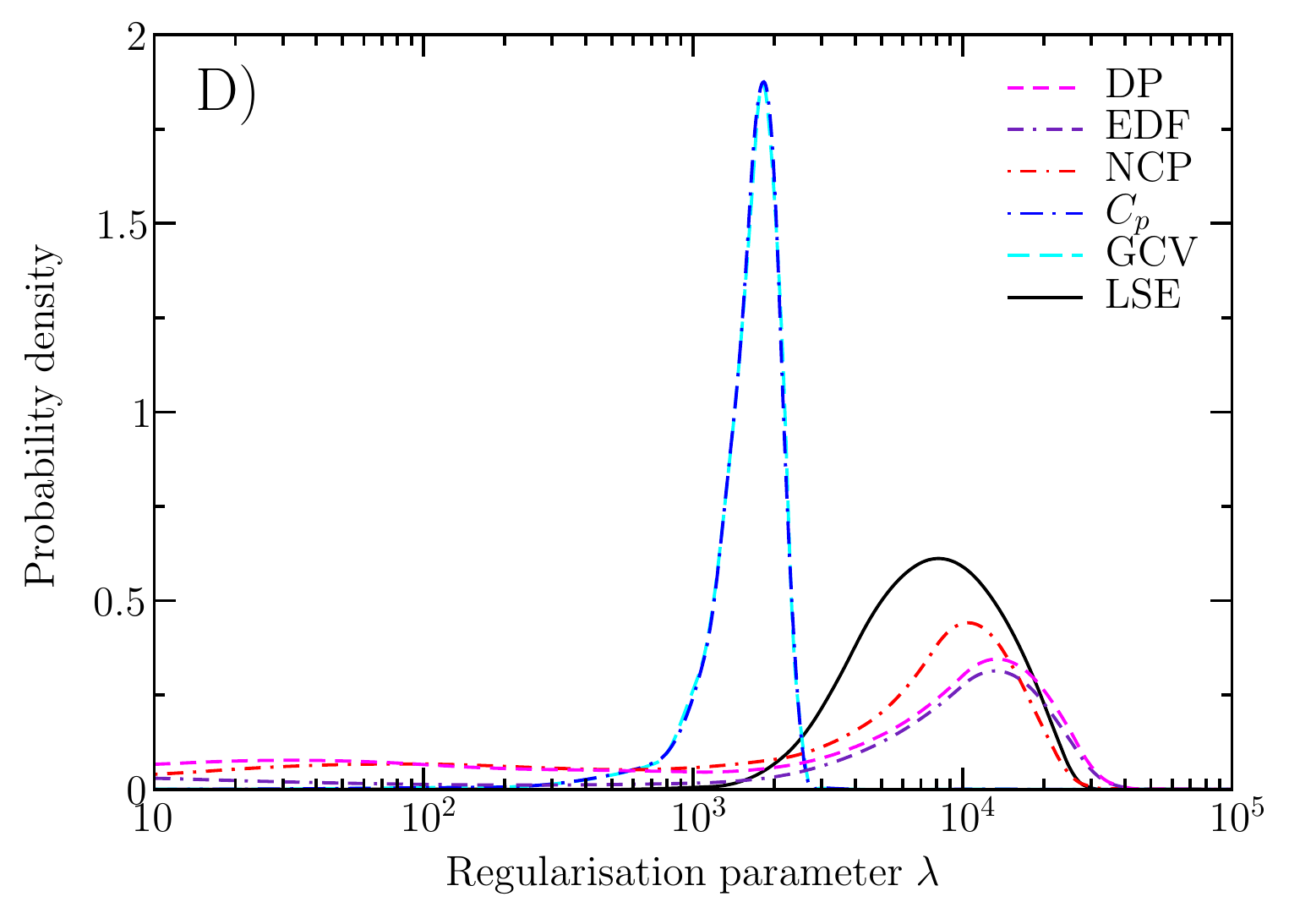}
\caption{\label{dist4} Same as Fig.\,\ref{dist1} for inversions
  using the WMAP-5, small-scale CMB and SDSS-4 LRG data
  (combination 4, Sec.\,\ref{spectra}), and for
  spectra A--D Sec.\,\ref{spectest}).}
\end{figure*}

We investigate the regularisation parameter selection methods by
applying them to mock data.  Minimisation of
$C_p\left(\lambda\right)$, $V_\mathrm{GCV}\left(\lambda\right)$ and
the SE is performed using Brent's method, while eqs.(\ref{chi}) and
(\ref{chi3}) are solved by Ridders' method \cite{press}.  Instead of
using $\chi^2\left(\hat{\pB}\right)$ as defined in eq.(\ref{chi}), in
practice we use the full log-likelihood function
$L\left(\pB,\dB\right)$ in eqs.(\ref{chi}), (\ref{chi3}), (\ref{cp})
and (\ref{gcv}), excluding the determinant terms. For each of the four
test spectra and four data combinations we generate $10^4$ mock
data realisations, to which the selection methods are applied. This
enables us to estimate the probability distributions of the chosen
parameter values by constructing histograms of the results. These are
shown in Figs.\,\ref{dist1}--\ref{dist4}.

The optimal values of $\lambda$ according to the LSE scheme are
consistently small for spectra B and C due to their infrared cutoffs
but are higher for spectra A and D. The peaks of the LSE distributions
closely follow the minima of the MSE functions shown in
Fig.\,\ref{mse}. The variation in the $\lambda_\mathrm{LSE}$ values is
related to the depth of the MSE minimum, e.g. the LSE distribution is
narrower for spectrum D than for spectrum A, since the MSE minimum of
spectra D is deeper. However, the ensemble mean of the
$\lambda_\mathrm{LSE}$ values equals the $\lambda$ value which
minimises the MSE only if the $\lambda$ versus SE curves of the
ensemble are symmetric about their minima \emph{and} translationally
symmetric copies of each other. Since these conditions are not met,
there is no absolute correspondence between the LSE distributions and
the MSE functions.

The distributions for Mallow's $C_p$ method follow the minima of the
MPE functions displayed in Fig.\,\ref{mpe}. This leads to relatively
successful selection of $\lambda$ for spectra C and D when the WMAP-5
data alone are used. However, with the addition of the small-scale CMB
and SDSS-4 LRG the $C_p$ method has a tendency to underestimate the
correct value of $\lambda$ for spectrum D and to overestimate it for
spectrum C. The $C_p$ method generally results in oversmoothed
reconstructions for spectrum A, and performs even more poorly when
applied to spectrum B, with extreme over smoothing.

\begin{figure*}
\begin{minipage}{150mm}
\includegraphics*[angle=0,width=0.5\columnwidth,trim = 32mm 171mm 23mm
  15mm, clip]{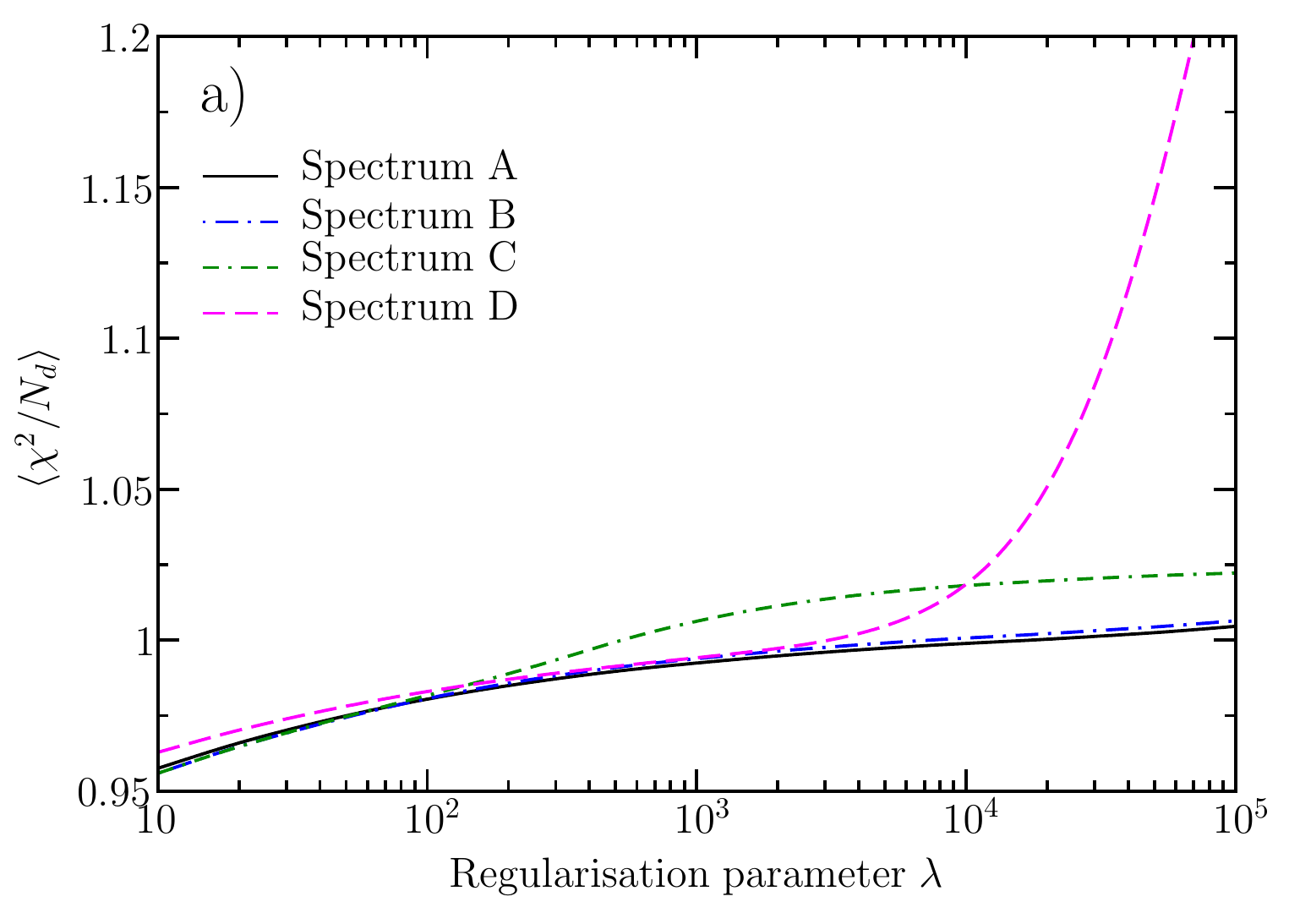}
\includegraphics*[angle=0,width=0.5\columnwidth,trim = 32mm 171mm 23mm
  15mm, clip]{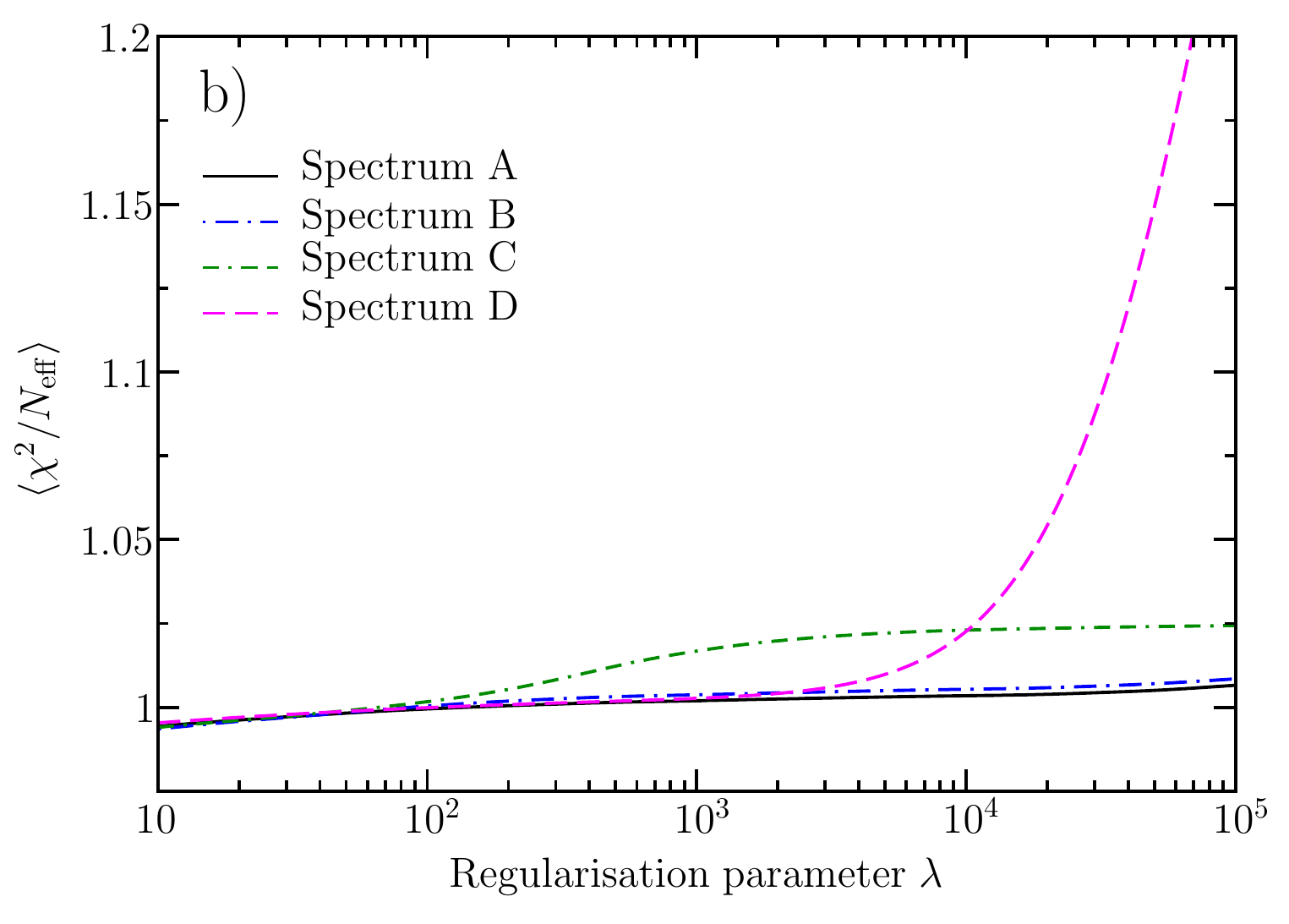}
\caption{\label{chind} The left panel shows the ensemble mean of
  $\chi^2\left(\pB\right)/N_d$ as a function of the regularisation
  parameter, for reconstructions of the four test spectra using the
  WMAP-5, small-scale CMB and SDSS-4 LRG data. The right panel shows
  the same for the quantity $\chi^2\left(\pB\right)/N_\mathrm{eff}$.}
\end{minipage}
\end{figure*}

The GCV distributions are almost identical to those of the $C_p$
method. This is due to the similarity of the $V_\mathrm{GCV}$ and
$C_p$ statistics.  In practice
$\sum_\mathbb{Z}\mathrm{Tr}\left(\mathsf{S}_\mathbb{Z}\right)/N_d$ is
of ${\cal O}(0.1)$ for reasonable reconstructions which permits the
Taylor expansion
\begin{equation}
V_\mathrm{GCV}\left(\lambda\right)=\chi^2\left(\hat{\pB}\right)
\left[1+2N_d^{-1}\sum_\mathbb{Z}\mathrm{Tr}\left(\mathsf{S}_\mathbb{Z}\right)+\ldots\right].
\end{equation} 
Since $\chi^2\left(\hat{\pB}\right)/N_d\simeq 1$ (see Fig.\,\ref{chind})
we have $V_\mathrm{GCV}\simeq C_p$, leading to the closeness of the
results.

The peaks of the DP distributions are in approximately the correct
location for spectra B and D when the WMAP-5, small-scale CMB and
SDSS-4 LRG data are used together, but not for the other
spectra. Moreover, the $\lambda$ values chosen by the DP method are
highly scattered. The DP results can be understood with reference to
Fig.\,\ref{chind}.  This shows the $\lambda$ dependence of the
ensemble means $\langle\chi^2/N_d\rangle$ and
$\langle\chi^2/N_\mathrm{eff}\rangle$, estimated from $10^5$
realisations of the WMAP-5, small-scale CMB and SDSS-4 LRG data
sets. The ensemble mean of the $\lambda_\mathrm{DP}$ values is
\textit{not} equal to the value of $\lambda$ for which
$\langle\chi^2/N_d\rangle=1$ because the $\chi^2/N_d$\,--$\lambda$
curves of the ensemble are not straight, parallel lines. The
probability density is greatest where the $\chi^2/N_d$-$\lambda$
curves intersect the $\chi^2/N_d=1$ line at the least acute
angle. Thus the peak of the distribution is located where the
$\chi^2/N_d$\,--$\lambda$ curves are steepest in the region where
$\chi^2/N_d$ is close to unity. For example, the
$\chi^2/N_d$\,--$\lambda$ curves are steepest for spectrum D at high
$\lambda$, flatten out, and then steepen again slightly at low
$\lambda$.  This leads to a well-defined peak at $\lambda\simeq 10^4$,
a trough at intermediate $\lambda$, and a small secondary peak at low
$\lambda$. The DP distributions are so broad because the change in
$\chi^2$ with $\lambda$ is small compared with the variation in
$\chi^2$ values at fixed $\lambda$ between different reconstructions
in the ensemble.

While most of the $\lambda_\mathrm{EDF}$ values are close to the
optimum for spectrum D when the WMAP-5, small-scale CMB and SDSS LRG
data are used, some greatly underestimate it. The effective number of
parameters decreases from
$\sum_\mathbb{Z}\mathrm{Tr}\left(\mathsf{S}_\mathbb{Z}\right)\simeq
60$ for $\lambda=10$ down to
$\sum_\mathbb{Z}\mathrm{Tr}\left(\mathsf{S}_\mathbb{Z}\right)\simeq 3$
for $\lambda=10^5$. Consequently the
$\langle\chi^2/N_\mathrm{eff}\rangle$\,--$\lambda$ curves of the test
spectra are higher at low $\lambda$ than the
$\langle\chi^2/N_d\rangle$\,--$\lambda$ curves but are almost
identical at high $\lambda$. This means that data realisations with
low $\lambda_\mathrm{DP}$ values have even lower
$\lambda_\mathrm{EDF}$ values, while realisations with high
$\lambda_\mathrm{DP}$ values have approximately the same high values
of $\lambda_\mathrm{EDF}$. This is reflected in the EDF distributions
which extend down to lower $\lambda$ than their DP counterparts.

The NCP method gives similar results to the DP method. This is because
for Gaussian distributed data the $\chi^2$ of the reconstruction has
the value $N_d\pm\sqrt{\left(2N_d\right)}$ when $\tilde{\yB}$ behaves
like white noise.

In summary, the $\lambda$ values of the DP, EDF and NCP methods are too 
widely scattered to be meaningful and bear little relation to the 
optimum figure. Mallow's $C_p$ method and the GCV method perform very well 
in minimising the PE of the reconstruction. Unfortunately this is of 
limited use as the PE is usually different to the SE for CMB data.

\section{Glossary}
\label{glossary}

We list below the notation and symbols used along with where they are
first mentioned:

\begin{longtable}[l]{p{65pt} p{45pt} p{290pt}} 

\textbf{Symbol}	& \textbf{Reference} & \textbf{Description} \\

$\bB$ & Sec.\,\ref{bivar}  & Second-order approximation to the bias of $\hat{\pB}$ \\ 

$\bB_1$, $\bB_2$, $\bB_3$, $\bB_4$ & Sec.\,\ref{bivar}  & Different components of $\bB$ \\ 

$\mathrm{C}\left(k_0;k\right)$ & eq.\,(\ref{equncorr})  & Correlation function of $\hat{\pB}$ \\

$\mathrm{d}_a^{(\mathbb{Z})}$ & eq.\,(\ref{int1}) & Data element $a$ of data set $\mathbb{Z}$ (whole vector is labelled $\dB_\mathbb{Z}$)\\

$\dB$ & Sec.\,\ref{tikreg}  & Vector of all data sets $\dB_\mathbb{Z}$\\

$\mathsf{D}$ & Sec.\,\ref{uncorr}  & Matrix whose square is the (diagonal) covariance matrix of $\tilde{\qB}$ \\

$\mathsf{G}$ & Sec.\,\ref{uncorr}  & Matrix relating $\tilde{\qB}$ to $\qB$ \\

$\hB$ & eq.\,(\ref{ncph}) & Normalised cumulative periodogram of $\yB$ \\

$\mathcal{K}_a^{(\mathbb{Z})}$ & eq.\,(\ref{int1}) & Integral kernel for element $a$ of data set $\mathbb{Z}$ \\

$L_\mathrm{bps}$ & eq.\,(\ref{eqbps})  & Beam and point source component of $L_\mathrm{TT}$ \\

$L_\mathrm{Gibbs}$ & Sec.\,\ref{wmap}  & Low-$\ell$ Gibbs sampler component of $L_\mathrm{TT}$ \\

$L_\mathrm{LRG}$ & eq.\,(\ref{eqlrg})  & SDSS LRG likelihood \\

$L_\mathrm{pol}$ & Sec.\,\ref{wmap}  & Total WMAP polarisation likelihood \\

$L_\mathrm{pix}$ & Sec.\,\ref{wmap}  & Low-$\ell$ pixel-based component of $L_\mathrm{pol}$ \\

$L_\mathrm{ss}$ & eq.\,(\ref{ssl})  & Small angular scale CMB likelihood \\

$L_\mathrm{pTE}$ & eq.\,(\ref{eqpte})  & High-$\ell$ component of $L_\mathrm{pol}$ \\

$L_\mathrm{pTT}$ & eq.\,(\ref{eqptt})  & High-$\ell$ component of $L_\mathrm{TT}$ \\

$L_\mathrm{TT}$ & Sec.\,\ref{wmap}  & Total WMAP TT likelihood \\

$L\left(\pB,\bm{\theta},\dB\right)$ & eq.\,(\ref{minq}) & $\equiv-2\ln\mathcal{L} \left(\pB,\bm{\theta}|\dB\right)$ \\

$\mathcal{L}\left(\pB,\bm{\theta}|\dB\right)$ & eq.\,(\ref{gauslike}) & Likelihood of $\pB$ and $\bm{\theta}$ given $\dB$ \\

$\mathsf{L}$ & eq.\,(\ref{firstl}) &  Matrix representation of the first-order derivative operator \\ 

$\mathrm{MPE}\left(\hat{\pB}\right)$ & eq.\,(\ref{eqmpe})  & Mean predictive error of $\hat{\pB}$ \\

$\mathrm{MSE}\left(\hat{\pB}\right)$ & eq.\,(\ref{eqmse})  & Mean squared error of $\hat{\pB}$ \\ 

$\mathrm{n}_a^{(\mathbb{Z})}$ & eq.\,(\ref{int1}) & Noise element $a$ of data set $\mathbb{Z}$ (whole vector is labelled $\nB_\mathbb{Z}$)\\

$\nB$ & Sec.\,\ref{tikreg}  & Vector of all noise sets $\nB_\mathbb{Z}$\\

$N$ & Sec.\,\ref{tikreg} & Number of data sets \\

$N_d$ & Sec.\,\ref{tikreg}  & Total number of data points \\

$N_\mathrm{eff}$  & Sec.\,\ref{pselm} & Number of degrees of freedom of $\hat{\pB}$ \\ 

$N_j$ & & Number of basis functions $\phi_i\left(k\right)$ \\

$N_\mathbb{Z}$ & Sec.\,\ref{tikreg} & Number of points in the set data set $\mathbb{Z}$ \\

$\mathsf{N}$ & Sec.\,\ref{tikreg}  & Matrix assembled from all $\mathsf{N}_\mathbb{Z}$ \\

$\mathsf{N}_\mathbb{Z}$ & Sec.\,\ref{tikreg} & Covariance matrix of $\mathrm{n}_\mathbb{Z}$ \\

$\mathcal{N}_\mathcal{R}\left(k\right)$ & eq.\,(\ref{null}) & Infinite-dimensional null space of functions \\

$\mathrm{p}_i$ & eq.\,(\ref{basis}) & Coefficients of the basis functions $\phi_i\left(k\right)$ (also labelled $\pB$) \\

$P_\mathrm{W1}$ & eq.\,(\ref{wish1})  & One-dimensional Wishart distribution \\

$P_\mathrm{W2}$ & eq.\,(\ref{wish2}) & Two-dimensional Wishart distribution \\

$\hat{\pB}$ & Sec.\,\ref{tikreg} & Estimator of $\pB$ \\

$\hat{\pB}_0$ & Sec.\,\ref{tikreg}  & Vector with the minimum $\mathrm{R}$ value of those that maximise the likelihood. Given by $\hat{\pB}_0 = \mathsf{W}_{\mathsf{N}\mathsf{\Gamma}}^{\dagger} \dB$ \\

$\hat{\pB}^\mathrm{(f)}$ & eq.\,(\ref{pf})  & First-order approximation to $\hat{\pB}$ \\

$\hat{\pB}_\mathrm{MAP}$ & Sec.\,\ref{bayesian}  & Maximum \textit{a posteriori} estimate of $\pB$ \\ 

$\hat{\pB}^\mathrm{(s)}$ & eq.\,(\ref{ps}) & Second-order approximation to $\hat{\pB}$ \\ 

$\pB_\mathrm{t}$ & Sec.\,\ref{error} & True value of $\pB$ \\

$\hat{\pB}^\mathrm{(z)}$ & eq.\,(\ref{pz})  & Zeroth-order approximation to $\hat{\pB}$ \\

$C_p\left(\lambda\right)$ & eq.\,(\ref{cp}) & Mallow's $C_p$ statistic \\

$\mathcal{P}_\mathcal{R}$ & eq.\,(\ref{2ptcorr}) & Primordial power spectrum (PPS) of $\mathcal{R}$\\ 

$\mathrm{PE}\left(\hat{\pB}\right)$ & Sec.\,\ref{performance}  & Predictive error of $\hat{\pB}$ \\ 

$\qB$ & Sec.\,\ref{uncorr}  & Correlated bandpowers of $\hat{\pB}$ \\

$\tilde{\qB}$ & Sec.\,\ref{uncorr}  & Uncorrelated bandpowers of $\hat{\pB}$ \\

$Q (\pB,\dB,\hat{\bm{\theta}},\lambda)$ & eq.\,(\ref{minq}) & $\equiv
  L\left(\pB,\hat{\bm{\theta}},\dB\right)+\lambda \mathrm{R}\left(\pB\right)$ \\ 

$R\left(k_0;k\right)$ & eq.\,(\ref{1storderkernel}) & First-order resolution function \\

$\mathrm{R}\left(\pB\right)$ & Sec.\,\ref{tikreg}  & Roughness function of $\pB$ \\

$\mathsf{R}$ & eq.\,(\ref{matrices})  & First-order resolution matrix \\

$\mathcal{R}$ & eq.\,(\ref{2ptcorr}) & Primordial comoving curvature perturbation\\ 

$\Re$ & Sec.\,\ref{bivar}  & Number of Monte Carlo simulations \\

$s_k$, $c_k$ & Sec.\,\ref{pselm} & Sine and cosine coefficients of the DFT of $\yB$ \\

$\mathrm{SE}\left(\hat{\pB}\right)$ & eq.\,(\ref{SE}) & Squared error of $\hat{\pB}$ \\ 

$\mathrm{T}_1$ & eq.\,(\ref{t1stat})  & Test statistic for a feature in the PPS (two-tailed) \\

$\mathrm{T}_2$ & eq.\,(\ref{t2stat})  & Test statistic for a feature in the PPS (one-tailed) \\

$\mathsf{T}$ & Sec.\,\ref{uncorr}  & Matrix relating $\qB$ to $\hat{\pB}$ \\

$\bm{\mathcal{T}}\left(\pB_\mathrm{t},\bm{\theta}_\mathrm{t},\hat{\bm{\theta}},\nB\right)$ & eq.\,(\ref{transfer}) & Transfer function relating $\hat{\pB}$ to $\pB_\mathrm{t}$ \\ 

$\mathrm{u}_\alpha$ & Sec.\,\ref{tikreg}  & Element $\alpha$ of uncertainty in $\hat{\bm{\theta}}$ (whole vector is labelled $\uB$)\\

$\mathsf{U}$ & Sec.\,\ref{tikreg}  & Covariance matrix for $\hat{\bm{\theta}}$ \\

$V_\mathrm{GCV}\left(\lambda\right)$ & eq.\,(\ref{gcv}  & Generalised cross-validation statistic \\

$W_{ai}^{(\mathbb{Z})}$ & eq.\,(\ref{rel1}) & Integral of  $\mathcal{K}_a^{(\mathbb{Z})}$ over wavenumber bin $i$ (also called $\mathsf{W}_\mathbb{Z}$)\\

$\mathsf{W}$ & eq.\,(\ref{rel2}) & Matrix assembled from all $\mathsf{W}_\mathbb{Z}$ \\

$\mathsf{W}_{\mathsf{N}\mathsf{\Gamma}}^{\dagger}$ & eq.\,(\ref{mp})  & Weighted Moore-Penrose inverse of $\mathsf{W}$ \\

$\yB$ & Sec.\,\ref{pselm}  & Vector of all data sets $\yB_\mathbb{Z}$\\

$\yB_\mathbb{Z}$ & Sec.\,\ref{pselm} & Data set $\dB_\mathbb{Z}$ whitened with a Cholesky decomposition\\

$\mathsf{Y}$ & eq.\,(\ref{matrices})  & Second-order resolution matrix \\

$Y\left(k_0;k_1,k_2\right)$ & eq.\,(\ref{2ndorderkernel}) & Second-order resolution function \\
\\
$\gamma$ & eq.\,(\ref{scale}) & SDSS LRG scaling factor \\
$\gamma_\mathrm{NCP}\left(\lambda\right)$ & Sec.\,\ref{pselm}  & Normalised cumulative periodogram statistic \\
$\delta\hat{\pB}^\mathrm{(f)}$ & eq.\,(\ref{deltapf}) & First-order approximation difference \\
$\delta\hat{\pB}^\mathrm{(s)}$ & eq.\,(\ref{deltaps}) & Second-order approximation difference \\ 
$\lambda$ & eq.\,(\ref{minq}) & Regularisation parameter \\
$\lambda_\mathrm{CP}$ & Sec.\,\ref{pselm} & Mallow's $C_p$ method $\lambda$ value \\
$\lambda_\mathrm{DP}$ & Sec.\,\ref{pselm} & Discrepancy principle method $\lambda$ value \\
$\lambda_\mathrm{EDF}$ & Sec.\,\ref{pselm} & Equivalent degrees of freedom method $\lambda$ value \\
$\lambda_\mathrm{GCV}$ & Sec.\,\ref{pselm} & Generalised cross-validation method $\lambda$ value \\
$\lambda_\mathrm{LSE}$ & Sec.\,\ref{pselm} & Least squared error method $\lambda$ value \\
$\lambda_\mathrm{NCP}$ & Sec.\,\ref{pselm} & Normalised cumulative periodogram method $\lambda$ value \\
$\nu_1$, $\nu_2$, $\nu_3$  & Sec.\,\ref{regularisation}  & Estimates of the effective number of free parameters of $\hat{\pB}$ \\ 
$\phi_i\left(k\right)$ & eq.\,(\ref{phidef}) & Basis functions for the reconstructed PPS \\
$\rho$ & Sec.\,\ref{tikreg}  & Rank of $\mathsf{W}$ \\
$\bm{\theta}$ & eq.\,(\ref{int1}) & Set of parameters defining background cosmology \\
$\hat{\bm{\theta}}$ & Sec.\,\ref{tikreg}  & Estimator of $\bm{\theta}$ \\
$\bm{\theta}_\mathrm{t}$ & Sec.\,\ref{error} & True value of $\bm{\theta}$ \\
$\sigma_i$ & Sec.\,\ref{tikreg}  & Weighted singular values of $\mathsf{W}$ \\
$\xi\left(k\right)$ & Sec.\,\ref{statsig})& Integral kernel specifying a feature in the PPS \\
\\
$\mathsf{\Gamma}$ & Sec.\,\ref{tikreg}  & Matrix specifying $\mathrm{R}$ \\
$\mathsf{\Sigma}_\mathrm{B}$ & eq.\,(\ref{sigb})  & Bayesian covariance matrix of $\hat{\pB}$ \\
$\mathsf{\Sigma}^{(\mathrm{f})}$ & eq.\,(\ref{sigf})  & Frequentist covariance matrix of $\hat{\pB}^\mathrm{(f)}$ \\ 
$\mathsf{\Sigma}^{(\mathrm{fs})}$ & eq.\,(\ref{sigfs}) & Frequentist covariance matrix of $\delta\hat{\pB}^\mathrm{(f)}$ and $\delta\hat{\pB}^\mathrm{(s)}$ \\  
$\mathsf{\Sigma}_\mathrm{F}$ & eq.\,(\ref{sigmaf})  & Frequentist covariance matrix due to data noise \\ 
$\mathsf{\Sigma}_\mathrm{N}$ & Sec.\,\ref{uncorr}  & Frequentist covariance matrix of $\qB$ \\ 
$\mathsf{\Sigma}_\mathrm{P}$ & eq.\,(\ref{sigmap})  & Frequentist covariance matrix due to background parameter uncertainties \\ 
$\mathsf{\Sigma}^{(\mathrm{s})}$ & eq.\,(\ref{sigs}) & Frequentist covariance matrix of $\hat{\pB}^\mathrm{(s)}$ \\
$\mathsf{\Sigma}^{(\mathrm{ss})}$ & eq.\,(\ref{sigss}) & Frequentist covariance matrix of $\delta\hat{\pB}^\mathrm{(s)}$ \\ 
$\mathsf{\Sigma}_\mathrm{T}$ & eq.\,(\ref{sigmat}) & Total frequentist covariance matrix of $\hat{\pB}$ \\

\end{longtable}

\end{appendix}

\end{document}